# Graph Neural Networks in Modern AI-aided Drug Discovery

Odin Zhang[1,#], Haitao Lin[2,#], Xujun Zhang[1,#], Xiaorui Wang[1,#], Zhenxing Wu[1], Qing Ye[1], Weibo Zhao[1], Jike Wang[1], Kejun Ying[3], Yu Kang[1], Chang-yu Hsieh[1,*], Tingjun Hou[1,*]

[1]College of Pharmaceutical Sciences, Zhejiang University, Hangzhou 310058, Zhejiang, China

[2]School of Engineering, Westlake University, Hangzhou 310024, Zhejiang, China

[3]T. H. Chan School of Public Health, Harvard University, Boston, MA, USA

[#]Equal Contributions

## Corresponding authors

**Tingjun Hou**

**E-mail:** tingjunhou@zju.edu.cn

**Chang-Yu Hsieh**

**E-mail:** kimhsieh@zju.edu.cn

# Abstract

Graph neural networks (GNNs), as topology/structure-aware models within deep learning, have emerged as powerful tools for AI-aided drug discovery (AIDD). By directly operating on molecular graphs, GNNs offer an intuitive and expressive framework for learning the complex topological and geometric features of drug-like molecules, cementing their role in modern molecular modeling. This review provides a comprehensive overview of the methodological foundations and representative applications of GNNs in drug discovery, spanning tasks such as molecular property prediction, virtual screening, molecular generation, biomedical knowledge graph construction, and synthesis planning. Particular attention is given to recent methodological advances, including geometric GNNs, interpretable models, uncertainty quantification, scalable graph architectures, and graph generative frameworks. We also discuss how these models integrate with modern deep learning approaches, such as self-supervised learning, multi-task learning, meta-learning and pre-training. Throughout this review, we highlight the practical challenges and methodological bottlenecks encountered when applying GNNs to real-world drug discovery pipelines, and conclude with a discussion on future directions.

# Contents

# 1. Introduction

## 1.1 Graph is a Natural Representation in Drug Discovery

Drug discovery is a resource-intensive and time-consuming endeavor. It is estimated that bringing a new drug to market requires over two billion US dollars in investment and typically spans more than a decade of research and development[1]. The high costs and risks associated with this process have limited the spectrum of diseases that can be effectively treated through pharmacological interventions. As such, any methodological breakthrough in drug discovery, particularly those that reshape foundational paradigms, has attracted considerable attention from both academic and industrial communities[2]. Over time, drug discovery has evolved from early reliance on serendipitous discovery to the advent of high-throughput screening and, more recently, to mechanism-based rational design, progressing steadily toward enhanced controllability and specificity[3].

In recent years, the rapid development of artificial intelligence (AI) has infused new momentum into drug discovery, accelerating the transition from rule-based heuristics to data-driven strategies. The 2024 Nobel Prize in Chemistry, awarded breakthroughs in protein design, highlights the transformative potential of incorporating AI into the standard practices of drug dsicovery[4]. Beyond algorithmic advances, the growing availability of high-quality experimental data, from structural biology to multi-omics, has further solidified the foundation for AI applications. Databases such as the Protein Data Bank[5] (PDB) and omics resources like The Cancer Genome Atlas (TCGA)[6] provide large-scale structural, transcriptomic, and proteomic data essential for data-driven modeling. In this context, AI-aided drug discovery (AIDD) is emerging as a system-level paradigm, encompassing target identification, hit discovery, lead optimization, and early risk mitigation[7]. AIDD is primarily deployed in the early stages of drug development, and its key objective is to find and improve the specificity and developability of candidate molecules while minimizing toxicity and the risk of clinical trial failure. Collectively, these advancements contribute to improved efficiency and reduced costs in the drug development pipelines[8].

To fully harness the potential of AI in drug discovery, molecular structures must first be encoded into machine-interpretable representations. To this end, a variety of molecular representation strategies have been developed, including string-based linearizations (e.g., SMILES[9] and SELFIES[10]), molecular fingerprints[11], and Graphs-based and 3D voxel grid representations[12], as shown in **Figure 1A**. When coupled with architectures like DNNs, CNNs[13], and Transformers[14], these representations have shown promising results in tasks such as quantitative structure–activity relationship (QSAR) modeling and scoring function development[15]. However, each representation has inherent limitations. Fingerprints and descriptors, which rely on expert-defined features, often suffer limited expressiveness due to their dependence on predefined statistical or physicochemical patterns[16]. String-based methods linearize molecular topology, facilitating compatibility with sequence-based models, but they struggle to capture structural symmetry, reversibility, and long-range dependencies[17].

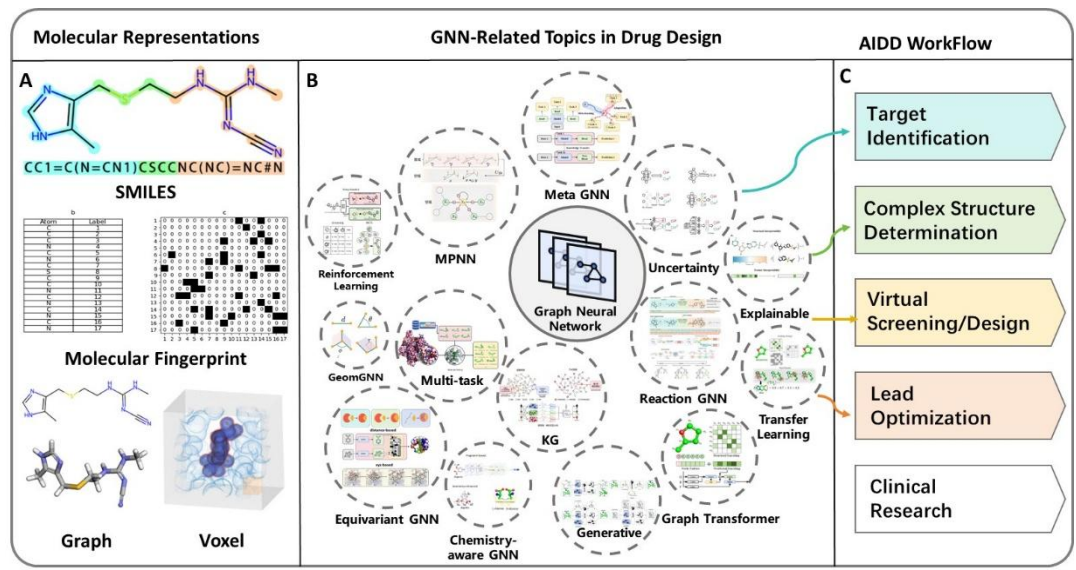

**Figure 1. Overview of molecular representations and graph neural network (GNN) applications in drug discovery.** (A) Common molecular representations; (B) GNN-related topics in AI-aided Drug Design; (c) AIDD workflow, with colord stages indicating where GNNs can be integrated.

Ontologically, molecules are inherently graph-structured objects, where xatoms constitute nodes and chemical bonds serve as edges, with additional three-dimensional (3D) coordinates defining their conformational geometry. This formulation naturally aligns with the modeling paradigm of graph neural networks (GNNs), which have shown exceptional promise in molecular modeling tasks. GNNs operate through local message-passing mechanisms to capture atomic coordination patterns and hierarchically aggregate them to encode molecular-level semantics. With the incorporation of equivariant or invariant operators (e.g., SO(3)-equivariant GNNs), GNNs can model stereochemistry, charge distributions, and spatial configurations in a symmetry-aware manner. Compared to traditional "flattened" representations, graph-based representations offer three key advantages: (i) physical consistency, preserving topological and geometric information; (ii) representational universality, enabling deployment across tasks such as property prediction, generative modeling, and pharmacological profiling; and (iii) enhanced interpretability through subgraph attention, edge attribution, and structural saliency maps.

As a result, graph representations are increasingly becoming the standard in molecular modeling. GNNs have been widely adopted by pharmaceutical companies (e.g., Roche, Novartis), AI-driven biotech startups (e.g., BenevolentAI, Insilico Medicine), and research communities supported by open datasets such as OGB-Mol[18] and GEOM[19]. These developments reflect not only the central role of GNNs in

AIDD but also the growing need for specialized GNN architectures tailored to diverse biomedical applications.

## 1.2. Scope and Organization of This Review

This review aims to provide a comprehensive and structured overview of GNN in modern AIDD, with a focus on their methodological foundations and practical challenges. In contrast to existing surveys that often center on specific tasks or empirical comparisons, we approach the subject from a model-centric perspective, examining the conceptual underpinnings of GNNs and their roles across diverse molecular design tasks, as shown in **Figure 1B**. Special attention is given to recent advances in large-scale GNNs, geometric deep learning, interpretability, uncertainty quantification, graph generative modeling, and reinforcement learning (RL). We further discuss how these architectures integrate with contemporary deep learning paradigms, such as self-supervised learning, multi-task learning, pretraining, and meta-learning, to improve data efficiency, generalization, and robustness in real-world drug discovery settings. This review highlights a unified modeling framework that harmonizes mathematical rigor with chemical intuition. It is intended for researchers in chemistry and biology applying AI to drug discovery, as well as those in computational sciences developing models for biochemical applications.

By highlighting the underlying logic and structural dependencies that connect seemingly disparate tasks, we aim to help readers develop a coherent conceptual framework that facilitates the effective adaptation of GNN-based techniques to address diverse challenges in drug discovery. The remainder of this review is organized as follows. Section 2 presents the methodological foundations of GNNs: (2.1) core principles of graph learning, including spectral and spatial frameworks; (2.2) symmetry-aware design principles, covering invariant and equivariant models for 3D molecular systems; (2.3) scalability challenges in large GNNs, such as oversmoothing, over-squashing, and pretraining strategies; (2.4) graph generation and structure editing methods; (2.5) graph editing and reinforcement learning techniques. Section 3 discusses molecular property prediction, including interpretability and uncertainty quantification in GNNs. Section 4 addresses virtual screening, including binding site prediction, protein-ligand docking, affinity scoring, and advances in structure prediction models such (e.g., AlphaFold). Section 5 presents GNN-based generative and reinforcement learning methods for molecular design and optimization. Section 6 explores the use of GNNs in knowledge graph construction and reasoning, particularly for biomedical entities. Section 7 focuses on chemical synthesis modeling, covering retrosynthesis, reaction condition prediction, and synthesis planning from a molecular generation perspective. Together, these sections are designed to connect theoretical models with practical applications, offering a two-way bridge between fundamental methodologies and their real-world deployment in drug discovery.

## 2. Advances in Graph Neural Networks

### 2.1 General Graph Theory

In principle, GNNs can be broadly categorized into two methodological paradigms based on their underlying theoretical foundations: **spectral-based** and **spatial-based** approaches, as illustrated in the **Figure 2**. Spectral methods extend the concept of convolution from Euclidean domains, which is used in traditional convolutional neural networks (CNNs), to arbitrary graph structures by leveraging spectral graph theory and graph Fourier analysis[20]. Representative works include SCNN[21], ChebNet[22], and GCN[23]. These models typically define graph convolution operations via the eigendecomposition of the graph Laplacian, leading to mathematically elegant yet relatively inflexible architectural designs. In contrast, spatial-based methods aggregate information directly from a node's local neighborhood using learnable functions that mimic the localized receptive fields of CNNs[24]. This approach enables efficient and intuitive modeling of local node interactions. Among spatial methods, the Message Passing Neural Networks (MPNN) framework[25] has emerged as a general abstraction, unifying the vast majority of modern GNN architectures, including GraphSAGE[26], GIN[27], and GAT[28]. Most contemporary graph deep learning libraries, such as PyG[29] and DGL[30], are built upon this framework, facilitating widespread adoption and practical implementation. Due to its conceptual clarity and ecosystem support, spatial-based graph convolution has become the predominant paradigm in chemistry-related tasks. In the first part of this section, we briefly review the core principles of spectral graph convolution to provide historical context. The remainder of this chapter focuses on spatial-based methods, with particular emphasis on MPNN-based frameworks and their applications in molecular modeling and drug design. The summary of these basic GNN architectures are provided in **Table 1**.

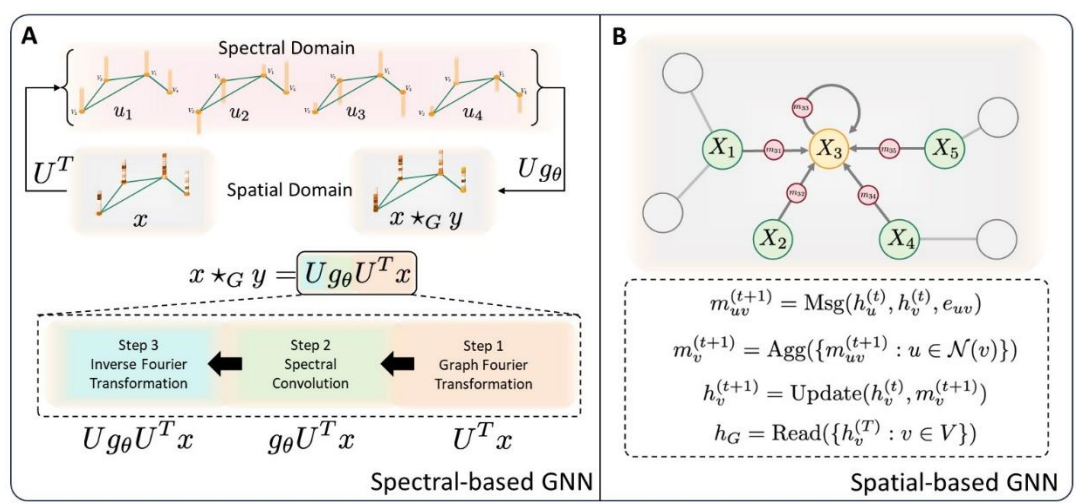

**Figure 2. Comparison between spectral-based and spatial-based GNNs.** (A) Spectral-based GNNs perform convolution in the spectral domain by applying learnable filters to the graph Laplacian

eigenbasis; (B) Spatial-based GNNs operate directly in the spatial domain by propagating and aggregating messages across graph neighborhoods. The formulas are presented in the MPNN framework.

Table 1. Overview of Spectral- and Spatial-based GNNs.

| Category | Models | Keywords |
| --- | --- | --- |
| Spectral-based | SCNN[21] | $g_\theta$ is a parameterized diagonal matrix |
|  | ChebNet[22] | Approximate $g_\theta$ with a fixed-order polynomial. |
|  | GCN[23] | Only keeping 1st and 2nd polynomial |
| Spatial-based | MPNN[25] | General Spatial-based GNN framework |
|  | GraphSAGE[26] | Sample and aggregate locally |
|  | GIN[27] | Discuss the upper bound of GNNs |
|  | GAT[28] | Introduce attention mechanism in GNNs |

### 2.1.1 Spectral-based GNN: Begin with Spectral Graph Theory

Convolution, broadly defined, describes how an input signal is modulated or filtered through the intrinsic characteristics of a given system, serving as a powerful mechanism for extracting informative features from raw input data. In classical signal processing, Fourier transforms are widely employed to project time-domain signals into the frequency domain, facilitating separate analysis of high- and low-frequency components, for instance, distinguishing between dolphin echolocation and high-pitched buzzes[31]. Similarly, in the context of graph-structured data, spectral methods seek to transform signals defined over nodes (i.e., signals in the spatial domain) into a spectral representation governed by the graph's topology. The ultimate goal is to design convolutional filters that encode the graph's frequency preferences, which can then be applied to extract and amplify meaningful node features from a given signal vector.

The Laplacian operator ($\Delta$) offers a natural mathematical formulation for characterizing graph structures. Conceptually, the (generalized) Laplacian quantifies the difference between the value of a function at a given node and the average value across its local neighborhood. In image processing, the Laplacian kernel is commonly employed to detect local variations with high sensitivity. A classic example is:

$$L = \begin{bmatrix} 0 & 1 & 0 \\ 1 & -4 & 1 \\ 0 & 1 & 0 \end{bmatrix},$$

When applied to a particular pixel, this operator effectively subtracts a weighted center value from the sum of its neighbors, emphasizing local intensity changes. Translating this concept to graphs, the graph Laplacian is defined analogously as:

$$L = D - A$$

where $A$ denotes the adjacency matrix (encoding connectivity) and $D$ the degree matrix (encoding local node degree). Linear algebra dictates that $L$ admits an eigendecomposition with $n$ orthonormal eigenvectors and non-negative eigenvalues. Let $U$ denote the matrix of eigenvectors (after normalization), satisfying:

$$UU^T = I$$

Each column of $U$ forms an orthogonal basis vector, which corresponds to the graph's harmonic components in the spectral domain. In this framework, multiplying a signal vector by $U$ projects it from the spatial domain into the spectral domain. To perform spectral graph convolution, a filter function $g$ is defined in the frequency domain to reweight (i.e., amplify or suppress) specific spectral components. In summary, the graph convolution of node features $x$ with filter $g$ can be decomposed into two steps: (1) projecting $x$ into the spectral domain, and (2) performing point-wise multiplication with the spectral representation of $g$, followed by projection back to the node domain:

$$x \star_G g = \mathcal{F}^{-1}\big(\mathcal{F}(x) \odot \mathcal{F}(g)\big) = U(U^T g \odot U^T x)$$

where $\star_G$ denotes graph convolution, $\mathcal{F}$ is graph Fourier transform, and $\odot$ is element-wise multiplication. To make the filters learnable, $g$ is parameterized as a function $g_w$, yielding:

$$x \star_G g_w = U(U^T g_w \odot U^T x)$$

For generalization, the spectral filter can be represented as a diagonal matrix in the eigenbasis of the Laplacian:

$$g_w{'} = \begin{pmatrix} g_{w_1}(\lambda_1) & & \\ & \cdots & \\ & & g_{w_n}(\lambda_n) \end{pmatrix}$$

Thus, the generalized spectral convolution can be formulated as:

$$x \star_G g_w' = U g_w' U^T x$$

This equation represents the most general and foundational formulation of spectral graph convolution. It was first introduced in its complete form by Shuman *et al.*[21]. The learnable spectral filter $g_w'$ is commonly denoted as $g_\theta$, where the subscript $\theta$ denotes the set of trainable parameters. By adopting different parameterizations or applying various constraints to $g_\theta$, a broad spectrum of spectral GNN architectures can be derived, each with distinct inductive biases and computational properties. In the following sections, we highlight three of the most representative models that exemplify these design strategies.

**SCNN**: The Spectral Convolutional Neural Network (SCNN) formulates the filter $g_\theta$ as a diagonal matrix of size $n \times n$, where each diagonal entry is a learnable parameter:

$$g_\theta = \begin{pmatrix} \theta_1 & & \\ & \ddots & \\ & & \theta_n \end{pmatrix}$$

This model represents one of the earliest attempts at leveraging spectral graph convolution and is widely regarded as the pioneering work in this domain. However, SCNN also suffers from several inherent limitations. First, computing the eigendecomposition of the graph Laplacian $L$ incurs a computational cost of $O(n^3)$, making it impractical for large-scale graphs. Second, the number of parameters in SCNN scales linearly with the number of nodes, i.e., $O(n)$, increasing the risk of overfitting on large graphs.

**ChebNet**: In SCNN, each entry of the filter matrix is learned independently, limiting the model's generalization across graphs of different sizes. ChebNet[22] addresses this challenge by expressing the spectral filter $g_\theta$ as a shared polynomial function over the graph spectrum. Among various polynomial bases, Chebyshev polynomials are particularly suitable due to their optimal approximation properties and numerical stability. Therefore, the spectral filter of ChebNet is approximated as:

$$g_\theta = \begin{pmatrix} \sum_{k=0}^{K} \beta_k T_k(\hat{\lambda}_1) & & \\ & \ddots & \\ & & \sum_{k=0}^{K} \beta_k T_k(\hat{\lambda}_2) \end{pmatrix}, \quad \hat{\lambda} = \frac{2}{\lambda_{max}}\lambda - 1$$

The compact expression is:

$$g_\theta(\Lambda) = \sum_{k=0}^{K} \beta_k T_k(\hat{\Lambda}), \quad \hat{\Lambda} = \frac{2}{\lambda_{max}}\Lambda - I_n$$

Thus, the graph convolution becomes:

$$x \star_G g_\theta = U g_\theta U^T x = \sum_{k=0}^{K} \beta_k T_k(\hat{L}) x$$

where $\hat{L} = \frac{2}{\lambda_{max}} L - I_n$. This formulation eliminates the need to explicitly compute the eigenvectors $U$, making the model scalable to large graphs. Furthermore, the number of learnable parameters depends only on the polynomial order $K$, rather than the graph size, thereby alleviating overfitting risks. ChebNet also enforces strict spatial locality: the polynomial order $K$ directly corresponds to the receptive field radius, ensuring that only nodes within $K$-hop neighborhoods are included in the convolution operation. This yields a localized filtering mechanism analogous to the fixed-size kernels used in conventional CNNs.

**GCN:** Building on ChebNet, Graph Convolutional Networks[23] (GCN) simplify the spectral filter even further by restricting the Chebyshev polynomial to the first two terms ($k = 0$ and $k = 1$):

$$g_\theta = \begin{pmatrix} \beta_0 T_0(\hat{\lambda}_1) + \beta_1 T_1(\hat{\lambda}_1) & & \\ & \ddots & \\ & & \beta_0 T_0(\hat{\lambda}_2) + \beta_1 T_1(\hat{\lambda}_2) \end{pmatrix},$$

Given that $T_0(\hat{L}) = 1$ and $T_1(\hat{L}) = \hat{L}$, the convolution reduces to:

$$x \star_G g_\theta = (\beta_0 + \beta_1 \hat{L})x$$

In the original GCN formulation, $\hat{L}$ is expressed by the normalized Laplacian $L_{sym}$, which is

$$L_{sym} = D^{-1/2} L D^{-1/2} = D^{-1/2}(D - W) D^{-1/2} = I_n - D^{-1/2} W D^{-1/2}$$

where $A$ is the adjacency matrix and $D$ the degree matrix. This normalization simplifies the spectral filter and ensures that the eigenvalues are bounded within a small interval, improving numerical stability and facilitating effective information propagation. To further streamline the model, the authors impose the constraint $\beta_0 = -\beta_1 = \theta$, yielding:

$$x \star_G g_\theta = (\theta(D^{-1/2}WD^{-1/2} + I_n))x$$

To mitigate gradient explosion in deep architectures, GCN introduces a renormalization trick by adding self-loops to the adjacency matrix, :

$$\widetilde{W} = W + I_n, \qquad I_n + D^{-\frac{1}{2}}WD^{-\frac{1}{2}} \to \widetilde{D}^{-\frac{1}{2}}\widetilde{W}\widetilde{D}^{-\frac{1}{2}}$$

leading to the final form:

$$x \star_G g_\theta = \theta(\widetilde{D}^{-1/2}\widetilde{W}\widetilde{D}^{-1/2})x$$

In this expression, $\widetilde{D}^{-1/2}\widetilde{W}\widetilde{D}^{-1/2}$ depends solely on the graph topology and defines a fixed operator, while $x$ represents node features and $\theta$ is the shared trainable weight. As in CNNs, this structure enforces a fixed, one-hop neighborhood receptive field and drastically reduces the number of parameters. Although GCN provides a clean and efficient bridge between spectral and spatial GNNs, some studies have noted that its simplicity may limit its expressiveness on more complex tasks[32].

In summary, SCNN, ChebNet, and GCN all fall within the spectral graph convolution framework, differing primarily in how they parameterize and constrain the spectral filter $g_\theta$. Fundamentally, they follow a common procedure: node features $x$ are projected into the spectral domain, filtered element-wise via $g_\theta$, and then transformed back to the spatial domain to yield updated node representations. These models provide a principled and mathematically grounded approach to capturing structural patterns in graph-structured data.

**2.1.2 Spatial-based GNN: Unified Framework with MPNN**

In contrast to spectral graph convolution, which relies on graph spectral theory and Fourier transforms, **spatial graph convolution** operates directly on a node's local neighborhood. It aggregates messages from neighboring nodes to update the central node's representation. The spatial approach is intuitive, adaptable, and particularly well-suited to dynamic graph structures. Beyond numerous model variants, the spatial GNN literature has also given rise to several **unified frameworks** that abstract common mechanisms among diverse architectures. Among these, the **Message Passing Neural Network**[25] (MPNN) framework stands out as one of the most widely adopted and foundational paradigms. Below, we use MPNN as a lens through which to introduce the core ideas of spatial GNNs, drawing comparisons to spectral methods (e.g., GCN) and highlighting representative models such as GraphSAGE, GAT, and GIN.

The MPNN framework formalizes GNNs through four key functions, whose specific implementations determine the model's behavior. These are:

**Message Function**: A message represents a pairwise interaction between two nodes. For a node $v$ with $k$ neighbors, there are $k$ messages generated from each neighbor $u \in \mathcal{N}(v)$. Formally, for edge $(u,v) \in E$, the message is computed as:

$$m_{uv}^{(t+1)} = \text{Msg}(h_u^{(t)}, h_v^{(t)}, e_{uv})$$

where $h_u^{(t)}$ and $h_v^{(t)}$ are the features of source and target nodes at step $t$, and $e_{uv}$ denotes edge features.

**Aggregation Function**: Aggregation summarizes all incoming messages for a node:

$$m_v^{(t+1)} = \text{Agg}(\{m_{uv}^{(t+1)} : u \in \mathcal{N}(v)\})$$

**Update Function**: The node updates its representation using its current state and the aggregated message:

$$h_v^{(t+1)} = \text{Update}(h_v^{(t)}, m_v^{(t+1)})$$

**Readout Function**: To obtain graph-level representations (for tasks such as graph classification), the node features are pooled:

$$h_G = \text{Read}(\{h_v^{(T)} : v \in V\})$$

For subgraph-level tasks, $V$ can be replaced with a subset $V_{sub}$.

The MPNN framework captures the essence of spatial GNNs: direct message exchange and iterative neighborhood aggregation. It provides a flexible abstraction adopted by most modern GNN toolkits (e.g., PyG[29], DGL[30]), allowing researchers to focus on architectural innovation rather than implementation complexity.

**GCN:** Although GCN[23] was originally derived from the spectral theory, its receptive field—limited to one-hop neighbors—aligns it more closely with spatial GNNs in terms of operational behavior. Within the MPNN framework, GCN can be described as:

$$\textbf{Message: } m_{uv}^{(t+1)} = h_u^{(t)}$$

$$\textbf{Aggregate: } m_v^{(t+1)} = \sum_{u \in \mathcal{N}(v)} \frac{1}{\sqrt{\hat{d}_u \hat{d}_v}} m_{uv}^{(t+1)} + h_v^{(t)}$$

$$\textbf{mUpdate: } h_v^{(t+1)} = \sigma(W^{(t)} m_v^{(t+1)})$$

Here, $\hat{d}_v$ denotes the degree of node $v$ including the self-loop, $W^{(t)}$ is a learnable weight matrix, and $\sigma$ is a nonlinear activiation function (e.g., ReLU). This can also be written more compactly as:

$$h_v^{(t+1)} = \sigma\left(\sum_{u \in \mathcal{N}(v) \cup \{v\}} \frac{1}{\sqrt{\hat{d}_u \hat{d}_v}} m_{uv}^{(t+1)} W^{(t)} h_u^{(t)}\right)$$

GCN's simplicity makes it a natural baseline for comparison with more expressive spatial GNNs.

**GraphSAGE**[26] (Sample and Aggregate) introduces two key innovations: (1) sampling a fixed-size neighborhood instead of using the entire graph, and (2) learning aggregation functions to combine neighbor features. Under MPNN, GraphSAGE is:

$$\text{Message: } m_{uv}^{(t+1)} = h_u^{(t)}$$

$$\text{Aggregation: } m_v^{(t+1)} = \text{Agg}(m_{uv}^{(t+1)} : u \in \mathcal{N}(v))$$

$$\text{Update: } h_v^{(t+1)} = \sigma(W^{(t)}[h_v^t || m_v^{(t+1)}])$$

The aggregation function can be **mean**, **LSTM-based**, or **pooling (e.g., max)**. A critical contribution of GraphSAGE is the emphasis on **permutation invariance**: the output should not depend on the order of neighbors. For LSTM aggregators, this is enforced by randomly permuting the input sequence.

**Graph Isomorphism Network (GIN)**[27] was developed to explore the theoretical expressive power of GNNs. Drawing inspiration from the Weisfeiler-Lehman (WL) test, GIN improves upon GraphSAGE by resolving cases where mean or max aggregation cannot distinguish different neighborhood structures with identical summaries. In MPNN form:

$$\text{Message: } m_{uv}^{(t+1)} = h_u^{(t)}$$

$$\text{Aggregation: } m_v^{(t+1)} = \sum_{(u,v) \in E} h_u^{(t)}$$

$$\text{Update: } h_v^{(t+1)} = \text{MLP}((1 + \epsilon^{(t+1)}) h_v(u) + m_v^{(t+1)})$$

where $\epsilon^{(t+1)}$ is a learnable scalar that balances the importance of a node's own features during update. Although simple in structure, GIN has been shown to match the WL test in its ability to distinguish graph structures, making it a strong candidate for tasks requiring high discriminative power.

**Graph Attention Network (GAT)**[28] integrates attention mechanisms into spatial aggregation, enabling the model to assign varying importance to each neighbor. The attention score between node $u$ and node $v$ is computed as:

$$e_{uv} = a^T \text{LeakyReLU}(W h_u + W h_v)$$

$$\alpha_{uv} = \frac{\exp(e_{uv})}{\sum_{k \in \mathcal{N}(v)} \exp(e_{vk})}$$

where and $\alpha_{uv}$ is the attention weights after normalization. Under MPNN, GAT can be expressed:

$$\text{Message: } m_{uv}^{(t+1)} = \alpha_{uv}^{(t)} h_u^{(t)}$$

$$\text{Aggregation: } m_v^{(t+1)} = \sum_{u \in \mathcal{N}(v)} m_{uv}^{(t+1)}$$

$$\text{Update: } h_v^{(t+1)} = \sigma(W^{(t)}[m_v^{(t+1)}])$$

Compared to GraphSAGE or GIN, GAT explicitly weighs each neighbor's contribution, offering a highly expressive, data-dependent aggregation strategy. Variants such as AttentiveFP[33] (which uses GRU-based readout) and EGAT[34] (which applies attention on edges) have further extended the utility of attention mechanisms in molecular property prediction and related tasks. Notably, GAT can be seen as a practical instantiation of models that achieve the expressiveness upper bound outlined by GIN.

## 2.2 Symmetric GNN: Incorporate Physical Laws

In traditional GNN applications, such as social networks, nodes typically possess limited attributes and relations. In contrast, molecular graphs in drug discovery and materials science are characterized by both their 2D topological structures and 3D geometric conformations. Specifically, each atom in a molecule is represented by a set of Cartesian coordinates in the 3D space. While these coordinates may vary under different reference frames, the molecular geometry, defined by the relative spatial arrangement of atoms, remains physically unchanged.

To make the model adhere to real-world physical constraints, it is essential to incorporate two types of symmetry principles into GNN design: **invariance** and **equivariance**[35]. **Invariance** applies to tasks where the output should remain unchanged under coordinate transformations. For example, when predicting scalar molecular properties such as thermodynamic stability or HOMO-LUMO energy gaps, the model's prediction must be independent of the molecule's orientation or position in space. **Equivariance**, on the other hand, is necessary for predicting vectorial or tensorial properties such as atomic forces or displacement directions. In these cases, if the input is rotated or translated, the output should transform in the same way. Conventional architectures like multilayer perceptrons (MLPs) do not inherently satisfy these symmetry constraints. To this end, it is necessary to embed appropriate physical inductive biases into the network architecture. Below, we review three representative approaches for incorporating geometric symmetries into GNNs within the broader message-passing framework.

Most symmetry-aware GNNs are formulated within the framework of **group theory**, which provides a rigorous mathematical formalism for defining invariance and equivariance. In the context of molecular systems, the most relevant symmetries lie in the **3D Euclidean space**, particularly:

- **SO(3)**: The group of all 3D rotation operations.
- **E(3)**: The Euclidean group, encompassing translations, rotations, and mirror reflections.
- **SE(3)**: The special Euclidean group, a subgroup of E(3), including only translations and rotations (excluding reflections).

These symmetry groups dictate how molecules can be transformed while preserving their physical identity. In the following section, we categorize invariant and equivariant GNNs into three classes, as summarized in **Table 2**.

Table 2. Overview of symmetric GNN models and their performance on several metrics. G is one of the scalar properties prediction tasks in QM9, and $F$ is the force prediction on the Aspirin molecule in MD17, the reported results are mean absolute error (MAE).

| Models | $G$ meV | F (Aspirin) meV/Å | Keywords |
|---|---|---|---|
| SchNet[36] | 14 | 58.5 | RBF basis function for handling $d$ |
| DimeNet[37] | 8 | 21.6 | Second-order MPNN considering $\theta$ |
| GemNet[38] | - | 10.3 | Third-order MPNN considering $\theta$ and $\tau$ |
| SphereNet[39] | 18 | - | Consider the $\phi$ between two planes rather than $\tau$ |
| ComENet[40] | 7.98 | - | Consider the $d, \theta, \tau, \phi$ |
| TFN[41] | - | - | Early CG-tensor product-based equivariant GNN |
| SE(3)-Transformer[42] | - | - | Introduce the dot product attention to the TFN |
| Cormorant[43] | 20 | - | Introduce non-linearity to TFN-like models |
| SEGNN[44] | 15 | - | Gated layer for equivariant non-linearity |
| EquiFormer[45] | 7.63 | 6.6 | Depth-wise tensor product, MLP-based attention |
| MACE[46] | - | 6.6 | Atomic Cluster Expansion, only 2-order interaction |
| GVP[47] | - | - | Isolate transformation and model multiplication |
| EGNN[48] | 12 | - | Atomic coord. difference multiplies invariant message |
| NewtonNet[49] | - | 15.1 | Extend EGNN to type, force, coords, and energy module |
| EQGAT[50] | 23 | - | Implement GVP equivariant mechanism in GAT |
| PaiNN[51] | 20 | 14.7 | Introduce Gated equivariant block to EGNN |

### 2.2.1 Internal Coordinate-based Invariant GNN

While Cartesian coordinates are reference-frame dependent, certain internal features, such as bond lengths, bond angles, and dihedral angles, remain invariant across reference frames. These internal coordinates are foundational in classical molecular mechanics and quantum chemistry (**Figure 3A**), as they represent E(3)-invariant quantities critical for energy and force calculations[52]. Consequently, one intuitive strategy is to convert Cartesian inputs into internal coordinates before feeding them into a GNN. This principle is adopted by many state-of-the-art molecular representation models.

**SchNet**[36]: One of the earliest and most influential models in this line of work is SchNet, which introduces continuous filter convolutions to model interactions between atoms. The core design centers around interatomic distances, ensuring SE(3) invariance when predicting molecular properties such as total energy. To enhance the representational capacity of distance-based features, SchNet employs Radial Basis Functions (RBFs), which is also called **Gaussian Smearing**, to map scalar distances into high-dimensional vectors, mitigating the limitation of using single scalar values as filter inputs. This RBF-based technique has since been widely adopted in subsequent models. Within the MPNN framework, SchNet can be abstracted as follows:

$$\text{Message function: } m_{uv}^{(t+1)} = W^{(t)} f_{\text{filter}}(d_{uv}) \odot h_u^{(t)}$$

$$\text{Aggregation: } m_v^{(t+1)} = \sum_{u \in \mathcal{N}(v)} m_{uv}^{(t+1)}$$

$$\text{Update: } h_v^{(t+1)} = \sigma\big(W^{(t)} h_v^{(t)} + m_v^{(t+1)} + b^{(t)}\big)$$

Here, the filter function is defined as:

$$f_{\text{filter}}(d_{uv}) = \exp\left(-\gamma(\| r_u - r_v \| - \mu)^2\right)$$

where $\gamma$ controls the width and $\mu$ the center of the Gaussian filter. Through this distance-based representation, SchNet ensures rotational and translational invariance for scalar predictions such as total energy $E$. The model also supports equivariant force prediction via the energy gradient:

$$E = \sum_{u \in V} h_u$$

$$F_i(Z_1, \ldots, Z_n, \bm{r}_1, \ldots, \bm{r}_n) = -\frac{\partial E}{\partial \bm{r}_i}(h_1, \ldots, h_n, \bm{r}_1, \ldots, \bm{r}_n)$$

Because $h_u$ is SE(3)-invariant, the energy $E$ is also SE(3)-invariant, and the force prediction $F_i$ transforms equivariantly under rotations:

$$F(Rr) = -\nabla E(Rr) = -R\nabla E(r) = RF(r)$$

This ensures that $F$ produces physically consistent outputs under coordinate transformations, i.e., SE(3)-equivariant predictions.

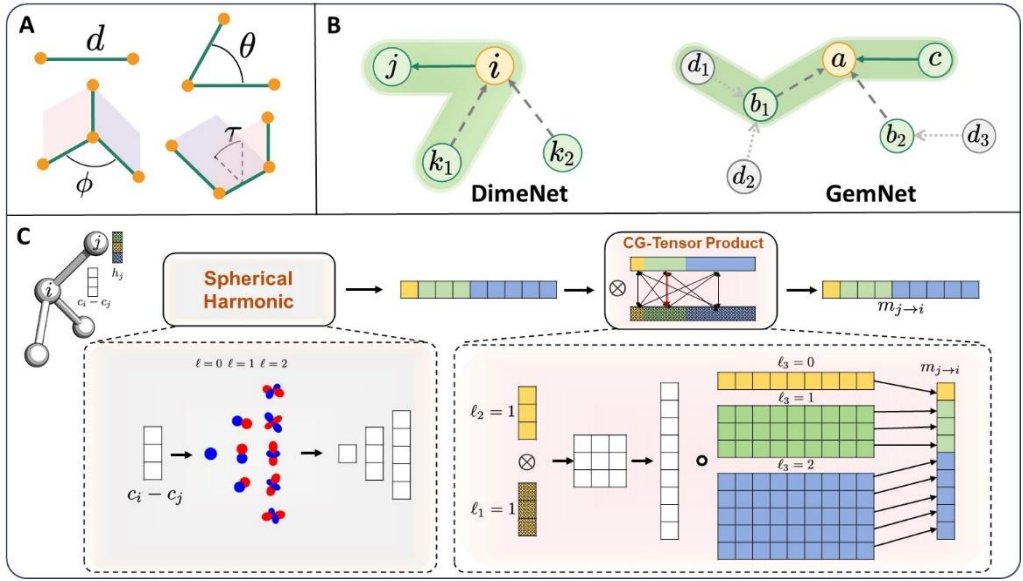

**Figure 3.** A) Internal coordinate system; B) The illustration of the DimeNet and GemNet message passing frameworks, which represent the second- and third-order interactions, respectively. C) The illustration of the CG-Tensfor Product method, where the CG-Tensor product operates between input geometric features and the spherical harmonics of relative position vectors.

**DimeNet**[37]: While SchNet captures pairwise interactions based solely on distance, it cannot distinguish molecular conformers with identical pairwise distances but different angular arrangements. To address this, **DimeNet** introduces directional message passing by incorporating angle-based features into the convolution process, enabling the model to capture richer geometric information, as illustrated in **Figure 3B**. Conceptually, its message function is defined as:

$$m_{ji}^{(t+1)} = f_{\text{update}}(m_{ji}^{(t)}, \sum_{k \in N(j) \setminus \{i\}} f_{\text{int}}(m_{kj}^{(t)}, e_{\text{rbf}}(d_{ji}), e_{\text{cbf}}(d_{ji}, \theta_{kji}))))$$

Here, $e_{\text{rbf}}(d_{ji})$ and $e_{\text{cbf}}(d_{ji}, \theta_{kji})$ denote radial and spherical basis embeddings of distances and angles, respectively. These embeddings are constructed using a combination of Bessel functions $j_l$, spherical harmonics $Y_l^0(\theta)$ and an envelope function $u(d)$ to map geometric inputs into high-dimensional latent spaces.

From the MPNN perspective, DimeNet captures second-order interactions: each message along edge $j \to i$ not only encodes the pairwise distance between $j$ and $i$, but also aggregates geometric context from a third atom $k$, forming angle $\theta_{kji}$. This higher-order aggregation improves the model's expressivity and aligns with a second-order Weisfeiler-Lehman test. Empirically, DimeNet outperforms SchNet by approximately 1.5× on benchmark datasets such as MD17 (aspirin data), which would be discussed later in the molecular property prediction section.

**GemNet:** While DimeNet accounts for angular information, a complete reconstruction of 3D molecular conformations also requires dihedral angles. To this end, the authors of DimeNet proposed **GemNet**[38], which extends directional message passing to include third-order geometric interactions via additional aggregation layers, as illustrated in **Figure 3B**. Here, we provide the conceptual message function in GemNet for better understanding:

$$m_{ca}^{(t+1)} = f_{\text{update}}(m_{ca}^{(t)}, \sum_{d \in N(b) \setminus \{a,c\}} \sum_{b \in N(a) \setminus \{c\}} f_{\text{int}}(m_{db}^{(t)}, e_{\text{rbf}}(d_{db}), e_{\text{cbf}}(d_{ba}, \theta_{abd}), e_{\text{sbf}}(d_{ca}, \theta_{cab}, \phi_{cabd})))$$

This formulation integrates a four-body interaction involving nodes $c \to a \leftarrow b \leftarrow d$, allowing GemNet to capture dihedral dependencies $\phi_{cabd}$, alongside distances $d_{db}, d_{ba}, d_{ca}$ and angles $\theta_{cab}, \theta_{abd}$, which aligns a third-order Weisfeiler-Lehman test. While GemNet improves expressivity, achieving approximately 2× performance improvement over DimeNet on datasets like MD17, it comes at a higher computational cost due to quadruplet enumeration, limiting its scalability for larger molecules.

**Other Models:** Beyond the aforementioned models, several other approaches pursue similar goals of capturing **multi-body geometric interactions**. For instance, **SphereNet**[39] models the angle between planes (dihedral-like) via a variable $\phi$. **ComENet** [6] combines multiple geometric terms, including distances, angles, and dihedrals, into a unified message formulation. These models aim to enhance expressivity by modeling higher-order interactions $f(h_1, h_2, ..., h_k)$ within $k$-hop neighbors. While SchNet, DimeNet, and GemNet capture first-, second-, and third-order interactions, respectively, SphereNet and ComENet explore alternative three-body definitions based on plane-plane angular system.

In summary, these invariant neural architectures leverage SE(3)-invariant features internally, from which they extract physically meaningful embeddings. Under the MPNN framework, these features are propagated and updated across graph layers, enabling the model to predict not only scalar properties (e.g., energy $E$) but also equivariant outputs (e.g., atomic forces $F(r) = -\nabla E(r)$). However, it is important to note that in such models, equivariant outputs are derived from gradients of invariant scalars, which inherently limits their flexibility. For tasks requiring directly learnable equivariant outputs, more general equivariant neural architectures are required.

Additionally, spherical harmonics play a dual role in these models: they embed angular information in a physically grounded way and also serve as angular basis functions. In the next section on tensor-product-based equivariant GNNs, we will see how these harmonics construct higher-order (type-$l$) equivariant features, enabling more expressive and generalizable models.

## 2.2.2 Tensor Product-based Equivariant GNN

An alternative strategy for achieving equivariance in GNN is to leverage group-invariant convolutional theory to operate directly on geometric features. While the associated mathematical tools are more complex, models based on these principles (e.g., SE(3)-Transformer) have been widely adopted in large biomolecular structure prediction tasks, such as RosettaFold[53] and AlphaFold[54], underscoring their strong research and application potential. The fundamental idea is to 1) extend common SO(3) equivariance to higher-order ones; 2) enforce operations equivariant via Glebsch-Gorden-tensor products (CG-tensor product), and 3) parameterize interactions across multiple representation orders using neural networks. The basic idea is illustrated in **Figure 3C**. Readers interested in the mathematical background may refer to **SI Part 1**.

The tensor product operation itself is inherently equivariant: when input features transform under rotations, the resulting tensor product preserves the same transformation behavior. According to the group theory discussed in the **SI Part 1**, scalar distances can be viewed as order $l = 0$ features, while coordinate vectors correspond to order $l = 1$. Most models begin with such inputs and enrich the geometric information by computing spherical harmonic embeddings $Y^l$ of relative coordinates $r_{ij}$, yielding higher-order equivariant type-$l$ features. These features are then passed through learnable CG-tensor products to generate output features across orders $\{0, \dots, m\}$. For scalar prediction such as energy or charge, the $l = 0$ output is used; for vector predictions such as force, the $l = 1$ output is used. On this foundation, we would discuss two representative models in detail: Tensor Field Networks (TFN) and SE(3)-Transformer.

**Tensor Field Networks**[41] (TFN) is among the first to apply group-invariant convolution theory to GNNs. The concept builds upon early developments in rotation-equivariant convolution, such as Harmonic Networks[55] and Spherical CNNs[56], and formalizes equivariant message passing through the use of CG tensor products. The message function is defined as:

$$m_{ij}^l = [h_i \otimes_{cg}^{\psi(d_{ij})} Y(e_{ij})]^l$$

where $h_i^l$ is the type-$l$ representation of node $i$, $d_{ij}$ is the interatomic distance, and $e_{ij}$ is the unit vector between nodes. The filter is constructed by modulating the spherical harmonic $Y$ with a learnable radial embedding $\psi$ and applying a CG product. The superscript $[\cdot]^l$ selects the type-$l$ component from the output. To be more specific, the message is:

$$m_{ij,m}^{(l)} = \psi_m(d_{ij}) \sum_{m_1=-l_1}^{l_1} \sum_{m_2=-l_2}^{l_2} C_{(l_1,m_1)(l_2,m_2)}^{(l,m)} h_{i,m_1}^{(l_1)} Y_{m_2}^{(l_2)}(e_{ij})$$

where $C^{(l,m)}_{(l_1,m_1)(l_2,m_2)}$ determine how the input features are coupled into output components of type-$l$. By construction, this operation ensures that the output message $m_{ij}^l$ remains equivariant. Aggregation and update functions are defined as:

$$\Delta h_i^l = \sum_{j \in \mathcal{N}(i)} m_{ij}^l$$

$$h_i^{l,(t+1)} = w^{l(t)} h_i^{l(t)} + \Delta h_i^{l,(t)}$$

where $w^{l(t)}$ is a learnable parameter, and $t$ is the layer index. The model output is split into the scalar ($l = 0$) and higher-order ($l > 0$) parts:

$$h_{i,\text{pred}}^0 = W_{\text{pred}} h_i^0$$

$$h_{i,\text{pred}}^l = \text{MLP}(h_i^l), \quad l > 0$$

Some works use a more general formulation of TFN's equivariant kernel:

$$h_{\text{out},i}^l = \sum_{k \geq 0} \sum_{j \in \mathcal{N}(i)}^n W^{lk}(x_j - x_i) h_{\text{in},j}^k$$

$$W^{lk}(\vec{r}_j - \vec{r}_i) = \sum_{J=|k-l|}^{k+l} \psi_J^{lk}(\|d_{ij}\|) W_J^{lk}(\frac{x_j - x_i}{\|d_{ij}\|}), \quad W_J^{lk} = \sum_{m=-J}^{J} Y_m^{(J)}\left(\frac{x_j - x_i}{\|d_{ij}\|}\right) C_{Jm}^{lk}$$

where $C_{Jm}^{lk}$ are the Clebsch–Gordan coefficients, and the spherical harmonics ensure angular dependency. This construction allows the kernel to transform type-$k$ inputs into type-$l$ outputs while preserving equivariance.

**SE(3)-Transformer**: Building upon TFN, the **SE(3)-Transformer** introduces a self-attention mechanism while maintaining the equivariance. Its message function is written as:

$$m_{ij}^l = \alpha_{ij}[h_i \otimes_{cg}^{\psi(d_{ij})} Y(e_{ij})]^l$$

where $\alpha_{ij}$ is an attention coefficient that modulates the influence of neighbor $j$ on node $i$. To preserve equivariance, $\alpha_{ij}$ must be invariant under rotation. This is achieved by defining $\alpha_{ij}$ via invariant queries and keys:

$$q_i = \bigoplus_{\ell \geq 0} \sum_{k \geq 0} W_Q^{\ell k} h_{\text{in},i}^k, \quad k_{ij} = \bigoplus_{\ell \geq 0} \sum_{k \geq 0} W_K^{\ell k}(x_j - x_i) h_{\text{in},j}^k$$

$$\alpha_{ij} = \frac{\exp(q_i^\top k_{ij})}{\sum_{j' \in \mathcal{N}_i \setminus i} \exp(q_i^\top k_{ij'})},$$

The attention-enhanced message is then aggregated and updated similarly to TFN, with an attention-based update coefficient:

$$\Delta h_i^l = \sum_{j \in \mathcal{N}(i)} m_{ij}^l$$

$$h_i^{l,(t+1)} = w_{\text{attn}}^{l(t)} h_i^{l(t)} + \Delta h_i^{l,(t)}$$

where $w_{\text{attn}}^{l(t)} = \text{MLP}(h_i^{l(t)} \cdot h_i^{l(t)})$ is a non-linear self-attention weight. Beyond TFN and SE(3)-Transformer, many related models refine CG-tensor product-based equivariant graph convolution. These architectures focus on improving either expressiveness or computational efficiency. **Cormorant**[43] introduces nonlinear interactions across tensors between different orders; **SEGNN**[44] introduces gating mechanisms for equivariant nonlinearity, where scalars are passed through standard activations and higher-order tensors are modulated via learned scalar gates:

$$(\bigoplus_i \phi_i(x_i)) \oplus (\bigoplus_j \phi_j(g_j) y_j),$$

where $x_i$ are input scalar features, $y_j$ are input type-$l$ ($l > 0$) tensors, and $g_j$ are gating scalars. This gating mechanism has become standard in subsequent works. **EquiFormer**[45] builds on SE(3)-Transformer by replacing dot-product attention with an MLP attention scheme and substituting full tensor products with depth-wise tensor products to reduce computational load while expanding model capacity. **MACE**[46] introduces Atomic Cluster Expansion (ACE) into TFN-like architectures. By using only pairwise convolutions of the form $h_i \otimes_{cg}^{\psi(d_{ij})} Y(e_{ij})$, MACE approximates many-body interactions through a complete basis, offering strong expressiveness with minimal cost.

### 2.2.3 Vector-based Equivariant GNN

Previous sections discuss invariant GNNs based on internal coordinates and equivariant GNNs leveraging CG tensor products. This section introduces a third paradigm: **vector graph neural networks**. In many practical tasks, the relevant physical quantities correspond to low-order rotational features, typically with angular momentum $l \leq 1$. Examples include atomic charges or molecular solubility ($l = 0$), and atomic velocities or force vectors ($l = 1$). While CG tensor products can represent higher-order equivariant features, such models often incur significant computational overhead. To improve efficiency, researchers have developed alternative equivariant operations tailored to scalar and vector features, avoiding the full complexity of CG algebra. These approaches, collectively referred to as vector GNNs, are introduced below through representative models.

**Geometric Vector Perceptron (GVP)**[47] employs an elegant equivariant design for updating scalar–vector pairs. The core formulation is as follows:

$$s', \vec{v}' = \text{GVP}(s, \vec{v})$$

$$s' = \sigma\left(W_m \begin{bmatrix} \| W_h \vec{v} \|_2 \\ s \end{bmatrix} + b\right)$$

$$\vec{v}' = \sigma\left(\left\| W_\mu W_h \vec{v} \right\|\right) \odot W_\mu W_h \vec{v}$$

Here, $\odot$ denotes element-wise multiplication. The vector feature $\vec{v} \in \mathbb{R}^{C \times 3}$ is designed such that linear transformations are applied from the left, and rotation matrices act from the right, ensuring equivariance under SO(3) rotations. While many implementations follow the PyTorch convention of applying linear layers to the final dimension (i.e., right-multiplication), the original GVP formulation adheres to the left-multiplication tradition of linear algebra. Readers may refer to the **SI Part 2** for a formal proof of GVP's equivariance. A GVP-based vector GNN typically follows this structure:

$$m_{ij}^{t+1}, \vec{m}_{ij}^{t+1} = \text{GVP}(s_i^t \| s_i^t \| e_{ij}, \vec{v}_i^t \| \vec{v}_j^t \| \vec{e}_{ij})$$

$$m_i^{t+1}, \vec{m}_i^{t+1} = \frac{1}{M} \sum_{j \in \mathcal{N}(i)} (m_{ij}^{t+1}, \vec{m}_{ij}^{t+1})$$

$$h_i^{t+1}, \vec{h}_j^{t+1} = (h_i^t, h_j^t) + \text{GVP}(m_i^{t+1}, \vec{m}_i^{t+1})$$

where $s_i, \vec{v}_i$ denote scalar and vector node features, and $e_i, \vec{e}_i$ represent edge features. A similar principle was adopted by Deng et al.[57] in the Vector Neurons framework, applied to point cloud data. Their model, VN-PointNet, demonstrated significant performance gains when SO(3) equivariance was enforced.

**Equivariant Graph Neural Network (EGNN):** Subsequent studies, such as Jing et al.[47], showed that GVP performs well for low-order geometric features. Victor et al. [48] later proposed **EGNN**, a model that became central to geometry-aware GNN research. Unlike GVP, which focuses on SO(3) equivariance, EGNN introduces a broader E(n)-equivariant formulation by using coordinate differences during message passing.

The model separates scalar and vector messages:

$$m_{ij}^t = \phi_e(h_i^t, h_j^t, d_{ij}, e_{ij}), \qquad \vec{m}_{ij}^t = (x_i^t - x_j^t)\phi_x(m_{ij}^t)$$

where $\phi_e, \phi_x$ are neural message functions, often implemented as MLPs. The design ensures that scalar messages remain invariant under transformations, while vector messages transform equivariantly. This can be formally verified as follows:

$$\vec{m}_{ij}^{t\,\prime} = \left(Rx_i^t + g - R_{x_j}^t - g\right)\phi_x(m_{ij}^t)$$

$$= R(x_i^t - x_j^t)\phi_x(m_{ij}^t)$$

$$= R\,\vec{m}_{ij}^t$$

Thus, equivariance is preserved under translation and rotation. The aggregation and update steps are defined as:

$$m_i^t = \sum_{j \in \mathcal{N}(i)} m_{ij}^t, \qquad \vec{m}_i^t = \frac{1}{M}\sum_{j \in \mathcal{N}(i)} \vec{m}_{ij}^t$$

$$h_i^{t+1} = \phi_h(h_i^t, m_i^t), \qquad \mathbf{x}_i^{t+1} = \mathbf{x}_i^l + \vec{m}_i^t$$

EGNN directly updates node coordinates, making it well-suited for structural prediction and dynamic simulations. For instance, **KarmaDock**[58] uses this mechanism to iteratively guide ligand fitting into protein pockets via force field-inspired optimization.

**NewtonNet**: Similar to EGNN, NewtonNet[49] uses coordinate differences $r_i - r_j$ to maintain equivariance, but introduces a modular architecture explicitly aligned with physical principles. The network is divided into four interacting modules: atomic features, forces, displacements, and energy.

Atomic features are updated as:

$$\phi_e(\vec{r}_{ij}) = \phi_{\text{rbf}}\left(e(\vec{r}_{ij})\right) e_{\text{cut}}(\|\vec{r}_{ij}\|), \qquad e_{\text{cut}}(\vec{r}_{ij}) = \sqrt{\frac{2}{r_c}} \frac{\sin(\frac{n\pi}{r_c}\|\vec{r}_{ij}\|)}{\|\vec{r}_{ij}\|}$$

$$m_{ij}^t = (a_j^t)\phi_e(\vec{r}_{ij}), \qquad a_i^{t+1} = a_i^t + \sum_{j \in \mathcal{N}(i)} m_{ij}$$

where $a_i^t$ is the scalar atomic feature at layer $t$, and $\phi_{\text{rbf}}, e_{cut}$ implement radial basis encodings, similar to those in SchNet. **Force vectors** are updated via:

$$\vec{F}_{ij} = \phi_F(m_{ij}^t)\vec{r}_{ij}$$

$$\mathcal{F}_i^{t+1} = \mathcal{F}_i^t + \sum_{j \in \mathcal{N}(i)} \vec{F}_{ij}$$

$$\Delta f_i = \sum_{j \in \mathcal{N}(i)} \phi_f(m_{ij})\vec{F}_{ij}$$

$$f_i^{t+1} = f_i^t + \Delta f_i^t$$

where $\vec{F}_{ij}$ represents latent forces and $f_i^t$ physical forces. If physical supervision is available, latent and physical forces can be regularized to align. **Displacement updates** follow a physics-inspired structure:

$$\delta r_i = \phi_r(a_i^{t+1}) f_i^{t+1}$$

$$dr_i^{l+1} = \sum_{j \in \mathcal{N}(i)} \phi_r'(m_{ij}) \, dr_j^l + \delta r_i$$

This step approximates coordinate evolution under force influence, though it does not strictly enforce Newtonian dynamics. **Energy changes** are computed by inner products between force and displacement:

$$\delta U_i = -\phi_u(a_i^{l+1})(f_i^{t+1} \cdot dr_i^{t+1})$$

$$a_i^{t+1} = a_i^t + \delta U_i$$

In classical mechanics $f_i^t \approx -\delta U/\delta r_i^t$, this term reflects the integration of force-driven energy change into atomic representations. Both NewtonNet and EGNN achieve equivariance via $r_i - r_j$, but NewtonNet explicitly separates force, displacement, and energy components, embedding more direct physical priors into the architecture.

In summary, vector-based GNNs can be broadly classified into two categories. The first, exemplified by GVP, ensures equivariance through the design of **vector neurons**. These methods are easily extended to 2D GNN variants. For instance, by incorporating attention into GVP-GNNs, one obtains models like GVP-GAT. The second category, represented by EGNN, ensures equivariance by scaling edge vectors $r_i - r_j$ with scalar messages. Models such as **EQGAT**[50] and **PaiNN**[51] adopt similar strategies. For example, EQGAT combines GAT with a simplified GVP-style mechanism, known as a value transformer:

$$s'_i = W_{sv}s_i + b_{sv}$$

$$\vec{v}'_i = \vec{v}_i W_{vv}$$

where $\vec{v}_i \in \mathbb{R}^{3\times c}$ is updated via right-multiplication, preserving SO(3) equivariance in the same spirit as GVP.

**PaiNN** extends EGNN's equivariant strategy but introduces a more intricate architecture. Though vector updates are limited to type-0 and type-1 representations, they are sufficient to construct high-order tensors through outer products:

$$T = \sum_{i=1}^{N} \sum_{k=1}^{R} \lambda(s_i)\, \vec{v}(\vec{v}_i)_{k,1} \otimes \cdots \otimes \vec{v}(\vec{v}_i)_{k,l}$$

For example, a rank-2 polarization tensor can be formed as:

$$\alpha = \sum_{i=1}^{N} \alpha_0(\mathbf{s}_i)I_3 + \vec{v}\left(\vec{\mathbf{v}}_i\right) \otimes \vec{r}_i + \vec{r}_i \otimes \vec{v}\left(\vec{\mathbf{v}}_i\right)$$

This demonstrates that even PaiNN, when restricted to type-0 and type-1 features, can recover higher-order physical quantities through tensor construction, retaining full expressive power for symmetric modeling tasks.

## 2.3 Large and Deep GNN: Challenges and Actions

The development of large-scale models in graph learning raises a central question: how can one construct GNNs that benefit from increased model capacity and training data, as observed in other domains? In principle, models with a larger number of parameters possess greater capacity for pattern recognition and representation learning, allowing them to extract deeper structures and relationships from complex data. As a result, such models often exhibit improved generalization and accuracy across a wide range of tasks.

Recent advances in neural scaling laws[59] have confirmed that scaling both model size and training data can substantially enhance performance in downstream tasks. More importantly, large models also demonstrate robust generalization in low-resource settings, such as **few-shot** and **zero-shot learning**. In natural language processing (NLP), **GPT-3**[60] achieved strong performance across diverse benchmarks after pretraining on massive text corpora. In vision tasks, large-scale training on datasets such as **Instagram images**[61] and **JFT-300M**[62] has also proven effective for transfer learning and low-shot scenarios. Even in zero-shot settings, models like **CLIP**[63], trained on 400 million image–text pairs using contrastive objectives, perform remarkably well.

Inspired by successes in NLP and computer vision, the graph learning community has begun exploring similar directions. However, training large GNNs presents unique challenges. From an architectural standpoint, increasing GNN depth does not always improve performance; in many cases, it can even degrade performance[64]. This phenomenon is attributed to the fundamental limitations of current GNN architectures, including **over-smoothing**, **over-squashing**, and **gradient instability**. On the data side, the scarcity of labeled data in real-world graph domains limits the training potential of large models. To overcome this, **self-supervised learning** has emerged as a promising pretraining strategy, leveraging large quantities of unlabeled data. However, if the pretraining task is poorly aligned with downstream objectives, it can result in the so-called **negative transfer**[65], where pretraining impairs rather than improves downstream performance. Therefore, this chapter focuses on three key aspects of successfully scaling graph models. Section 2.3.1 discusses the core architectural challenges, including over-smoothing, over-squashing, and gradient anomalies in deep GNNs. Section 2.3.2 explores the integration of Transformer architectures into GNN. Section 2.3.3 focuses on strategies for pretraining on unlabeled graphs, with an emphasis on generative, predictive, and contrastive objectives. The GNN solutions to these challenges are summarized in **Table 3**.

Table 3. Overview of strategies for solving problems when GNN goes deeper and larger.

| Problems | Method | Keywords |
| --- | --- | --- |
| Over-smoothing | DropGNN[66] | Node perturbation |
|  | DOTIN[67] | Attention-based node fusion |
|  | DropEdge[68] | Edge perturbation |
|  | SoftEdge[69] | Soft edge perturbation |
|  | Attention | Differentiate node features |
|  | DeepGCN[70] | Residual connection |
|  | GCNII[71] | Initial residual connect and identity mapping |
|  | JKNet[72] | Adaptive layer feature selection |
|  | DAGNN[73] | Gated adaptive layer feature selection |
| Over-squashing | SDRF[74] | Ricci-curvature-based rewiring |
|  | CurvDrop[75] | Gradual Layer rewiring |
|  | CBED[76] | Curvature-based edge dropping |
|  | Drew[77] | Dynamically densifying graph structures |
|  | CurvPool[78] | Curvature-based node merging and pooling |
|  | Transformer-like | Directly connect distant nodes |
| Gradient Anomaly | Residual Connect[79] | Direct gradient flow path |
|  | ReLU[80] et al. | Modified activation functions |
|  | Gradient Clip | As its name |
|  | Batch Norm[81] | Keep gradient within a normalized range |
|  | Initialization[82 82] | Prevent anomaly at the very beginning |

### 2.3.1 Over-Smooth, Over-Squashing, and Gradient Anomalies

#### 2.3.1.1 Problem Definition

Contemporary GNNs often achieve competitive results with only two or three layers. However, recent research has revealed that simply stacking more layers does not necessarily lead to better performance. This limitation has been traced to three interrelated phenomena: **over-smoothing**, **over-squashing**, and **gradient anomalies**, as illustrated in **Figure 4A**.

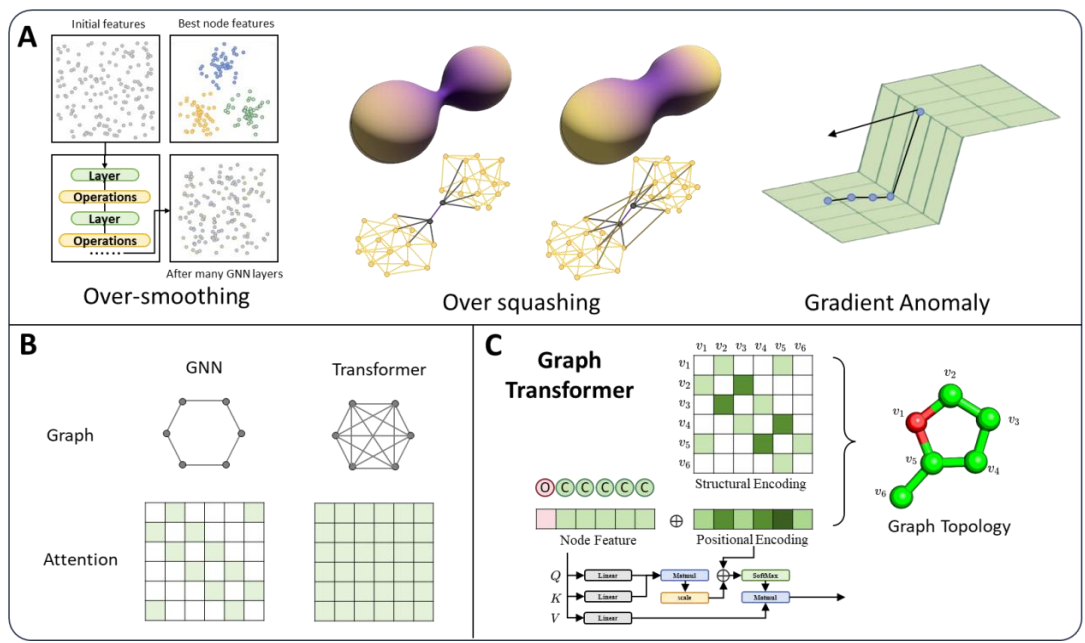

**Figure 4 A).** Challenges encountered when scaling GNN to deeper achitectures. B). The comparison between GNN and Transformer: GNN is a Transformer with sparse attention, wherase Transformer is a GNN with a fully connected graph. C) Illustration of a Graph Transformer architectures, which incorporates graph-structured inductive bias into both the input representation and the attention mechanism.

**Over-smoothing**[83] arises from the smoothing nature of graph convolution, where node representations are updated by aggregating features from their neighbors. As the number of layers increases, this process tends to homogenize the feature space, causing node representations to converge toward similar values. Consequently, discriminability across different nodes decreases, leading to diminished performance in downstream tasks. Li et al.[84] were among the first to formally characterize this phenomenon. They found that the message-passing mechanism of GCNs can be interpreted as a form of Laplacian smoothing[85], which tends to draw nodes of the same class closer in the embedding space. While this clustering can benefit classification tasks with well-separated communities, it also reduces feature diversity and model expressiveness.

**Over-squashing**[74] arises when a node aggregates messages from distant neighbors (multiple hops away), causing the receptive field to expand exponentially (e.g., $5^k$ paths for $k$ hops and 5 neighbors per node). This can compress or even lose critical information from distant but relevant nodes amid redundant messages, particularly when long-range dependencies are crucial for prediction. Recent work by Jake et al.[74] formalized this issue using edge-based combinatorial curvature, demonstrating that edges with negative curvature form bottlenecks in message passing, leading to over-squashing and poor information propagation.

**Gradient anomalies**, including both vanishing[86] and exploding gradients[87], further complicate deep GNN training. As network depth increases, the repeated application of linear transformations and non-linear activations during backpropagation causes gradients to either diminish or amplify uncontrollably. This prevents effective parameter updates and may lead to numerical instabilities (e.g., NaN outputs). In GNNs, these issues often emerge in the MLP layers responsible for feature transformation, especially when paired with insufficient normalization or residual connectivity.

It is worth noting that these three problems do not occur in isolation. Instead, they often interact in complex ways. Recent work[73] has proposed decomposing GNNs into two conceptual layers: the embedding propagation (EP) layer, which handles message aggregation across the graph, and the embedding transformation (ET) layer, which updates node representations. In their point of view, over-smoothing and over-squashing primarily affect the EP layer, while gradient anomalies are more likely to arise in the ET layer. Addressing these problems requires architectural innovations that balance depth, expressiveness, and stability.

### 2.3.1.2 Possible Solution

Significant research efforts have led to the development of various techniques to mitigate the key limitations of deep graph neural networks. These strategies have now been widely integrated into the design of modern GNN architectures. In what follows, we discuss methods organized according to the specific challenges they address.

#### 2.3.1.2.1 Mitigating Over-smoothing

**Topological Perturbations**: A common approach of mitigating over-sommothing is to perturb the graph's topology by modifying its node or edge structure, such as through random deletions. **Node perturbation** introduces randomness into local subgraphs, mimicking noise in real-world graphs. **DropGNN**[66] was among the first to propose random node dropout as a regularization method. **DOTIN**[67] later combined this idea with attention mechanisms, merging low-attention (i.e., less important) nodes into a single virtual node. This selectively preserves salient nodes while reducing smoothness in local representations. **Edge perturbation** focuses more directly on the structural inductive bias. **DropEdge**[68] randomly removes a subset of edges during each training epoch. **DropDEdge**[88] improves this by introducing signal-to-noise ratios and node-wise feature gain metrics to drop task-irrelevant or even harmful edges. **SoftEdge**[69] proposes a softer alternative by assigning each edge a random value between 0 and 1 in the adjacency matrix, resulting in a smoothed rather than binary connection structure.

**Anisotropic Message Passing**: Traditional GNNs typically aggregate neighbor messages using simple isotropic operations (sum or mean), treating all neighbors equally[89]. To enhance distinction between node features and prevent over-smoothing, anisotropic mechanisms are introduced. A prototypical example is **GAT**, which uses learned attention coefficients to weight neighbor contributions adaptively. In molecular graphs, incorporating geometric features such as distances and angles, as in **SchNet** and related models, also promotes anisotropy, enabling richer geometric constraints during message passing.

**Residual Connections**: Originally developed to mitigate gradient problems in CNNs, residual connections also alleviate over-smoothing and over-squashing in GNNs by modifying the receptive field. **DeepGCN**[70] adopts residual and dense connections inspired by ResNet[79] and DenseNet[90], enabling the stacking of deeper graph layers. **GCNII**[71] introduces initial residual connections and identity mapping, injecting original node features $H^0$ into each layer to preserve feature diversity. **DeeperGCN**[91] further extends this design with skip connections and batch normalization, achieving stable training for up to 100 layers. Other works[92] confirm the effectiveness of residual connections in preventing performance degradation in deep GNNs. Another work, **JKNet**[72], which concatenates the outputs from all layers into the final representation, allows the model to integrate features from multiple receptive field sizes. **DAGNN**[73] further introduces gating mechanisms to adaptively select important features across different layers. Both JKNet and DAGNN can be interpreted as adaptive multi-scale fusion strategies, which reduce the sensitivity of performance to model depth.

#### 2.3.1.2.2 Mitigating Over-squashing

**Graph Rewiring**: Rewiring involves targeted modifications to the graph structure to reduce bottlenecks and facilitate long-range message propagation. **FA**[93] introduces a fully connected adjacency matrix at the final layer, directly connecting distant nodes to mitigate over-squashing. **SDRF**[74] employs a Ricci curvature-based definition to identify negatively curved edges that are associated with structural bottlenecks, and adds supportive edges in their vicinity to improve information flow. **CurvDrop**[75] proposes randomly dropping negatively curved edges at each layer, effectively weakening bottlenecks without drastically altering the overall structure. **CBED**[76] follows a similar curvature-based strategy but with an alternative curvature metric. **Drew**[77] introduces a dynamic rewiring scheme that incrementally densifies the graph over layers, avoiding abrupt changes that may disrupt structural priors. **CurvPool**[78] clusters nodes based on curvature before applying pooling operations, simultaneously addressing over-smoothing and over-squashing.

Intuitively, over-squashing arises from long distances between nodes. By reducing shortest-path distances, rewiring naturally mitigates the problem. In this light, transformer-based models inherently avoid over-squashing by treating the graph as fully connected. These models connect all node pairs directly, ensuring a complete receptive field by design. We will explore such models in the next section.

#### 2.3.1.2.3 Mitigating Gradient Anomalies

Gradient explosion and vanishing are general problems in deep neural networks. In addition to residual connections, several techniques have proven effective in mitigating gradient pathologies in GNNs. Firstly, activation functions such as **ReLU**[80], **Leaky ReLU**[94], and **ELU**[95] ensure non-zero gradients and healthy gradient flow. Secondly, gradient clipping is often employed to constrain gradients within reasonable bounds. Thirdly, batch normalization[81] normalizes feature distributions within mini-batches, stabilizing training and preserving gradient magnitude. Lastly, weight initialization schemes such as **He**[82] and **Xavier**[82] are also critical for avoiding degenerate gradient behavior during the early training stage. Recent work[96] points out that gradient vanishing may occur even earlier than over-smoothing in deep GNNs. Consequently, the role of residual connections may be more central in addressing gradient instability than previously appreciated.

### 2.3.2 Transformer-like GNNs: Robust Large GNN models

According to the neural scaling law[97], the performance of deep learning models improves with increased data and model size. GNNs face similar demands when applied to large-scale tasks, such as pretraining on billion-scale molecular libraries (e.g., ZINC[98]), integrating multi-omics data in multimodal models, or training structure prediction models across the entire Protein Data Bank (PDB)[5] database, as seen in AlphaFold. Many downstream applications require larger and deeper models. In NLP, the success of large-scale transformer-based models such as GPT[99] (110M), BERT[100] (340M), GPT-3[60] (175B), and DeepSeed[101] (1T) has established the Transformer architecture[14] as the de facto foundation for large models. This has motivated research into building GNNs with Transformer architectures, aiming to replicate the transformative effects seen in NLP. In the following section, we would introduce an important perspective: a Transformer can be viewed as a fully connected graph, while a GNN can be interpreted as a Transformer with sparse (neighbor-only) attention. Based on this intuition, we would survey representative methods that inject graph inductive bias into Transformer-based architectures.

#### 2.3.2.1 GNNs vs. Transformers: A Structural Perspective

In MPNNs, a node aggregates information from its neighbors using the following two-stage mechanism:

$$m_{uv}^{(t+1)} = \text{Msg}(h_u^{(t)}, h_v^{(t)}, e_{uv})$$
$$m_v^{(t+1)} = \text{Agg}(\{m_{uv}^{(t+1)}: u \in \mathcal{N}(v)\})$$

where node $v$'s representation is updated by aggregating messages computed from its neighbors $u$, where $e_{uv}$ denotes edge features. This inherently induces a structural bias—only **local neighbors** affect a node's update.

In contrast, a standard Transformer operates over a fully connected graph defined over a sequence of tokens. It computes self-attention between every pair of nodes, regardless of their original adjacency:

$$Q_i = W_Q h_i^t, \qquad K_i = W_K h_i^t, \qquad V_i = W_V H_I^T$$

$$\alpha_{ij} = \text{softmax}\left(\frac{Q_i^T K_j}{\sqrt{d}}\right) = \frac{e^{\frac{Q_i^T K_j}{\sqrt{d}}}}{\sum_k e^{\frac{Q_i^T K_j}{\sqrt{d}}}}$$

$$H_i = \sum_j \alpha_{ij} V_j$$

where $Q_i, K_i, V_i$ are query, key, and value; $\alpha_{ij}$ are normalized attention weights; and $H_i$ are updated features. From the GNN perspective, the Transformer can be viewed as performing message passing on a complete graph, where all nodes attend to all others. Conversely, a GNN can be viewed as performing attention only on edges defined by the graph structure. The comparison between these arguments are illustrated in **Figure 4B**.

This duality has led to the rise of Graph Transformers, which combine the global receptive field of Transformers with structural priors from graphs. A typical formulation enhances the standard attention score with additional graph-specific bias terms:

$$A_{ij} = \frac{Q_i^T K_j}{\sqrt{d}} + b_{\phi(i,j)} + e_{ij}$$

where the bias terms $b_{\phi(i,j)}$ and $e_{ij}$ inject structural priors (such as adjacency, distance, or edge type) into the full attention map. This allows Graph Transformers to preserve the expressive power of global attention while emphasizing the inductive bias of graph topology. A popular way of integrating graph topology is illustrated in **Figure 4C**. In the following subsections, we review methods for incorporating positional and structural encodings into Transformer-based graph models.

### 2.3.2.2 Graph Transformer Models

#### 2.3.2.2.1 Graph Linearization and Structural Embedding

In NLP, input data naturally assumes a sequential form. However, Transformer architectures are inherently permutation-invariant, requiring additional position encodings to capture word order. Early approaches adopted absolute positional encodings based on sinusoidal functions[102], followed by relative encodings[103] that depend on pairwise token distances. In the case of graphs, directly serializing the node sequence (e.g., $[v_1, v_2, ..., v_n]$) destroys the structural inductive bias, as different node orderings lead to varied representations. To address this, **GraphBERT**[104] proposed injecting structural information into the serialized input using three encoding strategies: Weisfeiler–Lehman (WL) absolute encodings, intimacy-based encodings, and hop-based encodings. These strategies aim to preserve graph topology during sequence modeling. **Table 4** summarizes different Transformer-like architectures and their performance on the ZINC Test set.

**Table 4**. Overview of Transformer-like GNN models, ZINC Test MAE means the mean absolute error on the graph regression task of the ZINC Test set.

| Models | ZINC Test MAE | Keywords |
| --- | --- | --- |
| GAT[28] | 0.384±0.007 | Baseline that introduces the attention mechanism |
| GraphBert[104] | * | Fixed-siz e subgraph sampling |
| GraphTransformer[105] | 0.226±0.014 | Laplacian eigenvalues encoding |
| GraphTrans[106] | * | Virtual Node <cls> representing global feature |
| Graphormer[107] | 0.122±0.006 | Spatial-based distance positional encoding |
| GraphiT[108] | 0.211±0.010 | Graph kernel-based positional encoding |
| SAN[109] | 0.139±0.006 | Learned positional encoding |
| SAT[110] | 0.094±0.008 | GNN as pre-extractor for subgraph information |
| EGT[111] | 0.108±0.009 | Edge-enhanced |
| GRPE[112] | 0.094±0.002 | Topological and edge-awarded attention |
| GraphGPS[113] | 0.070±0.006 | General framework |
| Specformer[114] | 0.066±0.003 | Learned arbitrary filter bases |

**2.3.2.2.2 Laplacian-Dependent Graph Transformers**

**Laplacian Encodings and Spectral Bias**: The Laplacian matrix of a graph is defined as:

$$L = D - A$$

where $D$ is the degree matrix and $A$ is the adjacency matrix. In spectral graph convolution, the Laplacian serves as the foundation for defining graph Fourier transforms and designing convolution kernels. Thus, it inherently captures the complete structural inductive bias of the graph. **DGN**[115] demonstrates that Laplacian eigenvectors can distinguish graphs that 1-WL tests fail to differentiate. Therefore, encoding nodes using these eigenvectors can endow Graph Transformers with strong isomorphism recognition capabilities. **GraphTransformer**[105] adopts this idea by performing eigendecomposition on the Laplacian, using the top-k eigenvectors (sorted by eigenvalue magnitude) as node position embeddings. Specifically:

$$v_i^0 = C^0 v_i + c^0,$$
$$\hat{h}_i^o = A^0 h_i + a^0,$$
$$h_i^0 = \hat{h}_i^o + v_i^0,$$

where $h_i$ is the original node feature, $v_i$ is the position vector formed by Laplacian eigencomponents, and $A^0, a^0, C^0, c^0$ are learnable parameters. This integration embeds spectral position information into the Transformer input. Notably, GraphTransformer was among the first Graph Transformer models tested on molecular data.

**GraphiT: Kernel-Based Laplacian Encoding.** While GraphTransformer uses Laplacian eigenvectors for absolute position encoding, **GraphiT**[108] argues for a more expressive design by incorporating relative position information into attention. The model constructs a graph kernel $K_r$ from the Laplacian spectrum:

$$L = \sum_i \lambda_i u_i u_i^T, \qquad K_r = \sum_i r(\lambda_i) u_i u_i^T$$

Two choices of kernel functions $r(\lambda_i)$ yield well-known graph kernels:

**Diffusion kernel**: $K_D = \sum_{i=1}^{m} e^{-\beta \lambda_i} u_i u_i^T = e^{-\beta L} = \lim_{p \to +\infty} (I - \frac{\beta}{p} L)^p$

**p-step random walk kernel**: $K_{pRW} = (I - \gamma L)^p$

These kernels quantify pairwise node similarity and are injected into the attention computation:

$$\text{PosAttention}(Q, V, K_r) = \text{normalize}\left(\exp\left(\frac{QQ^T}{\sqrt{d_{out}}}\right) \odot K_r\right) V$$

**GraphiT** essentially introduces a structure-aware attention mechanism, where diffusion kernels with small $\beta$ promote localized attention.

**SAT: GNN-Augmented Structural Encoding.** SAT[110] (Structure-Aware Transformer) builds upon GraphiT by pointing out limitations in existing encodings, such as the inability of shortest-path encodings to distinguish certain graph motifs. Instead of relying solely on Laplacian kernels, SAT incorporates stronger structural embeddings derived from GNNs, i.e., obtaining node features encoding subgraph information $\phi_G(S(x_i))$ via k-subtree or k-subgraph GNN extractors. Here, the extractors compute subgraph-aware features from local neighborhoods up to depth $k$, enriching each node's representation. Empirical results on ZINC suggest that SAT, as a hybrid of GraphiT and GNN, achieves enhanced performance.

**SAN: Learnable Laplacian Encoding.** SAN[109] (Structure-Aware Network) proposes a learnable encoding scheme based on Laplacian decomposition. Instead of directly using the top eigenvectors, SAN feeds their combination into a Transformer encoder to learn node-wise positional encodings:

$$h_{lpe} = \sum_i^m w_i(\lambda_i, u_i) u_i$$

This allows the network to approximate node topologies using a Fourier basis. Moreover, SAN distinguishes between original graph edges and attention-induced edges, and controls their contributions via learnable parameters. Experiments on ZINC show that SAN outperforms previous GraphiT and SAT models[109].

**Specformer: Spectral Learning of Convolution Kernels.** Specformer[114] departs from prior work by placing attention not over node pairs, but directly over the spectral convolution kernels. The model aims to learn convolution filters in spectral space, motivated by the observation that traditional Transformers act as low-pass filters[116], limiting their capacity for localized representations.

**Spectral Embedding**: Eigenvalues $\lambda_i$ are embedded using sinusoidal functions:

$$\rho(\lambda, 2i) = \sin(\epsilon\lambda/10000^{2i/d})$$

$$\rho(\lambda, 2i+1) = \cos(\epsilon\lambda/10000^{2i/d})$$

**Transformer Encoding**: Let $Z = [\lambda_1 \| \rho(\lambda_1), \ldots, \lambda_n \| \rho(\lambda_n)]^T \in \mathbb{R}^{n \times (d+1)}$, then

$$\tilde{Z} = \text{MHA}(\text{LN}(Z)) + Z, \quad \hat{Z} = \text{FFN}(\text{LN}(\tilde{Z})) + \tilde{Z}$$

**Filter Construction**: Using another attention layer, learn $M$ convolution filters:

$$Z_m = \text{Attention}(QW_m^Q, KW_m^K, VW_m^V), \quad \lambda_m = \phi(Z_m W_\lambda)$$

$$S_m = U\text{diag}(\lambda_m)U^T, \quad \hat{S} = FFN([I_n \| S_i \| \ldots \| S_M])$$

**Convolution**:

$$x'_{ij} = \text{MLP}(x_i + e_{ij}), \quad h_i = \sum_{j \in \mathcal{N}(i)} x'_{ij} \hat{S}, \quad h'_i = h_i + h_i$$

Specformer enables dynamic adaptation of spectral filters, showing superior performance on both node-level and graph-level tasks. On ZINC, it achieves a notable test MAE of 0.066, suggesting that learning from Laplacian spectrum can significantly enhance transformer-based graph models[114].

**2.3.2.2.3 Laplacian-Free Graph Transformers**

While most of the previously discussed Graph Transformer models rely on Laplacian matrices to embed graph topology into Transformer architectures via spectral decomposition, computing such decompositions incurs a theoretical complexity of $O(n^3)$. Though the sparsity and symmetry of Laplacian matrices may alleviate this to some extent, the computational cost remains non-negligible for large-scale graphs. Therefore, recent work has sought computationally lighter alternatives for position and structure encoding that do not require Laplacian eigendecomposition.

**Graphormer: Fast Positional and Structural Encoding.** Graphormer[107] is a representative model in this category, specifically designed for molecular graph tasks. It introduces novel encoding mechanisms that are computationally efficient while preserving graph structure. Notably, it incorporates centrality encoding, spatial encoding, and edge encoding into both the input representation and attention mechanism.

**Centrality Encoding**: Graphormer augments each node's input features with learnable embeddings based on its in-degree and out-degree:

$$h_i^0 = x_i + z_{\deg^-(v_i)}^- + z_{\deg^+(v_i)}^+$$

where $x_i$ is the original node feature, and $z_{\deg^-(v_i)}^-$ and $z_{\deg^+(v_i)}^+$ are trainable embeddings for the in- and out-degrees of $v_i$, respectively.

**Spatial Encoding**: To reflect the intuition that closer nodes should exert stronger influence, Graphormer replaces the standard attention score computation with a topology-aware version:

$$A_{ij} = \frac{Q_i^T K_j}{\sqrt{d}}$$

$$A_{ij} = \frac{(h_i W_Q)(h_j W_K)^T}{\sqrt{d}} + b_{\phi(v_i, v_j)}$$

where $b_{\phi(v_i, v_j)}$ is a learnable bias based on the shortest path length between nodes $i$ and $j$.

**Edge Encoding** Graphormer also incorporates edge features, which are particularly important in molecular graphs (e.g., bond types). For each pair of nodes $ij$, the edge encoding $c_{ij}$ is defined as the weighted average of edge features along their shortest path:

$$c_{ij} = \frac{1}{N} \sum_{n=1}^{N} x_{e_n}^T w_n^E$$

where $x_{e_n}$ is the feature of the n-th edge along the path and $w_n^E$ is a trainable weight. Intuitively, $w_n^E$ is designed to decay with increasing $n$, reducing the influence of distant edges.

The final attention score becomes:

$$A_{ij} = \frac{Q_i^T K_j}{\sqrt{d}} + b_{\phi(i,j)} + c_{ij}$$

Through these efficient yet expressive encodings, Graphormer was the first to show that Transformer-like GNN can outperform traditional GNNs on graph-level tasks. For example, on the ZINC molecular dataset, Graphormer significantly outperforms GraphiT (0.122 vs. 0.211 in MAE) [107].

**EGT**[111] (Edge-augmented Graph Transformer) builds on Graphormer by updating edge features alongside node features during message passing. It also introduces degree-based gating mechanisms into attention to enhance structural awareness. On the ZINC LogP regression task, EGT improves over Graphormer (0.108 vs. 0.122). Then, GRPE (Graph Relative Positional Encoding) [112] further enhances Graphormer's attention mechanism by incorporating more fine-grained control over both spatial and edge encodings. Specifically, the attention scores are computed as:

$$a_{ij}^{\text{topology}} = q_i \cdot \mathcal{P}_{\psi(i,j)}^{\text{query}} + k_j \cdot \mathcal{P}_{\psi(i,j)}^{\text{key}}$$

$$a_{ij}^{\text{edge}} = q_i \cdot \mathcal{E}_{e_{ij}}^{\text{query}} + k_j \cdot \mathcal{E}_{e_{ij}}^{\text{key}}$$

$$a_{ij} = \frac{q_i \cdot k_j + a_{ij}^{\text{topology}} + a_{ij}^{\text{edge}}}{\sqrt{d_z}}$$

Here $\mathcal{P}_{\psi(i,j)}^{\text{query}}$ and $\mathcal{P}_{\psi(i,j)}^{\text{key}}$ denote position-dependent projection matrices based on a learned relative positional encoding function $\psi(i,j)$; $\mathcal{E}_{e_{ij}}^{\text{query}}$ and $\mathcal{E}_{e_{ij}}^{\text{key}}$ correspond to edge-specific projections conditioned on the edge type or attributes $e_{ij}$; $d_z$ is the dimension of the latent feature space, used for normalization. This fine-grained decomposition allows GRPE to better model structural dependencies, achieving further improvements on ZINC (MAE 0.094 vs. 0.108). **GraphTrans**[106] notices that tasks in ZINC are typically graph-level, where simple aggregation of atomic features may obscure important substructural patterns. Therefore, GraphTrans reinterprets the Transformer as a **global readout function**, akin to sequence classification in NLP. Inspired by the use of special <CLS> tokens in models like BERT, GraphTrans introduces a virtual node that aggregates global information from all graph nodes:

$$h_{cls} = h_{cls} + \sum_{v \in V} \alpha_{ij} h_v$$

The final graph-level prediction is obtained via:

$$y = \text{softmax}(W^{out} h_{cls})$$

This formulation allows GraphTrans to capture long-range dependencies and global graph semantics, leading to better performance in tasks where holistic understanding is crucial, such as molecular graph prediction or classification.

## 2.3.2.3 A Unified Framework for Graph Transformers

In the rapidly evolving landscape of Graph Transformer models, a recurring observation is that while the core mechanism for attention computation remains largely unchanged, most innovations lie in how graph inductive biases are incorporated into the architecture. Various methods have explored different strategies to inject structural information into the model. As discussed previously, **GraphTransformer** encodes node positions using the spectral features of the Laplacian matrix; **GraphiT** leverages Laplacian-based kernel methods to define relative positional relationships between nodes; **SAN** employs a Transformer encoder to enhance the expressivity of Laplacian spectral components, yielding learnable position encodings; **SAT** introduces a GNN module as a preprocessor to extract local structural information; and **Specformer** utilizes a Transformer to transform Laplacian decompositions into adaptive convolution kernels. On the other hand, **Graphormer**, **EGT**, and **GRPE** avoid the need for spectral decomposition entirely by relying on node degrees and shortest path distances to define positional and structural relations directly.

These diverse strategies suggest that it is possible and beneficial to systematize the construction of Graph Transformers into a unified, modular framework. **GraphGPS**[113] provides such a formulation, organizing graph-specific inductive biases **into two key components: positional encoding (PE) and structural encoding (SE).** Together with local and global update modules, this leads to a modular architecture that flexibly integrates both GNN-style locality and Transformer-style globality.

PE aims to represent where a node lies within the overall graph structure. Nodes that are spatially or topologically close should have similar positional embeddings. Depending on the implementation, PE methods vary in computational complexity: some approaches operate with $O(N^2)$ cost, such as those relying on pairwise distance matrices or attention scores across all node pairs [107, 117]. Others achieve linear complexity, either $O(N)$ or $O(E)$, by encoding node- or edge-level features, such as Laplacian eigenvectors or spectral gradients[109, 118].

SE is intended to inject higher-level graph topology. Nodes sharing similar subgraph structures, such as common motifs or patterns, are expected to have similar structural embeddings. One classic approach is to predefine a vocabulary of substructures (e.g., chemical fragments in molecular graphs) and then use one-hot encoding to assign structure-aware tokens to nodes within those subgraphs[119]. This method, however, requires strong domain knowledge and pre-curated fragment libraries. A more general and learnable approach is to derive structural signals directly from graph properties, such as the diagonal of the m-step random walk matrix node degree, or Ricci curvature[120]. These features encode local cyclicity, centrality, and geometric bottlenecks, respectively.

Building on this dual-channel encoding of graph inductive bias, GraphGPS introduces a hybrid architecture that combines a local MPNN module with a global Transformer module. While many Graph

Transformer models dispense with explicit message passing, GraphGPS preserves this element, motivated by findings from SAT that local GNN layers can enhance structural awareness, particularly in sparse graph regimes. Within the GPS framework, the MPNN block performs local aggregation over graph neighborhoods, whereas the Transformer module operates over a fully connected attention graph, capturing long-range dependencies and global structure. The resulting GraphGPS is as follows:

$$X^0, E^0 = \text{PE/SE}(X, E)$$

$$X_M^{t+1}, E^{t+1} = \text{MPNN}^t(X^t, E^t, A)$$

$$X_T^{t+1} = \text{GlobalAttention}^t(X^t)$$

$$X^{t+1} = \text{MLP}(X_M^{t+1} + X_T^{t+1})$$

Here, $X^0$ and $E^0$ are the initial node and edge features, enriched by positional and structural encodings. The updated features $X^{t+1}$ are derived by fusing the outputs from the local MPNN $X_M^{t+1}$ and global Transformer module $X_T^{t+1}$, followed by an MLP layer to produce the next-layer node representation. Importantly, to avoid the common issue of over-smoothing in deep GNNs, GraphGPS does not stack multiple MPNN layers. Instead, it alternates between single-layer MPNNs and Transformer layers, effectively balancing local and global feature extraction while mitigating over-smoothing and over-squashing effects. Thanks to this well-structured integration of encoding schemes and attention mechanisms, GraphGPS stands out as one of the first Graph Transformer models to achieve linear complexity $O(N + E)$ without sacrificing expressivity. It offers a flexible and scalable blueprint for constructing future Graph Transformers that can adapt across a wide range of graph-based learning tasks.

### 2.3.3 Pre-train Strategies: Utilizing Unlabeled Molecules

Unlike traditional fields such as CV and NLP, molecular representation learning faces severe challenges due to data sparsity and label noise. For instance, widely used datasets like PDBBind[121], which aim to model protein-ligand binding affinity, contain only tens of thousands of data points. Furthermore, key biological activity indicators such as $IC_{50}$ often suffer from batch effects introduced by differing experimental conditions and protocols across laboratories. For the same compound, the reported $IC_{50}$ can sometimes vary by orders of magnitude[122]. In such high-noise, low-resource regimes, supervised models are prone to overfitting and often fail to generalize out of the training data. Self-supervised learning offers a promising solution for mitigation. Instead of relying on human-annotated labels, self-supervised strategies aim to learn inherent properties of the data, such as structural regularities, topological similarities between graphs, or the importance of local substructures. These methods can be applied to large unlabeled molecular corpora, effectively providing the "fuel" for large models. Notably, even shallow models benefit from self-supervised pretraining: by initializing parameters closer to a favorable local minimum, they converge faster and generalize better[123]. In this section, we would explore generative pretraining for molecular graphs based on categorizing them into generative, predictive, and contrastive approaches, as illustrated in **Figure 5A**. The related methods are summarized in **Table 5**.

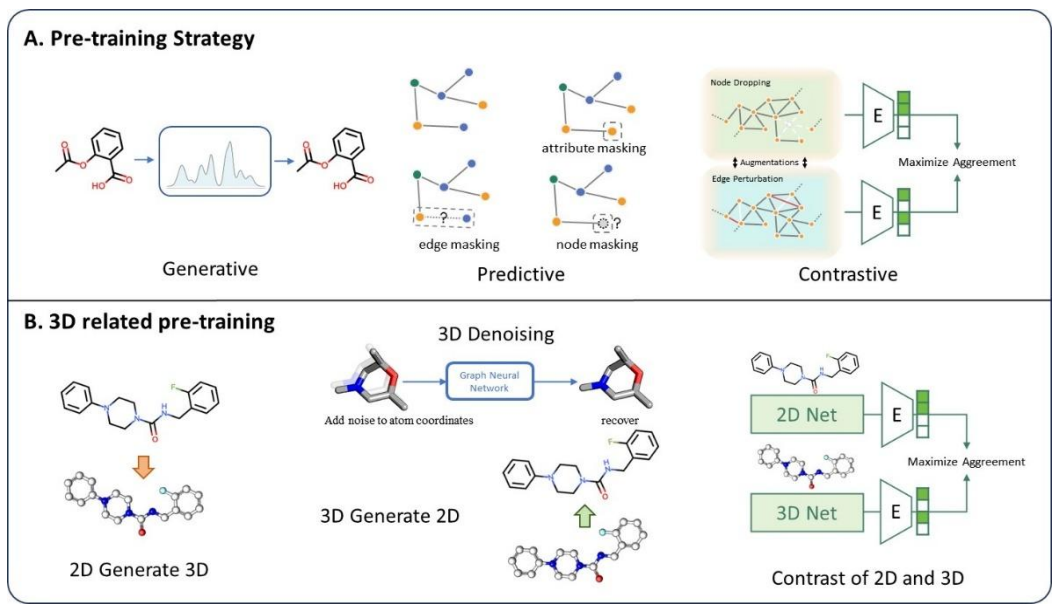

**Figure 5** A). Molecular pre-training strategies. B). 3D-related molecular pre-training strategies.

**Table 5**. Overview of molecular graph self-supervised learning strategies.

| Strategy | Method | Keywords |
|---|---|---|
| Generative | GPT-GNN[124] | Attribute and edge generation |
| | GraphMVP[125] | 2D->3D generation |
| | Unified 2D&3D[126] | 2D->3D/3D->2D generation |
| | FragVAE[127] | Fragment-based AE |
| | MISU[128] | Atom-, fragment-, and graph-level AE |
| Predictive | EdgePred[26] | Node and edge masking |
| | AttrMask[129] | Node and edge masking |
| | GROVER[130] | Graph Transformer, fragment existence prediction |
| | MGSSL[131] | Motif-based recovery |
| | Mole-BERT[132] | VQ-VAE tokenizer for environment-aware atoms |
| | MolCAP[133] | Chemical reactivity pretraining |
| | Noisy Nodes[134] | Noise adding and denoise of coordinates |
| | GeoSSL[135] | Transformation from one conformer to another |
| | Denoising[136] | Noise adding and denoise of coordinates |
| | Frad[137] | Adding torsional noise and denoise |
| Contrastive | GraphCL[138] | Four graph augmentation on topology change |
| | InfoGraph[139] | Treat subgraph as the augumentation |
| | GeomGCL[140] | Contrast 3D conformation and 2D graph |
| | 3D Infomax[141] | Contrast multi 3D conformation and 2D graph |
| | MICROGraph[142] | Treat fragments as the augumentation |
| | JOAO[143] | Automatic augmentation strategy selection |
| | MPG[144] | Classify half graph resources |
| | COATI[145] | Contrast 3D graph and SMILES |
| | SMICLR[146] | Contrast SMILES and graph |
| | CGIP[147] | Contrast 2D image and graph |
| | MoleculeJAE[148] | Treat diffusion trajectory as the augmentation |

### 2.3.3.1 Generative Pre-training

The central idea of generative pre-training is to model the underlying distribution $p(G)$ of real-world molecular graphs. Inspired by the notion that "what I cannot create, I do not understand," generative methods encourage models to construct meaningful representations by reproducing the structural statistics of molecular graphs. Rather than relying on external supervision, these approaches facilitate structural understanding through the process of self-construction and reconstruction, effectively embedding the statistical and geometric priors of molecular systems into the learned representation space.

#### 2.3.3.1.1 Autoregressive Generative Pre-training

Inspired by autoregressive language models such as GPT, **GPT-GNN**[124] extends the autoregressive paradigm to graph-structured data. Specifically, it formulates the probability of a graph $G$ as a joint distribution over node and edge attributes, denoted as $p(G) = p(X, E)$, where $X$ and $E$ represent node and edge features, repectively. To accommodate the inherently unordered nature of graphs, GPT-GNN first defines a total ordering $\pi$ over the nodes, effectively serializing the graph into a sequence. Under this framework, the model optimizes the expected likelihood over all possible permutations:

$$p(G; \theta) = \mathbb{E}_\pi[p_\theta(X_\pi, E_\pi)]$$

where $\theta$ denotes the learnable parameters, and $X_\pi, E_\pi$ are the node and edge features under permutation $\pi$. Assuming a uniform prior over permutations, the model factorizes the joint distribution autoregressively:

$$\log p_\theta(X, E) = \sum_{i=1}^{|\mathcal{V}|} \log p_\theta(X_i, E_i \mid X_{<i}, E_{<i})$$

meaning that at each step $i$, the model conditions on all previously generated nodes $X_{<i}$ and edges $E_{<i}$ to generate the attributes of the current node $X_i$ and its incident edges $E_i$. GPT-GNN is trained by maximizing the overall log-likelihood of the graph generation process, effectively learning to "write" molecular graphs in an autoregressive fashion. In doing so, it leverages the structural analogies between sequences and graphs to transfer the strengths of GPT-style pre-training into the molecular domain.

#### 2.3.3.1.2 Autoencoding-based Generative Pre-training

In addition to autoregressive modeling, auto-encoding approaches aim to learn compact and meaningful latent representations of molecular structures. The encoder maps the input graph to a low-dimensional latent space, while the decoder reconstructs the original graph from this latent code. The learned embedding, often referred to as the latent space, can then be used for downstream tasks such as regression or classification, or sampling to generate novel molecular graphs. In this category, **MISU**[128] proposes a multi-scale auto-encoding framework for molecular graphs, integrating supervision at the atomic, fragment, and molecule levels. Specifically, atomic-level structure is learned using a variational graph autoencoder (VGAE), fragment-level representation is captured using a junction tree VAE

(JTVAE[149]), and global graph features are distilled using fingerprint-based decoding. The encoder is then reused for downstream prediction tasks. Ablation experiments show that each level of supervision contributes to the model's performance, highlighting the effectiveness of hierarchical representation learning. Similarly, **FragVAE**[127] employs a fragment-based autoencoder and demonstrates strong performance in low-data virtual screening scenarios.

A key distinction between molecular graphs and generic graph data lies in their inherent 3D nature: real-world molecules are not only defined by their 2D topological structures but also by their physically meaningful 3D conformations. This motivates a more comprehensive generative formulation based on the joint distribution $p(G, R)$, where $G$ denotes the molecular graph and $R$ its conformation. To bridge this 2D–3D gap, recent works have introduced conformation-aware generative pre-training. For instance, GraphMVP[125] models the conditional distribution $p(R|G)$, encouraging the model to generate plausible 3D conformations conditioned on 2D molecular graphs. Conversely, Unified 2D&3D[126] explores the reverse process by modeling $p(G|R)$, reconstructing the molecular graph from atomic positions even in the absence of explicit atom-type labels. These approaches enrich learned representations with geometric priors and significantly enhance performance on both 2D and 3D molecular property prediction tasks.

### 2.3.3.2 Predictive Pre-training

While generative pre-training focuses on modeling the data distribution $p(G)$ to implicitly learn the underlying structure of molecular graphs, predictive pre-training seeks to define reliable, low-cost supervisory signals derived directly from the data. A straightforward approach involves masking and reconstructing either node attributes or edge features. Early methods, such as **EdgePred**[26] and **AttrMask**[129], adopt this strategy, where the model learns to recover masked atomic types or chemical bonds. However, these atom-level tasks often fail to produce transferable representations for graph-level downstream tasks and may even degrade performance, a phenomenon known as negative transfer[150]. To address this limitation, **MGSSL**[131] introduces a motif-based self-supervised learning approach, motivated by the intuition that fragments are more chemically meaningful for molecular property prediction. This assumption aligns with traditional fragment-based models for predicting properties such as solubility[151], which assume that molecular attributes are additive over constituent substructures. Accordingly, MGSSL treats molecular fragments as nodes in a coarse-grained graph and formulates pre-training objectives to recover both fragment identity and their topological relationships.

Complementing this line of reasoning, **Mole-BERT**[132] offers a different hypothesis for the observed limitations of AttrMask: the imbalance of the atomic vocabulary. As carbon atoms dominate most molecular graphs, a model trained to predict masked atoms may achieve high accuracy by simply favoring carbon, resulting in poor generalization. To mitigate this bias, Mole-BERT employs vector quantized variational autoencoding[152] (VQ-VAE) to construct a context-aware and enlarged atom vocabulary. This vocabulary reflects subtle variations in chemical environments, effectively partitioning a dominant atom type (e.g., carbon) into multiple subtypes and thus enriching the model's semantic capacity. Another prominent contribution is **GROVER**[130], which pioneers the applications of large-scale Graph Transformer architectures to molecular pre-training. Beyond traditional node and edge reconstruction, GROVER proposes a graph-level multi-label classification task to predict the presence of specific substructures within a molecule. Notably, ablation studies demonstrate that replacing the Graph Transformer with a shallow GNN leads to a noticeable drop in performance, highlighting the advantage of using large GNNs in pre-training settings.

Beyond leveraging topological information, recent works have proposed incorporating 3D molecular geometry as a source of supervision. Stable molecular conformations, which correspond to local minima on the potential energy surface, inherently encode physically meaningful inductive biases. Inspired by denoising diffusion models, **NoisyNodes**[134] introduces a conformation-aware pre-training paradigm by perturbing atomic coordinates and training the model to reconstruct the noise-free structures, thereby encouraging the model to learn local energy landscapes and embed spatial priors into the learned representations. Building upon this idea, **GeoSSL**[135] focuses on denoising between pairs of low-energy conformers of the same molecule. Instead of directly predicting 3D coordinates, it operates in distance-based latent space, thereby relaxing the constraints on equivariant model design. More recently, **Frad**[137] identified a key challenge in direct coordinate corruption: its strong dependence on noise scale. Small perturbations fail to explore the conformational landscape effectively, while large perturbations risk generating unrealistic molecular geometries. To address this, Frad proposed a two-step noise injection strategy: Gaussian noise is first applied to torsional angles to enable broad exploration of the conformational space, followed by traditional coordinate-level noise.

Beyond denoising objectives, cross-domain tasks can also be framed within the predictive pre-training paradigm. MolCAP[133], for instance, incorporates domain knowledge from chemical reactivity. By predicting bond reactivity and reconstructing product topology from reactants, the model learns to encode transformation patterns rooted in reaction chemistry. Additionally, MolCAP introduces simple reaction templates (e.g., hydration or protonation) to bridge the gap between pre-training objectives and downstream molecular property prediction tasks.

### 2.3.3.3 Contrastive Pre-training

The core principle of contrastive learning is to derive effective representations by training a model to discriminate between similar (positive) and dissimilar (negative) samples. In such a pre-trained embedding space, samples with similar semantics are drawn closer, while dissimilar ones are pushed apart. This structured space facilitates faster convergence in downstream tasks and often leads to improved generalization[153]. In the context of molecular graphs, constructing negative pairs is relatively straightforward because molecules from different structures naturally serve as negatives. Positive pairs, on the other hand, can be generated either manually or through automated data augmentations. Analogous to image-based contrastive learning, where augmentations such as rotation, translation, scaling, and color jitter are commonly applied[154], molecular graphs can undergo graph-specific augmentations to define positive views.

**2.3.3.3.1 Intra-model Contrative Learning**

**GraphCL**[138] was among the first to introduce contrastive learning into graph neural networks. The authors observed that directly using raw molecular graphs as contrastive inputs could lead to uniformly separated representations, which may degrade the quality of pre-trained embeddings. To address this, GraphCL proposed four types of graph augmentations: node dropping, edge perturbation, attribute masking, and subgraph sampling. By augmenting the input graph and maximizing agreement between the original and its augmented views, GraphCL effectively injected inductive biases into the model, promoting the learning of robust, perturbation-invariant features. Notably, empirical evaluations demonstrated that for molecular graphs, node dropping and subgraph sampling yielded particularly significant performance improvements.

It should be noted that contrastive pre-training and supervised fine-tuning may operate in distinct representation spaces. A mismatch between these objectives may result in negative transfer, where pre-training degrades downstream performance. For instance, the effectiveness of graph augmentations in GraphCL varied across datasets, necessitating manual tuning to optimize results. To address this, **JOAO**[143] (Joint Augmentation Optimization) introduces a bi-level optimization framework that dynamically selects the most beneficial augmentation strategies during training. This allows the model to adaptively learn an optimal augmentation policy per dataset, thereby effectively automating the construction of contrastive training pairs.

Inspired by Deep InfoMax[155], which maximizes mutual information between global and local features in images, **InfoGraph**[139] extends this paradigm to molecular graphs. It treats subgraphs as local views and the full graph as a global context, training the model to identify whether a subgraph originates from a given molecule. This naturally constructs positive pairs (subgraphs from same molecule) and negative pairs (subgraphs from different molecules). To reconcile the gap between unsupervised and supervised objectives, InfoGraph uses separate encoders for each task and applies task-specific losses at every layer, preventing representational interference and enhancing mutual information. This architectural decoupling helps achieve both effective contrastive training and improved downstream task performance. A related approach, **MPG**[144] (Molecular Pre-training Graph), introduces a subgraph-level pre-training task termed Pairwise Half-Graph Discrimination (PHD). MPG assumes that two halves from the same molecule are structurally compatible, while halves from different sources are not, thereby framing a discriminative contrastive task. Similarly, **MICROGraph**[142] leverages fragment-level contrastive learning, treating molecular fragments as natural augmentations of the parent molecule.

**2.3.3.3.2 Cross-model Contrastive Learning**

The molecular representation space is inherently multi-modal. Molecules can be described as 2D graphs, 3D conformations, or 1D strings (e.g., SMILES, SMARTS, SELFIES[156]). Each modality emphasizes different chemical characteristics and introduces distinct inductive biases. In this sense, cross-modal contrastive learning leverages one modality as an implicit augmentation of another, thereby broadening the scope of representation learning.

**GeomGCL**[140] was the first to apply contrastive learning between 2D graphs and 3D conformations. Utilizing SphereNet as the encoder, it constructs contrastive pairs by associating a molecule's 2D graph with its 3D conformation as a positive pair and using mismatched molecules as negatives. In contrast, **3D InfoMax**[141] explores intra-molecular conformational diversity by treating multiple low-energy conformers of the same molecule as positive pairs. Additionally, it employs the normalized temperature-scaled cross-entropy loss (NT-Xent[130]) to maximize mutual information between 2D and 3D representations, enabling a 2D GNN to implicitly learn geometric features. Remarkably, 3D InfoMax achieves state-of-the-art performance on geometry-sensitive tasks using only 2D GNNs, thus circumventing the computational cost of generating conformers during inference.

Inspired by vision-language pre-training methods[63], **COATI**[145] aligns SMILES sequences and 3D molecular graphs using contrastive learning. It introduces two input modalities: one processes SMILES directly via a Transformer, while the other uses 3D embeddings from an E(3)-equivariant neural network. The model learns both SMILES generation and cross-modal alignment by maximizing consistency between these two representations. Other works have similarly explored alternative modalities: **SMICLR**[146] performs contrastive learning between SMILES and graph embeddings, while **CGIP**[147] aligns molecular graph representations with 2D structural images.

Finally, **MoleculeJAE**[148] introduces a pseudo-modality derived from diffusion trajectories. These trajectories, generated via molecular diffusion processes, are treated as augmented views of the same molecule. Contrastive pairs are constructed within the same trajectory, while negative samples are drawn from distinct trajectories. By applying contrastive learning over both 2D topologies and 3D geometries along diffusion paths, MoleculeJAE outperforms diffusion-based predictive pre-training models such as GeoSSL. Intuitively, this framework is analogous to Gaussian-blurred augmentations in image-based contrastive learning[154], offering a new way to enrich graph representations by leveraging structured noise and temporal consistency.

## 2.4 Graph Generative Models and Probability Learning

Deep learning tasks are broadly categorized into two categories: predictive tasks, which aim to infer outputs given inputs, and generative tasks, which focus on modeling the underlying data distribution to generate new, plausible samples[157]. Compared with generative tasks in vision or language domains, molecular graph generation presents a unique challenge, that is, the variability in the number of atoms (nodes) and bonds (edges) in a molecule forces generative models to operate within a non-fixed-length decision space, which is a non-trivial modeling problem. There are roughly two ways to deal with this: 1) **Graph Embedding-based Approaches**: These methods encode entire molecular graphs into continuous latent representations and then decode them into graph structures, either in a single step or autoregressively. Such approaches are tightly linked with probabilistic generative modeling and often rely on principles from variational inference or density estimation. 2) **Graph Editing-based Approaches**: These operate directly on graph topology by iteratively modifying local structures (e.g., adding, removing, or replacing nodes and edges), often guided by reinforcement learning, to incrementally generate or optimize molecular structures. Among these, the majority of probabilistic generative methods fall under the graph embedding paradigm.

Drawing from the broader landscape of deep generative models, we can further categorize them into two primary classes: 1) **Likelihood-based Generative Models**: These explicitly model the data distribution and aim to maximize the likelihood of observed samples. New samples are generated by sampling from the learned distribution. A canonical example includes models like PixelRNN and PixelCNN[158], which sequentially model image pixels via conditional likelihood chaining. 2) **Latent Variable-based Generative Models**: These introduce latent variables to capture the data's underlying structure. The model then learns a mapping between the data and latent space, often using frameworks like the Variational Autoencoder (VAE), which jointly trains an encoder-decoder architecture to reconstruct data from latent representations. In the following sections, we would review the representative architectures under these two categories, with a particular focus on their adaptations for molecular graph generation. These methods are summarized in **Table 6** and four popular probability learning architectures are illustrated in **Figure 6**.

Table 6. Overview of probability-based generative models.

| Category | Methods | Keywords |
|---|---|---|
| Likelihood -based | Flow Model[159] | Invertible transformations; exact likelihood estimation |
| | EBM[160] | Unnormalized density, model potential functions |
| | Autoregressive | Sequential factorization; |
| Latent variable -based | VAE[161] | Variational inference |
| | GAN[162] | Adversarial training; Generator-Discriminator |
| | Diffusion[163] | Progressive noise injection and denoising |

| | |
|---|---|
| Diffusion Bridge[164] | Relax Gaussian prior constraints of diffusion |
| Flow Matching[165] | Direct path from noise to data |

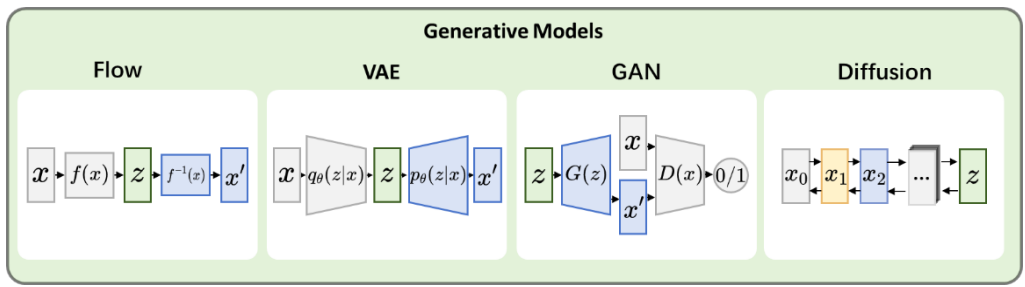

**Figure 6.** Four popular probability learning architectures.

### 2.4.1 Likelihood-based methods: Flow, EBM, AR

These models directly approximate the data distribution and are trained to maximize the likelihood of observed samples. In principle, one could fit a probability distribution to a molecular feature (e.g., interatomic bond length $d_{ij}$ by assuming it follows a known distribution (e.g., Gaussian), and estimating parameters via maximum likelihood:

$$\mathcal{L}(\theta) = \sum_{i,j} \log p(d_{ij}, \theta)$$

However, in practice, molecular data distributions are typically high-dimensional and highly nonlinear, making them intractable to model with simple analytic distributions. Likelihood-based generative models attempt to circumvent this by using deep neural networks to learn expressive approximations of arbitrarily complex distributions, enabling flexible modeling of high-dimensional data.

#### 2.4.1.1 Normalizing Flow Models

At the heart of normalizing flow models lies the idea of transforming a simple base distribution (e.g., a Gaussian) into a complex target distribution through a series of invertible transformations. Denoting $x$ as the molecular data and $z$ the latent variable, the transformation is defined as:

$$x = f^{-1}(z) \Leftrightarrow z = f(x)$$

where $f$ is a bijective (invertible) neural network, and the latent variable $z$ is typically assumed to follow a standard normal distribution:

$$z \sim \mathcal{N}(0,1), \quad q(z) = \frac{1}{(2\pi)^{D/2}} \exp\left(-\frac{1}{2} \parallel z \parallel^2\right)$$

By applying the change-of-variable formula, the likelihood of data $x$ under the model becomes:

$$q(x) = \frac{1}{(2\pi)^{\frac{D}{2}}} \exp\left(-\frac{1}{2} \parallel z \parallel^2\right) \left|\det\left[\frac{\partial f}{\partial x}\right]\right|$$

To ensure tractable training via maximum likelihood, flow models impose two constraints on the transformation $f$: 1) **Invertibility**: Every transformation must have a well-defined inverse to enable bidirectional mapping between latent and data space. 2) **Efficient Jacobian Determinant Calculation**:

The determinant $\det\left[\frac{\partial f}{\partial x}\right]$ must be computationally tractable (ideally scaling linearly in input dimension) to avoid bottlenecks during training and inference.

Different flow models propose various designs to meet these criteria. **NICE**[159] (Nonlinear Independent Components Estimation) is one of the earliest flow models, which employs a simple additive coupling layer. The input is split into two parts, and only one is transformed:

$$y_1 = x_1, \quad y_2 = x_2 + f(x_1)$$

This structure leads to a Jacobian determinant of 1, making likelihood computation trivial. However, it limits expressiveness due to the lack of scale transformations. Another model, **RealNVP**[166] (Real-valued Non-Volume Preserving), extends NICE by incorporating scaling and translation operations:

$$y_1 = x_1, \quad y_2 = x_2 + \exp(s(x_1)) + t(x_1)$$

Here, $s(\cdot)$ and $t(\cdot)$ are neural networks, and the Jacobian determinant becomes:

$$\det\left[\frac{\partial f}{\partial x}\right] = \exp\left(\sum_i s_i(x_i)\right)$$

This significantly improves flexibility while maintaining computational efficiency. **Glow**[167] further enhances RealNVP by introducing 1×1 invertible convolutions, enabling the model to mix information across channels more effectively and thereby increasing its expressive power.

### 2.4.1.2 Energy-based Model

Energy-based models are grounded in the principles of statistical physics, particularly the idea that any complex probability distribution can be approximated using a set of exponential functions. This notion is formalized by the Boltzmann distribution, which describes the probability of a system occupying a given state $x$ in thermal equilibrium:

$$p(x) = \exp\frac{-E(x)}{Z}, \quad Z = \exp\int\left(-E(x)\right)dx$$

where $E(x)$ denotes the energy associated with state $x$, and $Z$ is the partition function, a normalization constant integrating over all possible states. The key insight from this formulation is that by designing a suitable energy function $E(x)$, one can model arbitrarily complex data distributions. This serves as the theoretical foundation for **Energy-Based Models**[160] (EBMs).

In practice, EBMs assign an energy score to each data point $x$ through a neural network parameterization $E(x, \theta)$. Lower energy corresponds to higher likelihood, and model training seeks to minimize the energy of observed data while maximizing the energy of unobserved configurations. The log-likelihood objective becomes:

$$\mathcal{L}(\theta) = \sum_i \log p(x_i) = -\sum_i E(x_i, \theta) - \log Z$$

While conceptually elegant, computing the partition function $Z$ is generally intractable due to the integral over the high-dimensional data space. Consequently, EBMs are typically trained using approximate inference techniques, such as Contrastive Divergence[168] (CD) or Markov Chain Monte Carlo[169] (MCMC) methods. In molecular graph generation, GNNs are usually used to parameterize the energy function, allowing the model to capture structural dependencies within molecular graphs.

**2.4.1.3 Autoregressive Model**

Autoregressive (AR) models form a class of likelihood-based generative models in which the joint probability distribution is factorized into a product of conditional probabilities, enabling sequential sample generation. These models are widely used in NLP, where the probability of a word sequence is decomposed and each word is generated conditioned on all preceding tokens.

Formally, given a data instance $x = (x_1, x_2, ..., x_D)$, an autoregressive model defines its probability as:

$$p(x) = p(x_1)p(x_2|x_1)p(x_3|x_1, x_2) ... p(x_D|x_1, x_2, ..., x_{D-1})$$

This decomposition allows the model to capture complex, high-order dependencies among dimensions in a tractable manner, enabling both likelihood estimation and sample generation. In the context of graph generation, autoregressive models operate by sequentially constructing a molecular graph, including adding atoms, bonds, and substructures step by step. At each iteration, the model conditions on the existing partial graph and decides the next action, which may include introducing a new node (atom), forming a new edge (bond), or terminating the generation process. This paradigm aligns closely with **graph editing** and has the advantage of explicit control over generation steps. However, it may suffer from exposure bias and inefficient parallelization, motivating hybrid models that combine autoregression with latent or flow-based components.

**2.4.2 Latent variable methods: from VAE, GAN to Diffusion**

Latent variable methods aim to capture the underlying structure and dependencies within complex data by introducing latent representations. Unlike likelihood-based models that directly estimate the data distribution, latent models define a mapping between an observed data space and a lower-dimensional latent space. This strategy offers a more tractable way to approximate the data distribution, enabling efficient generation and reconstruction through operations in the latent domain.

One of the central challenges in training latent variable models lies in the intractability of the data likelihood, which often involves integrating over high-dimensional latent variables. To address this, researchers have developed a variety of approximation techniques, including variational inference and adversarial training, which seek to optimize a surrogate objective. In essence, these models trade exact likelihood estimation for a simplified and computationally feasible inference process. In this section, we would review three major classes of latent variable-based generative models that have demonstrated considerable success in molecular graph generation: Variational Autoencoders (VAE), Generative Adversarial Networks (GAN), and Diffusion Models.

#### 2.4.2.1 Variational Autoencoder (VAE)

VAEs[161] are a foundational class of latent generative models that introduce a probabilistic latent variable $z$ to model the underlying data distribution. A VAE consists of two components: an encoder, which maps the input data $x$ into the latent space by estimating a posterior distribution $q_\phi(z|x)$, and a decoder, which reconstructs $x$ from $z$ by modeling $p_\theta(x|z)$. Here, $\phi$ and $\theta$ denote the parameters of the encoder and decoder networks, respectively. The generative process is defined via the marginal likelihood:

$$p_\theta(x) = \int p_\theta(x|z)p(z)\,dz$$

where $p(z)$ is typically chosen as an isotropic Gaussian prior. However, computing this integral exactly is intractable for high-dimensional latent spaces. To overcome this, VAEs adopt variational inference by introducing a tractable approximation $q_\phi(z|x)$ to the true posterior $p_\theta(x|z)$, resulting in an evidence lower bound (ELBO) as the training objective:

$$\log p_\theta(x) \geq \mathbb{E}_{q_\phi(z|x)}[\log p_\theta(x|z)] - \mathrm{KL}\left(q_\phi(z|x) \,\|\, p(z)\right)$$

The first term represents the reconstruction loss, measuring how well the decoded sample $\tilde{x} \sim p_\theta(x|z)$ resembles the original input. The second term is the Kullback–Leibler divergence, which regularizes the approximate posterior to remain close to the prior $p(z)$, promoting structure and continuity in the latent space.

Thanks to this structured Gaussian prior, VAEs are capable of generating diverse and coherent samples. Compared to flow-based models, VAEs typically yield a more compact latent space, facilitating efficient learning and inference over high-dimensional molecular data. However, since the optimization is performed over a lower bound rather than the exact likelihood, VAEs may occasionally sacrifice sample fidelity in favor of tractability.

#### 2.4.2.2 Generative Adversarial Network (GAN)

GANs[162] offer an alternative approach to likelihood-free generative modeling by framing the learning process as a two-player game between a generator and a discriminator. The generator $G$ learns to map latent variables $z \sim \mathcal{N}(0,1)$ to data samples that resemble those in the training distribution. Meanwhile, the discriminator $D$ aims to distinguish between real and generated samples, thereby providing feedback that guides the generator toward producing more realistic outputs. The training objective of a GAN is formalized as a minimax problem:

$$\min_G \max_D V(D,G) = \mathbb{E}_{x \sim p_{\mathrm{data}}(x)}[\log D(x)] + \mathbb{E}_{z \sim p_z(z)}[\log(1 - D(G(z)))]$$

This adversarial framework allows GANs to implicitly learn the data distribution without requiring explicit likelihood estimation. As a result, GANs often produce high-quality samples with sharp details and rich structure. However, the adversarial training process can be unstable and highly sensitive to architectural choices and hyperparameter tuning. Additionally, GANs are prone to mode collapse, a phenomenon in which the generator produces a limited diversity of outputs, failing to capture the full variability of the target distribution. Despite these challenges, GANs have shown strong potential in molecular graph generation, particularly in settings where sharp structural fidelity is critical.

**2.4.2.3 Diffusion Models**

Diffusion models represent a rapidly advancing class of generative models known for their stable training and high-quality sampling. These models can be interpreted from several theoretical perspectives, including score-based generative modeling[170], denoising diffusion probabilistic models[171] (DDPMs), and stochastic differential equation (SDE)-based formulations[163]. They can even be understood as hierarchical extensions of VAE, where latent variables are gradually refined through sequential transformations.

At their core, diffusion models aim to bridge a simple latent distribution (e.g., standard Gaussian) and the complex data distribution via a multi-step stochastic process. Instead of learning this mapping directly, which is often highly nontrivial, diffusion models decompose the generation process into a sequence of incremental noising adding. This approach is conceptually analogous to numerical integration: breaking a complex transformation into finer-grained updates can lead to more accurate approximations of the target distribution. In what follows, we provide an intuitive overview of diffusion models using the SDE formulation, and discuss their extensions for improved efficiency.

Diffusion models consist of two complementary processes: a forward diffusion that gradually adds noise to the data, and a reverse process that denoises the signal to recover the original structure.

**Forward Process.** Starting from a data sample $x_0$, noise is incrementally added to reach a maximally perturbed latent state $x_T$. This process is governed by a stochastic differential equation of the form:

$$dx = f(x,t)dt + g(t)dW_t$$

where $f(x,t)$ is a drift term capturing deterministic trends, $g(t)$ scales the stochastic noise, and $dW_t$ is the increment of a Wiener process. By properly choosing $f$ and $g$, the terminal distribution $x_T$ can be ensured to approach a tractable prior, typically $\mathcal{N}(0,1)$.

**Reverse Process.** The generative objective is to invert the forward trajectory, reconstructing data $x_0$ from noise $x_T$. This reverse SDE takes the form:

$$dx = [f(x,t) - g(t)^2 \nabla_x \log p_t(x)]dt + g(t)d\overline{W}_t$$

Here, the gradient term $\nabla_x \log p_t(x)$ is known as the score function, representing the gradient of the log-probability at time $t$. The model learns to approximate this score function via a neural network $s_\theta(x,t) \approx \nabla_x \log p_t(x)$. Training is typically performed using a denoising objective:

$$L(\theta) = \mathbb{E}_{t,x_0,\epsilon}[\lambda(t) \parallel \epsilon - \epsilon_\theta(x_t,t) \parallel^2]$$

where $\epsilon$ is the added noise and $\epsilon_\theta$ is the network's prediction. This formulation connects the diffusion model with score matching, where the goal is to approximate the gradient of the evolving data distribution at each time step.

Notably, the SDE framework offers a unifying lens for understanding several previously independent formulations. For example, score-based models and DDPMs can both be recast as special cases under specific noise schedules:

Score Matching:

$$x_i = x_{i-1} + \sqrt{\sigma_i^2 - \sigma_{i-1}^2} z_{i-1}, \quad i = 1, \cdots, N, \quad \Rightarrow \quad dx = \sqrt{\frac{d[\sigma^2(t)]}{dt}} dw$$

DDPM:

$$x_i = \sqrt{1 - \beta_i} x_{i-1} + \sqrt{\beta_i} z_{i-1}, i = 1, \cdots, N \Rightarrow dx = -\frac{1}{2}\beta(t)xdt + \sqrt{\beta(t)}dw.$$

Both are interpreted as discretized trajectories within a continuous stochastic process.

**2.4.2.4 Extensions of Diffusion Models**

Since 2022, diffusion models have attracted growing attention across a wide range of domains, largely due to their ability to generate high-quality samples. However, their practical use still faces certain limitations. Notably, sample generation is relatively slow due to the stepwise denoising process, and training typically requires the initial data distribution to be Gaussian. These constraints have motivated the development of several extensions aimed at accelerating inference and relaxing assumptions on the initial distribution. Among recent directions, two lines of work have proven particularly influential: denoising diffusion bridge models[164] and flow matching frameworks[165]. Here, we focus on flow matching, which offers an intuitive alternative to the standard SDE-based formulation and provides insight into the dynamics of generative paths.

Rather than relying on stochastic differential equations, flow matching formulates data generation as the solution to an **ordinary differential equation (ODE)** of the form:

$$\frac{dx}{dt} = v(x, t)$$

where $v(x, t)$ denotes a velocity field that governs how the sample $x$ evolves over time $t$. Through this field, data points are transported from an initial distribution (typically noise) to the target data distribution. Interestingly, both score-based diffusion and DDPMs can also be rewritten under this framework, with their stochastic dynamics expressed as:

$$\frac{dx}{dt} = \sqrt{\frac{d[\sigma^2(t)]}{dt}} \frac{dw}{dt}$$

$$\frac{dx}{dt} = -\frac{1}{2}\beta(t)x + \sqrt{\beta(t)} \frac{dw}{dt}$$

These formulations highlight that in traditional diffusion models, the generative trajectory is nonlinear and the evolution speed varies over time. In contrast, the core idea of flow matching is to construct a linear path (e.g., a constant velocity flow) between two distributions. Specifically, given a sample pair $x_0$ and $x_1$, the transport path is defined as:

$$x_t = tx_1 + (1-t)x_0, \qquad \frac{dx}{dt} = x_1 - x_0$$

This linear evolution corresponds to an optimal transport path under the assumption of uniform velocity. The benefit of such a formulation is that it allows for larger integration steps during sampling, significantly reducing computational cost without sacrificing sample quality.

## 2.5 Graph Editing and Reinforcement Learning

In the previous section, we introduced a range of generative models grounded in graph embedding paradigms. We now turn our attention to graph editing methods and their close connection to reinforcement learning. In molecular design, graph editing involves sequentially modifying molecular graphs, through the addition, deletion, or substitution of atoms and bonds, to construct chemically valid structures that satisfy specific design objectives. This stepwise generation process naturally aligns with the framework of sequential decision-making, where each editing operation may have cascading effects on long-range molecular properties such as synthesizability or bioactivity.

Reinforcement learning offers a principled foundation for modeling this process. At each decision step, the model observes the current molecular state $s$ (i.e., the partial graph) and selects an editing action $a$ (e.g., attaching a functional group or forming a bond) with the goal of maximizing a cumulative reward. Over time, the model updates a policy $\pi = \{s, a\}$ to guide the construction toward high-quality molecules that meet predefined criteria.

Methodologically, RL-based approaches can be broadly classified into policy-based and value-based strategies. Policy-based methods (e.g., policy gradient algorithms) aim to directly optimize the action-selection policy to maximize expected returns. In contrast, value-based methods (e.g., Q-learning) focus on estimating the future reward associated with a given state-action pair, using this estimate to guide editing decisions. A third class of methods leverages Monte Carlo Tree Search (MCTS), which balances statistical sampling with heuristic exploration. Through repeated cycles of selection, expansion, simulation, and backpropagation over a search tree, MCTS can uncover promising action sequences in an exponentially large graph space. This section outlines the fundamental concepts behind these RL strategies, setting the stage for their application to molecular optimization and advanced editing strategies in subsequent discussions. The discussed methods are summarized in **Table 7**.

Table 7. Overview of reinforcement learning strategies.

| Category | Methods | Keywords |
| --- | --- | --- |
| Policy Gradient | REINFORCE[172] | Trajectory sampling, on-policy update |
| | TRPO[173] | KL-divergence constraint, trust region optimization |
| | PPO[174] | Clipped surrogate objective, efficient and stable |
| Value-based | Q-Learning[175] | Tabular update, greedy action selection |

| | DQN[176] | Use deep learning to approximate tabular Q |
| --- | --- | --- |
| | Actor-Critic[177] | Combined policy and value function |
| Others | IL[178] | Behavior cloning, expert trajectory supervision |
| | IRL[179] | Reward function inference |
| Search-based | MCTS[180] | Tree-based planning, rollout simulation |

## 2.5.1 Policy Gradient: Finetune Pretrained Graph Editing Models

Policy gradient methods aim to learn a parameterized policy $\pi_\theta$ that maximizes the expected cumulative reward $R$. In molecular generation, this policy is often represented as an autoregressive model that constructs molecules atom by atom or fragment by fragment, as illustrated in **Figure 7A**. At each step $t$, the model observes a partial molecular structure $s_t$ and samples an action $a_t$ according to a conditional distribution $p_\theta(a|s)$. Models pretrained on molecular datasets can be seen as encoding a "default policy" that captures the chemical priors present in the training set. Fine-tuning the model via policy gradient methods allows it to shift from general chemical plausibility toward task-specific optimization objectives (e.g., improving binding affinity or reducing toxicity).

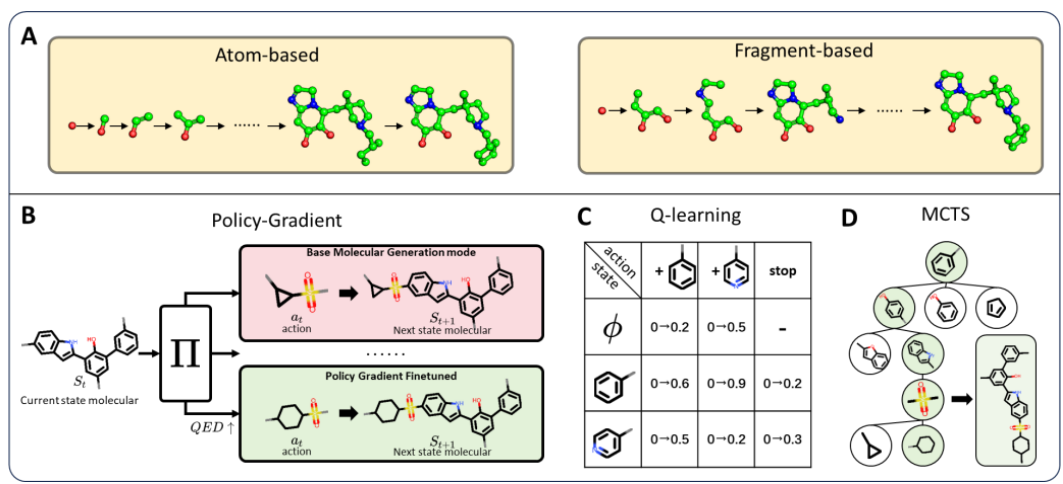

**Figure 7.** A). Atom- and fragment-based graph editing models. B) *Policy gradient* methods optimize a parameterized policy to guide graph editing actions toward desired objectives. C) *Q-learning* maintains a Q-table to estimate the expected reward of each state–action pair and select optimal strategies accordingly. D) MCTS performs rollouts to simulate possible future states, balancing exploration and exploitation during molecular generation.

### 2.5.1.1 REINFORCE: Basic Idea of Policy Gradient

The core of policy gradient learning is to adjust the model parameters so as to increase the expected reward received from interacting with the environment or an objective function, as illustrated in **Figure 7B.** The classic REINFORCE algorithm[172] offers a clean and intuitive instantiation of this idea. Assuming that the entire molecular generation process can be represented as a trajectory:

$$\tau = \{s_1, a_1, s_2, a_2, \ldots, s_T, a_T\}$$

where $s_t$ denotes the molecular state at step $t$, and $a_t$ is the editing action taken at that step. Once the full molecule is generated, a total reward $R_\pi$ is computed based on a predefined scoring function (e.g., drug-likeness or synthetic accessibility). The goal is to maximize the expected reward over all possible trajectories:

$$\overline{R}_\theta = \sum_\tau R(\tau) p_\theta(\tau)$$

Taking the gradient with respect to model parameters $\theta$, we obtain the policy gradient:

$$\nabla \overline{R}_\theta = \sum_\tau R(\tau) \nabla p_\theta(\tau) = \sum_\tau R(\tau) p_\theta(\tau) \nabla log p_\theta(\tau) = \mathbb{E}_{\tau \sim p_\theta(\tau)}[R(\tau) \nabla \log p_\theta(\tau)]$$

In practice, this expectation is approximated via Monte Carlo sampling. Given $N$ sampled trajectories $\tau^1, \ldots, \tau^N$, we have:

$$\nabla \overline{R}_\theta \approx \frac{1}{N} \sum_{i=1}^{N} R(\tau^i) \nabla \log p_\theta(\tau^i) = \frac{1}{N} \sum_{n=1}^{N} \sum_{t=1}^{T_n} R(\tau^n) \nabla \log p_\theta(a_t^n \mid s_t^n)$$

To reduce the variance of this estimator, a baseline $b$ (e.g., the average reward across the batch) is often introduced:

$$\nabla \overline{R}_\theta = \frac{1}{N} \sum_{n=1}^{N} \sum_{t=1}^{T_n} (R(\tau^n) - b) \nabla \log p_\theta(a_t^n \mid s_t^n)$$

This leads to the following update rule for the model parameters:

$$\theta \leftarrow \theta + \eta \nabla \overline{R}_\theta$$

The intuition is simple yet powerful: trajectories with positive rewards will have their log-likelihoods increased, making them more likely to be sampled again in the future. Conversely, poor trajectories with low or negative rewards will be suppressed. In molecular generation, this allows the model to evolve toward structures that optimize the target property while maintaining chemical plausibility.

### 2.5.1.2 TRPO & PPO: Improved Policy Gradient

The REINFORCE algorithm discussed above belongs to the family of **on-policy methods**, where the policy used to collect trajectories is the same as the one being updated. While straightforward, this constraint results in low data efficiency: each update requires new trajectories sampled from the latest policy. In contrast, **off-policy methods** allow updates to be performed using data collected under a different (possibly earlier) policy, significantly improving data utilization. Conceptually, this is akin to the ability to learn not only from one's own experiences but also from observing others: adjusting the policy without having to re-simulate entire trajectories under each new version of the model. To formalize this, we can rewrite the policy gradient as follows:

$$\nabla \overline{R}_\theta = \mathbb{E}_{\tau \sim p_\theta(\tau)}[R(\tau) \nabla \log p_\theta(\tau)] = \mathbb{E}_{\tau \sim p'_\theta(\tau)}\left[\frac{p_\theta(\tau)}{p'_\theta(\tau)} R(\tau) \nabla \log p_\theta(\tau)\right]$$

Here, trajectories $\tau$ are sampled from a behavior policy $\pi_{\theta'}$, while the target policy $\pi_\theta$ is the one being optimized. Expressing the reward $R(\tau)$ at the level of individual state-action pairs and subtracting a baseline $b$, we arrive at the advantage function $A(s, a) = R(\tau) - b$. The policy gradient becomes:

$$\nabla \overline{R}_\theta = \mathbb{E}_{(s_t, a_t) \sim \pi_{\theta'}} \left[ \frac{\nabla p_\theta(a_t|s_t)}{p_{\theta'}(a_t|s_t)} A(s_t, a_t) \right]$$

This form enables training a target policy using data collected from a different policy, leveraging importance sampling to correct for the distribution shift. However, when the divergence between the target and behavior policies is too large, the importance weights can exhibit high variance, destabilizing training. To mitigate this issue, the Trust Region Policy Optimization (TRPO) algorithm[173] introduces a constraint based on the Kullback-Leibler (KL) divergence, ensuring that the updated policy does not deviate excessively from the current one:

$$\max_\theta \quad \mathbb{E}_{(s_t, a_t) \sim \pi_{\theta'}} \left[ \frac{\nabla p_\theta(a_t|s_t)}{p_{\theta'}(a_t|s_t)} A(s_t, a_t) \right]$$

$$s.t. \quad KL(\theta, \theta') < \delta$$

This "trust region" constrains the update within a safe neighborhood around the behavior policy. Intuitively, it is akin to choosing mentors whose backgrounds or experiences are close to one's own, making their strategies more applicable and easier to internalize.

Despite its theoretical appeal, enforcing a hard KL constraint incurs substantial computational overhead during optimization. To address this, the famous Proximal Policy Optimization[174] (PPO) algorithm was proposed as a more efficient alternative. PPO relaxes the constraint in two ways: either by incorporating the KL penalty via a Lagrangian formulation or by clipping the update to avoid large policy shifts. The respective objective functions are:

- KL-penalized objective:

$$L^{KLPEN}(\theta) = \mathbb{E}_{(s_t, a_t) \sim \pi_{\theta'}}[r_t(\theta) A_t - \beta \mathrm{KL}[\pi_{\theta_{old}}(\cdot | s_t) \parallel \pi_\theta(\cdot | s_t)]]$$

- Clipped surrogate objective:

$$L^{CLIP}(\theta) = \mathbb{E}_{(s_t, a_t) \sim \pi_{\theta'}}[min(r_t(\theta) A_t, \mathrm{clip}(r_t(\theta), 1 - \epsilon, 1 + \epsilon) A_t)]$$

Here, $r_t(\theta) = \frac{\pi_\theta(a_t|s_t)}{\pi_{\theta'}(a_t|s_t)}$ is the importance weight, $A_t$ is the advantage estimate, $\beta$ is the KL penalty coefficient, and $\epsilon$ is the clipping parameter controlling the extent of allowable policy updates. In practice, PPO strikes a compelling balance between stability and sample efficiency, making it one of the most widely adopted algorithms in deep reinforcement learning. In molecular optimization, PPO has proven particularly useful for fine-tuning generation policies in large and complex chemical spaces, enabling efficient exploration while preserving chemical validity.

## 2.5.2 Value-based: Learn the Values of Each Graph Edit Step

In contrast to policy gradient methods that directly optimize the policy, value-based reinforcement learning focuses on estimating the expected utility of each decision, thereby guiding the selection of actions through learned value functions. The most representative method in this category is Q-learning[175]. Q-learning aims to learn an action-value function, denoted $Q^\pi(s, a)$, which estimates the expected cumulative reward of taking an action aaa in a given state $s$, and subsequently following a fixed policy $\pi$. Once this function is trained, it can be used to derive an improved policy $\pi'$ that favors actions with higher expected returns. Two key quantities are involved: 1) the action-value function $Q(s, a)$, which evaluates the long-term benefit of choosing action $a$ in state $s$, and 2) the state-value function $V(s)$, which evaluates the expected return starting from state $s$, following the current policy.

A common misconception when applying Q-learning to molecular design is to treat $Q(s, a)$ as a proxy for an immediate molecular property, such as QED[181] (quantitative estimate of drug-likeness), and greedily select the action that maximizes the local QED score at each step. However, such a greedy approach can be suboptimal. For instance, an intermediate molecule (state A) might exhibit a higher QED score than the starting structure, but fail to evolve into a more favorable final compound (state C) if only local gains are pursued. In contrast, Q-learning encourages learning future-aware value functions: ones that reflect not only the immediate structural benefits of a local modification but also its long-term impact on the eventual molecular objective.

To effectively train a Q-function for molecule optimization, one must sample sequences of edit steps from initial fragments to complete molecules, and evaluate the final rewards associated with those trajectories. These data are then used to iteratively update the Q-function so that it can assign meaningful future returns to each state-action pair. During inference, the trained Q-function guides molecule generation by selecting the action that maximizes long-term reward at each editing step. In practice, Q-learning maintains a **Q-table** or employs a neural network to approximate the function $Q(s, a; \theta)$, where $\theta$ denotes learnable parameters. The Q-function is updated iteratively using the Bellman equation:

$$Q_\pi(s, a) = Q_\pi(s, a) + \alpha [R(s, a) + \gamma \max_a Q_\pi(s, a) - Q_\pi(s, a)]$$

where $\alpha$ is the learning rate, $\gamma$ is the discount factor, and $R(s, a)$ denotes the immediate reward for executing action $a$ in state $s$. Despite its simplicity, classical Q-learning suffers from several limitations, such as limited exploration capacity and the tendency to overestimate Q-values. A variety of enhancements have been proposed to mitigate these issues, including Double Q-learning, Dueling Networks, and Noisy Nets. The Rainbow algorithm[182] provides a systematic comparison and unification of six key improvements to Q-learning, offering a comprehensive foundation for practitioners seeking more robust implementations.

In molecular design tasks, the Q-function serves as a value-based evaluator for guiding structure generation. After pretraining a generative policy to propose valid molecular edits, one can repeatedly update the Q-function to capture long-term optimization signals. During generation, this Q-function is then used to choose the next edit at each step by maximizing $Q(s,a)$, ultimately leading to more optimized molecules. **Figure 7C** illustrates the molecular generation process using Q-learning. Given the high-dimensionality of molecular state and action spaces, deep neural networks are typically used to parameterize the Q-function, resulting in the well-known Deep Q-Network[176](**DQN**), extending Q-learning to large, complex environments.

### 2.5.3 Hybrid Approaches

#### 2.5.3.1 Actor-Critic: Integrating Policy Gradients and Value Estimation

While policy gradient methods are well-suited for tasks with continuous action spaces, they often suffer from high variance in gradient estimation, which can destabilize training. A natural remedy is to combine them with value-based methods by using a learned value function to provide lower-variance estimates of expected returns. This leads to the **Actor-Critic**[177] framework, where the actor optimizes the policy by selecting actions, while the critic evaluates the quality of those actions by estimating the value function. In practice, the Actor-Critic algorithm operates in three stages:

1. **Action selection**: At each decision step, the actor samples an action $a \sim \pi_\theta(a|s)$ based on the current policy.

2. **Evaluation**: The critic computes the temporal-difference (TD) error to assess the action's quality:

$$\delta = r + \gamma V(s') - V(s)$$

where $r$ is the immediate reward, and $V(s)$ and $V(s')$ are the predicted values of the current and next states, respectively.

3. **Parameter update**: The actor is updated via policy gradients using the TD error as a surrogate reward:

$$\theta \leftarrow \theta + \alpha \delta \nabla_\theta \log \pi_\theta(a|s)$$

where $\alpha$ is the learning rate. This update encourages actions that lead to positive TD errors (actual returns higher than expected) and discourages those with negative TD errors.

By simultaneously optimizing both the policy (actor) and the value estimate (critic), the Actor-Critic framework benefits from more stable learning dynamics. Extensions such as **A2C** (**Advantage Actor-Critic**) and **A3C**[183] (**Asynchronous Advantage Actor-Critic**) further improve efficiency and generalization by employing the advantage function $A(s,a) = Q(s,a) - V(s)$ and by asynchronously collecting experience from multiple agents interacting with parallel environments, respectively. This parallelism accelerates convergence and increases robustness during training.

## 2.5.3.2 Imitation Learning and Inverse Reinforcement Learning

Although reinforcement learning does not require labeled data, its effectiveness depends heavily on the design of the reward function. In practice, defining meaningful and dense reward signals can be challenging, especially in environments with *sparse or delayed feedback*. For example, during SMILES string generation, molecular properties may only become computable after completing a syntactically valid sequence, such as when a closing bracket appears several steps after an opening bracket. These challenges motivate **imitation learning** and **inverse reinforcement learning**, both designed to address such sparse reward settings.

**Imitation learning (IL)** bypasses the need for an explicit reward function by directly mimicking expert behavior. In the context of molecular design, expert behavior typically corresponds to complete generation trajectories. A widely used approach is **Behavior Cloning**[178] **(BC)**, which formulates trajectory imitation as a supervised learning task: given expert state-action pairs $(s_t, a_t)$, the model learns a mapping $\hat{a} = f(s_t)$. BC is often used for pretraining, enabling models to reach a reasonable performance level before further refinement via reinforcement learning. For example, Meldgaard et al.[184] pretrain a molecular generator on GDB-11 using BC and use it to initialize a subsequent Q-learning phase. Their loss function combines energy, force, and Q-value prediction errors:

$$l = \rho_e \|E - \hat{E}\|^2 + \rho_f \frac{1}{3N} \sum_{i=1}^{3N} \text{Huber}(F_i, \hat{F}_i) + \rho_q (-\sum_{i=1}^{M+1} Q_i \log \hat{Q})$$

**Inverse reinforcement learning (IRL)**, on the other hand, attempts to infer the underlying reward function by observing expert trajectories. This is especially valuable in molecular design, where it is difficult to construct a comprehensive reward function from first principles. By analyzing how expert actions lead to desirable molecular outcomes (e.g., improved drug-likeness), IRL aims to recover reward signals that drive such behavior. A representative algorithm is **Apprenticeship Learning**[179], which assumes the reward function is a linear combination of features:

$$R(s, a) = \omega^T \phi(s, a)$$

where $\phi(s, a)$ are predefined feature vectors, and $\omega$ are the learnable weights. The objective is to minimize the distance between the expected feature counts under the expert policy $\pi_E$ and the learned policy $\pi$:

$$\min_{\theta} \| \mathbb{E}_{(s,a) \sim \pi}[\phi(s, a)] - \mathbb{E}_{(s,a) \sim \pi_E}[\phi(s, a)] \|$$

Despite its conceptual appeal, IRL has seen limited application in molecular generation due to computational challenges and the difficulty of designing suitable feature representations. One of the few notable attempts is the work by Agyemang et al.[185], who apply **maximum entropy IRL** to infer reward functions that bootstrap subsequent reinforcement learning for molecular generation.

## 2.5.4 Search-based: MCTS for Searching Discrete Graph Spaces

**Monte Carlo Tree Search (MCTS)** is essentially a heuristic search algorithm that, more broadly, falls under the umbrella of reinforcement learning. It has proven effective for navigating large, discrete decision spaces by combining random sampling with statistical evaluation to approximate near-optimal action sequences. A landmark success of MCTS was its integration into **AlphaGo**[180], which famously defeated human world champions in the game of Go, sparking renewed enthusiasm in reinforcement learning research.

The central idea of MCTS is to treat the decision space as a vast, intractable search tree, where each node represents a possible state (e.g., a partially constructed molecule) and edges correspond to actions (e.g., adding or modifying atoms or bonds). Rather than relying on a predefined value function, MCTS evaluates the utility of each action by performing repeated rollouts, which are stochastic simulations of possible futures, and incrementally refines its policy based on the outcomes. In the context of molecular generation, MCTS can be used to explore the chemical space by iteratively selecting graph editing actions, as illustrated in **Figure 7D**. At each step, a surrogate model may be used to stochastically complete the rest of the molecule, after which the resulting structure is evaluated using a predefined scoring function (e.g., drug-likeness or binding affinity). Over time, MCTS builds a search tree that prioritizes actions and molecular fragments leading to promising outcomes. The algorithm comprises four key phases:

1. **Selection**: Traverse the tree from the root node to a leaf by selecting child nodes that maximize an upper confidence bound.
2. **Expansion**: Add one or more child nodes to the tree by exploring new actions from the selected leaf.
3. **Simulation**: Perform a randomized rollout from the new node to estimate potential rewards.
4. **Backpropagation**: Propagate the result of the simulation upward, updating statistics (e.g., average reward) along the traversed path.

In summary, this section has introduced three primary reinforcement learning paradigms for molecular graph optimization: policy-based, value-based, and search-based methods. Additionally, we discussed hybrid strategies such as Actor-Critic, as well as imitation learning and inverse reinforcement learning, which offer practical alternatives in settings with sparse or undefined reward functions. In practice, policy-based methods are particularly effective for fine-tuning pretrained generative models, while value-based methods and MCTS are often deployed as standalone optimization frameworks with their own independent training protocols.

# 3. Molecular Property Prediction

## 3.1 Tasks and Challenge

Molecular property prediction represents one of the most fundamental tasks in AIDD, and it was also among the earliest applications of GNNs in this domain. Approximately two-thirds of drug candidates fail during clinical trials, primarily due to insufficient in vivo efficacy or the emergence of toxicity and adverse effects[186]. Therefore, the ability to assess drug-likeness at the early stages of drug development is critical for filtering low-potential compounds and reducing overall Research & Development (R&D) costs. In early implementations of AIDD, researchers commonly employed handcrafted molecular fingerprints, such as ECFP4, to represent molecular structures, followed by conventional machine learning algorithms for property prediction, e.g., random forests and support vector machines[187]. While these approaches achieved moderate success and demonstrated utility in practice, their reliance on human-engineered descriptors often introduced subjective biases, which in turn limited both predictive accuracy and model generalizability[188]. In contrast, GNNs offer a data-driven alternative by learning task-specific molecular representations directly from the graph-based molecular structures. This capacity to automatically infer relevant features without manual intervention has garnered increasing attention from the community. Moreover, the flexible architecture of GNNs enables seamless integration with advanced techniques such as pretraining, multitask learning, and transfer learning, which can mitigate data sparsity issues by incorporating chemical and physical knowledge into model training. These advantages have made GNNs a promising direction for molecular property prediction. This section provides a comprehensive overview of the applications and remaining challenges of GNNs in this context.

### 3.1.1 Task Descriptions and Datasets Available

#### 3.1.1.1 ADMET Properties

Absorption, distribution, metabolism, excretion, and toxicity (ADMET) properties are basic pharmacokinetic and pharmacodynamic descriptors for evaluating drug-likeness. Each aspect reflects a distinct stage of in vivo behavior:

**Absorption (A):** Absorption describes the process by which a drug enters systemic circulation after administration, typically quantified by intestinal absorption, oral bioavailability or the percentage of drug detected in plasma[189]. This process is influenced by membrane transport mechanisms in the gastrointestinal tract. In vitro assays such as the Caco-2 cell model[190] and P-glycoprotein (P-gp) transporter systems [191] are commonly employed to evaluate permeability and efflux characteristics.

**Distribution (D):** After absorption, a drug must be distributed to its site of action. The apparent volume of distribution (VD) is a commonly used pharmacokinetic parameter[192]. A related property is blood-brain barrier (BBB) permeability[193], which is of particular importance for drugs targeting the central nervous system (CNS).

**Metabolism (M):** Many drugs undergo metabolic transformation to become pharmacologically active or to be eliminated from the body. Cytochrome P450 (CYP) enzymes play a dominant role in drug metabolism[194], and prediction tasks in this category often involve modeling compound activity against specific CYP isoforms.

**Excretion (E):** Excretion is the process by which a drug is cleared from the body, commonly described by parameters such as clearance rate (CL) and half-life ($t_{1/2}$) [195].

**Toxicity (T):** Toxicity evaluation is essential for determining drug safety and includes endpoints such as teratogenicity, hepatotoxicity, and dermatotoxicity. **Table 8** summarizes commonly used datasets for toxicity prediction. In addition, the ChEMBL database[196] provides a comprehensive collection of physicochemical and ADMET-related information, which is frequently referenced by independent datasets. For an extensive compilation of ADMET-related endpoints, the ADMETlab platform[197] serves as a valuable resource.

Table 8: Overview of datasets for molecular property prediction. The C and R in the Task columns denote the classification and regression tasks.

| Domain | Dataset | Task | # mols | Keywords |
|---|---|---|---|---|
| ADMET | Human Bio[198] | C | 1013 | Oral bioavailability for absorption |
| | Caco-2[190] | R | 100 | In vitro simulation of SIECs for absorption |
| | P-gp[191] | | 1302 | Receptors associated with transporter |
| | BBBP[193] | C | 2039 | Blood-brain barrier |
| | Tox21[199] | C | 7831 | Toxicology in the 21st Century |
| | ToxCast[200] | C | 8675 | Toxicology data |
| | SIDER[201] | C | 1427 | Side effect resource |
| | ClinTox[202] | C | 1478 | Clinical approved/failed drugs |
| | IUPAC pKa[203] | R | 10,624 | Experimental pKa |
| Physical Chemistry | ESOL[204] | R | 1128 | Water solubility (LogS) |
| | FreeSolv[205] | R | 642 | Hydration free energy |
| | Lipophilicity[206] | R | 4200 | Membrane permeability and solubility (LogP) |
| | AQSOL[207] | R | 9982 | Aqueous solubility values from 9 sources (LogS) |
| | QM7[208] | R | 7165 | Molecules up to 7 atoms and their quantum properties |
| | QM8[209] | R | 21,786 | Molecules up to 8 atoms and their quantum properties |
| | QM9[210] | R | 133,885 | Molecules up to 9 atoms and their quantum properties |
| | MD17[211] | R | >100,000 | Eight organic molecule dynamics trajectories calculated from DFT. |
| | PCQM4M[212] | R | 3,803,453 | HOMO-LUMO energy gap |
| QSAR | PCBA[213] | R | 437,929 | PubChem BioAssay, containing 128 target tasks |
| | MUV[214] | R | 93,087 | Refined PCBA, containing 17 target tasks |
| | HIV[215] | C | 41,127 | Molecules with 0/1 active labels on HIV |
| | BACE[216] | C/R | 1,513 | Molecules with 0/1 active labels on β-secretase enzyme |
| | NCI1[217] | C | 4110 | Molecules with 0/1 active labels on cell lung |

Molecular property prediction tasks can be broadly formalized as: $p_{\text{ADMET}} = f(m, o)$, where $m$ denotes the molecular structure and $o$ represents external interactions such as drug-drug or drug-food interactions. Thus, molecular structure-based prediction alone is inherently limited in fully capturing ADMET behaviors. While, in theory, large-scale datasets could allow models to implicitly learn the impact of $o$ from molecular features, most available ADMET datasets contain only thousands to tens of thousands of compounds. As such, direct inference of ADMET endpoints from $m$ alone remains a highly challenging task. To enhance the reliability of ADMET prediction, advanced GNN techniques are needed. For example, models can use pretraining strategies to promote structurally meaningful molecular clustering, as well as use knowledge graphs, which incorporate explicit biological knowledge, to better model the multifaceted nature of ADMET behavior.

**3.1.1.2 Physicochemical Properties**

Compared to ADMET, physicochemical property prediction is more straightforward. Theoretically, these properties depend solely on the molecular structures, and can be formulated as: $p_{\text{phychem}} = f(m)$. Although environmental conditions (e.g., solvent effects, pH) may influence certain endpoints, their impact is typically more limited than in ADMET modeling. Common physicochemical properties relevant to drug design include $pK_a$[203], aqueous solubility[204], and lipophilicity[206]. These properties aid both early-stage screening for drug activity and downstream ADMET filtering. Lower-level descriptors such as polarizability, HOMO–LUMO gap, and dipole moment[210], while often weakly correlated with macroscopic pharmacokinetics, are frequently used as benchmarks for evaluating model expressivity. We would focus on three representative endpoints with high relevance to drug design:

**$pK_a$:** Protonation state at physiological pH affects molecular behavior and binding. Accurate prediction of site-specific $pK_a$ values is foundational for downstream tasks including ADMET and activity modeling.

**Aqueous solubility:** Typically expressed as logS (mol/L), solubility influences a compound's bioavailability, absorption kinetics, and distribution profile.

**Lipophilicity:** This property governs molecular partitioning between hydrophilic and hydrophobic environments and is critical for both membrane permeability and solubility. Non-ionizable compounds are described by LogP, whereas ionizable compounds are characterized by LogD.

Notably, $pK_a$ prediction focuses on local atomic environments and is usually modeled as an atom-level task, while solubility and lipophilicity are global properties better captured by graph-level prediction. Several GNN-based approaches have been proposed for physicochemical property modeling. For instance, **Graph-pKa**[218] employs a GAT-like architecture to predict $pK_a$ values. Follow-up methods such as

**MolGpKa**[219] and **Epik**[220] leverage GCN-based frameworks. Notably, **MolGpKa** utilizes two separate networks for acidic and basic sites to prevent signal interference. **MF-SuP-pKa**[221] further improves prediction by employing pretraining on noisy ChEMBL data and fine-tuning on the high-quality DataWarrior dataset[222], achieving state-of-the-art accuracy via effective transfer learning.

**3.1.1.3 Quantitative Structure-Activity Relationship (QSAR)**

QSAR modeling seeks to infer biological activity against specific targets directly from molecular structures. Formally, a QSAR task maps a molecular graph to its activity label under a given biological target. From a biophysical standpoint, accurate modeling of molecular-target interactions ideally requires knowledge of the binding complex. However, in a data-driven context, small molecules targeting the same protein often share substructures or pharmacophores. This allows machine learning models to learn implicit binding patterns from large-scale data, forming the conceptual foundation of QSAR modeling.

The **PubChem BioAssay (PCBA)** database provides over 1.25 million bioactivity records derived from high-throughput screening (HTS) experiments, covering 125 biological targets. These data can be used to construct and evaluate a corresponding set of 125 QSAR models. The **Maximum Unbiased Validation (MUV)** dataset[214], a curated subset of PCBA, was designed to mitigate the bias introduced by structural analogs through nearest-neighbor filtering, and retains 17 particularly challenging targets for model validation. In addition, the **NCI1** dataset[217] contains activity screening results across various cancer cell lines and is commonly used as a benchmark in chemoinformatics. The **HIV** dataset[215], which is part of the NCI screening effort, comprises antiviral virtual screening data against HIV and has been widely adopted in molecular property prediction studies. The **BACE** dataset[216], which focuses on β-secretase 1 (BACE-1) inhibitors, is another popular benchmark for evaluating regression-based molecular property prediction models. Among these datasets, **HIV** and **NCI1** provide categorical molecular labels (typically classified as inactive, moderately active, or active), whereas most of the other datasets are labeled with **continuous activity values**, such as $IC_{50}$ values against specific targets, making them suitable for regression tasks.

**3.1.2 Chemistry-enhanced GNNs: Fragments, Conformers, and Chirality**

In molecular property prediction, a straightforward strategy involves directly applying standard GNN architectures to specific tasks. For example, GAT has been employed in QSAR modeling[33], while EGNN[48] has been used to predict quantum mechanical properties in the QM9 dataset. Although these general-purpose models have seen widespread application, a growing body of research focuses on embedding domain-specific chemical knowledge into GNN architectures to further improve performance. This section introduces several representative classes of chemistry-enhanced GNNs (**Table 9**), illustrating how chemical priors can be integrated into GNNs.

Table 9. Overview of chemical-enhanced GNN in molecular property prediction.

| Category | Methods | Keywords |
|---|---|---|
| Edge-enhanced | D-MPNN[223] | Bonds as nodes; while atoms as edges |
| Fragment-based | FragGAT[224] | Extract fragment as nodes |
|  | HiGNN[225] | Hierarchical atom graph to fragment graph layers |
|  | RG-MPNN[226] | Pharmacophore-based graph |
| Geom-enhanced | - | Refer to Table 4 |
| Chirality-enhanced | ChIRo[227] | d, θ, φ message passing layers |
|  | MolKGNN[228] | Build chiral convolutional kernel |

### 3.1.2.1 D-MPNN: Bond-as-node

One of the earliest and most influential GNN architectures in molecular property prediction is the Directed Message Passing Neural Network (D-MPNN)[223]. Unlike conventional GNNs that treat atoms as nodes and bonds as edges, D-MPNN models chemical bonds as nodes during message passing, with the two atoms connected by the bond treated as directed edges. This reconfiguration enables more chemically intuitive information flow across the molecular graph. The message update scheme can be formalized as:

$$m_{uv}^{t+1} = \sum_{k \in N(v) \setminus u} \text{Mes}(x_u, x_v, h_{uv}^t)$$

$$h_{uv}^{t+1} = \text{Upd}(h_{uv}^t, m_{uv}^{t+1})$$

where $h_{uv}^t$ represents the feature on the bond (edge) from atom $u$ to atom $v$ at step $t$. Subsequent models such as EMNN[229] and GEA-D-MPNN[230] further extended this edge-centric message passing paradigm. The widely used Chemprop framework[231], which builds on D-MPNN, has been applied across a variety of molecular property benchmarks with consistent success.

### 3.1.2.2 Fragment GNNs: Fragment-as-node

In traditional GNNs, molecules are typically represented as graphs where atoms are nodes and bonds are edges. While this approach works well for many tasks, it may be insufficient to capture higher-level chemical features such as ring systems or functional groups, which play critical roles in determining molecular properties. To address this limitation, several recent models have explored coarse-grained representations, treating molecular fragments as the graph nodes, as illustrated in **Figure 8A**.

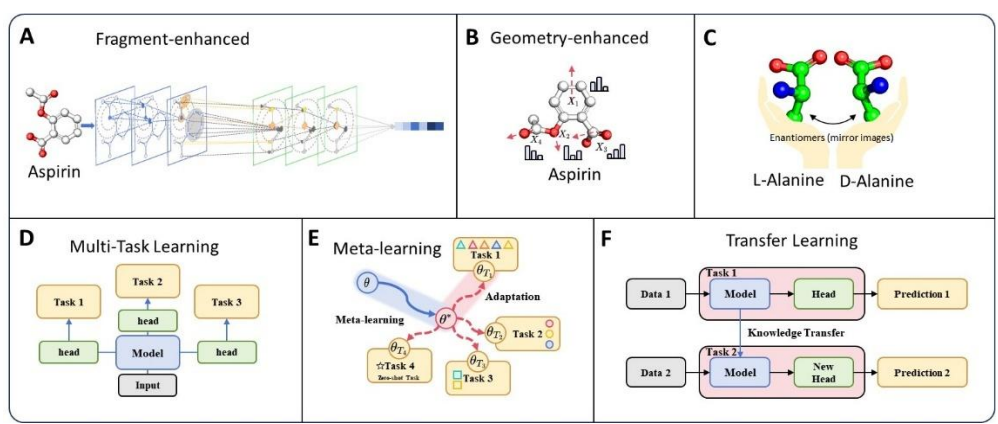

**Figure 8**. A,B,C) Chemistry-enhanced GNN models that incorporate domain knowledge to improve molecular representation learning. D,E,F) Representative strategies for addressing data sparsity in molecular property

prediction: D) Hard parameter sharing framework, where all tasks are jointly trained using a shared backbone (trunk) and task-specific output heads. E) MAML, which learns an initialization of model parameters that can rapidly adapt to new tasks with limited data. F) Transfer learning, which leverages knowledge from a source domain (Data 1) to improve performance in a target domain (Data 2).

**FraGAT**[224] exemplifies this direction by first breaking molecular graphs along acyclic single bonds to generate fragments, and then constructing a fragment graph in which each node represents a substructure. The architecture employs a graph attention mechanism inspired by AttentiveFP, with fragment-level features computed as the sum of the constituent atom features: $h_f = \sum_{i \in G_f} h_i$, where $G_f$ denotes the subgraph corresponding to fragment $f$. Across multiple prediction benchmarks, FraGAT demonstrated consistent improvements over atom-level models such as AttentiveFP. Similarly, **HiGNN**[225] (Hierarchical Informative Graph Neural Network) adopts a hierarchical learning strategy, with fragments derived using the BRICS decomposition rules[232]. One of HiGNN's key innovations lies in its explanation mechanism: by comparing the learned representations of fragment graphs and the full molecular graph, the model can provide fragment-level explanation for property predictions. Another approach, **RG-MPNN**[226], incorporates pharmacophoric information by combining the atom-level molecular graph with a pharmacophore graph. Message passing is performed at both levels, enabling the model to capture both fine-grained atomic interactions and higher-level functional motifs. This two-scale framework enhances the expressiveness of the molecular representation and improves predictive performance in tasks where pharmacophore context is essential.

**3.1.2.3 Conformation-Aware GNNs**

Molecules in nature do not exist as static 2D structures; rather, they adopt dynamic 3D conformations that evolve over time. Under reasonable assumptions, these conformations can be assumed to follow the Boltzmann distribution: $P(R) \propto e^{E(R)/kT}$, where $E(R)$ is the energy of conformation $R$, $k$ is the Boltzmann constant, and $T$ is the absolute temperature. Nearly all molecular properties are influenced, directly or indirectly, by 3D conformation, though the degree of dependence varies. For example, quantum mechanical properties are highly conformation-sensitive, while macroscopic properties such as solubility can often be approximated from 2D structures alone. In principle, learning from molecular conformations provides a more physically grounded modeling approach than relying solely on 2D graphs.

In GNN frameworks, a natural extension is to represent molecules as 3D graphs, denoted by $G = (V, E, X)$, where $V$ and $E$ are the node and edge, respectively, and $X$ denotes the Cartesian coordinates of each atom, as illustrated in **Figure 8B**. As discussed in the **Section 2.2**, many geometric GNNs are explicitly designed to operate on such 3D structures, often incorporating equivariance properties to respect spatial transformations. A growing body of work has demonstrated that incorporating conformational information via geometric GNNs significantly enhances performance in predicting quantum properties[233], as well as in QSAR and ADMET tasks[234]. Several pretraining strategies also incorporate geometric signals. For instance, noisy node masking[134] involves denoising perturbed atomic coordinates; GhemRL-GEN[235] and GROVER[130] trains models to predict bond lengths and angles. These methods have shown superior performance in downstream tasks, benefiting from richer spatial encodings.

Notably, recent studies have explored the use of conformational ensembles rather than single conformers as input to GNNs[236]. This so-called "4D-GNN" approach, where the fourth dimension denotes temporal or ensemble variability, aligns more closely with the physical reality of molecular motion and flexibility. However, the computational cost of generating and processing large conformer sets remains prohibitive. A promising direction for future research is the development of architectures that can encode ensemble-level information implicitly from a single conformer, thereby balancing physical realism and computational efficiency.

**3.1.2.4 Chirality-Aware GNNs**

Chirality is a crucial yet often underrepresented factor in molecular property prediction. Chiral molecules, or enantiomers, share identical molecular formulas and connectivity but differ in their spatial arrangement such that they are non-superimposable mirror images. A classic example is a carbon center bonded to four different substituents, as illustrated in **Figure 8C**. Chirality is especially important in drug development: although two enantiomers may share similar physicochemical properties (e.g., solubility or electronic energy levels), their biological activities can differ dramatically[237]. For instance, the R-enantiomer of thalidomide exhibits relatively safe sedative effects, while the S-enantiomer is teratogenic and causes severe birth defects[238]. Therefore, incorporating chirality into GNN architectures is a natural and necessary enhancement for chemically informed modeling. Conventional GNNs typically cannot distinguish between enantiomers. A naive strategy is to append chirality labels (e.g., R/S) to atomic features. However, in deep networks, such signals may be overwhelmed or diluted during message passing. A more robust approach involves architectural modifications that enable the model to explicitly resolve chiral distinctions. Ideally, a chirality-aware GNN should satisfy:

$$f_\chi(m_R) \neq f_\chi(m_L)$$

That is, the learned representations of R- and L-enantiomers should be distinct in latent space. Theoretically, this requires the network to be equivariant to SE(3) but not to E(3), as it must break equivariance under reflections to encode mirror asymmetry. Some equivariant GNNs naturally capture chirality through their treatment of 3D coordinates. For example, GVP-GNNs, which operate directly on Cartesian vectors, and SphereNet, which performs third-order message passing using internal coordinates. In addition, several architectures have been specifically designed for chirality modeling. ChIRo[227] (Chiral Invariant-to-Reflection Neural Network) incorporates distance, angle, and dihedral updates from third-order atomic neighborhoods, yielding a conformation-sensitive representation of each atom:

$$\hat{y} = f_{out}(\sum_{i \in \mathcal{G}} h_i, z_d, z_\theta, z_\phi)$$

where $(z_d, z_\theta, z_\phi)$ represent distance, angle, and dihedral encodings, respectively. These encodings are concatenated with atom features and aggregated to produce the final graph-level output. For details on the computation of $z_d, z_\theta, z_\phi$, refer to the original paper. Another approach, **MolKGNN**[228], introduces a chiral convolution kernel to distinguish enantiomers. Its core idea is to use a tetrahedral numerical signature derived from the kernel: if the neighborhood orientation matches the kernel's chirality, the convolution yields a positive value; if mismatched, the output is negative, thereby enabling the network to differentiate molecular mirror images at the architectural level.

### 3.1.3 Strategies for Mitigating Data Sparsity: Multi-task, Meta-, and Transfer Learning

Data sparsity remains one of the most critical challenges in molecular property prediction. As shown in Table 8, many property labels, particularly those requiring wet-lab measurements, are available for only a few hundred or thousand molecules. Such limited data volume makes it difficult to train predictive models with strong generalization capabilities. To mitigate this issue, researchers have proposed a variety of data-efficient learning strategies, many of which are particularly well-suited to GNN frameworks due to their architectural flexibility. This section focuses on three representative directions: multi-task learning, meta-learning, and pretraining with transfer learning, with a summarized **Table 10**.

**Table 10.** Overview of GNN methods focusing on alleviating the data sparsity issue in molecular property prediction

| Category | Methods | Keywords |
| --- | --- | --- |
| Multi-task Learning | MGA [239] | Hard parameter sharing on ADMET |
| | DeepMolNet[240] | Hard parameter sharing on QM9 |
| | SST-GCN[241] | Add STL results p as additional MTL features |
| | MTGL-ADMET[242] | Adaptive selection of beneficial side tasks |
| Meta-Learning | Nguyen et al[243] | Prove three meta-learning tasks work on CHEMBEL |
| | Meta-MGNN[244] | Adaptive task in outer loop |
| | GS-Meta[245] | Sampling tasks based on molecular relation graph |
| Transfer Learning | - | Refer to Table 5 |

STL: single task learning, MTL: multi task learning.

#### 3.1.3.1 Multi-Task Learning

Multi-task learning (MTL) aims to simultaneously train a model on multiple related tasks, enabling it to leverage shared patterns and improve generalization compared to single-task learning (STL). In the context of molecular property prediction, MTL helps alleviate the requirement for large datasets on any individual task by jointly training with auxiliary tasks.

The most basic form of MTL extends an STL model by replacing its single-task output head with a multi-task head. However, this naive implementation often suffers from gradient conflicts between tasks, leading to negative transfer, where performance degrades rather than improves[246]. A widely used solution is hard parameter sharing, in which the model is partitioned into shared modules and task-specific heads: the shared layers learn generalizable representations, while each task head focuses on its respective output[247], as illustrated in **Figure 8D.** For example, **MGA**[239] employs a shared GNN backbone for all tasks and uses separate MLPs as task-specific heads, enabling joint classification and regression modeling. Similarly, **DeepMolNet**[240] adopts a hard sharing architecture to predict quantum chemical properties. Within this framework, the design of multi-task loss functions is critical. Han et al.[248] systematically explored the impact of different multi-task loss formulations:

$$\mathcal{L} = \sum_{k=1}^{K} \mathcal{L}_k, \qquad \mathcal{L}_{\text{un}} = \sum_{k=1}^{K} (\frac{1}{2\sigma_k^2}\mathcal{L}_k + \log \sigma_k)$$

$$\mathcal{L}_{un_r} = \sum_{k=1}^{K} (\frac{1}{2\sigma_k^2}\mathcal{L}_k + \log (1 + \sigma_k^2))$$

$$\mathcal{L}_{\text{DWA}}(t) = \sum_{k=1}^{K} \lambda_k(t)\mathcal{L}_k(t)$$

$$\text{where } \lambda_k(t) = \frac{Ke^{w_k(t-1)/T}}{\sum_j e^{w_j(t-1)/T}}, w_k(t-1) = \frac{\mathcal{L}_k(t-1)}{\mathcal{L}_k(t-2)}$$

where $\mathcal{L}, \mathcal{L}_{\text{un}}, \mathcal{L}_{un_r}, \mathcal{L}_{\text{DWA}}(t)$ are the uniform multi-task lass, uncertainty-based multi-task loss, corrected uncertainty-based multi-task loss and dynamic weight averaging multi-task loss, respectively; $\mathcal{L}_k$ denotes the loss for task $k$, and $\sigma_k$ denotes task uncertainty. Their experiments revealed that while adapted STL performs best when tasks are highly correlated, it fails under task conflict. In contrast, uncertainty-based losses consistently outperformed naive averaging, highlighting the importance of task weighting in MTL.

Beyond hard sharing, more innovative MTL frameworks have emerged. For instance, the Stacked Single-Target (SST) approach[241] first trains individual STL models and then uses their predictions $\hat{p}$ as extended features, concatenated with the original molecular representation $x$ to form an augmented feature space $\{x, \hat{p}\}$. The intuition is that $x$ encodes core chemical information, while $\hat{p}$ captures cross-task signals. The resulting **SST-GCN**[241] outperformed standard MTL baselines on multiple benchmarks. It is worth noting that not all auxiliary tasks are beneficial. The **MTGL-ADMET**[242] framework addresses this by selectively identifying helpful auxiliary tasks for a given main task. It constructs a task-relatedness graph and applies a max-flow algorithm to identify optimal task bundles. A gated multi-task learning module is then trained on each bundle. Empirical results demonstrated that MTGL-ADMET achieved state-of-the-art performance in multi-property ADMET prediction.

### 3.1.3.2 Meta-Learning

Meta-learning, often referred to as "learning to learn", aims to train models that can quickly adapt to new tasks using only a small number of labeled samples[249]. For example, humans can distinguish between an armadillo and a pangolin after seeing only a few images of each, despite their visual similarity, and this kind of rapid generalization exemplifies the goal of meta-learning. In molecular property prediction, where labeled data for specific endpoints may be extremely limited, meta-learning offers a promising avenue for improving few-shot performance. Meta-learning strategies are generally categorized into two major classes: **metric-based** methods[250] and **gradient-based** methods[251]. Metric-based approaches construct a latent space in which new, unlabeled samples can be matched to a small set of labeled examples. In contrast, gradient-based methods aim to learn a set of model parameters that are **easily adaptable** to new tasks with minimal gradient updates. The most widely used gradient-based method in molecular modeling is **Model-Agnostic Meta-Learning (MAML)** [251], as illustrated in **Figure 8E**.

Nguyen et al.[243] first applied MAML and its two simplified variants, **FO-MAML**[252] (First-Order MAML) and **ANIL**[253] (Almost No Inner Loop), on ADMET prediction using the GGNN[254] as a backbone and ChEMBL as the training corpus. Their results showed that meta-learned models significantly outperformed baselines in low-data regimes. Building on this foundation, Guo et al. introduced **Meta-MGNN**[244], which augments the MAML framework with **task-dependent attention mechanisms** during the outer loop updates, allowing dynamic reweighting of tasks. Additionally, it introduces **auxiliary self-supervised losses** within each task, such as atom and bond masking and recovery, to enhance structural representation learning. Meta-MGNN achieved strong performance gains on low-shot toxicity prediction benchmarks like Tox21 and SIDER under 1-shot and 5-shot settings.

Borde et al.[255] extended meta-learning to **ensemble models**, confirming that meta-learned initialization benefits both individual and aggregated predictors in quantum property prediction. More recently, Zhuang et al.[245] highlighted a unique characteristic of few-shot learning in molecular science: the same molecule can appear across multiple datasets with different endpoints, and many molecular properties exhibit biological or mechanistic correlations. For example, predictions of cardiotoxicity may benefit from known endocrine-related side effects due to overlapping biological pathways. To exploit this cross-property dependency, Zhuang et al.[245] proposed **GS-Meta** (Graph Sampling-based Meta-Learning), which constructs a molecular-property relationship graph and samples tasks accordingly for meta-training. This task sampling strategy leverages molecular overlap and property interdependence to improve generalization. Despite these advances, most current studies still evaluate meta-learning under artificial conditions, typically 1-shot or 5-shot settings on curated datasets. In real-world applications, further evaluation and integration with domain knowledge are necessary for meta-learning to realize its full potential in chemistry.

### 3.1.3.3 Pretraining and Transfer Learning

With the widespread success of foundation models such as ChatGPT[256], pretraining has become a mainstream strategy for addressing data scarcity in molecular property prediction. Transfer learning, broadly defined, encompasses various paradigms including pretraining–fine-tuning, feature transfer, and parameter transfer. In molecular property prediction, researchers typically focus on the pretraining–fine-tuning strategy, as illustrated in **Figure 8F**. Depending on the use of labeled data, pretraining can be classified into supervised and unsupervised approaches. In supervised pretraining, large volumes of low-confidence but cheaply generated data, like simulation-generated labels are used to pretrain a model, which is then fine-tuned on high-confidence but scarce experimental data[257]. In unsupervised pretraining, models extract self-supervised signals from molecular structures without relying on any labels, making it possible to train on millions of unlabeled compounds to explore chemical space in a task-agnostic and unbiased manner. Common unsupervised objectives also include generative modeling, predictive tasks (e.g., masked atom prediction), and contrastive learning. These strategies have been discussed in detail in Chapter 2 of this review.

However, transfer learning is not always effective. In some cases, pretrained models may even underperform task-specific baselines, a phenomenon known as **negative transfer**. Several factors may contribute to this issue: **1). Oversimplified pretraining objectives.** For example, if the pretraining task is masked node prediction and the dataset is highly imbalanced (e.g., dominated by carbon atoms), the model might achieve excellent pretraining performance by trivially predicting carbon atoms without learning meaningful chemistry. As a result, the learned representation may be biased or misleading in downstream tasks. **2). Inappropriate architecture.** As discussed in our section on deep GNNs, oversmoothing and overcompression are common pitfalls in graph-based models. These issues necessitate careful architecture design when adopting pretraining strategies. For instance, GROVER[130] demonstrated that replacing a graph transformer backbone with a simpler GNN led to performance degradation. Similarly, Buterez et al.[257] showed that the choice of readout function significantly impacts transferability: simple summation or averaging may obscure differences between active and inactive molecules. To address this, they proposed an attention-based readout to better preserve discriminative information. **3). Mismatch between pretraining and downstream tasks.** For example, node-level pretraining objectives often have limited benefit for graph-level downstream tasks.

Overall, the complexity of chemical tasks far exceeds that of conventional domains such as computer vision or NLP. Therefore, universal models with coarse pretraining objectives are unlikely to be sufficient. A more promising approach involves combining large-scale pretraining with carefully tailored fine-tuning strategies, integrating molecular representations with domain-specific knowledge. This direction holds great potential not only for AIDD, but also for the broader field of AI4Science.

### 3.1.4 Challenges in Molecular Property Prediction

In addition to the sparsity problem discussed earlier, molecular property prediction faces a range of other challenges, including data imbalance, ambiguous task definitions, and high levels of data noise.

### 3.1.4.1 Data Imbalance

Imbalanced datasets can significantly degrade the performance of predictive models in both classification and regression tasks[258]. In many chemical datasets, the distribution of active vs. inactive compounds (or positive vs. negative samples) is highly skewed. For example, if 99% of a dataset consists of negative samples, a model that predicts all instances as negative could achieve 99% accuracy, but such a model would be practically useless. This issue is especially prevalent in chemistry, particularly in datasets derived from literature curation. For instance, in the FreeSolv dataset[205], most compounds are recorded as active, while in large-scale screening datasets like HIV[215], the vast majority of compounds are inactive. This imbalance causes models to primarily learn how to identify the dominant class, rather than truly understanding the determinants of molecular activity, thus limiting their effectiveness and reliability in real-world applications.

### 3.1.4.2 Incomplete Modeling

Many molecular property prediction tasks suffer from inherently incomplete modeling. As discussed earlier in the data description section, physicochemical properties are often directly inferred from molecular structures. In contrast, properties like ADMET are influenced by complex in vivo conditions, inter-individual variability, and drug-drug or drug-food interactions. For example, the SIDER dataset[201] contains records of side effects for 1,427 marketed drugs across 27 organ systems. The occurrence of these side effects depends not only on the drugs' chemical structures, but also on patient-specific physiological conditions, concomitant medications, and food interactions. This multifactorial nature makes it extremely challenging to predict side effects solely from molecular structures. Moving forward, researchers may need either extremely large-scale datasets to allow models to implicitly learn such physiological influences, or explicitly incorporate "human system states" into modeling, an idea that will be further explored in the section on knowledge graphs.

### 3.1.4.3 High Data Noise

Real-world data inevitably contains noise due to human and technical factors. In chemistry, this includes not only experimental variability introduced by researchers but also inherent errors in measurement techniques. Mistakes during database curation can also introduce label inconsistencies. For example, the same compound may have conflicting activity annotations in databases like PDBBind and BindingMOAD. High noise levels can degrade model training and lead to a "Garbage In, Garbage Out" scenario. To mitigate this, researchers have explored strategies such as data cleaning[259] and uncertainty quantification[260] to improve data quality and enhance model robustness. A detailed discussion on building more trustworthy GNN-based prediction models through uncertainty quantification will follow in a later section.

## 3.2 Uncertainty Quantification in GNN Property Prediction: Control Risk

In real-world scenarios, molecular property prediction often faces inherent data noise and limited chemical space coverage caused by data sparsity. As a result, models are confronted with two key limitations: generalization ability and awareness of predictive boundaries. On the one hand, we expect that models can make reasonably reliable predictions for compounds that fall outside the chemical space covered by the training data. On the other hand, we expect that models can quantify the uncertainty of predictions to avoid overconfidence in false positive cases. Consequently, incorporating uncertainty quantification into molecular property prediction has emerged as a critical direction for building more robust and trustworthy models. According to the classical Bayesian theory, uncertainty generally arises from two sources: aleatoric uncertainty, which originates from the inherent noise in the data, and epistemic uncertainty, which stems from the model's limited knowledge. To model uncertainty, a common approach is to make model output a variance alongside the predictive mean during inference. **Table 11** summarizes the uncertainty quantification methods we would discuss below.

Table 11. Overview of uncertainty quantification methods. The (B) in the Methods column denotes the Bayesian method, A and E in the Uncertainty column refer to aleatoric and epistemic uncertainty.

| Methods | Keywords | Uncertainty |
|---|---|---|
| HR[261] | Only change loss function | A |
| MC-Dropout (B)[262] | Random dropout sampling model configurations | A+E |
| Ensembling (B)[263] | Random initial states sampling model configurations | A+E |
| Bootstrapping (B)[264] | Resampling dataset | A+E |
| SVGD (B)[264] | Sampling model configurations via gradient descent | A+E |
| SGLD (B)[265] | Langevin dynamics sampling model configurations | A+E |
| Backprop (B)[266] | Sampling model configurations via backpropagation | A+E |
| Evidential DL[267] | Estimate Gaussian prediction (low-order) from Normal Inverse-Gamma (high-order) distribution | A+E |
| Temperature Scalling[268] | Post-hoc calibration adjusting output probability | A |
| Conformal Prediction[269] | Two-step modeling: error and confidence | A+E |

### 3.2.1 Aleatoric (Data) Uncertainty

#### 3.2.1.1 Heteroscedastic Regression

In traditional regression tasks, aleatoric (or data) uncertainty is often assumed to be homoscedastic, i.e., the noise level is constant across data points. Under this assumption, models typically minimize the mean squared error (MSE) without explicitly modeling uncertainty. However, in real-world applications, the noise may vary with the input, which is known as heteroscedasticity. For example, in molecular property prediction, variations in experimental conditions, measurement instruments, or intrinsic chemical properties can lead to different measurement noise for different compounds.

To address this, heteroscedastic regression (HR)[261] has been proposed, allowing models to predict both the output mean $\mu$ and variance $\sigma^2$, as illustrated in **Figure 9A**. The implementation is straightforward: one adds an auxiliary output head for variance prediction alongside the mean, and the model is trained by minimizing the following loss:

$$\mathcal{L}(\theta) \propto \frac{1}{N} \sum_{i=1}^{N} \frac{1}{2\sigma_\theta^2(\mathbf{x}_i)} \| \mathbf{y}_i - \mu_\theta(\mathbf{x}_i) \|_2^2 + \frac{1}{2} \log \sigma_\theta^{\,2}(\mathbf{x}_i)$$

Intuitively, this loss encourages the model to learn input-dependent noise: data points with higher predicted variance receive lower weight during training, reducing their influence on the final model. HR is highly flexible and can be combined with other uncertainty quantification methods, such as Monte Carlo dropout and ensembling, to produce a more comprehensive uncertainty estimate in molecular property prediction.

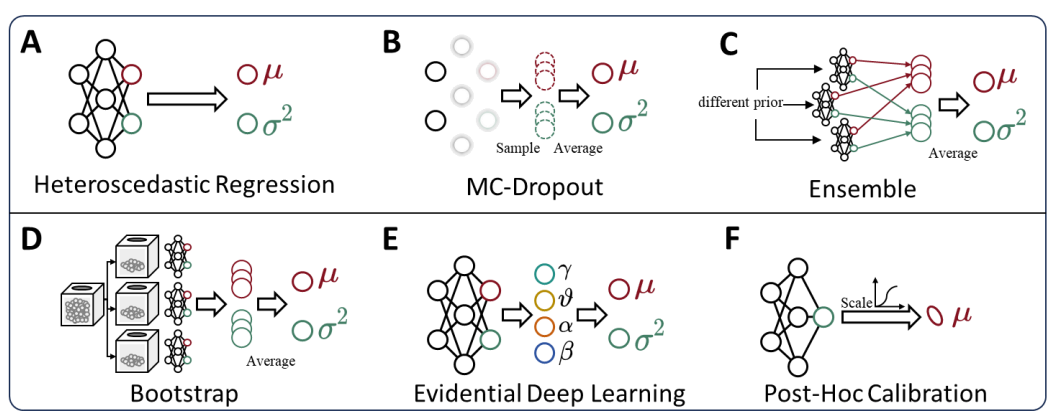

**Figure 9 A-F).** Illustration of popular uncertainty quantification methods.

### 3.2.2 Epistemic (Model) Uncertainty

Epistemic uncertainty reflects the model's incomplete knowledge due to the limited data or architectural ambiguity, arising from sources such as model initialization, structure, and training dynamics. According to some research[71], Bayesian inference provides a principled way to jointly describe both aleatoric and epistemic uncertainties. Readers of interest can refer to the appendix for a comprehensive explanation.

**3.2.2.1 Variational Inference Methods**

One major category of epistemic uncertainty estimation involves sampling from the model posterior via variational inference. Widely adopted methods include MC Dropout, ensembling, and bootstrapping, which are briefly described below.

**MC Dropout**[262] **(Figure 9B)** is straightforward and can be easily integrated into any model architecture. During training, dropout is applied as usual; during inference, dropout remains active to generate $M$ stochastic forward passes. These outputs approximate samples from the model's posterior $q(\theta)$. The eptistemic uncertainty can be obtained through the analysis of these $M$ samples:

$$\tilde{y} = \frac{1}{M}\Sigma y^i, \qquad \sigma_e^2 = \text{var}(y^i), \qquad \sigma_a^2 = \frac{1}{M}\sum \sigma_a^i$$

where $\tilde{y}$ is the consensus prediction, $\sigma_e^2$ captures epistemic uncertainty, and $\sigma_a^2$ reflects aleatoric uncertainty. A major limitation of MC-dropout is the sensitivity to the dropout rate $p$, which is a hyperparameter and should be carefully tuned. To address this, Concrete Dropout[270] uses gradient-based optimization to adaptively select dropout probabilities.

**Ensembling**[263] and **Bootstrapping** (or **bagging**)[264] are other two statistical approaches for modeling epistemic uncertainty, illustrated in **Figure 9C,D**. Ensembling involves training multiple models with different random initializations and aggregating their predictions. Similar to MC Dropout, the mean and variance of the outputs represent the consensus prediction and epistemic uncertainty, respectively. Despite their simplicity, ensembles often exhibit strong empirical performance and can be interpreted as implicit variational approximations of the posterior[271]. Bootstrapping follows a similar principle but trains ensemble models on resampled datasets to capture variability in the training data..

Beyond these classical methods, more advanced approaches include Stein Variational Gradient Descent (SVGD)[272], Stochastic Gradient Langevin Dynamics (SGLD)[265], and Bayes by Backprop[266]. In molecular property prediction, these techniques are often evaluated in combination. For instance, Zhang et al.[273] decomposed total predictive uncertainty into aleatoric and epistemic components using Bayesian inference, and compared MC Dropout and SVGD on a GCN-based framework, concluding that SVGD offered greater robustness. Ryu et al.[274] showed, via logP ablation studies, that increasing data noise elevated aleatoric uncertainty while epistemic uncertainty remained stable, consistent with Bayesian theory. A comparative study by Carbriele et al.[275] found that ensembles outperformed bagging and MC Dropout across multiple datasets. The recently proposed **MUBen benchmark**[276] unified four major UQ methods and demonstrated that pretrained models generally achieve both better accuracy and more reliable uncertainty estimates than standard GNNs.

**3.2.2.2 Evidential Deep Learning**

Unlike previous sampling-based approaches, Evidential Deep Learning (EDL)[267] enables the joint estimation of aleatoric and epistemic uncertainties in a single forward pass. As discussed earlier, heteroscedastic regression (HR) treats the target $y$ as a realization of a Gaussian distribution $\mathcal{N}(\mu, \sigma^2)$, and directly optimizes the negative log-likelihood to estimate $\mu$ and $\sigma^2$.

$$\mathcal{L}(\theta) \propto \frac{1}{N} \sum_{i=1}^{N} \frac{1}{2\sigma_\theta^2(\mathbf{x}_i)} \parallel \mathbf{y}_i - \mu_\theta(\mathbf{x}_i) \parallel_2^2 + \frac{1}{2}\log \sigma_\theta^2(\mathbf{x}_i)$$

However, $\sigma$ in HR captures only aleatoric uncertainty, neglecting epistemic aspect. To model both uncertainties simultaneously, EDL (**Figure 9E**) introduces a higher-order distribution over the parameters of the Gaussian $\mathcal{N}(\mu, \sigma^2)$, defined as:

$$p(\mu, \sigma^2 | \gamma, \nu, \alpha, \beta) = \frac{\beta^\alpha \sqrt{\nu}}{\Gamma(\alpha)\sqrt{2\pi\sigma^2}} (\frac{1}{\sigma^2})^{\alpha+1} \exp\{-\frac{2\beta + \nu(\gamma - \mu)^2}{2\sigma^2}\}.$$

Sampling from this higher-order distribution yields a Gaussian $\mathcal{N}(\mu, \sigma^2)$, from which both uncertainties can be derived:

$$\tilde{y} = \gamma, \qquad \sigma_a^2 = \frac{\beta}{\alpha - 1}, \quad \sigma_e^2 = \frac{\beta}{\nu(\alpha - 1)}$$

where $\tilde{y}$ is the consensus output, $\sigma_a^2$ denotes aleatoric uncertainty, and $\sigma_e^2$ denotes epistemic uncertainty. In implementation, this requires modifying the HR output heads from $(\mu, \sigma^2)$ to $(\gamma, \nu, \alpha, \beta)$. In molecular property prediction, Soleimany et al.[277] used EDL to guide data acquisition in active learning. For QSAR tasks in antibiotic discovery, uncertainty estimates improved the hit rate by focusing on high-confidence predictions. The approach was also integrated into Chemprop[231].

**3.2.2.3 Post-Hoc Calibration**

In classification tasks, models typically produce probability distributions over classes, with the highest-probability class selected as the prediction. However, models often exhibit overconfident predictions, even when their accuracy is high[274]. This mismatch between predicted probabilities and actual correctness undermines decision-making. Post-hoc calibration techniques aim to align predicted probabilities with empirical accuracies. A representative method is temperature scaling[268] (**Figure 9F**), which introduces a learnable scalar $T$ to soften the softmax outputs:

$$\mathrm{softmax}_T(z_i) = \frac{e^{z_i/T}}{\sum_j e^{z_j/T}}$$

When $T > 1$, the output distribution becomes less sharp, mitigating overconfidence. While calibrated probabilities partially reflect aleatoric uncertainty, they are limited in capturing epistemic uncertainty. In molecular property prediction, Busk et al.[278] applied post-hoc calibration to MPNN-based models and found it effective in reducing overconfident errors.

**3.2.2.4 Conformal Prediction**

Conformal prediction[269] provides an alternative framework that generates statistically valid prediction intervals or confidence sets under minimal assumptions. It assumes that the prediction error is independent and identically distributed (i.i.d.) across samples; more generally, exchangeability suffices. Data is split into three sets: training, calibration, and test. A model is trained on the training set and used to compute conformal scores on the calibration set, often defined as the absolute residual:

$$\alpha_i = |y_i - \hat{y}_i|$$

where $\hat{y}_i$ denotes the predicted value, $y_i$ is the true label, and $\alpha_i$ is the conformal score. All conformal scores are then sorted, and a threshold $d$ is determined based on the selected confidence level $1 - \epsilon$ (e.g., for a 95% confidence level, $\epsilon = 0.05$). This threshold $d$ represents the uncertainty estimated by the model. During inference, the predicted interval at confidence level $1 - \epsilon$ is given by $(\hat{y}_i - d, \hat{y}_i + d)$.

Although conformal prediction has been well-developed in the field of property prediction[279], most existing methods rely on traditional machine learning. Conformal prediction based on GNNs remains underexplored, with only a few works emerging in recent years, such as **CoDrug**[280] and **CF-GNN**[281]. Some methods, while not explicitly referencing conformal prediction, have adopted its core ideas. For example, the pLDDT (predicted Local Distance Difference Test) scoring system used in AlphaFold2[54] to evaluate the local confidence of protein structure predictions follows a principle very similar to conformal prediction, in which uncertainty is inferred jointly from the model's output and the associated error.

## 3.3 Explainable GNN for Reliable Prediction: Uncovering the Blak-box GNN

In the preceding sections, we explored the use of GNNs for molecular property prediction and introduced techniques for uncertainty quantification. However, uncertainty estimation remains fundamentally a "black-box" approach. In high-stakes decision-making scenarios such as drug design, incorporating human expert judgment is crucial. If a model can clearly highlight which molecular features or substructures drive its predictions, domain experts can better evaluate, modify, or trust the model's outputs. This motivation forms the foundation of interpretable GNNs in chemistry.

### 3.3.1 Feature Attribution and Substructure Attribution

In the broader field of neural network explanation, feature attribution is a widely used technique. Approaches such as LIME[282], DeepLIFT[283] (Learning Important Features), and SHAP[284] (Shapley Additive Explanations) aim to identify which input features contribute most significantly to a prediction. For instance, in QSAR models, where molecular descriptors such as hydrophilicity, lipophilicity, and hydrogen bond counts are used as inputs, feature attribution can help pinpoint which physicochemical properties are most influential, thus assisting medicinal chemists in evaluating model reliability.

In GNNs, however, explanation must extend beyond feature-level interpretations to the graph structure itself. Specifically, substructure attribution focuses on identifying molecular fragments or localized patterns that are most influential for a prediction. Chemically meaningful substructure attribution typically requires satisfying three criteria: 1) **Sparsity**: The explanation should involve as few atoms/bonds as possible; 2) **Connectivity**: The substructure should form a connected subgraph; 3) **Sufficiency**: Retaining only the substructure should yield similar model predictions. Substructure attribution not only reflects causal reasoning at the graph level but also aligns better with chemical intuitions. The illustration of substructure and feature attribution is provided in **Figure 10A**.

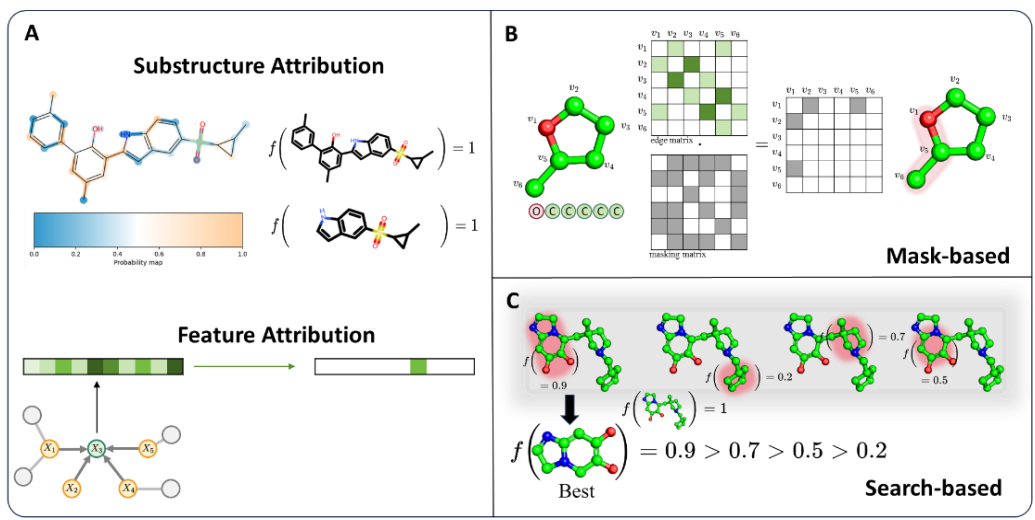

**Figure 10.** A) Comparison between substructure and feature attribution. B,C) Two examples of explainable GNN methods: mask-based methods, which learn masks over molecular graphs to identify important substructures; search-based methods, which iteratively evaluate candidate substructures to select the most interpretable one.

### 3.3.2 Factual Explanation: Post-Hoc and Self-Explanation

Explanation techniques can be broadly classified into two categories: post hoc and self-explanation approaches. Post hoc methods generate explanations after a model has been trained. A secondary "explainer" is trained separately to interpret the predictions of the original black-box model. The main advantage is modularity—the predictor and explainer are decoupled, making this approach model-agnostic, flexible, and non-intrusive. However, it may suffer from spurious correlation, where the explanation reflects artifacts of the explainer rather than the true decision process of the original model.

Self-explanation methods are designed with explanation built into the model architecture. In these methods, the explanation module is jointly trained with the predictor, ensuring that explanation is an intrinsic part of the prediction pipeline. This tight integration often leads to better causal alignment, but comes at the cost of increased complexity: such models must be trained from scratch, and require careful architectural design to ensure that the explanation module does not compromise predictive performance.

Explanation techniques can also be categorized by model visibility: model-agnostic and model-accessible methods. Model-agnostic methods require access only to model inputs and outputs. Model-accessible methods rely on internal model information (e.g., gradients, attention weights, or hidden states), and may require specific architectural components. **Table 12** summarizes representative explanation methods across these axes. In general, model-agnostic post hoc methods offer the greatest flexibility and ease of implementation, while model-accessible self-explanation methods provide the most faithful reflection of the model's internal reasoning but are the most technically demanding.

**Table 12**. Overview of explainable GNNs. F denotes factual interpretable methds; while CF denotes

counter-factual interpretable methods.

| / | Sub Category | Methods | Keywords |
|---|---|---|---|
| F | Gradient/ Parameter -based | Grad-SM[285] | Attribute to the gradient norm |
| | | CAM[286] | Limited to FC and AVA Pool as the final layer |
| | | Grad-CAM[287] | Replace weights CAM's final layer with gradients |
| | | Excitation-BP[286] | Attribute to the neuron activation |
| | | IG[288] | Gradient integration between v and v' |
| | Masking | GNNExplainer[289] | Attribute to the subgraph |
| | | PGExplainer[290] | Use Gubmel-Softmax for hard mask |
| | | GraphMask[291] | Replacing base edges instead of dropping edges |
| | Search/Generative -based | SubGraphX[292] | Attribute to connected subgraphs with MCTS |
| | | SME[293] | Attribute to chemically intuitive subgraphs |
| | | Gem[294] | Use VAE to generate explainable subgraphs |
| | Local Structure Interpretation | GAT | Attribute to attention scores |
| | | GraphLIME[295] | Use local graph as input for feature attribution |
| | | RelEX[296] | Use local graph as input for subgraph attribution |
| | | GraphSVX[297] | Unified framework including MASK, GEN, EXP |
| | Self-explain | GSAT[298] | Introduce stochastic attention for better causality |
| | | ProtGNN[299] | Attribute to the prototype closest to the data point |
| CF | Masking | CF-GNNExplainer[300] | CF extension of GNNEXplainer |
| | | CF2[301] | Combining factual and counter-factual causality |
| | | RCExplainer[302] | Decision-making boundaries enhanced |
| | Search/Generative -based | MEG[303] | Counter-factual reward guided RL generation |
| | | MMACE[304] | Generate candidates and filter them with CF loss |
| | | GCFExplainer[305] | Use random-walk to explore the CF candidates |
| | | CLEAR[306] | CF-VAE |

### 3.3.2.1 Post Hoc Methods: Training an Explainer

### 3.3.2.1.1 Gradient and Parameter-Based Methods

The earliest explanation techniques in GNNs were adapted from CNNs, including gradient-based saliency maps[285], class activation mapping (CAM)[286], gradient-based CAM[287], excitation backpropagation[286], and integrated gradients (IG)[288]. The core intuition behind these methods is that the most important parts of the input should correspond to larger gradients or parameter weights. By backpropagating from the model output, one can trace the influence of input features and identify which components contribute most to the prediction. Since all these methods require access to internal model weights or gradients, they are classified as model-accessible approaches.

Grad-SM computes the gradient of the model output with respect to its input, producing a heatmap where the norm of the gradient at each input reflects its importance. Typically, negative gradients are zeroed out to highlight features that positively activate the output neuron. The attribution score is computed as:

$$L^c_{\text{Gradient}} = \| \text{ReLU}\left(\frac{\partial y^c}{\partial x}\right) \|$$

where $y^c$ denotes the logit corresponding to class $c$ (the value before applying the softmax function), and $x$ is the input feature. While this method is intuitive and easily implemented via automatic differentiation, it has been shown to be sensitive to noise and less reliable than more advanced attribution techniques[307].

Unlike Grad-SM, CAM does not rely on gradients. Instead, it uses weights from the final fully connected layer, combined with the output of the global average pooling layer, to compute importance scores for node $n$ with respect to class $c$:

$$L^c_{CAM}[n] = \text{ReLU} \left( \sum_k w^c_k F^L_{k,n} \right))$$

where $F^L_{k,n}$ is the $k^{th}$ feature of node $n$ at layer $L$, and $w^c_k$ is the weight of $k^{th}$ feature corresponding to class $c$. However, CAM requires specific network architecture, namely, a global average pooling followed by a fully connected layer. **Grad-CAM** generalizes CAM by computing the gradient of $y^c$ with respect to feature maps $F^l_{k,n}$, thus relaxing the architectural constraints and enabling visualization at intermediate layers:

$$\alpha^{l,c}_k = \frac{1}{N} \sum_{n=1}^{N} \frac{\partial y^c}{\partial F^l_{k,n}}$$

$$L^c_{Grad-CAM}[l, n] = \text{ReLU} \left( \sum_k \alpha^{l,c}_k F^l_{k,n} \right)$$

where $\alpha^{l,c}_k$ represents the sensitivity of the prediction for class $c$ with respect to the feature $F^L_{k,n}$, and $L^c_{Grad-CAM}$ quantifies the contribution of atom $n$ at layer $l$ to the prediction of class $c$. In fact, CAM can be seen as a special case of Grad-CAM at the final layer, where the gradients directly correspond to output weights.

**Excitation-BP** attributes importance to each neuron by recursively decomposing activation values from the output layer down to the input. At each layer, the activation of a neuron can be interpreted as a weighted sum of the activations from neurons in the preceding layer. It treats the network as a probabilistic chain rule:

$$P(a^{(l-1)}_j) = \sum_i P(a^{(l-1)}_j \mid a^l_i) P(a^l_i)$$

where $a^{(l-1)}_j$ is the activation of neuron $j$ at layer $(l-1)$. Excitation-BP traces how each upstream neuron contributes to the final output, offering a probabilistic view of neural importance. Details of the conditional distributions $P(a^l_i)$ can be found in the original work.

**IG**[288] estimates the importance of an input feature $x_v$ by integrating the gradients along a straight-line path from a baseline input $x'_v$ to the actual input:

$$\text{IG}(x_v) = (x_v - x'_v) \int_\Omega \frac{\partial f(x'_v + \alpha(x_v - x'_v))}{\partial x_v} d\alpha$$

where IG($x_v$) denotes the integrated gradient, and the integration domain Ω is typically set to [0,1]. In practice, the continuous integral is approximated using Riemann summation, yielding:

$$\text{IG}(x_v) \approx \frac{(x_v - x'_v)}{m} \sum_{r=1}^{m} \frac{\partial f(x'_v + \frac{r}{m}(x_v - x'_v))}{\partial x_v}$$

Unlike simple difference methods $f(x_v) - f(x'_v)$, IG provides a more nuanced measure of feature importance by aggregating gradients along the entire input path.

These foundational methods form the basis for many explainability studies in molecular property prediction. For instance, Sanchez-Lengeling et al.[308] benchmarked multiple attribution methods across four GNN architectures and found that CAM achieved the best performance across variants, particularly with GCN models. Later, Jiménez-Luna et al.[309] extended this work to investigate activity cliffs, using random forest (RF) as a baseline. Rao et al.[310] constructed domain-specific benchmarks involving hepatotoxic substructures, Ames mutagenicity motifs, and CYP3A4-related cliffs, demonstrating how explainability tools can be grounded in chemical relevance. Despite these efforts, most studies to date have applied standard attribution techniques to GNNs, rather than designing explanation mechanisms intrinsic to the GNN architecture itself. This may partly explain why Jiménez-Luna et al. observed that RF combined with explainability often outperformed GNNs paired with traditional attribution methods.

### 3.3.2.1.2 Mask-Based Methods

**GNNExplainer**[289] is one of the earliest interpretation frameworks specifically designed for GNNs, focusing on the substructure attribution. It formulates the task of identifying informative subgraphs as an information maximization problem, aiming to find a subgraph that retains the most mutual information with the model prediction:

$$\max_{G_S} \text{MI}(Y, G_S) = \text{H}(Y) - \text{H}(Y \mid G = G_S).$$

where MI denotes the mutual information between the predicted label $Y$ and the selected subgraph $G_S$, and H is the entropy function. Since the predictive model is fixed, $H(Y)$ is constant, and the optimization objective reduces to minimizing conditional entropy:

$$\min_{\mathcal{G}} \mathbb{E}_{G_S \sim \mathcal{G}} H(Y \mid G = G_S)$$

In this formula, the objective becomes minimizing the predictive uncertainty conditioned on a given subgraph $G_S$. To obtain such a subgraph, GNNExplainer adopts a mean-field variational inference approach, where the graph is modeled as a set of independent edges governed by Bernoulli distributions: $P(G) = \Pi_{(i,j) \in \mathcal{E}} P(e_{ij})$, with $P(e_{ij} = 1) = \theta_{ij}$. A learnable mask matrix $\sigma(M_E)$ is introduced, from which the subgraph is generated as $G_s = A \odot \sigma(M_E)$. Accordingly, the optimization objective becomes:

$$\min_{\mathcal{G}} \mathbb{E}_{G_S \sim \mathcal{G}} H(Y \mid G_s = A \odot \sigma(M_E))$$

Building on GNNExplainer, **PGExplainer**[290] points out that the output of $\sigma(M_E)$ lies in the continuous range [0,1], forming a soft mask that still requires post-processing (e.g., thresholding) to obtain a binary hard mask. To overcome this limitation, PGExplainer employs the Gumbel-Softmax[311] trick, a differentiable approximation to sampling discrete variables, enabling end-to-end training of binary edge selections. The new objective becomes:

$$\min_{\mathcal{G}} \mathbb{E}_{G_S \sim \mathcal{G}} H(Y \mid G = G_S \odot \text{Gumbel} - \text{Softmax}(M_E))$$

Notably, although PGExplainer leverages random Gilbert graphs to model subgraph distributions, whereas GNNExplainer uses mean-field variational modeling, the essential ideas are similar: both methods reduce substructure search to learnable mask optimization. The following **GraphMask**[291] argues that direct binary masking may break the graph's topological integrity. To address this, it proposes edge replacement instead of removal: when $z_{u,v} = 0$, the message from edge is replaced by a learnable vector $b$. The modified edge message becomes: $\widetilde{m}_{uv} = z_{u,v} \cdot m_{u,v} + b \cdot (1 - z_{u,v})$, $z_{u,v} \in \{0,1\}$. This strategy preserves information flow and mitigates the destructive effects of hard masking on graph connectivity.

In this class of mask-based methods (**Figure 10B**), sparsity and connectivity are key constraints for generating chemically meaningful explanations. Effective rationales are expected to be compact and localized, ideally corresponding to functional groups rather than dispersed atomic patterns. While GNNExplainer introduces sparsity through soft regularization to promote subgraph compactness, it does not explicitly enforce topological connectivity, which limits its ability to produce chemically realistic substructures. Despite this limitation, GNNExplainer and its derivatives have been widely adopted in drug discovery and molecular property prediction, providing a flexible framework for interpretable graph learning.

**3.3.2.1.3 Search- and Generation-Based Methods**

In addition to mask-based approaches for identifying explanatory substructures, researchers have also developed search-based strategies. In principle, one could enumerate all possible subgraphs and select the one that contributes most to the model's prediction as the explanation. However, this strategy faces a well-known combinatorial explosion in graph-structured data, making exhaustive search computationally infeasible. To address this, recent work has explored efficient graph search algorithms (e.g., MCTS) and domain-constrained substructure generation to make the task tractable.

**SubGraphX**[292] is a representative method that uses graph search to identify important subgraphs, as illustrated in **Figrue 10C**. It leverages Shapley values[312], a concept from cooperative game theory, to quantify the contribution of each subgraph to the model's prediction. In this formulation, nodes are treated as players, and subgraphs (node combinations) are treated as coalitions. The Shapley value provides a principled and fair way to assess the importance of each coalition. Based on this, SubGraphX uses the Shapley value as a scoring function within MCTS[180] to efficiently explore the subgraph space. As

MCTS incrementally expands the search from an initial node, each candidate subgraph is guaranteed to be connected, thereby satisfying the connectivity requirement for chemical explanation. However, while SubGraphX forces the explanations connected, SME[293] (Structure-aware Molecular Explanation) argues that such subgraphs may lack chemical interpretability. To address this, SME incorporates domain-informed decomposition techniques, such as Murcko scaffolds[313] and BRICS fragmentation[232], to generate chemically meaningful substructures that serve as candidates for explanation. This approach not only narrows the search space but also grounds the rationale in functional motifs. For each candidate subgraph, SME estimates its contribution via a simplified Shapley value, computed as the change in prediction when the substructure is removed, $f(G) - f(G_s)$. A complete enumeration is then performed to select the fragment that best aligns with the model's decision.

Beyond mask- and search-based techniques, directly generating explanations is also a hot topic in GNN explanation. For example, XGNN[314] treats the model to be explained as a feedback function within a RL framework. It learns to generate graph structures that are assigned high confidence scores by the target model, thereby revealing global patterns associated with specific classes or predictions. However, this approach is geared more toward global model-level explanation, uncovering general rules or motifs, rather than explaining individual predictions. As such, it is less commonly applied in practical tasks related to drug discovery, where instance-level explanations are typically preferred. In contrast, Gem[117] focuses on instance-level explanation by quantifying the causal influence of individual edges based on a Granger Causality-inspired metric[315]. Specifically, for a trained model $f$, a true label $y$, and a loss function $\mathcal{L}$, the influence of an edge $e_{ij}$ is defined as: $\Delta_{e_{ij}} = \mathcal{L}(y, f(G)) - \mathcal{L}(y, f(G \setminus e_{ij}))$. Using this criterion, Gem identifies target explanatory substructures over the training set. Then, a graph autoencoder is trained to learn a mapping from the input graph to the computed explanatory substructure. In summary, the core idea of Gem is to use causality-inspired attribution to define ground-truth explanations in a data-driven manner, and then train a model to reproduce them. In drug design, a similar philosophy can be applied by incorporating chemically meaningful criteria, e.g., known pharmacophores or toxicophores, to define explanation ground truth. These targets could then be learned by generative models that map full molecular graphs to these small, interpretable substructures.

### 3.3.2.1.4 Local Structural Methods

In large-scale graphs such as knowledge graphs, it is often infeasible to perform global explanation analysis due to the graph's sheer size and complexity. This challenge has motivated the development of local structural explanation strategies. Since many graph reasoning tasks, such as node classification, are inherently local, some studies argue that explanation can be achieved by focusing on the neighborhood of the target node. To do so, GraphLIME[295] extends the well-known LIME method[282], originally developed for tabular feature attribution, to the context of subgraph explanations. LIME uses the Hilbert-Schmidt

Independence Criterion (HSIC) to quantify nonlinear dependence between input features and model predictions, and employs Lasso regression for sparsity, retaining only the most informative features to reduce overfitting. GraphLIME adapts this idea by replacing the original feature space with the *k*-hop neighborhood of the target node $v_i$, applying the LIME procedure to select the top $n$ most important nodes as the explanation. **RelEx**[296] builds upon the mask-based explanation paradigm. It begins by performing Breadth-First Search (BFS) to sample subgraphs with a fixed probability, thereby constructing a local explanation dataset. A joint loss function is then defined based on the prediction target $Y$ and the distributional difference induced by masking parts of the subgraph:

$$\arg\min_{M} L = L(Y_i, g(A_i \odot M)) + \Psi(M)$$

where $\Psi(M)$ is a regularization term to promote sparsity in the learned mask matrix $M$. To generate hard masks (i.e., binary-valued masks), RelEx employs the Gumbel-Softmax technique, enabling discrete optimization within a differentiable framework. Among local explanation methods, **GraphSVX**[297] proposes a unified framework that jointly handles both feature attribution and structural explanation. It decomposes the explanation process into three modular operations: MASK, GEN and EXP. This design enables a holistic interpretation of node-level predictions by considering both attribute and topology perturbations.

### 3.3.2.2 Self-Explaining Methods: Predictors Explain

As discussed earlier, post-hoc methods operate by training an explanation model $g_\phi$ separately on top of a pre-trained predictor $f_\theta$, aiming to generate an explanatory subgraph $G_s = g_\phi(G)$ while preserving the model's original prediction. However, from a theoretical standpoint, such approaches are suboptimal, as the explanation model and the predictor are optimized in isolation. Without joint training, the explanation model cannot guarantee proximity to the global optimum of a combined predictive-explanatory objective. This decoupling explains a recurring phenomenon observed in post-hoc methods such as PGExplainer and GraphMask: these methods often demonstrate strong explanation in the first few (e.g., 5–10) training iterations, but their explanation quality deteriorates as training progresses, even though the prediction loss continues to decrease. This disconnect highlights the need for self-explaining methods, where $f_\theta$ and $g_\phi$ are optimized jointly to ensure consistency between prediction and explanation.

Self-explaining methods aim to co-train the predictor and the explainer, thereby addressing the consistency gap between explanation and prediction. Some GNN architectures inherently support structure-aware explanation, as they internally compute edge importance scores. Examples include GAT, Top-k Pool[316], and SAG Pool[317]. Take GAT as an example: it computes attention coefficients on edges, which naturally highlight the importance of different parts of the graph. High attention scores correspond to influential edges, while low scores indicate less relevant connections. However, such attention-based explanations have epistemological limitations. Specifically, attention weights do not always align with causal importance, and may reflect learned associations that do not generalize well across data distributions[289, 318]. As a refinement, **GSAT**[298] (Graph Stochastic Attention) introduces a new framework based on stochastic attention mechanisms. GSAT first computes edge attention scores $p$, and then samples discrete hard masks $M_\alpha$ from a Bernoulli distribution based on $p$. The masked subgraph is then passed through a GNN for downstream prediction. This mechanism combines the discrete explanation of hard masking with the flexibility of learned attention.

Another representative self-explaining model is **ProtGNN**[299], which adopts a prototype learning paradigm. It introduces a set of $k$ learnable prototypes $h_i$, and computes the similarity between the graph embedding $h$ (obtained from a GNN) and each prototype. These similarity scores are then used to generate the final prediction. Intuitively, this approach mirrors classical fragment-additive models in cheminformatics[319], where predictions are composed from predefined functional groups—except here, the "fragments" are learned prototypes in the latent space. To extract interpretable substructures corresponding to each prototype, ProtGNN employs MCST and mask generation techniques to localize the most relevant regions in the original graph. In summary, self-explaining GNNs is an emerging and active research direction. A key open question is how to leverage the causal insights revealed by self-explaining mechanisms to enhance predictive performance in real-world tasks.

### 3.3.3 Counterfactual Explanation Methods: More Natural to Drug Discovery

Until now, most explanation methods for GNNs have focused on identifying a subgraph $G_s$ that explains the model's prediction $f(G)$, which follows the factual reasoning perspective. From a causal inference standpoint, however, there exists a complementary approach which asks: *"What would happen if a certain part of the graph did not exist?"* This is counterfactual reasoning. Whereas factual reasoning considers *whether B would happen given A*, counterfactual reasoning examines *whether B would still happen if A did not*. In the context of GNN explanation, this translates into identifying subgraphs $G_s$ whose removal significantly alters the model prediction. Such counterfactual explanations are particularly well-suited for tasks like activity cliff detection in molecular modeling. In this section, we would review counterfactual explanation methods that offer another causal perspective on model behavior.

#### 3.3.3.1 Mask-Based Counterfactual Generation

**CF-GNNExplainer**[300] extends the idea of GNNExplainer by modifying the optimization objective to reflect counterfactual logic. The objective combines a prediction divergence term with a structural similarity constraint:

$$\mathcal{L} = \mathcal{L}_{\text{pred}}(f(G), f(G_s)) + \beta \mathcal{L}_{\text{dist}}(G, G_s)$$

Here, $\mathcal{L}_{\text{pred}}$ aims to maximize the difference between the predictions on the original graph $G$ and the subgraph $G_s$, while $\mathcal{L}_{\text{dist}}$ penalizes structural differences between them. Ideally, the optimal subgraph $G_s$ should differ minimally from $G$ yet cause a significant change in prediction, making the residual graph $G - G_s$ a meaningful counterfactual explanation. Similar to GNNExplainer, CF-GNNExplainer employs a learnable masking model $g$ to generate the explanatory subgraph via:

$$G_s = A \odot \sigma(M_E) = A \odot \sigma(g(G))$$

The loss function is defined as:

$$\mathcal{L}_{\text{pred}}(f(G), f(G_s)) = -\mathbb{1}[f(G) = f(G_s)] \odot \mathcal{L}_{\text{NLL}}(f(G), f(G_s))$$

where $\mathbb{1}$ is an indicator function that evaluates whether $f(G)$ equals $f(G_s)$, and $\mathcal{L}_{\text{NLL}}$ denotes the negative log-likelihood, quantifying prediction divergence.

**CF2**[301] (Counterfactual and Factual) generalizes CF-GNNExplainer by incorporating both factual and counterfactual loss terms, such that $G_s$ explains why the model made a certain prediction (factual), while $G \backslash G_s$ explains how that prediction would change if certain components were absent (counterfactual). Similarly, **RCExplainer**[302] adopts this dual optimization approach. The overall objective is:

$$\mathcal{L}(\theta) = \sum_{G \in D} \{\lambda \mathcal{L}_{\text{same}}(G_s, G) + (1 - \lambda)\mathcal{L}_{\text{opp}}(G, G \backslash G_s) + \beta \mathcal{R}_{\text{sparse}}(M) + \mu \mathcal{R}_{\text{discrete}}(M)\}$$

Unlike earlier methods that use simple prediction agreement (e.g., $-\mathbb{1}[f(G) = f(G_s)]$) to quantify explanation quality, RCExplainer proposes a more geometric approach: it explicitly models decision regions in the final GNN layer as convex polytopes[320]. That is to say, each class is associated with a distinct decision region; graphs that map into the interior of region $i$ are classified as class $i$. To this end, RCExplainer employs a progressive peeling strategy to identify the decision boundaries of each class and rewrites the loss functions as:

$$\mathcal{L}_{\text{same}}(G_s, G) = \frac{1}{|\mathcal{H}_G|} \sum_{h_i \in \mathcal{H}_G} \sigma(-\mathcal{B}_i(f(G)) * \mathcal{B}_i(f(G_s))),$$

$$\mathcal{L}_{opp}(G, G \backslash G_s) = \min_{h_i \in \mathcal{H}_G} \sigma(\mathcal{B}_i(f(G)) * \mathcal{B}_i(f(G \backslash G_s))),$$

$$\mathcal{R}_{\text{sparse}}(M) = \|M\|_1$$

$$\mathcal{R}_{\text{discrete}}(M) = -\frac{1}{|M|} \sum_{i,j} (M_{ij} \log(M_{ij}) + (1 - M_{ij}) \log(1 - M_{ij}))$$

Here, $\mathcal{B}_i = w_i^T x + b_i$ represents the decision boundary for class $i$, and the sign of $\mathcal{B}_i(x_i)$ indicates on which side of the hyperplane a sample lies. $\mathcal{H}_G$ is the set of decision boundaries. The loss $\mathcal{L}_{\text{same}}$ encourages the subgraph $G_s$ to lie on the same side of the decision boundary as $G$ (i.e., preserving the class), while $\mathcal{L}_{opp}$ encourages $G_s$ to cross the boundary (i.e., switch class). $\mathcal{R}_{\text{sparse}}$ enforces sparsity in the mask, and $\mathcal{R}_{\text{discrete}}$ promotes binarization of the mask elements.

### 3.3.3.2 Search- and Generation-Based Counterfactual Methods

In addition to mask-based approaches, counterfactual reasoning in GNNs has also been explored through search-based and generative methods. Analogous to factual explanation techniques, search-based methods employ RL or other efficient algorithms to navigate the combinatorial explosion of subgraph possibilities.

**MEG**[303] (Molecular Explanation Generator) is a representative RL-based search approach. It adopts a counterfactual loss as the reward signal, enabling the agent to explore the space of plausible substructures. Similar to CF-GNNExplainer, the objective function is defined as:

$$\mathcal{L} = \mathcal{L}_{\text{pred}}(f(G), f(G_s)) + \beta \mathcal{L}_{\text{dist}}(G, G_s)$$

The agent's actions in RL consist of adding or removing atoms and bonds. After multiple iterations, the model converges on a substructure that serves as a locally optimal solution to the counterfactual objective. **MMACE**[304] (Molecular Model-Agnostic Counterfactual Explanations) generates the candidate subgraphs first, followed by counterfactual scoring. It uses the STONED[156] graph generation method to produce a batch of perturbed molecules $\{G_s\}$, and applies the following counterfactual selection criterion:

$$\min_{G_s} d(G, G_s), \text{ s.t. } f(G) \neq f(G_s).$$

Given the large size of the search space, MMACE clusters the generated candidates and selects the cluster centroids as representative counterfactuals. One potential limitation is that rare but informative counterfactuals may be excluded during clustering, affecting the coverage and diversity of explanations. **GCFExplainer**[305] (Global Counterfactual Explainer) follows a similar structure to MMACE but introduces a more efficient exploration strategy based on vertex-reinforced random walks (VRRW)[321]. It first constructs an edit map centered on the input graph $G$, where each node represents a candidate molecule and edges represent single-step graph edits. VRRW is then used to explore the neighborhood of $G$ in the edit map, using the same counterfactual selection criterion as MMACE. The method ultimately identifies a counterfactual that optimizes both proximity and predictive divergence.

CLEAR[306] is a generative-based counterfactual method that integrates counterfactual loss directly into the training of a VAE. Unlike GEM, which uses generation for factual explanation, CLEAR is tailored to produce counterfactual graphs as the output of the VAE decoder. The standard VAE loss is given by:

$$\mathcal{L}_{VAE} = \mathcal{L}_{recon} + \mathcal{L}_{prior}$$
$$= \mathbb{E}_Q[\ln P(G_{recon} \mid Z, Y, G)] - \text{KL}(Q(Z \mid G, Y) \parallel P(Z \mid G, Y^*))$$

To generate a counterfactual graph $G_{CF}$, the decoder is encouraged to produce outputs that are structurally close to the original graph while yielding different predictions. The modified reconstruction term becomes:

$$\mathbb{E}_Q[\ln P(G_{recon} \mid Z, Y, G)] = \mathbb{E}_Q[d(G, G_{CF}) + \alpha \mathcal{L}(f(G_{CF}), f(G))]$$

where $d(G, G_{CF})$ enforces similarity to the original graph, while $\mathcal{L}(f(G_{CF}), f(G))$ ensures prediction divergence. While counterfactual methods are still in thee early stages of application to molecular problems, they hold great promise for enabling causal inference in drug discovery. Interestingly, the formalism of counterfactual reasoning closely mirrors that of activity cliffs, where small molecular changes lead to sharp changes in activity. This conceptual alignment suggests that counterfactual GNNs may offer a more principled framework for understanding critical structure-activity relationships. We anticipate that these techniques will become increasingly influential in rational drug design and beyond.

### 3.3.4 Challenges of XAI in Drug Discovery

Conceptually, explainable AI (XAI) methods have wide-ranging applications in drug discovery. For example, when applied to the molecular property prediction tasks, explainability naturally evolves into techniques for identifying substructures[322], toxicophores[323], or pharmacophores[323]. In the context of protein-ligand binding affinity prediction, explainability methods help characterize protein–ligand interactions[324]. When applied to reaction prediction, they enable mechanistic interpretation of reaction pathways[325].

However, in practice, the majority of explainability applications still rely on **classical attribution techniques**, such as CAM, Grad-CAM, and Integrated Gradients. In contrast, the GNN-specific explainability methods introduced in this review are often limited to conceptual demonstrations: frequently evaluated on small benchmark datasets like MUTAG[326], which consists of only 4,337 molecules and distinguishes mutagenic activity based on the presence of functional groups such as $NH_2$ or $NO_2$. Explanatory models are typically assessed by their ability to highlight these predefined motifs. This highlights a gap between method development and real-world deployment, where the chemical complexity far exceeds that of simplified benchmarks.

In the current era of foundation models and leaderboard-driven molecular benchmarks[327], property prediction tasks are increasingly being solved by powerful pre-trained architectures. As these models grow in scale and complexity, explainability techniques must evolve to augment them. This evolution should go beyond providing post-hoc insights: it must also uncover causal relationships and correct learned biases, thereby enhancing both robustness and generalizability. Furthermore, researchers should explore the use of XAI not merely as a tool for explanation but as a means to improve predictive performance. Instead of trading accuracy for explanation[328], the goal should be to leverage causality-aware explanations to enhance model reasoning. After all, a model that learns the correct causal structure should inherently generalize better to unseen data.

## 4. Virtual Screening

### 4.1 Binding Site Prediction

Binding site identification serves as the starting point in virtual screening, as illustrated in **Figure 11**. While binding sites are often well-characterized in later stages of drug development, their prediction remains a critical and non-trivial challenge in earlier phases such as target identification and validation. Traditional binding site prediction approaches largely rely on template-based methods, where known ligand-binding information is transferred to target proteins through structural alignment. However, these approaches are inherently constrained by the coverage and diversity of existing template libraries[329]. More recently, language models applied directly to protein sequences have been explored for this task[330], yet these methods often lack explicit incorporation of 3D structural information, limiting their predictive accuracy. In contrast, GNNs offer a promising alternative by operating directly on protein conformations, enabling the learning of intricate spatial geometries and energetic features that are critical for accurate binding site identification.

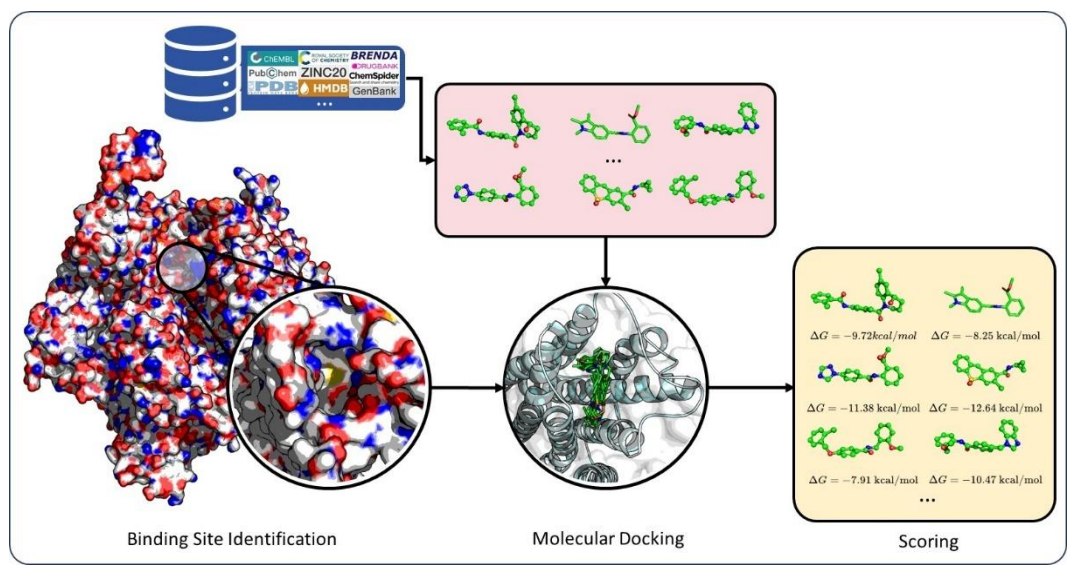

**Figure 11**. Virtual screening workflow.

### 4.1.1 Task Descriptions and Datasets Available

Proteins are capable of interacting with a diverse range of molecular partners, including macromolecules such as DNA, RNA, and other proteins, as well as small molecules such as chemical compounds, peptides, and metal ions. Accordingly, the identification of binding sites on proteins can be categorized into several distinct tasks, including protein-small molecule binding site prediction, protein-protein interaction interface prediction, and others. In the context of drug discovery, the primary focus lies in predicting protein-small molecule binding sites. Unless otherwise specified, the term "binding site" throughout this review refers specifically to protein-small molecule interactions. **Table 13** summarized the related datasets.

**Table 13**: Overview of datasets for protein-ligand interaction, including binding site prediction, docking, and scoring problem.

| Dataset | Keywords |
| --- | --- |
| BioLip[331] | Protein-ligand complex with binding site annotation |
| PDBBind[121] | For Docking and Scoring, 3D crystals and affinity labels |
| PoseBuster[332] | An external benchmark for Docking |
| BindingMoad[333] | Similar to PDBBind but contains data without affinity labels |
| CASF[121] | The core dataset of PDBBind |
| DEKOIS2.0[334] | 81 targets, each with 30 actives and 1200 decoys for screening |
| DUD-E[335] | 102 targets, each with varying actives and 50x decoys. |
| LIT-PCBA[336] | 15 targets, with 10030 actives and 2m decoys in total (from PubChem) |
| Merck FEP[337] | 8 targets, 264 actives, with FEP+ energy computation (high quality) |
| BindingDB[338] | 1265 targets, 542126 exp-affinity, but without binding structures |

A widely used dataset for training binding site prediction models is BioLiP[331], which provides curated information on protein-ligand complexes, including ligand-binding residues, binding affinities, and catalytic site annotations. In BioLiP, a residue is defined as a binding residue if at least one of its atoms lies within a certain distance threshold of any atom in the ligand. Mathematically, this is defined as:

$$d_{\min}(r, l) < r_{\text{vdW}}(r) + r_{\text{rdW}}(l) + 0.5\text{Å}$$

where $d_{\min}(r, l)$ denotes the minimum distance between any atom in residue $r$ and any atom in ligand $l$, and $r_{\text{vdW}}(r)$, $r_{\text{rdW}}(l)$ represent the van der Waals radii of the atoms in the residue and ligand, respectively. Under this formulation, binding site prediction is generally treated as a binary classification problem, where the goal is to determine, given a protein structure, whether specific nodes (residues or atoms) constitute a ligand-binding site.

### 4.1.2 Method Development

Using GNN for binding site prediction is an emerging area of research, which is summarized in **Table 14**. Specifically, **GraphSite**[339] represents protein residues as nodes and utilizes a JK-Net architecture to capture inter-layer message passing. It employs focal loss as the objective function to mitigate the issue of class imbalance between binding and non-binding residues. A similar architecture is adopted by **GrASP**[340] (Graph Attention Site Prediction), which incorporates a 12-layer GATs, along with the residual connections to alleviate the over-smoothing problem. Furthermore, GrASP integrates the geometry-based pretraining method, i.e., Noisy Node, which we previously discussed, to enhance the model's robustness to structural noise and improve generalization capabilities. In addition to residue-level representations, surface-based approaches such as **MaSIF**[341] discretize solvent-accessible surfaces (SAS) into triangular meshes and directly tackle the task of surface binding site prediction. From a modeling standpoint, these representation strategies primarily differ in the type of nodel-level information they encode, though all leverage GNN-based message passing for feature learning. For surface-based modalities, models such as MaSIF, SurfGen[342], and GeoBind[343] have introduced geodesic-aware message passing schemes specifically designed to capture the geometric properties of protein surfaces.

Table 14. Overview of binding site predition methods.

| Methods | Keywords |
|---|---|
| GraphSite[339] | Residue-based protein graph + JKNet |
| GrASP[340] | Residue-based protein graph + GAT+ Noisy-node |
| PeSTo[344] | 3D Graph Transformer |
| ScanNet[345] | Frame averaging-based 3D GNN |
| GPSite [346] | Multi-task learning of different binding entities |
| ZeroBind[347] | MAML-based meta-learning for binding site preditction |

From a physical standpoint, protein binding site prediction is inherently an SE(3)-invariant problem: the prediction outcome should remain unchanged under arbitrary translations and rotations of the input protein structure. Consequently, incorporating equivariant architectures into binding site prediction has emerged as a key research direction. **PeSTo**[344], for instance, employs a geometric Graph Transformer framework for generalized binding site prediction, including protein-small molecule interactions. **ScanNet**[345] adopts an alternative model-agnostic strategy for incorporating invariance, known as Frame Averaging[348], to endow GNNs with SE(3)-invariance, applying it to protein-protein interface prediction. **GPSite**[346] (Geometry-aware Protein Binding Site predictor) frames binding site prediction as a multi-task learning problem. It combines pre-trained embeddings from the protein language model ProtTrans[349] with geometric features extracted via an SE(3)-equivariant GNN, and performs predictions on ESMFold-predicted structures[350]. By integrating both sequence-level and structural information, and by training on predicted rather than experimentally resolved structures, GPSite achieves strong generalization performance even on low-confidence conformations derived from structure prediction models.

A notable recent development is **ZeroBind**[347], which formulates the binding site prediction task as a zero-shot learning problem. Specifically, ZeroBind treats each protein-ligand pair as a distinct task and employs MAML, a zero-shot learning strategy also discussed in Chapter 2 Property Prediction, to make the model generalize well on new protein-ligand pairs. To account for variation across tasks, ZeroBind incorporates a task-level self-attention mechanism that quantifies the relative importance of each protein-ligand pair during meta-optimization. For further details on meta-learning, readers are referred to Chapter 2.

## 4.2 Protein-Ligand Binding Structure Prediction: Docking

A second topic in virtual screening is protein-ligand binding conformation prediction, commonly known as molecular docking. Molecular docking aims to simulate the interactions between small molecules and target proteins in order to predict their most probable binding poses and associated binding affinities. Traditional docking approaches typically follow a two-step process: conformation sampling followed by pose scoring. First, a set of potential binding conformations is generated using a conformational search algorithm, and these conformations are then ranked via a scoring function, with the top-scoring poses selected as the final predictions. However, this process faces two major limitations. First, the conformational search is computationally expensive due to the high complexity of the protein-ligand potential energy surface. Although a range of heuristic algorithms have been employed to accelerate this process, such as the genetic algorithm in AutoDock[351] and the simulated annealing technique in LeDock[352], molecular docking remains time-consuming. Screening large-scale compound libraries such as

ZINC or ChEMBL typically requires CPU time on the scale of years. Second, the accuracy of scoring functions is limited. Most conventional scoring functions rely on empirical parameters and simplified physical models, which often fail to generalize well in realistic biological settings. In recent years, the rapid development of GNNs has inspired a new wave of research aimed at improving both the accuracy and efficiency of molecular docking. As a natural representation of molecular structures, GNNs can effectively model protein-ligand interactions and predict atomic coordinates, thereby enabling docking pose prediction in an end-to-end learning framework. In this section, we would review recent advances in molecular docking powered by GNNs. **Table 13** provides a summary of the relevant datasets, while **Table 15** categorizes docking methods into three classes.

Table 15. Overview of docking methods, SR refers to the success rate (the ratio of RMSD<2.0Å) on PoseBuster, which is extracted from the literature report.

| Category | Methods | SR(%) | Keywords |
|---|---|---|---|
| Semi-Flexible Docking | DeepDock[353] | 18 | Train a pose scoring function followed by searching |
| | EDM[354] | - | Predict DM and reconstruct it to 3D coords |
| | TankBind[355] | B-15.0 | Use triangular attention to predict DM |
| | UniMol[327] | 22.9 | A large pre-training model to predict DM |
| | CarsiDock[356] | 79.6 | Multi-startegy in DM protocol |
| | Equibind[357] | B-2.6 | Use EGNN to directly update 3D coords |
| | LigPose[358] | - | Use EGNN with recyclingto optimize 3D coord. |
| | KarmaDock[58] | 30.4 | EGNN and MDN for docking and scoring |
| | DiffDock[359] | B-37.9 | Diffuse internal coordinates |
| | HelixDock[360] | 85.2 | Diffuse atomic coord., PL docking scaling law |
| | FABind[361] | - | Combine binding site and conf. prediction |
| | FABind+[362] | - | Permutation-invariant loss & dynamic pocket radius |
| Co-Folding | RF-AA[363] | B-42.0 | Extend RoSettaFold to all molecules |
| | AlphaFold3[364] | B-76.4 | Diffuse atomic coord. with a non-equivariant model |
| | UMol[365] | B-21.0 | Extend AF2 to the molecular docking, DM protocol |
| | NeuralPLexer[366] | B-54.9 | AF2's condition to guide the PL atomic diffusion |
| Flexible Docking | FlexPose[367] | - | LigPose v2, add sidechain orientation with GVP |
| | DiffBindFR[368] | 50.2 | Co-predict torsion angle of protein side-chain |
| | DynamicBind[369] | - | Extend DiffDock on protein internal coord. update |
| | Re-Dock[370] | - | Diffusion bridge to model the apo to holo conf. |

DM is the distance matrix; PL is protein-ligand; conf. is conformation.

### 4.2.1 Task Descriptions and Datasets Available

**PDBBind:** PDBBind[121] is one of the most widely used benchmark datasets for protein-ligand binding affinity prediction and is also commonly employed for evaluating binding pose prediction tasks. It is constructed by collecting experimentally resolved 3D structures of biomolecular complexes with binding affinity measurements (e.g., $K_i$, $K_d$, $IC_{50}$) from the PDB[371]. The dataset encompasses a broad range of complex types, including protein-ligand and nucleic acid-ligand interactions. PDBBind is organized into three subsets: General Set, Refined Set, and Core Set. The curation criteria of PDBBind are primarily tailored toward binding affinity prediction, excluding crystal structures that lack experimental activity data. Consequently, for docking pose prediction tasks, which may benefit from structural diversity over affinity labels, researchers may consider constructing larger datasets by incorporating additional structure-only entries. The detailed curation process for PDBBind will be discussed further in the section on protein-ligand binding affinity prediction.

**PoseBuster:** PoseBuster[332] is a recently proposed dataset that curates high-quality protein-ligand complex structures released after 2021. Importantly, PoseBuster has no overlap with PDBBind v2020, making it a suitable independent test set for evaluating docking models trained on PDBBind. This setting exemplifies a practical application of temporal data split. In addition, PoseBuster introduces a set of conformation validation protocols to rigorously assess the physical plausibility of predicted docking poses, thereby enhancing the reliability of pose evaluation.

### 4.2.2 Semi-flexible Docking: Distance Geometry and Direct XYZ

Semi-flexible docking (**Figure 12A**), often referred to as rigid docking in machine learning literature, typically assumes that the protein binding pocket remains static during the docking process while allowing conformational changes in the ligand. This approach must be distinguished from historical rigid docking, which further constrains the ligand to a fixed internal geometry, disallowing any internal flexibility. Semi-flexible docking has gained widespread use in virtual screening due to its favorable balance between computational cost and predictive accuracy. By avoiding the complexity of modeling receptor flexibility, it enables high-throughput evaluation of ligand binding poses against a fixed protein conformation. Early GNN-based docking frameworks predominantly adopted this semi-flexible setting, wherein ligand coordinates are updated during simulation, while the protein structure remains unchanged.

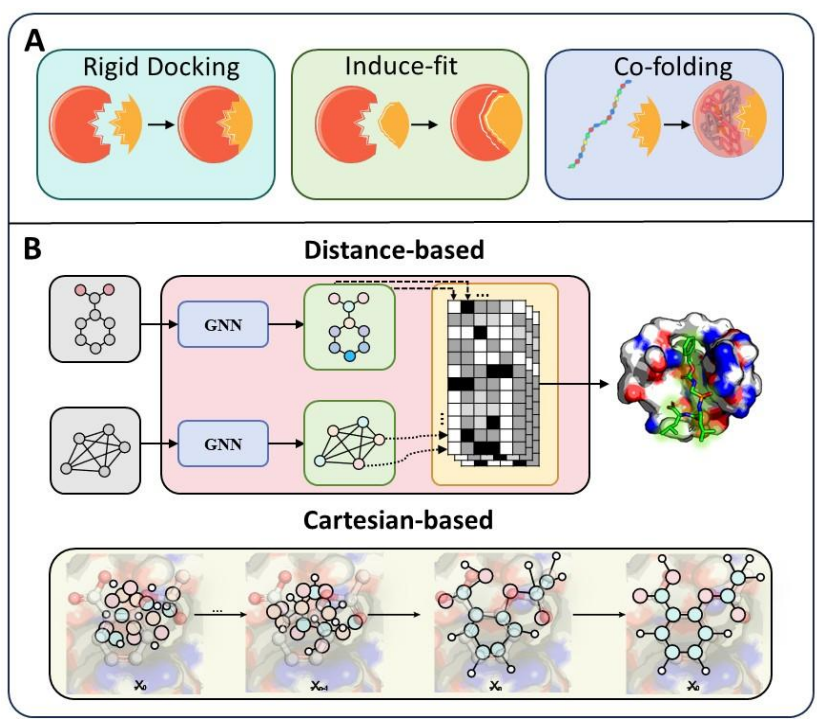

**Figure 12.** A) Illustration of three protein-ligand conformation prediction strategies. B) Distance-based and Cartesian-based protocols for GNN-based docking methods.

**4.2.2.1 Regressing Distance Matrices**

Historically, molecular conformation prediction has primarily relied on distance geometry-based algorithms, which first sample pairwise distance matrices prior and then reconstruct Cartesian coordinates **(Figure 12B)**. Early GNN-based methods largely followed this distance-driven paradigm, with **DeepDock**[353] representing an early exploration in this direction. Rather than directly predicting molecular coordinates, DeepDock constructs a scoring function using a GNN combined with a Mixture Density Network (MDN). Specifically, DeepDock employs MaSIF to encode protein surfaces, while small molecules are represented as 3D molecular graphs. A deep residual GNN is then trained to learn the interaction potential between protein-ligand pairs, which is modeled as a multivariate Gaussian mixture distribution via the MDN. The MDN parameterizes the interaction distribution using a set of learnable variables $\{\mu, \sigma, \pi\}$, corresponding to the means, variances, and mixture coefficients of the Gaussian components. During training, the model minimizes the **negative log-likelihood** of the observed interatomic distances, thereby encouraging the predicted distribution to approximate empirical interaction patterns. The parameters $\{\mu, \sigma, \pi\}$ are optimized via gradient descent, enabling iterative refinement of the interaction landscape. Once trained, DeepDock uses the learned interaction potential as a scoring function during inference. That is to say, the learned MDN is combined with traditional conformational search algorithms to explore the conformational space of the protein-ligand complex, ultimately predicting binding conformations.

An alternative strategy is to directly predict the protein-ligand distance matrix, from which Cartesian coordinates can be reconstructed. **EDM**[354] is a representative approach of this kind. **TankBind**[355] adopts a similar idea. It first identifies potential binding sites using P2Rank[372], then predicts a distance matrix $D^{pred}$ between the ligand and either protein residues or atoms, followed by ligand coordinate refinement. This optimization can be formalized as:

$$\min_{x^l} \| D^{opt} - D^{pred} \|_2,$$

$$\text{where } D^{opt}_{ij} = \|x^l_i - x^l_j\|$$

where $x^l_i$ denotes the optimized 3D coordinates of the ligand atoms, and $D^{opt}$ is the distance matrix of the current complex conformation.

**UniMol**[327] extends this methodology by leveraging a 3D version of Graphormer for pretraining separately on protein and ligand conformations. It predicts the interatomic distance matrix and reconstructs Cartesian coordinates to complete the docking process. Its successor, **UniMol-Docking V2**[373], improves the coordinate reconstruction step by applying constraints in internal coordinates (e.g., global translation, rotation, and local torsion angles) rather than directly optimizing Cartesian coordinates, ensuring physical plausibility of predicted conformations.

**CarsiDock**[356] achieves state-of-the-art performance in this line of work by incorporating various architectural innovations and large-scale pretraining on physics-based docking datasets. It employs separate encoders for proteins and ligands to project them into a shared feature space, followed by a cross-interaction encoder that performs message passing on the protein–ligand complex graph. Notably, a triangle attention mechanism is introduced to capture geometric relationships among atomic triplets. CarsiDock has two prediction objectives: (1) the complex distance matrix $D$, and (2) a distance-based potential energy modeled by an MDN, denoted as $N(\mu, \sigma, \pi)$. The former is used to generate candidate conformations, while the latter is employed to rank them. During inference, the model first predicts an approximate distance matrix $\widetilde{D}$ for the protein-ligand pair. It then randomly initializes 10 ligand conformations using RDKit, followed by conformation refinement under internal coordinate constraints. Among these 10 docking poses, the final output is selected based on a scoring function, CarsiScore, derived from the MDN-based distance potential $N(\mu, \sigma, \pi)$.

#### 4.2.2.2 Regressing Cartesian Coordinates

The aforementioned class of methods, which first predict interatomic distance matrices and subsequently reconstruct molecular conformations from those distances, require the model to be SE(3)-invariant, implying that the global rotations or translations of the input structure do not affect the predicted distances. While this formulation simplifies model design by decoupling geometry from prediction, the mapping from pairwise distances to 3D coordinates is inherently non-unique, potentially introducing reconstruction errors. To address this, more recent approaches directly regress the 3D coordinates of ligand atoms (**Figure 12B**). In the context of GNNs, this involves producing node-level outputs that correspond to predicted atomic positions, with models typically designed to be SE(3)-equivariant, ensuring that output coordinates transform consistently with the input.

**EquiBind**[357] was one of the earliest models to adopt this formulation, and it applies SE(3)-equivariant coordinate regression layers (inspired by EGNN) to directly predict ligand docking poses. It further aligns RDKit-generated initial conformations with the predicted binding poses, making the final output more physically plausible. Building upon this, **LigPose**[358] introduced a recycling mechanism, where the predicted conformation is iteratively fed back into the model for refinement, significantly improving docking accuracy. Conceptually, this recycling process can be interpreted as a denoising diffusion over atomic coordinates, and the trained LigPose network can be viewed as a learned denoising model operating in coordinate space. Subsequently, **KarmaDock**[58] adopted a similar iterative refinement strategy, further improving docking accuracy under the condition of a given binding pocket. In addition, KarmaDock incorporated protein-ligand interaction energies into the modeling process by introducing a MDN-based scoring function, enabling simultaneous prediction of docking poses and binding affinities. Moreover, **FABind**[361] adopts a two-stage generation process: it first predicts potential binding regions within the protein pocket, and then estimates ligand coordinates within the identified region. This effectively decomposes the docking problem into binding site localization followed by local pose generation. **FABind+**[362] further extends this approach by introducing adaptive pocket sizes and a permutation-invariant loss, as proposed in LigPose, thereby enhancing the robustness of binding site identification. Furthermore, FABind+ utilizes MC-Dropout to extend single-conformation regression into multi-conformer generation, and ranks these conformers using a confidence score trained in parallel with the model.

**4.2.2.3 Generative Approaches**

In contrast to the regression-based models, another line of research employs generative models for molecular docking. Among these, the most prominent is **DiffDock**[359], which leverages a diffusion-based generative framework. DiffDock constrains coordinate updates to the internal coordinate system, ensuring the conformational plausibility of small molecules. Specifically, the model predicts the global translation, rotation, and local torsional angles of the ligand. It builds upon a tensor product network architecture (see Chapter 2), ensuring that each geometric quantity respects the corresponding physical symmetries. By operating in internal coordinates, DiffDock significantly reduces computational demands and enables blind docking across entire protein surfaces. However, this approach introduces a challenge: there is no natural ordering among the predicting variables (translations, rotations, and torsions). As a result, the model must iteratively update these variables until convergence. To address the limitations of diffusion in internal coordinate space, **HelixDock**[360] proposes a diffusion process directly over Cartesian atomic coordinates. While this enhances the model's flexibility, it also requiressubstantially more training data and computational resources. Consequently, HelixDock is primarily designed for targeted pocket-specific docking tasks and mitigates data scarcity by pretraining on conventional docking simulation datasets.

### 4.2.3 Flexible Docking: Co-Folding and Induce-Fit Docking

Flexible docking refers to modeling protein-ligand interactions where both the ligand and the protein are permitted to undergo conformational changes, thereby capturing the induced fit phenomenon more realistically. Unlike semi-flexible docking, which explores only the ligand's conformational space, flexible docking requires joint prediction of atomic coordinates for both binding partners. Ideally, flexible docking could start from an arbitrary protein conformation and, through induced fit, relax into a stable bound state of the complex. This would overcome the limitations of rigid methods that neglect protein flexibility. However, practical implementation of flexible docking remains highly challenging. First, modeling all atoms of the protein in Cartesian space significantly increases the degrees of freedom, resulting in prohibitive computational demands. Second, most benchmark datasets such as PDBBind contain only static protein-ligand complexes. While suitable for training semi-flexible docking models, they fail to represent the dynamic range of conformational changes associated with protein flexibility. As a result, models trained on such datasets often struggle to generalize to real-world scenarios where significant protein rearrangements occur during binding. The lack of dynamic structural data thus remains a key bottleneck in advancing flexible docking approaches.

#### 4.2.3.1 Co-folding: Protein Sequence Input

Despite the aforementioned challenges, several pioneering efforts have been made to explore flexible docking. Among them, two particularly noteworthy advances are **RosettaFold-All-Atom** (RF-AA)[363] and **AlphaFold3** (AF3)[374], which mark major milestones by shifting the focus from classical docking to whole complex structure prediction. These models take protein sequences and ligand SMILES as input and directly generate the 3D structure of the protein-ligand complex, transitioning from single-molecule structure prediction to holistic biomolecular modeling. Co-folding discards the assumption of a pre-defined static protein conformation. Instead, the protein and ligand are folded simultaneously into a bound state, enabling implicit modeling of induced fit and conformational flexibility. While this introduces additional sources of error related to protein structure prediction, it effectively removes the reliance on fixed input geometries. This approach is particularly advantageous in cases where the protein structures are unknown, highly dynamic, or exhibits conformational heterogeneity. By jointly exploring the high-dimensional energy landscape, co-folding methods may also help overcome local minima and discover novel binding modes inaccessible via conventional docking.

The RF-AA model builds upon **RosettaFold**[53] to support all-atom structure prediction across diverse biomolecules. For proteins, RF-AA preserves residue-level rigid-body representations using N-C-C coordinate frames. For nucleic acids, an analogous scheme is adopted, with nucleotide nodes defined by the OP1-P-OP2 axis, enabling seamless extension to protein-nucleic acid complexes. For small molecules and metal ions, where coarse-grained definitions are unavailable, RF-AA directly models atomic coordinates. Within this framework, RF-AA predicts translation and orientation vectors for coarse-grained protein and nucleotide nodes, while only translation vectors are predicted for atoms. The model integrates 1D (sequence), 2D (chemical topology), and 3D (chirality and spatial geometry) information through separate channels, employing a SE(3)-Transformer to preserve equivariance under physical transformations. Coordinate refinement is performed iteratively using a recurrent update scheme. As one of the earliest attempts to unify all-atom precision with biomolecular generalization, RF-AA represents a key step forward in AI-driven flexible docking.

Compared to RF-AA, AF3 introduces more extensive architectural innovations. Building upon the insights gained during the development of **AlphaFold2** (AF2)[54], the AF3 team significantly streamlined various model components. For example, the 48-layer Evoformer used for multiple sequence alignment (MSA) processing in AF2 is replaced with a lightweight 4-layer PairFormer module. Moreover, AF3 eliminates rigid-body frame prediction and instead directly regresses all-atom coordinates for residues and ligands. Another major innovation lies in AF3's transition from the traditional recycling mechanism to diffusion models. As discussed earlier, both recycling and diffusion serve similar roles: they iteratively refine partially optimized structures by denoising, gradually transforming low-quality conformations into accurate ones. During training, to mitigate model hallucination, AF3 incorporates both cross-distillation and self-distillation strategies. Specifically, it leverages pretraining on predicted structures generated by AF2[54] and **AF-Multimer**[375], inheriting the strengths of existing state-of-the-art models. For self-distillation, AF3 introduces partially trained models to predict intrinsically disordered protein regions and incorporates these outputs into the training set. This exposure ensures that the model recognizes disordered structures during generation and avoids erroneously folding all proteins into artificially compact conformations. Beyond changes in generative methodology, AF3 departs from SE(3)-equivariant architectures in favor of a standard Graph Transformer, significantly reducing computational overhead. Empirical evaluations suggest that this simplification does not compromise performance, prompting renewed debate on whether equivariance is a necessary inductive bias in molecular structure prediction.

In the technological trajectory of protein-ligand co-folding, beyond the most prominent representatives (RFAA and AF3), two contemporary efforts, **UMol**[365] and **NeuralPLexer**[366], have emerged. These models share architectural similarities with RFAA and AF3, respectively, and can be broadly viewed as earlier-stage prototypes or precursors. UMol extends the core design principles of RosettaFold and AlphaFold2 by adapting local coordinate systems and iterative recycling mechanisms for protein-ligand complex prediction. Unlike AlphaFold2, UMol eliminates the MSA component, instead directly leveraging protein sequence information and 2D molecular graphs of small molecules. Through iterative recycling, it refines a distance histogram representing interatomic distances within the complex. This histogram is then decoded into 3D docking poses, followed by physics-based energy minimization with molecular force fields to generate the final complex structure. UMol supports both blind docking (from sequence alone) and pocket-guided docking, where binding site information is provided as an input constraint to bias ligand pose generation toward the specified protein pocket. NeuralPLexer, on the other hand, adopts a generative framework akin to AF3, extending AlphaFold2 into a diffusion-based model for small molecule docking. It comprises two main components: a Contact Map Prediction (CMP) module and a molecular coordinate diffusion module. Analogous to the PairFormer block in AF3, the CMP module predicts interatomic distances between the protein and ligand, which serve as conditioning inputs for the coordinate generation stage. To align small molecules within the residue-level local coordinate systems of the protein, NeuralPLexer decomposes the ligand into rigid fragments, predicting their rigid-body centers and orientation angles. Both the protein and ligand are thus coarse-grained into residue/fragment-level rigid bodies. Based on this representation, the CMP module autoregressively constructs the inter-molecular distance map frame by frame. Subsequently, the diffusion module generates the full-atom coordinates of the protein-ligand complex, conditioned on the predicted distance constraints.

**4.2.3.2 Induce-fit Docking: Protein Structures Input**

Induce-fit docking refers to traditional flexible docking strategies where, starting from a given protein structure, the model evolves the protein structure to better accommodate a small molecule ligand. This approach partially mimics the regulatory effects of small molecules on protein conformational states. To train such models, databases containing paired apo (ligand-free) and holo (ligand-bound) protein structures are constructed, enabling the model to learn conformational plasticity associated with ligand binding. For instance, ApoBind[376] identifies potential apo structures corresponding to known protein-ligand complexes in PDBBind by searching the PDB using the following criteria: (1) sequence identity > 80% with the holo protein, (2) no other ligands within 15 Å of the binding pocket, and (3) Cα backbone RMSD < 15 Å upon alignment with the holo structure. However, aligning protein structures at the all-atom level remains non-trivial. As an alternative, some studies utilize protein conformations predicted by AlphaFold2 or **ESMFold**[350] as proxies for apo states. In practice, these two strategies are often combined to enrich the conformational diversity and flexibility landscape available for model training.

Building on this idea, the LigPose team proposed **FlexPose**[367], which introduces protein flexibility by simultaneously updating the rigid-body orientation and position of amino acid residues while predicting the full-atom coordinates of the ligand. This approach captures major backbone-level flexibility but lacks fine-grained modeling of side-chain torsions. To facilitate orientation updates, FlexPose replaces the EGNN framework used in LigPose with a vector-based message passing architecture, namely the Geometric Vector Perceptron (GVP). This is because EGNN can only update Cartesian coordinates of nodes, lacking the ability to handle geometric vector features such as residue orientation, whereas GVP is inherently designed to process both scalar and vector features of graph nodes:

$$\Delta x, s' = EGNN(x, s)$$

$$v', s' = GVP(v, s)$$

Here, $x$ denotes the 3D coordinates of the node, while $v$ and $s$ represent its vector and scalar features, respectively. For further architectural details, readers are referred to the vector-based GNN discussion in Chapter 2. During inference, FlexPose takes as input a noisy ligand conformation and a fixed protein structure. Through iterative recycling, the model updates side-chain orientations, atomic coordinates of the protein, and the ligand conformation until convergence, yielding a final complex that reflects the induced fit process. An alternative strategy, inspired by DiffDock, applies similar translation, rotation, and torsional sampling, which is originally designed for small molecules, to protein residues. For example, **DynamicBind**[369] and **DiffBindFR**[368] extend DiffDock by incorporating protein flexibility: the former generalizes the transformation space to include amino acid-level translations, orientations, and side-chain torsions, while the latter focuses solely on modeling the angles of side chains. **Re-Dock**[370], on the other hand, employs a Diffusion Bridge model[164] that relaxes the requirement for diffusion to start from random noise, allowing initial sampling from any complex distribution. This enables the model to capture the transition from the apo-state distribution $P_{apo}(R)$ to the holo-state distribution $P_{holo}(R)$. In contrast to DynamicBind and DiffBindFR, which start from a randomly perturbed protein conformation $P_{rand}(R)$, Re-Dock initiates from a realistic apo conformation $P_{apo}(R)$, embedding more structural priors into the diffusion process and potentially reducing errors associated with intermediate-state inference.

### 4.2.4 Challenges of Molecular Docking Models

#### 4.2.4.1 Generative or Regression?

From a modeling perspective, molecular docking can be framed either as a generative task or a regression problem. Rooted in physical principles, protein-ligand interactions often exhibit not a single unique binding pose but rather multiple local minima on the energy landscape. Generative modeling is thus advantageous for capturing the intrinsic conformational diversity of binding modes. However, in practical settings, experimental techniques such as X-ray crystallography or cryo-electron microscopy typically provide only a single binding conformation, and scoring functions are usually designed to evaluate static structures. As such, regression-based frameworks that directly predict the "optimal" pose may offer a more targeted and computationally efficient approach. Nevertheless, recent studies suggest that the most effective AI-driven docking pipelines often adopt a hybrid strategy: generating multiple candidate poses followed by scoring-based selection. This scheme can be regarded as a form of ensemble prediction, which enhances the robustness and fault tolerance of the model. For example, in regression-based docking, pose diversity can be achieved by perturbing model inputs or leveraging stochastic architectural components. CarsiDock generates distinct docking poses by providing multiple ligand conformations, while FABBind+ utilizes dropout layers to sample different sets of model parameters, thereby producing diverse predictions for the same input. Collectively, these approaches demonstrate that multi-pose generation combined with scoring-based selection consistently outperforms single-shot regression models in terms of accuracy and reliability.

**4.2.4.2 Challenges in Physical Plausibility**

Although AI-based docking models have advanced and even surpassed traditional docking methods in predicting ligand-protein binding poses, substantial challenges remain in generating physically and chemically plausible conformations. That is to say, conformations generated by current AI models sometimes exhibit unrealistic ligand geometries, including distorted local structures, non-planar aromatic rings, and abnormal bond lengths. These issues are particularly pronounced in models that directly predict Cartesian coordinates, such as KarmaDock, LigPose, and FABind. Even models that utilize internal coordinates, e.g., DiffDock, frequently produce steric clashes between ligand atoms and protein residues. To address this, many docking pipelines incorporate a post-prediction refinement step using classical force fields, which relaxes predicted structures toward a local minimum on the protein-ligand potential energy surface (PES). The rationale is that deep learning models can approximate low-energy regions, but classical optimization is still needed to reach chemically valid minima. However, force field refinement does not guarantee improvements in docking accuracy and introduces additional computational overhead. Consequently, there is growing interest in developing end-to-end differentiable models that enforce physical plausibility during learning. Proposed strategies include incorporating steric clash penalties into the loss function or predicting internal coordinates rather than Cartesian positions. In the long term, as training datasets expand in scale and fidelity, the problem of conformational implausibility may diminish, enabling AI docking models to produce physically consistent predictions without post hoc correction.

**4.3 Binding Affinity Prediction**

Binding affinity prediction, often referred to as scoring, aims to assign a score to each candidate small molecule interacting with a target protein, thereby identifying the most promising binders. Early studies typically relied on two main strategies: (1) computing handcrafted descriptors from protein-ligand complexes and applying classical machine learning algorithms such as SVMs to fit activity labels, and (2) converting protein-ligand complexes into voxelized 3D grids and training CNNs to fit affinity labels. The former approach heavily depends on manually designed descriptors, which may introduce human bias and limit generalizability. The latter, while more automated, suffers from high memory consumption that scales cubically with grid resolution, making it less practical for large-scale virtual screening. With the advent of GNNs, molecular graphs have emerged as a compact and intuitive representation for modeling protein-ligand interactions. GNN-based models have been widely adopted for binding affinity prediction, mitigating key limitations of earlier methods.

From a methodological perspective, binding affinity prediction using GNNs can be framed as a property prediction task, in which the model outputs an affinity value based on the input protein-ligand graph. However, unlike conventional property prediction, the scoring problem entails richer physical and chemical implications. On one hand, there is a strong desire to explicitly capture chemical interactions embedded in protein–ligand complex graph. On the other hand, beyond achieving numerical agreement with experimental binding affinities of few tested compounds, scoring models are also expected to effectively discriminate between active and inactive compounds in a vast chemical space, which is at the core of virtual screening. These characteristics introduce unique challenges and opportunities in both model design and evaluation protocols for binding affinity prediction. **Table 16** summarizes the GNN-related scoring methods we would discuss.

Table 16. Overview of GNN-based scoring methods, RMSE refers to root mean square error, $R_p$ refers to person coefficient, two metrics are evaluated on PDBBind-core and extracted from the literature report.

| Category | Methods | RMSE | $R_p$ | Keywords |
|---|---|---|---|---|
| Interaction-Free | GraphDTA*[377] | 1.470 | 0.743 | Interaction-free, CNN-sequence and GNN-mol |
| | MGraphDTA[378] | 1.439 | 0.753 | Deep GNN version of GraphDTA |
| | PotentialNet[379] | 1.503 | 0.772 | Two-stage nonco-/co-valent conv. |
| | 3DEmbDTA[380] | - | - | Interaction GNN minus non-interaction GNN |
| Interactin Graph | IGN[381] | 1.220 | 0.827 | Map intermolecular edges to affinity |
| | SIGN[382] | 1.316 | 0.797 | Angle-aware GNN and matrix regularizer |
| | CurvAGN[383] | 1.217 | 0.831 | Curvature-based GNN for topological learning |
| | GIGN[384] | 1.190 | 0.840 | Hetero graph for nonco-/co-valent conv. |
| | EHIGN[385] | 1.150 | 0.854 | Hetero graph and pairwise addition correction |
| | PIGNet[386] | - | 0.761 | Physical layer and introduce decoys |
| | LGN[387] | 1.177 | 0.842 | Fuse molecular descriptor GNN |
| Fusion Model | FAST[388] | 1.308 | 0.810 | Fusion model on grid (CNN) and graph (GNN) |
| | HAC-Net[389] | 1.210 | 0.846 | Advanced model variants of FAST |
| | GraphLambda[390] | - | - | Fuse 3 GNNs and interaction descriptpr |
| | FGNN[391] | 1.140 | 0.873 | Fuse spatial GNN and spectral GNN |
| Ranking | PBCNet[392] | - | - | Siamese GNN for relative binding |
| Docking | DeepDock[353] | - | 0.460 | Pose scoring potential with MDN |
| Screening | RTMScore[393] | - | 0.455 | Superior screening power with MDN and GT |
| | PIGNet2[394] | - | 0.747 | More data enhancement on PIGNet |
| General Scoring | GenScore*[395] | - | 0.834 | Add affinity-correlated term to RTMScore |
| | EquiScore[396] | - | - | More decoys: 2D similar but 2D dissimilar mols. |
| | ConBAP[397] | 1.237 | 0.864 | Contrastive learning for balanced 4 powers |

| | | | |
|---|---|---|---|
| IGModel[398] | | 0.831 | Affinity and RMSD prediction |
| PointVS[399] | - | 0.805 | Affinity and active conf. prediction |
| PLANET[400] | 1.311 | 0.824 | Affinity, mol. conf., and contact map prediction |

GraphDTA* is the GIN variant. GenScore* is GatedGCN_ft_1.0 version; GT is Graph Transformer, MDN is Mixture Density Network.

### 4.3.1 Task Descriptions and Datasets Available

#### 4.3.1.1 Scoring, Ranking, Docking and Screening Powers

Many models formulate binding affinity prediction as a regression task: given the molecular graphs of a protein and a ligand, the model outputs a single scalar value representing their interaction strength. However, in real-world drug discovery, the expectations for scoring functions go far beyond numerical accuracy. A reliable scoring function should not only correlate well with experimental affinities but also support critical tasks such as compound prioritization, pose discrimination, and virtual screening. To meet these broader objectives, an effective scoring function is typically expected to demonstrate four key capabilities[401]:

**Scoring Power**: This refers to the ability of the model to produce binding affinity predictions that are linearly correlated with experimentally determined values across different protein-ligand complexes. It is typically quantified by the Pearson correlation coefficient $R_p$:

$$R_p = \frac{\sum_{i=1}^{n} (X_i - \overline{X})(Y_i - \overline{Y})}{\sqrt{\sum_{i=1}^{n} (X_i - \overline{X})^2 \sum_{i=1}^{n} (Y_i - \overline{Y})^2}}$$

where $X_i$ and $Y_i$ denote the predicted and experimental binding affinities, respectively. $R_p$ ranges from –1 to 1, with values closer to 1 indicating stronger linear correlation.

**Ranking Power**: This measures the model's ability to correctly rank a series of ligands for a given target, independent of absolute affinity values. It is commonly assessed using the Spearman rank correlation coefficient $R_s$:

$$R_s = 1 - \frac{6 \sum_{i=1}^{n} d_i^2}{n(n^2 - 1)}$$

where $d_i$ is the difference between the ranks of the predicted and experimental values for the $i$th sample. $R_s$ focuses on ranking consistency rather than numerical accuracy, with $R_s = 1$ indicating perfect agreement and –1 indicating a completely reversed order.

**Docking Power**: A robust scoring function should effectively distinguish correct binding poses from incorrect ones. Docking power is often assessed by the success rate of identifying a predicted binding pose with a root-mean-square deviation (RMSD) of less than 2 Å from the experimentally determined pose.

**Screening Power**: This reflects the model's effectiveness in identifying active compounds from a large virtual library. Screening power is typically quantified by the enrichment factor (EF), defined as:

$$\text{EF}_\alpha = \frac{\text{NTB}_\alpha}{\text{NTB}_\text{total} \cdot \alpha}$$

where $\text{NTB}_\alpha$ is the number of true binders ranked within the top $\alpha$ of the screened compounds, and $\text{NTB}_\text{total}$ is the total number of known active compounds for a given target. A higher EF value indicates greater enrichment of true actives in the top-ranked subset, highlighting the model's practical utility in virtual screening.

**4.3.2 Datasets**

**PDBBind:** PDBBind is one of the most widely used benchmark datasets for developing and evaluating scoring functions. As introduced in the molecular docking section, PDBBind collects protein-ligand complex structures with experimentally measured binding affinities from the PDB. The dataset comprises three subsets: general set, refined set, and core set. The refined set is a curated subset of the general set, selected based on the following criteria to ensure data quality: (1) crystallographic resolution ≤ 2.5 Å to ensure structural accuracy; (2) activity measurements restricted to $K_d$ or $K_i$ values to reduce label noise. In more recent versions, additional filtering criteria have been introduced for the Refined set, including exclusion of complexes with missing backbone or side chain atoms within 8 Å of the ligand, removal of complexes with extreme binding affinities, exclusion of non-standard amino acids within 5 Å of the ligand, and omission of ligands bound in shallow binding pockets. The core set originates from the Comparative Assessment of Scoring Functions (CASF)[402] benchmark, which aims to evaluate the performance of scoring functions across diverse protein targets. To ensure diversity, proteins in the refined set are first clustered based on ≥90% sequence identity; then, for each cluster, three representative complexes are selected, corresponding to ligands with high, medium, and low binding affinities, respectively. Newer versions of CASF retain this clustering strategy while refining ligand selection criteria. Typically, deep learning-based scoring functions are trained on the general or refined sets and evaluated on the core set. Some alternative protocols adopt a time-split strategy[357], where the dataset is partitioned into training, validation, and test sets on publication dates to simulate real-world scenarios.

**CASF:** CASF is the core set of PDBBind, serving as a standard benchmark for evaluating four key aspects of scoring functions.

**Binding MOAD:** The Binding MOAD[333] (Mother Of All Databases) dataset is another curated collection of protein-ligand complexes, widely used in virtual screening and scoring function evaluation. Similar to PDBBind, it includes complexes sourced from the PDB with measured binding affinities and resolution ≤ 2.5 Å. However, it excludes covalent inhibitors, nucleic acids with more than four nucleotides, and peptides longer than ten amino acids. Due to the overlapping data collection protocols between Binding MOAD and PDBBind, entries common to both datasets can be cross-referenced in practical applications. This cross-validation enables the correction of potentially misannotated or inconsistent entries, enhancing dataset reliability.

**DEKOIS 2.0, DUD-E, and LIT-PCBA**: DEKOIS 2.0[334] and DUD-E[335] are widely used external benchmark datasets designed to evaluate the virtual screening performance of scoring functions. Both datasets contain multiple protein targets, experimentally validated active compounds, and carefully constructed decoy sets—molecules that are chemically similar to actives but presumed to be inactive. DEKOIS 2.0 comprises 81 protein targets, each associated with 30 actives and 1,200 decoys, while DUD-E includes 102 targets, with variable numbers of actives and a fixed decoy-to-active ratio of 50:1. Since most protein-ligand complex structures are not available for these datasets, docking tools (i.e., Glide) is used to generate the corresponding binding poses. LIT-PCBA[336] is a more recent and challenging benchmark designed to better reflect real-world virtual screening conditions. It integrates active and inactive compounds from 149 PubChem BioAssays, all supported by experimental validation. After rigorous filtering and bias removal, the final dataset includes 15 protein targets, 10,030 true actives, and 2,798,737 decoys, resulting in a highly imbalanced active-to-inactive ratio of approximately 1:1000. Each target is associated with multiple PDB-derived structural templates for docking, providing a rich source of conformational diversity. As with DEKOIS 2.0 and DUD-E, docking tools are employed to generate 3D binding poses.

**Merck FEP Benchmark:** The Merck FEP Benchmark[337] was originally developed to evaluate the accuracy of rigorous free energy perturbation (FEP) calculations. It is a smaller but higher-quality benchmark for assessing the predictive power of scoring models. This dataset includes eight drug targets and a total of 264 experimentally validated active compounds. The binding poses are generated either using Flexible Ligand Alignment or Glide core-constrained docking based on known reference structures, under the assumption that molecules with similar scaffolds tend to adopt similar binding modes. Compared to standard docking-based pose generation, the conformations provided in this dataset are of higher quality. Additionally, this benchmark includes the reference results from the state-of-the-art FEP+[403] method, enabling direct comparison and stricter evaluation of machine learning-based scoring functions.

**BindingDB:** BindingDB[338] is a large-scale repository of protein-ligand binding data, compiled from peer-reviewed publications, patents, PubChem BioAssays, and ChEMBL. It includes data for 2,685 proteins and 542,126 experimentally measured binding affinities, categorized into 1,265 unique targets. Unlike PDBBind, BindingDB does not provide 3D structures of protein-ligand complexes, making it primarily suitable for training 2D GNNs or 3D models after generating poses via molecular docking. A unique feature of BindingDB is its target-specific validation set, where each target is associated with a series of homologous active compounds, making this subset particularly valuable for training models to predict relative binding affinities within congeneric series.

### 4.3.2 Mining Protein-Ligand Interactions with GNN

In the field of binding affinity prediction, early data-driven approaches often involved separate encoding of protein and ligand features, followed by feature concatenation as input to a regression model. While such methods capture coarse interaction patterns, they often suffer from limited generalizability due to the scarcity of high-quality labeled data. To overcome this limitation, recent studies have increasingly adopted GNNs to model protein-ligand complexes directly, enabling the extraction of intermolecular interaction patterns and the learning of underlying physical principles. In this section, we would review the evolution of GNN-based scoring methods, from early interaction-free GNNs, which do not explicitly model inter-molecular contacts, to more recent interaction-aware GNNs, which are designed to capture fine-grained physical interactions. We highlight how these models progressively uncover deeper physical insights into protein-ligand binding.

#### 4.3.2.1 Protein-Ligand Complex GNNs

One of the early attempts to employ GNNs in drug-target affinity prediction was **GraphDTA**[377], which utilized a GNN to extract features from the 2D molecular graph of a small molecule, and combined these features with protein features encoded via a one-hot representation followed by a 2D-CNN. However, **MGraphDTA**[378] later pointed out that shallow GNNs struggle to capture global molecular characteristics, such as cyclic substructures. To address this limitation, MGraphDTA adopted a deeper GNN architecture based on the DenseGNN design (Chapter 2), extending the network depth to 27 convolutional layers to better model the global context of small molecules. Notably, both GraphDTA and MGraphDTA primarily focused on encoding molecular structures, without explicitly modeling the interactions between proteins and small molecules.

The first representative interaction-aware GNN was **PotentialNet**[379], which introduced a spectral GNN to encode both intra- and intermolecular interactions within a protein-ligand complex through a block adjacency matrix:

$$A = \begin{bmatrix} A_{L:L} & A_{L:P} \\ A_{P:L} & A_{P:P} \end{bmatrix}$$

Here, L and P denote ligand and protein atoms, respectively. $A_{L:L}$ and $A_{P:P}$ encode covalent bonds within the ligand and protein, respectively, while $A_{L:P}$ and $A_{P:L}$ capture spatial distance-based intermolecular interactions (e.g., defined by a distance threshold, such as <3Å). PotentialNet first performs convolutions over $A_{L:L}$ to learn covalent intra-ligand features, and then applies convolutions over the full matrix A to incorporate noncovalent protein-ligand interactions. The final readout is used to predict binding affinity. Building upon this idea, **3DEmbDTA**[380] proposed a learnable protein-ligand interactions:

$$A_{ij} = e^{-(d_{ij} - \mu)/\sigma}$$

where $d_{ij}$ is the Euclidean distance between ligand atom $i$ and protein atom $j$, and $\mu, \sigma$ are learnable parameters. Edges are only constructed between atom pairs within a 5Å distance cutoff. An interesting innovation in 3DEmbDTA is its final prediction strategy: the output of a non-interaction-aware GNN is subtracted from that of the interaction-aware GNN, thereby isolating noncovalent contributions. This strategy aligns with the understanding that binding free energy is predominantly driven by noncovalent interactions.

Subsequent studies have explored more explicit modeling of intermolecular edges. For example, **IGN** (Interaction Graph Network)[381] performs readout over intermolecular edges rather than nodes, directing learning toward contact-specific representations. However, IGN exhibited limited generalizability across datasets, likely due to the simplicity of early GNN architectures and the naive edge feature definition $e_{ij} = [h_i \| h_j]$, which lacks fine-grained interaction modeling. To address these limitations, **SIGN** (Structure-aware Interactive Graph Neural Networks)[382] introduced two architectural innovations: a polar-inspired graph attention layer (PGAL), which incorporates angular information through edge-neighbor aggregation, and a pairwise interaction pooling (PiPool) module that uses an empirically derived atom-type interaction matrix[404] $Z$ as a regularization ($z^{pred} - z$), enforcing physical priors during training. A related approach, **CurvAGN**[383], retained the overall architecture of SIGN but replaced the base GNN with a curvature-based model to better capture the geometric topology in protein-ligand complexes.

While these studies collectively suggest that interaction-aware modeling improves prediction performance, empirical validation through quantitative ablation studies is essential. **GIGN** (Geometric Interaction Graph Neural Network)[384] explicitly conducted such experiments, demonstrating that models ignoring protein-ligand interactions consistently yield lower accuracy. GIGN further introduced two enhancements: (1) separating covalent and noncovalent interactions via heterogeneous message passing over spatial graphs, reducing RMSE from 1.460 to 1.380); and (2) comparing SE(3)-invariant models based on interatomic distances with coordinate-based models, showing clear advantages for equivariant modeling, (RMSE from 1.518 to 1.380). While these ideas were previously explored, GIGN provided quantitative evidence of their utility for interaction learning. Building on this, **EHIGN**[385] introduced an additional correction term to predict pairwise atomic interaction forces, further reducing RMSE to 1.297. This result underscores the growing consensus that embedding physical priors and explicitly modeling interactions are critical for improving model accuracy and generalizability in protein-ligand affinity prediction.

### 4.3.2.2 Physics-inspired Interaction Learning

While many interaction-aware GNNs implicitly incorporate physical priors, for example, through architectural designs or adjacency definitions that distinguish intra- and intermolecular interactions, such strategies often lack interpretability in terms of alignment with established physical laws. To address this limitation, some researchers have explored integrating explicit physical models into scoring networks, thereby imposing stronger constraints on predictions. A representative work in this direction is **PIGNet** (Physically-Informed Graph Neural Network)[386], which incorporates various energy terms in the final layer of the GNN to mimic traditional force field-based scoring functions. Specifically, PIGNet accounts for van der Waals interactions, hydrogen bonding, metal-ligand interactions, and hydrophobic effects, by computing pairwise atomic energy contributions. For example, the van der Waals interaction is modeled as follows:

$$E^{\text{vdW}} = \sum_{i,j} c_{ij} \left[ \left(\frac{d'_{ij}}{d_{ij}}\right)^{12} - 2\left(\frac{d'_{ij}}{d_{ij}}\right)^{6} \right]$$

where $d_{ij}$ is the distance between atoms $i$ and $j$, $d'_{ij}$ is the equilibrium distance constants, and $c_{ij}$ is a learnable coefficient. For the other three types of interactions (hydrogen bonds, metal coordination, and hydrophobic interactions), PIGNet adopts a unified functional form:

$$E^{hbond,metal,hydro} = \sum_{ij} c_{ij} e_{ij} = \begin{cases} w & \text{if } d_{ij} - d'_{ij} < c_1, \\ w\left(\frac{d_{ij} - d'_{ij} - c_2}{c_1 - c_2}\right) & \text{if } c_1 < d_{ij} - d'_{ij} < c_2, \\ 0 & \text{if } d_{ij} - d'_{ij} > c_2 \end{cases}$$

In addition, PIGNet incorporates a rotatable bond penalty term inspired by AutoDock Vina to approximate entropic effects:

$$T = 1 + C_{tor} \cdot N_{tor}$$

where $N_{tor}$ is the number of rotatable bonds in the ligand, and $C_{tor}$ is a learnable parameter. With all these components, the final predicted energy is given by:

$$E = \frac{E^{vdW} + E^{hbond} + E^{metal} + E^{hydro}}{T}$$

Since a ligand typically binds to a protein in a conformational state corresponding to a local minimum on the potential energy surface, PIGNet imposes the constraint $\nabla E = 0$ to guide the network towards toward physically plausible binding poses.

### 4.3.2.3 Enhancing Interaction Learning via Fusion Models

Different data sources or feature representations are typically associated with distinct inductive biases. By integrating these complementary features, models can more effectively capture diverse aspects of molecular interactions, thereby improving prediction accuracy, a concept known as fusion strategies[405]. In the context of GNN-based scoring functions, fusion learning enables the model to simultaneously learn different physicochemical interactions, such as electrostatic forces and van der Waals interactions, through separate feature channels. Fusion strategies can be broadly categorized into three types: feature-level, decision-level, and model-level. Below, we outline these approaches and their applications in GNN-based affinity prediction

**Feature-Level Fusion**: Feature-level fusion, also known as early fusion, refers to the strategy of concatenating multiple feature representations into a unified vector prior to model input. For instance, **LGN**[387] merges handcrafted molecular descriptors with GNN-derived embeddings before feeding the combined features into a predictive model. This design allows the model to simultaneously leverage domain knowledge and learned representations. **FAST**[388] (Fusion models for Atomic and molecular Structures), for example, jointly uses 3D grid-based inputs (via CNNs) and molecular graphs (via GNNs), and these features are fused and passed to a DNN for affinity prediction. **HAC-Net**[389] further improves upon FAST by integrating more expressive network backbones, reducing RMSE from 1.308 to 1.205.

**Decision-Level Fusion**: Decision-level fusion, or late fusion, involves aggregating predictions from multiple independent models (or model variants). **GraphLambda**[390] exemplifies this strategy by training GAT, GCN, and GraphSAGE models separately on the same protein-ligand complex graph, then combines their predictions via a weighted voting scheme, capturing diverse modeling perspectives.

**Model-Level Fusion**: In contrast to decision-level fusion, model-level fusion involves interaction between sub-models or modules during the feature extraction phase, enabling deeper integration of different representation learning techniques. For example, **FGNN**[391] (Fusion GNN) fuses spatial-domain graph representations from AttentiveFP with spectral-domain features from SignNet, integrating local and global structural information. This architecture achieves state-of-the-art performance on the PDBBind core benchmark.

### 4.3.3 Four Powers of GNN Scoring Functions: Beyond Regression

As previously discussed, scoring functions are expected to demonstrate strong performance across four key tasks: scoring, ranking, docking, and screening. However, most existing approaches primarily optimize scoring performance by minimizing regression loss to approximate experimental binding affinities. For example, some interaction-aware GNNs, such as IGN, achieve an RMSE of 1.2 on the PDBBind core set, but their performance in ranking, docking, or screening tasks remains limited. Recently, GNN-based models have emerged to address these additional capabilities. For instance, **PBCNet**[392] focuses on ranking, **DeepDock**[353] aims to improve docking, and **RTMScore**[393] targets screening. This section reviews the design principles behind these models and discusses strategies for balancing all four capabilities.

#### 4.3.3.1 Ranking Power

Ranking power can be viewed as an extension of scoring power, emphasizing relative comparisons among ligands targeting the same protein. Traditional regression-based models predict absolute binding affinities, which do not inherently guarantee correct ordering of compounds. A direct solution is to predict relative affinities, as demonstrated by PBCNet[392]. PBCNet adopts a Siamese Network[406] framework to perform pairwise comparisons between ligands binding to the same target, thereby learning to capture differences in molecular activity. Specifically, PBCNet creates multiple "same-target, multi-ligand" series based on the BindingDB dataset, and uses Glide to generate 3D protein-ligand complexes. For each series, it constructs paired samples, labeling them with the activity difference $\Delta_i$ between two ligands. The loss consists of a regression loss on $\Delta_i$ and an entropy-based ranking loss. PBCNet achieves ranking accuracy comparable to **FEP+**[403], while requiring only 0.9 seconds to predict the relative binding free energy between two compounds, making it suitable for high-throughput virtual screening. In case studies, PBCNet, combined with SME-based interpretability methods, has shown that its learned key interactions (e.g., on JNK-1) align well with Schrödinger-calculated results.

**4.3.3.2 Docking Power**

Docking power evaluates a scoring function's ability to distinguish between different poses of the same ligand and identify the one closest to the native (crystal) conformation. A scoring function with strong docking power can serve as a post-filter for docking or generative models that output multiple poses. Docking ability can be modeled as a classification (e.g., native vs. non-native pose) or regression (e.g., predicting RMSD). **DeepRMSD**[407] follows the regression route, using 3D voxel input to predict RMSD directly. Other methods, like PIGNet, incorporate force field constraints (e.g., $\nabla E = 0$) to favor stable poses. A notable departure from these paradigms is offered by **DeepDock**, which adopts a probabilistic framework using MDNs to model the likelihood of protein-ligand conformations. It learns the probability distribution of interatomic distances and sums the likelihoods across all atom pairs to evaluate each pose. During training, it maximizes the likelihood of known crystal structures. At inference time, it selects the pose with the highest total likelihood as the most plausible native conformation.

**4.3.3.3 Screening Power**

Screening focuses on identifying active compounds from large-scale libraries, requiring models to effectively distinguish actives from a vast majority of inactives. This necessitates sufficient exposure to negative samples during training. Target-specific approaches such as **ParaVS**[408] are trained on both active and inactive compounds for each target in DUD-E, enabling rapid activity judgments for new molecules. However, these models need retraining for new targets. More generalized approaches integrate screening capability into the training of general scoring-power models. For instance, **PIGNet** incorporates many inactive compounds during training, assigning them extremely low affinity labels, enhancing the model's discriminative power. **RTMScore**, which builds on DeepDock's likelihood-based strategy and uses a Graph Transformer architecture, significantly improves screening performance without explicitly including inactive molecules during training. Instead, its exposure to native conformations of active compounds appears sufficient to differentiate decoy poses of inactives, a hypothesis supported by CarsiDock. The CarsiScore model, a variant of RTMScore trained with a large number of inactive molecules, showed improved docking performance but degraded screening capability, suggesting that excessive inactive data can blur the distinction between actives and inactives by making the model less sensitive to the unique features of actives.

**4.3.3.4 Balancing Four Powers**

Most task-specific models discussed above excel in one area but often underperform in others. Developing a scoring function with balanced performance across all four tasks requires deliberate multi-task design and training strategies. In principle, scoring, ranking, docking, and screening can be framed as distinct tasks, each with its own loss function under a multi-task learning framework. However, obtaining sufficient training data and designing differentiable loss functions for each task remain major challenges. As a result, most recent works enhance baseline models by integrating auxiliary objectives or curated data augmentations, thereby extending model generalization across tasks while preserving core performance in their primary domain.

**4.3.5.4.1 Empower Additional Abilities on Baseline Models**

**GenScore**[395] represents a unified scoring framework that balances all four capabilities. It builds upon RTMScore (which excels in docking and screening) by introducing an affinity correction loss to enhance scoring and ranking. The training objective includes:

A likelihood-based loss derived from a MDN, modeling atomic pairwise distances:

$$\mathcal{L}_{MDN} = -\log P(d_{u,v} \mid h_u^{prot}, h_v^{lig})$$

An affinity correction loss that encourages correlation between the MDN-derived scores and experimental binding affinities:

$$E_{(x)} = \sum_{u=1}^{U} \sum_{v=1}^{V} \mathcal{L}_{MDN} = -\text{Score}$$

$$\mathcal{L}_{affi} = Corr(\text{Score}, y_{affi})$$

From a physical standpoint, the MDN in RTMScore learns a distance-based potential function similar to traditional pairwise atomistic scoring functions, which is effective at identifying stable poses within the protein pocket. However, such atomic-level interactions may not fully capture molecular-level binding energies. GenScore addresses this by using a correction loss to empower RTMScore with scoring and ranking powers.

**4.3.5.4.2. Data Augmentation Strategies**

As previously discussed, PIGNet achieves a reasonable balance among all four capabilities. Its successor, **PIGNet2**[394], further enhances performance via data augmentation, including both positive and negative sample augmentation. For positive augmentation, PIGNet2 uses RDKit to generate 1000 ligand conformers per compound, which are then overlaid onto the corresponding crystal structures and refined using Smina[409] to resolve steric clashes. Conformers with RMSD < 2 Å and Smina score < 1 kcal/mol are retained, thereby capturing the conformational flexibility of active ligands within the binding pocket. The original activity labels are uniformly assigned to all conformers derived from the same compound, reinforcing the notion of a shared binding energy landscape. For negative augmentation, PIGNet2 utilizes three strategies: (1) redocking, where ligands are re-docked and poses with RMSD > 4 Å are selected as non-native; (2) cross-docking, in which ligands are docked into unrelated protein targets; and (3) random docking, pairing random ligands with targets. Negative samples are assigned low affinity labels (e.g., > −6.8 kcal/mol) using a hinge loss, enforcing separation from actives. Builing on the PIGNet's physically inspired framework, PIGNet2 achieves performance comparable to GenScore across all four tasks.

**EquiScore**[396] further increases the difficulty of negative sample discrimination by introducing decoys that are structurally similar but chemically distinct. Using **DeepCoy**[410], 500 physicochemically similar molecules are generated, and the five most pose-similar candidates are selected via Schrödinger's shape-based screening module[411]. **ConBAP**[397] adopts a contrastive learning strategy with augmented data to balance all four tasks. It combines 3D protein pocket information with 2D ligand structures as anchors. Ligand poses with RMSD > 2 Å are treated as decoys (negatives), while those with RMSD < 2 Å are positives. These are embedded into a shared representation space using triplet contrastive loss, which pulls positives closer to anchors and pushes negatives away. Pretraining in this manner yields improved generalization in docking, ranking, and screening tasks. Fine-tuning on affinity prediction further augments scoring performance.

### 3. Multi-task Learning

Given the distinct yet complementary nature of scoring, docking, ranking, and screening, these tasks naturally lend themselves to multi-task learning frameworks. **IGModel**[398] exemplifies this by jointly training two tasks: binding affinity prediction and RMSD prediction. Specifically, for each protein-ligand complex in the PDBBind dataset, IGModel generates 15 docked conformations using AutoDock Vina and LeDock. During training, the model simultaneously predicts the binding affinity label and the RMSD between the docked and crystal conformations. The overall loss function is formulated as:

$$\mathcal{L} = \alpha \cdot mse\,(\text{RMSD}_{\text{real}}, \text{RMSD}_{\text{pred}}) + \beta \cdot mse\,(pK_d^{label}, pK_d^{pred}) + \gamma \cdot 1/N \sum pK_d^{pred}$$

where $\alpha, \beta, \gamma$ are task-specific weights. Importantly, IGModel models the inverse correlation between predicted binding affinity and RMSD by adjusting the affinity label of each docked pose as:

$$pK_d^{label} = pK_d^{nat} - w \cdot RMSD_{real}, \qquad w = \sigma(WV_{pkd})$$

where $pK_d^{nat}$ denotes the affinity of the crystal pose, and $V_{pK_d}$ is the learned affinity representation. This coupling allows the model to improve docking and ranking in addition to scoring, while data augmentation contributes to screening capability.

A similar multi-task formulation is adopted in **PointVS**[399], which formulates RMSD prediction as a binary classification problem: it classifies conformations as active (RMSD < 2 Å from the crystal pose) or inactive. Subsequently, **PLANET**[400] extends the multi-task paradigm by jointly training three tasks: binding affinity prediction, ligand conformation prediction, and protein-ligand distance matrix prediction. To mitigate the negative impact of decoy conformations, PLANET restricts negative samples (high-RMSD poses) to the affinity prediction task only, echoing the training strategy of Uni-Mol (previously discussed in molecular property prediction). Although Uni-Mol has not been explicitly benchmarked accross all four capabilities, it is reasonable to expect that Uni-Mol-based scoring functions would exhibit comparable generalization across the four tasks.

### 4.3.4 Challenge in GNN-based Scoring Functions: Overfitting

#### 4.3.4.1 Overfitting and Dataset Biases

Overfitting is a well-recognized issue in deep learning, and GNNs are no exception. Due to their heightened sensitivity to local graph structures, GNNs employed for protein-ligand binding affinity prediction are particularly prone to a "memorization effect": rather than learning the underlying physical principles of molecular interactions, models may memorize associations between specific molecular scaffolds or functional groups and higher activity in the training set. As a result, such models often fail to generalize when exposed to novel chemical spaces or protein conformations. To mitigate overfitting, researchers have explored architectural innovations that encourage the learning of physically meaningful features and the incorporation of auxiliary tasks to broaden the model's learning scope. In addition to model design and training objectives, overfitting is also significantly influenced by dataset properties. Biases embedded in datasets can contaminate the learning process, undermining the generalizability of trained models. Sieg et al.[412] identified three predominant types of bias in datasets in affinity prediction: artificial bias, domain bias, and label bias. Each of these can severely affect model training and predictive performance.

**Artificial bias** often arises during dataset construction, where experimental design choices or manual curation introduce an imbalanced distribution of physicochemical properties between positive (active) and negative (inactive) samples. For instance, highly active compounds are frequently selected as positives, while clearly inactive compounds are labeled as negatives. This artificial enrichment leads to a disproportionate concentration of training data at the two extremes of the activity spectrum. As a result, the model may learn to distinguish these extremes effectively but fails to generalize to compounds with intermediate activity labels, which are critical for real-world applications. Instead of capturing meaningful patterns of molecular recognition, the model may rely on superficial correlations that do not hold in unseen chemical spaces.

**Domain bias** reflects the uneven representation of chemical scaffolds or functional groups within a dataset. For instance, if the actives in the training data predominantly share similar scaffolds, the model may generalize well to compounds within that chemical space but fail on structurally distinct molecules.

This bias typically originates from selective sampling in experimental workflows or research interests, leading to overrepresentation of certain molecular classes.

**Label Noise and Bias** concerns inaccuracies or inconsistencies in binding affinity labels, stemming from experimental noise, annotation errors, or class imbalance. For instance, incorrect measurements or data recording errors can directly degrade dataset quality. Additionally, severe label imbalance may cause the model to overfit the dominant class while underperforming on the minority class, a phenomenon known as shortcut learning. Here, the model relies on easy-to-learn discriminative features rather than acquires a deeper understanding of intermolecular interactions.

Volkov et al.[413] demonstrated that GNNs trained solely on ligand or protein information often outperform those trained on full protein-ligand complex graphs for affinity prediction. This finding suggests that models may be leveraging ligand structure memorization rather than learning true interaction mechanisms. To further support this hypothesis, the authors constructed a simple k-nearest-neighbor (kNN) model that predicts binding affinity by averaging the activities of the most similar ligands in the training set. Remarkably, this baseline approach achieved performance comparable to state-of-the-art GNNs, further underscoring the overfitting issue in scoring tasks. Subsequently, Mastropietro et al.[414] applied explainable methods (e.g., EdgeSHAPer and GNNExplainer) to analyze feature contributions, and their findings revealed that approximately 60% of the predictive signal originated from ligand features, 20% from protein features, and only 20% from protein-ligand interactions. These results underscore the tendency of current GNNs to memorize ligand-specific features rather than learn robust, generalizable interaction patterns.

**4.3.4.2 Strategies to Mitigate Overfitting**

One straightforward strategy to mitigate overfitting is to explicitly incorporate features that represent protein-ligand interactions. For instance, the LGN model integrates composite representations including compound fingerprints (CFP), structural interaction fingerprints (SIFP), and extended connectivity interaction features (ECIF) into a unified complex graph, thereby enhancing generalization capability. Similarly, as discussed in previous sections on interaction mining, methods that explicitly encode physical principles into the model architecture or training objective can further improve generalization.

Another effective strategy is dataset expansion. While the PDBBind dataset offers high-quality structures with experimentally determined binding affinities, its relatively small size ($\sim 10^4$ samples) is insufficient for training deep models with millions of parameters. In such data-sparse scenarios, GNNs are naturally inclined to exploit "cheap" shortcuts, such as memorizing ligand features, to minimize loss on the validation set. To address this, PIGNet introduces a broader set of negative samples and leverages docked poses as contrasting inputs to enhance discriminative learning. Uni-Mol employs large-scale pretraining on billions of small molecules and 180,000 protein conformations to substantially increase model exposure to diverse chemical and structural patterns. Moreover, given that PDBBind contains predominantly positive (active) samples, incorporating a larger number of negative (inactive) examples can help models better differentiate between active and inactive molecular states. Therefore, many methods like Equiscore and ConBAP adopt different neative sampling strategies to enhance the model's ability to generalize across diverse chemical spaces.

## 5. Molecular Generation

In **Section 4**, we discussed how molecular representations can be leveraged to identify candidate compounds from existing chemical libraries, with a focus on how GNNs capture both intra- and intermolecular interactions. Beyond mining known molecules, an increasingly active direction is to generate novel compounds directly using graph-based generative models. Given the enormous size of chemical space[415] (estimated at over $10^{60}$) compared to the limited coverage of current compound databases ($10^{11}$), generative approaches offer not only efficiency in candidate discovery but also the potential to reveal unexplored chemical structures with favorable pharmacological and pharmacokinetic properties. As introduced in Chapter 2, molecular generation builds upon the core paradigms of generative modeling, including likelihood-based and latent-variable-based methods. However, molecular graph generation poses unique challenges that distinguish it from other graph generation tasks (e.g., social networks or geometric modeling). In chemistry, even a subtle perturbation, such as adding or deleting a bond, or altering an atom or bond type, can lead to invalid molecular structures that violate basic chemical rules. This stands in stark contrast to other domains (e.g., image generation), where flipping a pixel rarely compromises semantic validity.

Beyond structural validity, molecular graph generation often involves domain-specific constraints tailored to drug discovery applications. For example, structure-based generation conditions the model on a protein binding site to generate target-specific ligands; 3D molecular generation is required to capture spatial conformations critical for activity; fragment-based generation focuses on assembling known chemical moieties to ensure synthetic feasibility; property-driven optimization aims to improve pharmacokinetic properties such as clearance or blood-brain barrier permeability; and scaffold-constrained design attempts to modify existing lead compounds while preserving their core structural components. These real-world constraints greatly enrich the generative landscapebut also increase modeling complexity.

In this chapter, we distinguish between two core tasks in molecular design: **molecule generation** and **molecule optimization**. While traditionally addressed separately, these two problems share overlapping formulations and can be unified under a common modeling framework. We aim to clarify this connection and discuss how GNN-based generative models can be effectively applied to diverse drug discovery scenarios, offering conceptual clarity and practical insight into model behavior and design trade-offs.

## 5.1 Free Molecular Generation: The Foundation
## 5.1.1 Atom-wise Molecular Generation

Molecular graph generation (**Figure 13A**) methods can be broadly divided into two categories: **one-shot generation** and **autoregressive (AR) generation**, which are summarized in **Table 17**. Both approaches aim to model the probability distribution over valid molecular graphs, typically formulated as:

$$p(G) = p(n, X, A) = p(X, A|n) \cdot p(n)$$

where $n$ is the number of atoms, $X \in \mathbb{R}^{n \times F_a}$ represents atomic features, and $A \in \mathbb{R}^{n \times n \times F_b}$ is the adjacency tensor encoding bond types between atom pairs. The dimensions $F_a$ and $F_b$ denote the number of atom types and that of bond types, respectively. A key modeling challenge lies in the variable nature of $X$ and $A$, since their shape depends on the sampled number of atoms $n$. Handling such variable-length data remains nontrivial for standard deep learning architectures.

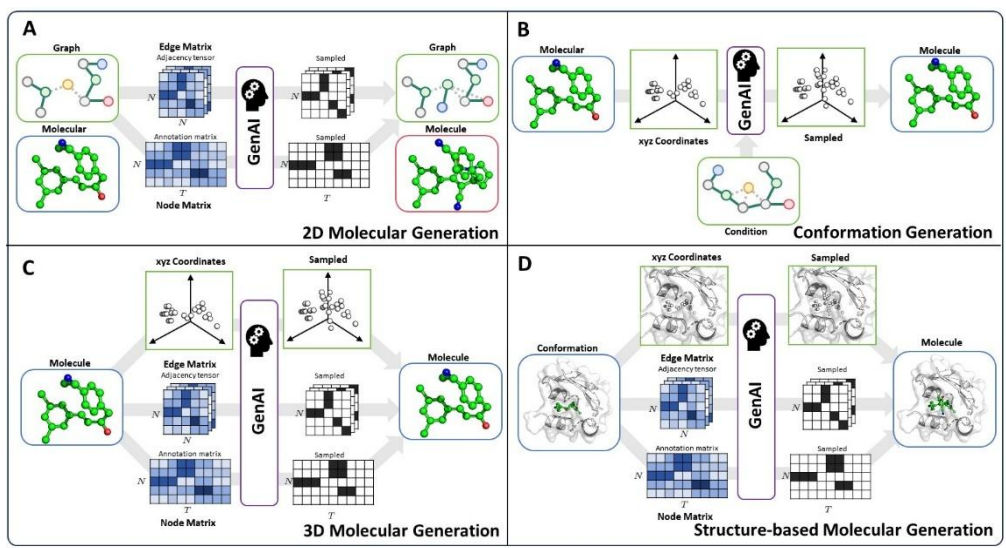

**Figure 13** A) 2D molecular generation, which learns the distribution of nodes and edges; B) Conformation generation, which learns the conditional distribution of 3D coordinates given a fixed molecular graph topology; C) 3D molecular generation, which jointly models the distribution of nodes, edges, and coordinates; D) Structure-based molecular generation, a conditonal form of 3D generation conditioned on the structure of a target protein.

**Table 17**. Overview of free molecular generation models, which learn $p(G)$ in an atom manner.

| Category | Methods | Model | Keywords |
|---|---|---|---|
| OS | GNF[416] | Flow | Normalizing Flow for $X = \text{Flow}(z_x)$, $A = \text{Decoder}(X)$ |
|  | GraphNVP[417] | Flow | RealNVP for $A = \text{Flow}(z_A)$, $X = \text{Flow}(A, z_X)$ |
|  | GraphVAE[418] | VAE | Use VAE to obtain a compact hidden space |
|  | MolGAN[419] | GAN | Ues GAN to sidestep graph matching |
|  | GDSS[420] | Diff | Diffusion for one-shot molecular generation |
|  | DiGress[421] | Diff | Discrete Diffusion for one-shot molecular generation |
| AR | MolecularRNN[422] | Edit | Classical AR molecular generation model |
| | GraphAF[423] | Flow | Use flow to model each step in the AR approach |
| | GraphCNF[424] | Flow | Categorical Normalizing Flows variant of GraphAF |
| | GraphDF[425] | Flow | Discrete Flows variant of GraphAF |

OT refers to one-shot, AR refers to autoregressive, Diff refers to diffusion, and Edit refers to graph editing.

### 5.1.1.1 One-Shot Strategy: Determine Length First

The one-shot approach addresses this challenge by generating the entire molecular graph structure in a single step. Instead of modeling the full joint distribution $p(n, X, A)$, it focuses on the conditional distribution $p(X, A|n)$, which simplifies generation by fixing the atom count in advance. The overall generation process can then be written as:

$$p(G) = p(n, X, A) = p(X, A|n) \cdot p(n)$$

where $p(n)$ is a prior over atom counts, often derived empirically from training data. A sample $n$ is first drawn, and the model proceeds to generate corresponding feature and adjacency matrices of size $n \times n$, representing atom types and connectivity.

Notable examples of one-shot molecular graph generators include Graph Normalizing Flows[416] (GNF) and GraphNVP[417], which are based on flow-based generative models. GNF models the node features $X$ using normalizing flows, followed by a decoder that infers the adjacency structure $A$. The GNF can be formulated as:

$$X = \text{Flow}(z_x), \qquad A = \text{Decoder}(X)$$

where the decoder uses a soft connection function based on Euclidean distance.

$$A_{ij} = \frac{1}{1 + \exp\left(C(\|x_i - x_j\|_2^2 - 1)\right)}$$

where $x_i$ and $x_j$ are latent representations of atoms $i$ and $j$, and $C$ is a temperature parameter (typically set to 10). In contrast, GraphNVP employs RealNVP-style coupling layers to jointly model both node and edge variables. Specifically, it first generates adjacency tensors $A$ from latent noise $z_A$, and then conditions on $A$ and additional noise $z_X$ to generate node features $X$:

$$A = \text{Flow}_A(z_A), \qquad X = \text{Flow}_X(A, z_X)$$

Beyond flow-based models, VAE is also widely used in molecular generation. The continuous latent space induced by VAE is particularly useful for downstream optimization tasks. In VAE-based models, generation is often restricted to the node features $X$, and the number of atoms must typically be specified or fixed beforehand, limiting flexibility in molecular size.

To address this limitation, GraphVAE[418] introduced a placeholder atom strategy. The decoder input is expanded to a fixed size of $k \times F_a$ for node features and $k \times k \times F_E$ for bond types, where $k$ is the maximum number of atoms considered. The model predicts full-size outputs, with unused positions filled with zero-valued placeholder atoms. This approach allows the model to support variable molecule sizes, though it may suffer from memory inefficiencies and is generally applicable only to small-molecule datasets such as QM9. A further complication in one-shot generation is the graph matching problem. Because graph representations are permutation invariant, the same molecule may correspond to multiple adjacency matrices depending on atom ordering. Accurate computation of reconstruction loss thus requires alignment between predicted and reference graphs—a process known as graph matching. While computationally intensive $O(n^3)$, most implementations sidestep this step by assuming consistent atom ordering. However, GraphVAE, due to its placeholder mechanism, must explicitly align generated graphs with reference structures. The alignment is achieved via the Hungarian algorithm, which uses an alignment matrix $X \in \mathbb{R}^{k \times n}$ to transform the original adjacency matrix $A \in \mathbb{R}^{n \times n}$ into a new form:

$$A' = XAX^T$$

where $A' \in \mathbb{R}^{k \times k}$ matches the generated graph size and enables training via RMSE loss. Subsequent studies, such as that by Kwon et al.[426], proposed approximate matching schemes that leverage statistical similarities between input and output graphs, e.g., atom types or bond counts, to reduce computational cost while preserving alignment quality.

In addition to flow- and VAE-based approaches, **MolGAN**[419] emerged as a prominent example of applying GAN to one-shot molecular graph generation. Unlike previous likelihood-based methods, MolGAN avoids the graph matching step required for reconstruction loss computation. Instead, it adopts an implicit generative framework to increase efficiency. Specifically, the generator samples node features $X$ and adjacency tensors $A$ from latent noise $z$, and a discriminator is trained to distinguish real from generated molecular graphs. This setup eliminates the computational overhead of explicit graph alignment, making MolGAN significantly faster in practice. Beyond adversarial learning, MolGAN incorporates an additional RL loss to guide the generator toward molecules with desirable chemical properties, additionally conditioning the generative distribution $p(G|z)$ on target attributes $c$, resulting in $p(G|z, c)$. The overall training objective becomes a weighted sum:

$$L(\theta) = \lambda \cdot L_{\text{GAN}}(\theta) + (1 - \lambda) \cdot L_{\text{RL}}(\theta)$$

This design inspired a wave of follow-up studies[426, 427] that adopt similar strategies by incorporating chemically meaningful feedback scores to enhance the quality of one-shot molecular generation. However, GAN-based models are notoriously unstable during training, with mode collapse being a persistent issue. Although MolGAN introduces Wasserstein GAN[428] (WGAN) architecture to partially alleviate this, instability remains a concern. Subsequent work explored retraining strategies that incorporate high-quality molecules into the generator's learning process to further stabilize training[429]. Despite its early promise, GAN-based models for molecular graph generation have yet to yield broadly adopted, production-grade frameworks, largely due to the intrinsic difficulty of GAN optimization.

Recent efforts have also explored applying diffusion models to one-shot graph generation. Architecturally analogous to flow models, these methods replace the invertible transformations in flows with diffusion-based transformations. For instance, GDSS[420] leverages continuous-time stochastic differential equations for graph diffusion, while DiGress[421] adopts a discrete denoising diffusion framework. Although current applications in 2D molecular graphs remain limited, diffusion models have shown particular promise in structure-based and 3D-aware molecular generation, which will be discussed in detail in later sections.

**5.1.1.2 Autoregressive Strategy: Grow Molecules Gradually**

While one-shot methods simplify generation by sidestepping sequential modeling, they often struggle to produce chemically valid molecules, especially in complex design tasks. **Autoregressive (AR) models** address these shortcomings by modeling molecular generation as a stepwise, sequential decision-making process. Rather than producing the entire graph in a single pass, these models build molecules incrementally, generating a sequence $\{s_0, s_1, \ldots, s_n\}$, where each $s_i$ corresponds to a partial molecular graph at step $i$. The generative probability is expressed as:

$$p(G) = p(s_0) \cdot p(s_1|s_0) \cdots p(s_n|s_0, \ldots, s_{n-1})$$

AR models can be further categorized into **atom-wise** and **fragment-wise** methods. Atom-wise methods sequentially add atoms and predict their connections to the existing graph, while fragment-wise methods operate at a coarser granularity, adding predefined substructures at each step.

In atom-wise modeling, each step $p(s_i|s_{<i})$ is decomposed into two sub-probabilities: one for the atom type $x_i$ and the other for its bonding pattern $A_i$. This decomposition can be ordered in two ways:

$$p(s_i|s_{<i}) = p(A_i|x_i, s_{<i}) \cdot p(x_i|s_{<i}),$$
$$\text{or } p(s_i|s_{<i}) = p(x_i|A_i, s_{<i}) \cdot p(A_i|s_{<i}).$$

The first formulation predicts the atom type before its connectivity, while the second does the reverse. Some models further include prediction of attachment points $f$, and in fragment-based generation, additional variables such as geometric configuration or fragment identity complicate the learning targets at each step.

Despite their more complex structure, AR models often produce more chemically plausible molecules. They can adaptively account for intermediate substructures. For example, they may tend to generate additional rings when cyclic motifs are already present, leading to more stable and realistic outputs. AR models also naturally handle the variable-length generation process. Termination can be modeled explicitly with a "stop" prediction module, where generation halts once $(f|\cdot) < \delta$ for some threshold $\delta$.

One of the early representatives of autoregressive molecular generation is **MolecularRNN**[422], which adopts the "atom-then-bond" decomposition. During generation, the model computes node features through a NodeRNN, followed by atom type prediction via a feedforward layer:

$$h_i^{node} = \text{NodeRNN}(h_{i-1}^{node}, s_{i-1}), \qquad x_i = \text{NodeMLP}(h_i^{node})$$

Then, bond types are predicted through an EdgeRNN operating over previously generated atoms:

$$h_{i,j}^{edge} = \text{EdgeRNN}(h_{i,j-1}^{edge}, s_{i-1}), \qquad A_{ij} = \text{EdgeMLP}(h_{i,j}^{edge})$$

where each $x_i$ denotes an atom type, and each $A_{ij}$ represents the connectivity pattern between atom $i$ and exisiting atom $j$, $h_i$ and $h_{ij}$ are the hidden features of atom $i$ and edge $ij$. The overall likelihood of the generated molecule is computed as the average of per-step probabilities:

$$p = \frac{1}{T} \sum_{t=1}^{T} p(x_i \mid x_{<i}, A_{<i}) p(A_i \mid x_i, x_{<i}, A_{<i})$$

where $A_{<i}$ is the bonding information before the $i$th state. This per-atom log-likelihood can serve as a proxy for evaluating the quality of generated graphs. Building upon the similar autoregressive paradigm, **GraphAF**[423] (Graph Autoregressive Flow) replaces the feedforward prediction layers with flow-based models to improve expressiveness. The model predicts atoms and bonds as follows:

$$p(x_i | s_1, \ldots, s_{i-1}) = \text{Flow}_X(h_i)$$
$$p(A_{ij} | x_i, s_1, \ldots, s_{i-1}) = \sum_j \text{Flow}_A(h_i, h_j)$$

Although flow models are inherently designed for continuous variables, GraphAF adapts them for discrete molecular graphs via dequantization, adding Gaussian noise to discrete inputs to enable continuous modeling. However, this introduces two drawbacks: potential distortion of the original distribution and increased instability due to noise sensitivity. To address these challenges, follow-up work such as **GraphCNF**[424] (Graph Normalizing Flow) and **GraphDF**[425] (Graph Discrete Flow) proposed tailored flow-based solutions for discrete data. GraphCNF learns continuous latent representations for each discrete variable, while GraphDF discards the continuous latent space altogether, using modulo shift transformations within the flow architecture to directly model discrete variables.

In summary, one-shot and autoregressive models represent two complementary paradigms in 2D molecular graph generation. While autoregressive methods typically yield more valid and structured molecules, they suffer from slower inference due to their sequential nature. On the other hand, one-shot methods are more efficient, but often produce less chemically robust molecules. Importantly, the sequential decision-making structure of AR models makes them amenable to **reinforcement learning**, enabling direct integration with policy gradient algorithms for molecular property optimization, a direction we will explore in subsequent sections.

### 5.1.2 Fragment-wise Molecular Generation

Although atom-wise molecular graph generation models have achieved substantial progress, c sequentially adding atoms and bonds often leads to generation quality issues and poor synthetic accessibility. These issues stem from the fact that even subtle errors in chemical structure (e.g., extra bonds, improperly closed rings, or distorted aromatic systems) can render the entire molecule invalid. To mitigate these problems, fragment-based molecular generation has been proposed: by restricting the generation space to the assembly of synthetically feasible fragments, models can improve the validity and synthesizability of the resulting molecules. Mathematically, this approach can be regarded as a form of multi-scale graph generation:

$$p(G) = p(X_F, A_F) \cdot P(X_a, A_a | X_F, A_F)$$

where a coarse-grained fragment graph $(X_F, A_F)$ is first generated to determine the number, types, and connectivity of fragments, followed by decoding into a full atom-level graph $(X_a, A_a)$. This ensures that each generated substructure is chemically meaningful, significantly reducing the synthetic challenge of the final molecule. Under this formulation, the molecular representation shifts from atom-level $p(G) = p(n, X, A)$ to fragment-level $p(G) = p(n_F, X_F, A_F)$, where $n_F$ is the number of fragments, $X_F$ their types, and $A_F$ their connectivity. As with atom-level generation, fragment-based approaches can be either one-shot or autoregressive, while most recent models adopt autoregressive strategies. **Table 18** summarizes these fragment-based generation methods.

**Table 18**. Overview of fragment-based molecular generation models, which learn $p(G)$ in a fragment manner.

| Methods | Model | Keywords |
|---|---|---|
| JTVAE[149] | VAE | Generate junction tree followed by graph decoding |
| HierVAE[430] | VAE | Graph editing actions as the VAE decoder |
| MARS[431] | MCMC | Graph editing to model the Markov chains |
| MoleculeChef[432] | VAE | VAE for a bag of reactants then complete them with reaction predictor |
| PS-VAE[433] | VAE | Mining fragments using the principal subgraph concept |
| MiCAM[434] | VAE | Connection-aware motif-enhanced principal subgraph |

#### 5.1.2.1 Junction Tree VAE and Hierarchical Models

Junction Tree Variational Autoencoder[149] (JTVAE) was one of the earliest models to incorporate the notion of molecular fragments into molecular generation. It decomposes molecular generation into two stages: generating a fragment-based tree structure $T(V, E)$ and decoding it into a molecular graph $G(V, E)$. Specifically, JTVAE first transforms the molecular graph into a junction tree, where each node corresponds to a structural motif such as a ring or a bridgehead atom. Although the construction may appear complex, the junction tree essentially reflects the molecule's topology from a fragment perspective. JTVAE employs a tree decoder and a graph decoder. The tree decoder initiates from a leaf node and incrementally generates the tree using breadth-first traversal. It comprises two main components: a topology predictor $p_t = \sigma(h_{x_i}, h_T)$ that determines whether to grow a new child node, and a label predictor $q_j = \text{softmax}(h_{x_i}, h_T)$ that assigns a fragment type to the new node. Once the full junction tree $T(V, E)$ is generated, the graph decoder translates it into an atom-level graph $G(V, E)$. This is done by enumerating all chemically valid attachment configurations using RDKit and selecting the highest-scoring candidate:

$$\{G_1, G_2, \ldots, G_n\} = \text{RDKit}(\mathcal{T}(\mathcal{V}, \mathcal{E}))$$

$$G = \arg\max_G f(\{G_1, G_2, \ldots, G_n\})$$

Thus, JTVAE encodes molecular features at both atomic and fragment levels into a shared latent space, from which the model samples tree structures and reconstructs molecular graphs. HierVAE[430] extends JTVAE by simplifying the representation of fragments and directly generating fragment graphs without explicitly using junction trees. It uses three autoregressive decoders to predict: the next fragment type ($p_{S_t}$), the attachment site on the new fragment ($p_{A_t}$), and the attachment site on the existing molecule ($p_M$). All three predictions are based on multi-layer perceptrons acting on latent features and the current graph representation $z_G$. To train the latent representation, HierVAE performs hierarchical message passing to encode the initial molecule into $z_G$, which is then used for fragment-by-fragment decoding. Another work, **MARS**[431], takes a different route by modeling the molecule construction process as a Markov chain. Starting from an initial molecule $G_0$, it applies fragment addition or deletion steps iteratively, and after sufficient iterations, the distribution $G_n \sim p(G)$ approximates the target. Both operations are parameterized by GNNs. The model computes probabilities for fragment addition, selection, and deletion via learned neural modules. Each step applies either an addition or deletion with 50% probability, effectively sampling from a learned transition distribution $q(G'|G^t)$. While HierVAE decodes from a latent variable using a VAE framework, MARS operates directly on the molecule structure using discrete fragment editing.

**5.1.2.2 Fragment Mining and Structural Vocabulary Learning**

While JTVAE and HierVAE represent two major lines of fragment-based generation, either of them predicts a full fragment graph before connecting or assembling fragments one by one. Subsequent research has focused on how to better define and utilize fragments themselves. We now discuss a few representative studies in this direction.

**MoleculeChef**[432] argued that most previous methods ignore explicit synthetic constraints. It proposes a two-stage model: first, a VAE generates a set of "synthons", fragment-like precursors; then, a retrosynthesis-inspired model assembles them into a valid molecule. While its encoder is based on GNNs, its decoder relies on RNNs to predict reaction sequences. This design enables the model to generate not only synthesizable molecules but also their corresponding synthesis routes, thus enhancing practical reliability. **PS-VAE**[433] (Principal Subgraph VAE) and **MiCaM**[434] (Mined Connection-aware Motif) adopt a data-driven approach to discover and optimize fragment vocabularies. They point out that JTVAE's decomposition method can oversimplify chemical structures, sometimes breaking fused rings incorrectly. They also note that MoleculeChef's BRICS-based rules[232] are inherently limited by their reliance on predefined cleavage points, restricting the accessible chemical space. PS-VAE addresses this by mining frequent and chemically meaningful subgraphs from large molecular databases to construct a vocabulary, and then follows a VAE-style generation pipeline similar to MoleculeChef, with an additional neural linker to reconnect subgraphs. MiCaM goes further by incorporating not only fragment frequency but also connection patterns between subgraphs. Its "connection-aware motifs" embed both structural and attachment site information, allowing the model to enforce more precise attachment constraints during generation. Furthermore, MiCaM can perform ring closures when predicted attachment points belong to internal atoms, enabling the formation of fused or bridged ring systems. This flexibility allows MiCaM to generalize beyond its predefined motif vocabulary.

## 5.2 Constrained Molecular Generation: Various Application Domains

As previously discussed, unconditional molecular generation aims to learn from distributions of known chemical structures to generate valid molecules. However, such models typically lack the ability to control or design molecules toward specific biochemical goals. In drug discovery, the primary interest lies in conditional generation, where the model is expected to incorporate prior constraints such as target proteins, binding pockets, or pharmacophoric features. Mathematically, this corresponds to modeling the distribution $p(G|c_0, \ldots, c_n)$, where $\{c_i\}_{i=0}^{n}$ represent various conditioning factors. In this section, we outline how to extend unconditional models toward practical, constraint-aware drug design.

### 5.2.1 Molecular Conformation Generation: Foundation of 3D Generation

In the real world, molecules exist in 3D space, and their conformational arrangements directly impact physicochemical properties and biological activities. Consequently, learning to generate valid 3D geometries is of substantial value, especially for downstream applications such as structure-based drug design (SBDD). As discussed in earlier sections, 3D coordinates are inherently equivariant to physical transformations, and thus, most 3D molecular generation models incorporate symmetry-aware architectures. Common choices include equivariant graph networks such as EGNN or SE(3)-Transformers, which directly output coordinates, and invariant networks such as SchNet, which model distance matrices and apply coordinate updates through physically consistent mechanisms.

To begin, we focus on the problem of 3D conformation generation (**Figure 13B**), a task closely related to but distinct from 3D molecular graph generation (**Figure 13C**). While the latter involves simultaneously generating molecular topology and geometry, the former assumes the molecular graph $G$ is given and seeks to predict a physically plausible 3D arrangement of atoms, i.e., modeling $p(R \mid G)$. As one of the earliest applications of geometric deep learning in chemistry, conformation generation has become a cornerstone for many subsequent 3D molecular modeling approaches. Since atom types and counts are fixed, the task is typically framed as a regression or generative problem over a fixed-dimensional coordinate space. Existing methods fall into two main categories, depending on how they handle geometric data: distance-based approaches and direct Cartesian generation, as summarized in **Table 19**.

Table 19. Overview of molecular conformation generation models, which learn $p(R|G)$.

| Methods | Model | Keywords |
|---|---|---|
| GraphDG[435] | VAE | VAE for generating Distance matrix |
| CGCF[436] | Flow | CNF for generating Distance matrix |
| ConfVAE[437] | VAE | VAE for generating Distance matrix |
| ConfGF[18] | Diff | Score matching for D score first, then differentiate to that of XYZ |
| SDEGen[438] | Diff | SDE for generating D, followed by geometry relaxation |
| DMCG[439] | VAE | VAE for generating XYZ, using conformer alignment for equivariance |
| GeoDiff[440] | Diff | DDPM on XYZ; EGNN architecture for equivariance |
| Torsional Diff[441] | Diff | SDE on rotation variable; TFN architecture for equivariance |

D denotes distance, XYZ denotes the Cartesian coordinates, EGNN is E(n)-GNN, and TFN is Tensor-field Network, Diff is the diffusion model.

**5.2.1.1 Distance Geometry-Based Methods**

To preserve physical equivariance, early approaches avoid directly predicting Cartesian coordinates and instead focus on modeling the distance matrix $D$, followed by recovering 3D structures using distance geometry techniques[442]. For example, GraphDG[435] is a representative example of this paradigm, and it employs a GNN encoder to process node and edge features of the molecular graph and uses these to predict pairwise distances via a variational latent space:

$$(h_i', e_{ij}') = \text{GNN}_{\text{enc}}(h_i, e_{ij} \| d_{ij}), \qquad z_i = \text{MLP}_{\text{enc}}(h_i')$$

where $z_i$ represents latent node embeddings aligned to a Gaussian prior. The decoder reconstructs edge distances as follows:

$$(h_i'', e_{ij}'') = \text{GNN}_{\text{dec}}(h_i' \| \hat{z}_i, e_{ij}'), \qquad \hat{z}_i \sim \mathcal{N}(0,1)$$

The model is trained by minimizing a reconstruction loss on $d_{ij}$ and KL divergence in the latent space. Since distance prediction is edge-based and symmetric by nature, this framework is highly flexible and can incorporate a wide variety of GNN architectures (e.g., GIN, GAT) or generative mechanisms (e.g., flows, diffusion models). Notable extensions include **CGCF**[436] (flow-based), **ConfVAE**[437] (CNF-based), **ConfGF**[443] (score matching), and SDEGen[438] (SDE-based diffusion). However, one limitation of this paradigm is that a distance matrix does not uniquely determine a Cartesian embedding, potentially introducing ambiguity during reconstruction.

### 5.2.1.2 Direct Cartesian Coordinate Generation

To avoid the potential loss of geometric fidelity associated with distance-based reconstruction, several methods directly generate 3D coordinates, i.e., they learn $p(R \mid G)$ under SE(3)-equivariant constraints. DMCG[439], for example, introduces physical equivariance through its loss function by minimizing the Frobenius distance after optimal rigid alignment:

$$\ell_{\text{DMCG}}(R, \hat{R}) = \min_{\rho} \| R - \rho(\hat{R}) \|_F^2$$

Here, $\rho$ denotes the optimal rigid-body transformation (via the Kabsch algorithm). Another class of methods utilizes equivariant neural architectures to ensure symmetry-preserving coordinate prediction. For instance, GeoDiff[440] and Torsional Diffusion[441] build on EGNNs to generate coordinates within a diffusion framework:

$$x_i', h_i', e_{ij}' = \text{EGNN}(x_i, h_i, e_{ij}), \qquad \hat{x}_i = \text{Diffusion}(x)$$

It is worth noting that although ConfGF was initially introduced as a distance-based model, its coordinate refinement mechanism approximates an equivariant update:

$$s_\theta(R_i) = \sum_{j \in N(i)} \frac{1}{d_{ij}} \cdot s_\theta(d_{ij}) \cdot (r_i - r_j)$$

This formulation closely mirrors the force field update scheme in SchNet, where gradient-based derivations of distance-informed invariants yield equivariant forces.

### 5.2.2 3D Molecular Generation: Co-Design of Molecular Graphs and Coordinates

Compared with 3D conformation generation, 3D molecular generation atims to model the joint prediction of both molecular topology and the atomic coordinates, written as:

$$p(G, R) = p(n, X, A, R)$$

where $n$ is the number of atoms, $X$ denotes atom types, $A$ the bond adjacency matrix, and $R$ the atomic coordinates. Similar to 2D molecular generation, this task can be approached via one-shot or autoregressive strategies. One-shot models typically extend a 2D generator with an additional coordinate-generation branch, parameterized by equivariant networks. Autoregressive models, by contrast, factor the generation process step-by-step and incorporate coordinate prediction into each intermediate state $s_i$:

$$p(s_i | s_{<i}) = p(R_i | A_i, x_i, s_{<i}) \cdot p(A_i | x_i, s_{<i}) \cdot p(x_i | s_{<i})$$

where $s_i$ may represent either an atom-level or fragment-level state. Compared to 2D models, the key difference lies in the prediction of spatial coordinates, which can be either internal (in terms of bond length $d$, bond angle $\theta$, and torsion angle $\phi$) or Cartesian. Below, we discuss both modeling paradigms in detail, while the methods are summarized in **Table 20**.

**Table 20**. Overview of 3D molecular generation models, which learn $p(G, R)$.

| Category | Methods | Model | Keywords |
| --- | --- | --- | --- |
| OS | E-NF[444] | Flow | Normalizing Flow for 3D MG with EGNN |

| | EDM[445] | Diff | Diffusion Model for 3D MG with EGNN |
| | EMDS[446] | Diff | Edge-involved version of EDM |
| | JODO[447] | Diff | Diffusion Graph Transformer version of EDM |
| | MDM[448] | Diff | Introduce variational noise in diffusion models for 3D MG |
| | MolFlow[449] | FM | Continuous/Discrete flow matching for 3D MG |
| AR | G-SchNet[450] | Edit | Using Schnet autoregressively add atoms to the molecule |
| | G-SphereNet[451] | Edit | Using SchereNet autoregressively add atoms to the molecule |

MG is molecular generation, XYZ denotes the Cartesian coordinates, and EGNN is E(n)-GNN.

### 5.2.3.1 One-Shot Models for 3D Molecular Generation

A typical one-shot approach for 3D molecular generation builds upon a 2D graph generator by adding an equivariant coordinate decoder. For example, E-NF[444] (E(3)-Equivariant Normalizing Flows) leverages EGNN to jointly model atomic types and coordinates via:

$$z_x, z_h = f(x, h) = [x(0), h(0)] + \int_0^1 \phi(x(t), h(t)) dt.$$

where $(x(1), h(1)) = (z_x, z_h)$ are the latent representations of coordinates and atom types, initialized from $x(0), h(0)$. The function $\phi$ is implemented as an EGNN. This paradigm was later advanced by EDM[445] (Equivariant Diffusion Model), which replaces the normalizing flow with a score-based diffusion model and achieves state-of-the-art performance. However, both models sidestep explicit graph generation; bond connectivity is inferred post hoc using rule-based heuristics from Open Babel[452], potentially introducing inconsistencies.

To address this, EMDS[446] and JODO[447] extend EDM by jointly generating both bond matrices and 3D coordinates. EMDS employs stochastic differential equations (SDEs) for simultaneous modeling, while JODO introduces a graph transformer backbone and an adaptive normalization scheme[453] for conditional generation $p(G, R|c)$, where $c$ could represent protein binding pocket information. In parallel, MDM[448] builds on the distance-based strategy of ConfGF and applies diffusion in distance space, followed by coordinate updates via score-based gradients:

$$s_\theta(d_{ij}) = \text{MLP}\left(\left[h_i \cdot h_j, h_{e_{ij}}\right]\right), \qquad s_\theta(R_i) = \sum_{j \in N(i)} \frac{1}{d_{ij}} \cdot s_\theta(d_{ij}) \cdot (r_i - r_j)$$

Here, $d_{ij}$ denotes pairwise distances, and $s_\theta$ is the learned score function. Additionally, MDM incorporates a VAE-style reparameterization of noise variables to improve sample diversity. MolFlow[449] explores the application of flow matching to 3D molecular generation, extending EDM into the discrete domain. It introduces a SimplexFlow model for discrete variables (atoms and bonds) but reports diminished performance, highlighting the gap between intuitive design and practical stability.

### 5.2.3.2 Autoregressive Models for 3D Molecular Generation

Analogous to 2D autoregressive models, 3D autoregressive methods predict the type, position, and bonding of each new atom conditioned on the existing subgraph. Representative examples include G-

SchNet[450] and G-SphereNet[451] developed by the same group, where the former considers bond lengths and the latter also incorporates angles and torsions. In G-SchNet, the generative process is factorized as:

$$p(s_i|s_{<i}) = p(d_{ij}|A_i, s_{<i}) \cdot p(A_i|s_{<i})$$

where $s_i$ is the status of $i$, i.e., all information of the previously generated subgraph, $A_i$ is the type of atom $i$, and $d_{ij}$ is the distance between atoms $i$ and $j$. That is, given the atom type, the model predicts distances to previously placed atoms. The new coordinate $x_i$ is then obtained by solving a constrained minimization problem:

$$\min_{x_i} \sum_j \|\|x_i - x_j\|_2^2 - d_{ij}\|_2^2$$

where $x_j$ are the known coordinates and $d_{ij}$ are predicted distances. Similar to E-NF, these models rely on empirical rules for assigning bond types after coordinate generation, rather than learning connectivity directly.

### 5.2.3 Structure-based Molecular Generation: Directly Design Molecules to Targets

With the foundation of 3D molecular generation, models are now capable of generating molecules in 3D space. By conditioning the generation process on protein pocket information, one can perform structure-based molecular generation (SBMG, **Figure 13D**), which is a central task in real-world drug design. SBMG mirrors how medicinal chemists rationally design ligands: by analyzing the geometric features and interaction hotspots of a protein pocket, they design compatible molecular structures to achieve favorable binding. Formally, this corresponds to learning the conditional distribution $p(G|p)$ or, for more interaction-aware formulations, $p(G, R|p, p_R)$, where $G$ denotes the molecular graph, $R$ its 3D conformation, and $p, p_R$ represent the protein sequence and its structure, respectively. In low-data regimes, modeling protein-ligand interactions usually improves convergence and generalization.

Protein structure encoding can be implemented in several ways: one strategy leverages sequence-based language models (e.g., Transformers) to extract token-level embeddings[350], while another discretizes the 3D structure into voxel grids for processing with CNNs. However, the cubic computational cost of 3D voxel-based CNNs often becomes prohibitive at high resolutions[454]. GNNs, by contrast, offer a more efficient and chemically intuitive framework by directly operating on the protein's geometric and topological information. Accordingly, most SBMG models adopt GNN-based encoders and pair them with symmetry-aware generative models. Both autoregressive and one-shot generation strategies can be employed in this context, as summarized in **Table 21**.

Table 21. Overview of Structure-based molecular generation models, which learn $p(G, R|p)$.

| Category | Methods | Keywords |
| --- | --- | --- |
| AR | GraphBP[455] | $\{d, \theta, \phi\}$ for coordinate generation |
|  | 3D-SBDD[456] | Using MCMC for sampling a joint distribution $p(a_i, r_i|s_{<i})$ |
|  | Pocket2Mol[457] | Decompose the AR to focal, xyz, atomic type, and bond generation |
| | ResGen[458] | Multi-model for reducing the computational budget in 3D SBMG |
| | DeepICL[459] | Integrating interaction profile of reference molecules for generation |
| | SurfGen[342] | Lock-and-key analogy with Geodesic- and Geoattn- GNN |
| | PocketFlow[460] | Flow model in each step in AR-state |
| | FLAG[461] | Fragment-based model with five steps in each AR-state |
| | FragGen[462] | Complex fragment-based model with seven steps in each AR-state |
| | DiffSBDD[463] | Diffuse on continuous xyz and discrete atom types |
| | DiffBP[464] | Diffuse on continuous xyz and discrete atom types |
| OS | TargetDiff[465] | Diffuse on continuous xyz and discrete atom types |
| | PIDiff[466] | Introducing L-J potential in the training loss for tighter binding |
| | KGDiff[467] | Using docking score guidance in diffusion sampler |
| | DecompDiff[468] | Achieving multi-Gaussian prior (fragment locations prior) |

AS is autoregressive; OS is one-sot.

**5.2.4.1 Autoregressive SBMG Models**

Compared to standard 3D generation, autoregressive SBMG faces two key challenges: how to inject protein information into the generation process, and how to define the initial generation point for the ligand. For the first issue, a common solution is to construct a compound graph by merging ligand and protein nodes, enabling GNN-based message passing to integrate structural signals. This is often realized using: (1) bipartite graphs connecting all protein-ligand node pairs, (2) radius graphs that connect only spatially proximal nodes, or (3) k-NN graphs to ensure degree regularization. Each formulation introduces different priors-global context (bipartite), local interactions (radius), or topological stability (k-NN). These edge-building methods allow ligand nodes to absorb multiscale protein information, guiding the generative model toward biologically plausible outputs.

The second challenge is the initialization of the first ligand atom. In standard 3D models, the first atom is often randomly selected, but this lacks biological interpretability in SBMG. A widely adopted compromise is to identify the protein-ligand atom pair $(a_p, a_l)$ with the shortest interatomic distance in a known complex. The ligand atom $a_l$ is then set as the starting point, and the generation path is constructed via breadth-first traversal. While not optimal, this scheme enforces consistency between training and inference and has demonstrated stable performance in practice.

Implementation details vary across models, depending on their GNN backbone, generation order, and handling of geometric variables. For example, GraphBP[455] adopts an autoregressive sequence of center prediction → atom type → coordinate generation. Coordinates are generated in internal coordinates $\{d, \theta, \phi\}$ using a VAE to model the joint distribution of $\{a, d, \theta, \phi\}$. While internal coordinates are rotation- and translation-invariant, they introduce error accumulation due to the need to define second- and third-order atom relations early in the process, limiting GraphBP's empirical performance.

An alternative strategy is to directly predict Cartesian coordinates $r$, often reformulated as relative displacements $\Delta r_i = r_i - r_{i-1}$ to minimize absolute error. To satisfy SO(3) equivariance, these displacements are typically parameterized using equivariant networks such as EGNN or GVP. 3D-SBDD[456] was among the first to jointly model atom types and positions with:

$$p(a_i, r_i | s_{<i}) = \frac{\exp(f[a_i])}{Z}, \qquad f[a_i] = \text{MLP}(h_{r_i})$$

where $h_{r_i}$ is the node embedding at $r_i$, and $Z$ is an implicit normalization term. The conditional formulation is:

$$p(a_i | r_i, s_{<i}) = \frac{\exp(f[a_i])}{1 + \sum_{a_i} \exp(f[a_i])}, \qquad p(\emptyset | r_i, s_{<i}) = \frac{1}{1 + \sum_{a_i} \exp(f[a_i])},$$

The sampling of 3D-SBDD is done via MCMC. Pocket2Mol[457] further decomposes this distribution and adds bond-type prediction. PocketFlow[460] leverages normalizing flows to model each autoregressive component. However, autoregressive inference scales heavily with the ligand size. For example, GraphBP requires $4n$ inference steps for $n$ atoms, which makes inference expensive when combined with protein GNN encoding. ResGen[458] mitigates this by coarse-graining the molecular system, enabling parallel modeling on the same GPU.

Some methods incorporate stronger physical priors during generation. For instance, DeepICL[459] annotates protein nodes with key interaction points and autoregressively generates ligand atoms to satisfy these contacts. While effective in encoding interaction priors, it struggles with multi-ligand systems. SurfGen[342] treats the protein as a topological surface, using Geodesic-GNN and GeoAttn-GNN to extract rich geometry-aware signals, offering a conceptual interpretation akin to "designing keys for a given lock".

Fragment-based autoregressive models, such as FLAG[461] and FragGen[462], represent another line of work. Their key challenge lies in constructing physically realistic poses for the next fragment. FLAG, for example, includes the steps for center prediction, fragment classification, connectivity prediction, torsion angle sampling, and pose refinement. During attachment prediction, FLAG adopts a strategy similar to JTVAE: for each new fragment, the model exhaustively enumerates all chemically and spatially feasible attachment points on the current molecular scaffold. For each candidate attachment mode $G$, a neural network computes a compatibility score $f(G^c)$. The best configuration is selected via:

$$G_i = \arg\max_{G^c} f(G^c)$$

where $G^c$ denotes all possible combinations of the new fragment and the current molecule, and $G_i$ is the configuration with the highest score. If there are 10 candidate attachment sites, the model performs 10 forward passes for scoring, highlighting the computational cost associated with exhaustive evaluation. Following attachment, FLAG predicts the optimal torsion angle for the newly placed fragment using:

$$\Delta\alpha = \text{MLP}(h_{cnt_1}, h_{cnt_2}, h_G)$$

Here, $h_{cnt_1}, h_{cnt_2}$ are the local embeddings of the connecting atoms from the scaffold and the new fragment, respectively. $h_G$ is a globally pooled representation of the existing molecular graph, enabling the model to incorporate global structural context. Next, FLAG refines the geometry of the molecule through force-based position updates. The neural network predicts pairwise interatomic forces $f_{i,j}$ as a function of node features and spatial displacements:

$$f_{i,j} = g(h_i, h_j, r_i - r_j)$$

These forces are then aggregated to update the coordinates of each atom:

$$r'_i = r_i + \frac{1}{n}\sum_{j \neq i} f_{i,j}$$

where $n$ is the total number of atoms. This iterative refinement is inspired by physics-based force fields but learned purely from data. While elegant, it introduces sensitivity to inaccurate force predictions: small errors may compound and distort the final geometry.

FragGen[462] takes this paradigm further by systematizing the entire pipeline of geometry-aware fragment placement. Its autoregressive generation involves seven dependent sub-tasks: (1) center atom prediction, (2) identification of permissible pocket regions, (3) fragment selection, (4) prediction of attachment points on the scaffold, (5) bond type assignment, (6) fragment geometry initialization, and (7) torsion angle optimization. Although this fine-grained decomposition increases model complexity and inference time, it leads to significant improvements in molecular validity, binding relevance, and physical plausibility. In summary, fragment-wise generation in the structure context provides a promising yet technically demanding approach to protein-aware molecule design. The critical open problem is how to reduce inference latency and model complexity while maintaining high-quality generation under geometric and biochemical constraints.

**5.2.4.2 One-shot SBMG Models**

In one-shot SBMG, atomic coordinates are often treated as an additional feature channel and integrated into the diffusion-based generation process, similar to unconstrained 3D molecule generation. The main distinction lies in the need to incorporate protein context at the very beginning by constructing a protein-ligand complex graph. Representative approaches, such as **DiffSBDD**[463], **DiffBP**[464], and **TargetDiff**[465], follow this paradigm, enabling the direct generation of ligand conformations within binding pockets. For example, DiffBP utilizes EGNN to extract features from atomic coordinates and types:

$$x'_i, h'_i = \text{EGNN}(x_i, h_i)$$

To accommodate the nature of each variable, Gaussian noise is applied to the continuous coordinate channel, while categorical noise is used for discrete atomic types. The diffusion steps are defined as follows:

$$q(x \mid x_0) = \mathcal{N}(\alpha_t x_0, \sigma_t^2 I), \qquad q(h_t \mid h_0) = \text{Cat}(h_0 \bar{Q}_t),$$

Here, $\alpha_t$ and $\sigma_t^2$ denote the scale of Gaussian noise at step $t$, and $\bar{Q}_t$ is the cumulative transition matrix in the categorical diffusion process. Building on this foundation, TargetDiff introduces an additional binding affinity prediction module. Instead of using equivariant coordinates, the model uses invariant features $h'_i$ along with an MLP to estimate protein-ligand binding strength, guiding the generation toward more biophysically favorable molecules. PIDiff[466] further refines this approach by explicitly incorporating a Lennard-Jones-like potential into the one-shot generation loss:

$$E_{LJ} = \sum_{j \in P} \sum_{i \in M} \epsilon [(\frac{d'_{ij}}{d_{ij}})^{12} - 2(\frac{d'_{ij}}{d_{ij}})^6]$$

where $\epsilon$ denotes the dielectric constant, while $P$ and $M$ correspond to the protein and ligand atoms, respectively. The gradient of this potential is added to the training loss:

$$L = \sum_{j \in P} \sum_{i \in M} \frac{\partial E_{LJ}}{\partial d_{ij}}$$

This incorporation of physical priors effectively improves the stability of generated molecules, as indicated by reduced RMSD in pre- and post-docking comparisons.

Another work, DecompDiff[468], addresses a different aspect of the diffusion process: the choice of prior distribution. Rather than sampling from a simple Gaussian, the model uses a Gaussian mixture model (GMM), with each mode corresponding to an atom cluster derived from known ligands. During training, these clusters are extracted from decomposed molecular fragments; during inference, atom clusters are predicted based on the binding pocket, enriching the generation with chemically meaningful priors. This design improves both the structural diversity and physical plausibility of the generated molecules.

### 5.2.4 Real-World Challenges in Applying Molecular Generation

#### 5.2.4.1 Challenges in 3D Molecular Generation

We have thus far described how unconstrained generative models can be extended to generate 3D molecular conformations while incorporating protein binding pocket interactions. This enhancement highlights the critical importance of training data quality. A representative example is Torsional Diff, a model focused on torsional angle optimization. During inference, it relies on starting from conformations with physically plausible bond lengths and angles, often generated via the ETKDG algorithm in RDKit. However, training utilizes conformations from semi-empirical quantum calculations, leading to a severe mismatch between training and inference distributions. To mitigate this issue, Torsional Diff aligns training data with RDKit-generated conformations, reducing distributional shifts and improving inference robustness.

Regarding data sources, conformational generation models often use the GEOM[19] dataset, which provides approximately 37 million conformers for 450,000 molecules using the CREST[469] method. Although GEOM enables learning diverse thermodynamically stable conformers, these structures may not correspond to bioactive conformations relevant for drug design[470]. In SBMG tasks, most models utilize the CrossDock[471] dataset, an extension of PDBBind, which augments protein-ligand complexes through pocket similarity clustering and cross-docking. This yields a dataset of over 22 million structures, yet many conformations arise from docking algorithms and may reflect biases in software rather than physical reality[472].

Moreover, current models underutilize available data. For instance, SDEGen is trained on only a subset of 40,000 molecules from GEOM, each contributing five conformers (totaling 200,000 samples). Models like Pocket2Mol or ResGen only use a fraction of the CrossDock dataset. Fully leveraging large-scale conformational data may enhance model generalizability and robustness. Another key issue is conformational diversity within individual molecules. Single-conformer datasets are suitable for regression tasks, where only one stable structure is needed. However, generative models aiming to reproduce the full spectrum of plausible conformations must be trained on datasets reflecting one-to-many mappings. Otherwise, the models may struggle to generate diverse and realistic geometries.

**5.2.4.2 Biases in Evaluating Molecular Generation**

Evaluation of molecular generation remains a complex and multi-faceted challenge. Broadly, assessment metrics can be grouped into three categories. First, statistical metrics evaluate the population of generated molecules based on diversity and novelty. This includes internal diversity (inter-molecular similarity), uniqueness (fraction of non-redundant molecules), and external diversity (similarity to the training set). These are typically computed using the ECFP[473] fingerprints or learned embeddings such as ChemNet[474]. Second, drug-likeness metrics assess individual molecules using criteria such as QED[181], synthetic accessibility[475] (SA), Lipinski's rule of five[476], and logP[477]. These indicators suggest whether a compound is chemically viable and pharmacologically promising. Third, binding affinity metrics measure interactions between generated ligands and target proteins, using docking scores[478], MM-PBSA/GBSA[479] calculations, or QSAR-based models.

Each metric class is susceptible to inherent biases and may conflict with others. For instance, maximizing diversity may reduce similarity to the training distribution, which could either enhance or impair generalization depending on context. Similarly, molecules that score well on SA tend to be small and easily synthesizable, while those with strong predicted binding affinity are often large and synthetically challenging. These trade-offs reflect the multifactorial nature of molecular design and cannot be resolved by optimizing a single metric. Furthermore, individual metrics can be fundamentally flawed. Docking scores, for example, are known to produce high false positive rates in virtual screening campaigns. Thus, many chemists remain skeptical when model performance is evaluated solely using biased or non-validated criteria.

Nonetheless, these metrics are necessary, if not sufficient, for benchmarking model progress. At least, models should achieve baseline performance across key drug-likeness and affinity metrics. For more rigorous validation, wet-lab experiments or expert-driven qualitative assessments are essential. CBGBench[480] was developed to address this by re-implementing existing SBMG methods and providing generated molecules on a shared CrossDock-based benchmark, enabling community-wide comparisons under standardized conditions.

### 5.2.4.3 Why Molecular Generation Has Yet to Scale in Practice

Despite increasing academic interest, the real-world impact of AI-driven molecular generation remains limited due to two fundamental barriers. First, molecular syntheziability is highly sensitive to specific chemical sites, where even minor structural modifications can render a compound unsynthesizable. As noted by Gisbert Schneider, machine-generated molecules are rarely perfect and often require expert-guided modifications[481]. However, identifying editable regions that improve synthetic tractability without sacrificing activity remains an expert-dependent task. Second, most generative models operate effectively only in the preclinical stage. While some metrics such as SA provide coarse estimates of synthetic feasibility but offer little insight into downstream challenges such as toxicity, pharmacodynamics, or ADMET properties. SBMG models, in particular, are optimized to design ligands that bind tightly to a known target, but they provide limited guidance for later-stage clinical considerations.

To overcome these limitations, future molecular generation frameworks must incorporate expert-in-the-loop paradigms. By interacting directly with domain experts, models can leverage human intuition to guide learning in data-scarce settings and prioritize compounds with higher translational potential. In domains where expert knowledge is indispensable, AI should be viewed not as a replacement but as a tool that augments and accelerates human decision-making.

### 5.3 Molecular Optimization: Generate Better Molecules

In the preceding sections, we reviewed how generative models can construct molecular graphs from scratch and how structural and functional constraints can be incorporated into the generative process for different drug discovery tasks. These efforts largely fall within the paradigm of *de novo* molecular design. However, in practical medicinal chemistry, another critical problem is **molecular optimization**, which is the process of modifying existing molecules to improve their drug-likeness, bioactivity, or other properties of interest.

From a modeling perspective, molecular optimization can be formulated as a conditional generation problem, where the goal is to generate a molecule $G_f$ conditioned on an initial molecule $G_0$, such that $h(G_f) > h(G_0)$. Here, $h(\cdot)$ denotes a target property function to be maximized. Based on this formulation, one may either design models specifically for optimization or adapt general-purpose molecular generative models to fulfill this objective. While molecular generation models are often categorized into one-shot and autoregressive classes based on their generation strategy, in the context of molecular optimization, we categorize approaches based on how they tackle the **optimization challenge** itself. Specifically, we divide them into three major groups, as summarized in **Table 22**.

Table 22. Overview of molecular optimization models.

| Category | Methods | Keywords |
|---|---|---|
| RL-Driven | GCPN[482] | First PPO finetuned an atom-wise graph editing model |
| | FREED[483] | Soft AC with fragment edits strategy |
| | Rational-RL[484] | MCTS to identify key substructures, then complete them. |
| | MolDQN[485] | Q-learning for estimating the values of each graph edit |
| | MORLD[486] | Introduce docking score in MolDQN |
| | DeepFMPO[487] | Fragment embedding for fragment edits, trained with AC |
| | DeepFMPO v3D[488] | 3D similarity score version of DeepFMPO |
| | DrugEX v3[489] | Pareto rewards in PPO with a Graph Transformer agent |
| | MolGAN[419] | PPO for finetuning one-shot GraphAF |
| | GraphAF[423] | PPO for finetuning autoregressive GraphAF |
| | GraphDF[425] | PPO for finetuning autoregressive GraphDF |
| | GraphCNF[424] | PPO for finetuning autoregressive GraphCNF |
| Embed-based | VTJNN[149] | Style transform vector on the JTVAE space |
| | HierG2G[430] | Style transform vector on the HierVAE space |
| | MoDof[490] | Introduce delete fragment action based on VTJNN |
| | GraphVAE[418] | Make optimization problem a condition generation |
| | DEL-JTVAE[491] | Evolutionary algorithm on JTVAE embedding space |
| Search-Based | GraphMCTS[492] | MCTS with pre-trained atom-wise graph editing |
| | MolSearch[492] | Pareto rewards MCTS with MMP-defined fragment editing |
| | GraphGA[492] | Graph genetic algorithm on |
| | MoGADdrug[493] | Graph genetic algorithm with fragment edits |
| | MEGA[494] | Multi-objective graph genetic algorithm with fragment edits |

AC refers to the Actor-Critic algorithm.

**Search-based methods**, which directly explore the discrete graph space to identify molecules that improve the target property.

**Embedding-based methods**, which encode molecules into a continuous latent space, perform optimization therein, and then decode optimized latent points back to molecular graphs.

**RL-based methods**, which either fine-tune pre-trained generative models using policy gradients or directly model the graph construction process through value-based learning.

### 5.3.1 Search-based Methods: From GA to MCTS

A natural and intuitive strategy for molecular optimization is to directly search the discrete chemical space by editing molecular graphs and evaluating candidates according to a target property. However, the discrete nature of graph structures renders many classical optimization methods, such as simulated annealing or particle swarm optimization, ineffective. This limitation has led researchers to explore discrete search strategies combined with graph-editing operations to not only approach objectives of interest but also ensure chemical validity of optimized molecules. Two prominent discrete search paradigms have been utilized in the search-based methods, i.e., Genetic Algorithms (GA) and MCTS.

### 5.3.1.1 Genetic Algorithm-Based Search

In the GA framework, **GraphGA**[495] builds upon early efforts, such as Nathan Brown's work[496] and the ACSESS[497] (Chemical Space Exploration with Stochastic Search), to perform direct optimization over molecular graphs. It defines a set of mutation operators at both the node level (e.g., appending, trimming, inserting, deleting, or replacing atomic fragments) and the edge level (e.g., adding, deleting, or modifying chemical bonds). These operations allow for flexible structural transformations while preserving graph connectivity and chemical validity. Moreover, GraphGA introduces advanced crossover strategies, such as multi-point crossover and subgraph exchange, enabling the recombination of molecular substructures from parent molecules to produce offspring with desirable features. By analyzing the ZINC database, GraphGA assigns chemical priors to various operations, guiding the evolution toward property-improving molecules over multiple iterations.

Similar ideas have been extended in **MoGADdrug**[493] and **MEGA**[494]. MoGADdrug incorporates multiple objectives like similarity to a reference compound and drug-likeness through a weighted scoring scheme. Its initial population consists of acidic and amine fragments, which are evolved through standard GA operations. MEGA further enriches the chemical space exploration by using graph-based chromosome representations. It leverages fragment mining tools to extract substructures annotated with synthetic sites and weights. During evolution, fragments with higher weights, which are termed "advantageous fragments", are preferentially used, enabling efficient navigation of the molecular space.

### 5.3.1.2 MCTS-Based Search

In contrast to the population-based paradigm of GAs, GraphMCTS[495] applies the MCTS strategy to molecular graph optimization. As we discussed in Chapter 2, MCTS operates by iteratively selecting, expanding, simulating, and backpropagating within a search tree, continuously updating the statistics associated with each decision branch. When applied to molecular design, each node in the search tree represents a candidate molecule. Expansion corresponds to chemical modifications such as atom addition, ring formation, or bond type alteration, which yield new molecular structures. If a simulated modification leads to an improved property value, such as better binding affinity or increased QED score, the corresponding path in the tree receives a positive update, steering the search toward more promising regions of chemical space.

**MolSearch**[492] builds upon GraphMCTS with several key innovations. In the expansion phase, it incorporates Matched Molecular Pair (MMP) transformations mined from chemical databases. These transformations introduce diverse and chemically valid editing operations that support broader exploration of the molecular space, including fragment substitution and bond rearrangement. To further align with practical drug development pipelines, MolSearch introduces a two-stage optimization scheme. The HIT-MCTS stage focuses on affinity-related objectives, such as docking scores, while the LEAD-MCTS stage targets drug-likeness metrics including QED and synthetic accessibility (SA). This separation mirrors the typical workflow in medicinal chemistry, where initial lead identification is followed by lead optimization. In addition, MolSearch leverages Pareto-based ranking to balance multiple objectives during optimization. This multi-objective framework allows the algorithm to maintain diversity in the candidate pool and to avoid bias toward any single metric. Such design enables more flexible and robust discovery of high-quality molecules in complex chemical spaces.

### 5.3.2 Graph Embed Methods: GNN-encoded Latent Molecules

A key challenge in search-based molecular optimization lies in the difficulty of performing gradient-based or continuous optimization over the discrete space of graph structures. To overcome this, a natural strategy is to encode molecular graphs into a continuous latent space and then perform optimization within that space. Early models such as GraphVAE introduced this idea by using a VAE to learn continuous molecular representations. However, directly formulating molecular optimization as a conditional generation problem, in the spirit of conditional VAEs (cVAEs) in computer vision, is often less effective in the context of drug design. This is mainly because molecular properties are typically represented by single scalar values, lacking the rich semantic content of image-level conditions. As a result, the molecular optimization filed has developed alternative approaches that better accommodate the discrete nature of graphs and scalar conditions, including graph-to-graph translation and latent space optimization.

#### 5.3.2.1 Graph-to-Graph Translation

The concept of graph-to-graph translation is analogous to style transfer in images, where a source molecule $G_0$ is transformed into a target molecule $G_f$ through operations in a latent space. A representative method is VTJNN[149] (Variational Junction Tree Neural Network), which is built upon the JTVAE framework. VTJNN maps a molecule into both junction tree-level and graph-level latent representations, and then learns a latent translation vector from $G_0$ to $G_f$.

For a pair of molecules $X$ and $Y$ (corresponding to $G_0$ and $G_f$), the model defines transformation vectors as follows:

$$\delta_T^{X,Y} = \sum c_T^Y - \sum c_T^X, \quad \delta_G^{X,Y} = \sum c_G^Y - \sum c_G^X$$

$$z_T \sim \mathcal{N}\left(\boldsymbol{\mu}(\delta_T^{X,Y}), \boldsymbol{\Sigma}(\delta_T^{X,Y})\right), \quad z_G \sim \mathcal{N}\left(\boldsymbol{\mu}(\delta_G^{X,Y}), \boldsymbol{\Sigma}(\delta_G^{X,Y})\right),$$

Here, $c_T, c_G$ denote latent features at the tree and graph levels, respectively, and the sampled latent shifts $z_T, z_G$ represent the transformation from the source to the target molecule. During inference, these sampled shifts are added to the original latent vectors of $G_0$, producing transformed representations:

$$\tilde{x}_T = x_T + \delta_T^{X,Y}, \quad \tilde{x}_G = x_G + \delta_G^{X,Y}$$

These are then decoded using the JTVAE decoder to reconstruct the optimized molecule $G_f$. Building on VTJNN, MoDof[490] (Modifier with One Fragment) introduces more stringent chemical constraints by focusing on local transformations. It assumes that an optimized molecule should largely preserve the original molecular structure, with modifications limited to specific substructures. To this end, MoDof adds a bond disconnection prediction module during decoding, allowing local changes to be made precisely on the input molecule, thereby enabling fine-grained and chemically interpretable optimization. A related development is HierG2G[430], an extension of HierVAE that incorporates graph-to-graph optimization into a fragment-level hierarchical framework. HierG2G defines translation vectors at both the fragment and graph levels and uses a MLP to parameterize the corresponding Gaussian distributions:

$$\delta_S^{X,Y} = \sum c_S^Y - \sum c_S^X, \quad \delta_G^{X,Y} = \sum c_G^Y - \sum c_G^X$$

$$[\mu_{X,Y}, \sigma_{X,Y}] = \text{MLP}(\delta_S^{X,Y}, \delta_G^{X,Y})$$

where $c_S$ and $c_G$ represent fragment-level and graph-level latent features, respectively. This design enables seamless continuous transformations between molecular states at multiple levels of abstraction.

**5.3.2.2 Latent Space Optimization**

Unlike graph-to-graph translation, which requires explicit training on molecular pairs $(G_0, G_f)$, latent space optimization seeks to directly manipulate the encoded representation of a molecule to improve desired properties. Although this process still relies on optimization algorithms, it operates entirely within the latent space and is thus categorized under embedding-based methods. For instance, in property-guided optimization, a QSAR model is first trained to predict a target property from the latent representation $e(G)$:

$$\text{QSAR}_i = h(e(G_i))$$

Given a source molecule $G_0$, optimization algorithms such as GA or particle swarm optimization (PSO) are used to search for an optimal latent vector $e(G_f)$ that maximizes the QSAR score. The optimized vector is then decoded back to a molecular graph using the inverse mapping $e^{-1}(e(G_f))$. This approach does not require molecular pairs and allows flexible reuse across different optimization tasks by simply changing the QSAR objective.

A prominent example is **DEL**[498] (Deep Evolutionary Learning), which combines a variational autoencoder with an evolutionary search strategy. DEL first trains a FragVAE model to learn a continuous fragment-level latent space. Crossover and mutation operations are performed on latent vectors, enabling smooth transitions and avoiding the challenges associated with editing discrete chemical structures. DEL-JTVAE extends this idea by replacing FragVAE with JTVAE, thereby enhancing the structural diversity of the generated molecules. Several other approaches adopt similar optimization strategies in the latent space, often based on SMILES representations. These are not discussed here in detail, as the underlying optimization methodology remains consistent across different molecular encodings.

### 5.3.3 RL-based Methods: Modify Structures as a Human

RL-based approaches for molecular optimization are closely related to autoregressive molecular generation. In principle, any autoregressive generative model can be repurposed as a molecular optimization model using RL. The core idea lies in how the generation process is structured: an autoregressive model seeks to learn the joint distribution over a molecular graph $G$ via sequential factorization:

$$p(G) = p(s_0) \cdot p(s_1|s_0) \cdots p(s_n|s_0, \ldots, s_{n-1})$$

During maximum likelihood training, the objective is to maximize the likelihood of observed molecules from the dataset. In optimization scenarios, however, the objective is redefined to update this generation process so that the resulting molecules better align with a given target function $h$. In other words, the goal is to learn a conditional policy $p(s_i|s_{j<i}, h)$ that maximizes the expected reward associated with the generated molecule $G$ under constraint $h$. While the autoregressive model provides the structural framework to ensure syntactic validity, RL introduces preferences to guide generation toward chemically meaningful and property-specific directions. Two major paradigms are commonly employed in this context: policy gradient methods and value-based methods.

### 5.3.3.1 Policy Gradient Methods

Policy gradient-based models directly learn the generation policy $p(s_i|s_{j<i}, h)$ to bias autoregressive models toward producing optimized molecules. A representative example is GCPN[482] (Graph Convolutional Policy Network), which uses GCNs to parameterize the generation policy. The model encodes the molecular graph using GCN and defines the action probabilities as follows:

$$h'_i = \text{GCN}(h_i, e_{ij})$$

$$f_{\text{first}}(s_t) = \text{softmax}(\text{MLP}(h'_i))$$

$$f_{\text{second}}(s_t) = \text{softmax}\left(\text{MLP}\left(h'_{i_{a_{\text{first}}}}, h'_i\right)\right)$$

$$f_{\text{edge}}(s_t) = \text{softmax}\left(\text{MLP}\left(h'_{i_{a_{\text{first}}}}, h'_{i_{a_{\text{second}}}}\right)\right)$$

$$f_{\text{stop}}(s_t) = \text{softmax}(\text{MLP}(\sum_i h'_i))$$

These components respectively represent the probabilities of selecting the first atom $f_{\text{first}}$, the second atom to connect $f_{\text{second}}$, the bond type to form $f_{\text{edge}}$, and whether to terminate generation $f_{\text{stop}}$. Unlike earlier generative models trained via maximum likelihood, GCPN is trained using the PPO algorithm. Its objective function is given by the clipped surrogate loss:

$$L^{CLIP}(\theta) = \mathbb{E}_{(s_t, a_t) \sim \pi_{\theta'}}[\min(r_t(\theta)A_t, \text{clip}(r_t(\theta), 1-\epsilon, 1+\epsilon)A_t)]$$

The exact definition of the clipped surrogate loss $L^{CLIP}(\theta)$ can be found in Section 2.6. For readers less concerned with the mathematical derivation, it suffices to understand that this loss function enables the model to incorporate target-specific constraints $h$ into the originally unconditional policy $p(s_i|s_{j<i})$, yielding a constrained generation process $p(s_i|s_{j<i}, h)$. In this way, RL acts as a preference model, guiding the autoregressive generation pathway toward regions of chemical space aligned with the desired molecular properties.

Building upon the GCPN framework, several studies have introduced improvements in both the model architecture and training dynamics. For example, **DrugEx v3**[489] replaces the graph convolutional encoder with a Graph Transformer to enhance the representation power of molecular structures. Other models such as **GraphAF**[423], **GraphDF**[425], and **GraphCNF**[424] follow a similar paradigm by integrating PPO into their respective autoregressive generation backbones. These efforts exemplify a widely adopted strategy in molecular optimization: rather than training a generative model from scratch under new constraints, it is often more practical to fine-tune an existing model through RL. This not only saves computational resources but also leverages previously learned structural priors. It is also important to highlight that MolGAN[419], although categorized as a one-shot generative model, employs a policy gradient framework to enhance its capacity to generate chemically valid molecules. This is achieved by conceptualizing the one-shot generation as a special case of sequential decision-making with a sequence length of one.

Within the policy gradient framework, various improvements have been proposed. These include the use of actor-critic algorithms and the shift from atom-level to fragment-level editing. **DeepFMPO**[487] is a representative model in this category. It introduces a binary tree encoding of molecular fragments based on the Tanimoto and Levenshtein similarity, enabling fragment-level operations that resemble molecular fingerprints. The model combines policy-based and value-based optimization using an actor-critic architecture, achieving faster convergence and enhanced stability. Inspired by the medicinal chemistry philosophy of Murcko[499], DeepFMPO emphasizes local modifications near a given scaffold rather than global diversity. **DeepFMPO v3D**[488] extends this framework by incorporating 3D electronic similarity into the reward function, enabling spatially constrained molecular optimization. However, the reliance on the predefined fragment templates in DeepFMPO may limit its flexibility. **FREED**[483] addresses this limitation by introducing a more general set of graph editing operations and employing soft Actor-Critic for training. The policy network is composed of three components:

$$p^{act1} = \text{softmax}(MI(h_g, h_i))$$
$$p^{act2} = \text{softmax}(z_{act1}, \text{ECFP}(h_{cand}))$$
$$p^{act3} = \text{softmax}(z_{act2}, h_{att})$$

These components sequentially determine the attachment point on the current molecule $p^{act1}$, the type of fragment to be attached $p^{act2}$, and the exact site on the fragment to perform the connection

$p^{act3}$. The integration of these three networks enables effective and flexible fragment-based optimization in a chemically meaningful manner.

### 5.3.3.2 Value-Based Methods

In contrast to policy gradient methods, value-based approaches estimate the expected return $Q(s, a)$ of a given state-action pair. Instead of directly learning the generation probability $p(s_i|s_{j<i}, h)$, these methods learn a value function that evaluates the long-term reward of taking a particular action in a given molecular state. The policy is then derived by selecting the action with the highest $Q$-value. A seminal example is MolDQN[485], which highlights several advantages of this framework: it does not require pretraining on large molecular datasets and tends to be more sample-efficient and stable due to the well-understood properties of Q-learning. In MolDQN, each state corresponds to a molecular graph, and actions are defined as modifications such as adding atoms, altering bond types, or deleting existing connections. The following are representative bond editing operations defined in the model:

- No atom → one of the allowed atom types;
- No bond → single, double, or triple bond;
- Single bond → double or triple bond;
- Double bond → triple bond;
- Triple bond → single, double, or no bond;
- Double bond → single or no bond;
- Single bond → no bond.

To improve training stability, MolDQN adopts a double DQN architecture, which mitigates overestimation of Q-values. It also employs bootstrapped DQN and epsilon-greedy exploration to enhance policy diversity. Several subsequent studies have built upon MolDQN by incorporating additional property-specific objectives. For example, MORLD[486] incorporates docking scores into the reward function, thus enabling optimization in structure-based design tasks.

### 5.3.4 Challenges: Multi-objective and Many-objective Optimization

In realistic drug discovery scenarios, molecular optimization must account for a wide range of factors beyond efficacy alone. A successful therapeutic agent must balance efficacy with safety, pharmacokinetics, administration route, off-target effects, and manufacturability[500]. Consequently, molecular design is inherently a **multi-objective optimization** (MOO) problem. In broad terms, the objectives include maximizing (1) efficacy, (2) structural novelty, and (3) favorable pharmacokinetic properties, while minimizing (4) synthetic complexity and cost, and (5) toxicity or side effects. A central challenge of lead optimization is how to extend single-objective optimization methods to multi-objective contexts without compromising the model's ability to navigate conflicting criteria. One straightforward approach is to combine multiple objectives into a weighted sum, forming a scalar surrogate objective such as $h_{MMO} =$

$\sum w_i h_i$, where $w_i$ are task-specific weights. However, this approach offers limited flexibility in adjusting the relative importance of each objective and may obscure trade-offs when objectives are inherently conflicting. For example, synthetic accessibility tends to favor smaller molecular structures, while drug-likeness often prefers molecules within a moderate molecular weight range. As a result, effective MOO requires more nuanced techniques capable of decomposing and managing such trade-offs.

**5.3.4.1 Pareto Optimality: A More Flexible Optimization Paradigm**

To address these conflicts between competing objectives in molecular optimization, the concept of **Pareto optimality** has been widely adopted. A solution is said to be Pareto-optimal if no objective can be improved without deteriorating at least one other. The set of all such non-dominated solutions forms the **Pareto front**, which provides a spectrum of trade-offs across the objective space. In drug design, the Pareto front enables chemists to explore and select from a portfolio of candidate molecules that exhibit distinct yet balanced trade-offs between critical properties, such as binding affinity, synthetic feasibility, safety profile, pharmacokinetics, and cost.

Among search-based approaches, Pareto optimization has been effectively implemented. Notably, **non-dominated sorting genetic algorithms**[501] (NSGA) sort each generation of candidate molecules based on dominance relationships and selectively retain those approximating the Pareto front. Over successive iterations, this process incrementally improves the population toward a more balanced set of trade-offs. In contrast, RL-based molecular optimization has not yet fully embraced the Pareto framework. Most current RL-based methods still collapse multiple objectives into a single scalar function via fixed weightings. Developing more flexible RL paradigms that can simultaneously maintain solution diversity and optimization balance across multiple objectives remains an open and important research direction.

**5.3.4.2 Many-Objective Optimization (ManyMO): The High-Dimensional Frontier**

Within the broader context of MOO, it is important to distinguish between **Multi-objective optimization** (typically involving up to three objectives) and **Many-objective optimization (ManyMO)**, which involves four or more. As the number of objectives increases, the Pareto front evolves from a curve in two dimensions to a surface in three, and ultimately to a high-dimensional manifold in the ManyMO setting. The size of the non-dominated solution set grows exponentially with the number of objectives, making it difficult to visualize, evaluate, and converge upon representative solutions.

To address the complexity of ManyMO, a number of algorithmic strategies have been proposed. For instance, Angelo et al.[502] categorize solutions into five methodological families: relaxed dominance, indicator-based selection, objective decomposition, dimensionality reduction, and hybrid frameworks. These strategies are designed to navigate the high-dimensional solution space more effectively, balancing the need for diversity with the requirement for convergence.

However, it is worth noting that real-world drug discovery rarely requires optimizing more than a handful of objectives at once. In practice, most tasks focus on three or four core criteria, enabling chemists to maintain intuitive control over the optimization process. For example, Zhang et al.[503] employed the Delete model, which is designed for lead optimization, to optimize lead compounds with respect to only three parameters: binding affinity, half-life, and clearance rate. This low-dimensional optimization not only improves interpretability but also facilitates visual selection and comparative analysis. Nonetheless, many MO research continues to push the frontier of automated molecular design. If efficient and robust optimization can be achieved in high-dimensional settings, it would substantially advance the automation and scalability of rational drug discovery pipelines.

# 6. Knowledge Graph

While previous sections have explored GNN applications on molecular graphs for tasks such as property prediction, virtual screening, and molecular generation/optimization, these models often fall short of capturing the system-level complexity of drug discovery. From preclinical research to clinical trials, therapeutic development involves multi-scale processes spanning genes, proteins, pathways, tissues, and phenotypes. Focusing solely on molecular structures risks overlooking the broader biological context essential for understanding drug mechanisms, adverse effects, and combinatorial therapies[504, 505]. In this context, knowledge graph (KG) has emerged as a powerful tool in biomedical research. By organizing heterogeneous and multi-relational biological data within a unified graph framework, KG enables the integration of information spanning molecular, cellular, and clinical levels. This enables a more holistic understanding of the complex relationships among drugs, diseases, genes, and other biomedical entities, thereby facilitating more informed and mechanistically grounded decision-making in drug discovery and development[506].

## 6.1 Knowledge Graph Construction

### 6.1.1 Basic Concepts of Knowledge Graph

A KG is a graph-based data structure used to represent structured and conceptual knowledge across one or multiple domains. It consists of nodes, commonly referred to as entities or concepts, and edges that define the relationships among them[507]. Formally, a KG can be denoted as:

$$G = (V, E, R)$$

where $V$ is the set of entities, $R$ is the set of relation types, and each edge $e \in E$ is typically represented as a triplet $(h, r, t)$ corresponding to a "head entity" $h$, a relation $r$, and a "tail entity" $t$. Biomedical KGs are inherently heterogeneous, including a wide variety of entity types such as genes, proteins, drugs, diseases, symptoms, phenotypes, biological pathways, clinical trials, and even patient records. The relations among these entities are also diverse in their semantics. For example, a KG may encode facts such as "gene $X$ encodes protein $Y$", "drug $A$ inhibits protein $B$", "mutation in gene $Z$ leads to disease $D$", or "drug $C$ interacts with drug $D$". These relationships can be either directed (e.g., gene mutation causes disease) or undirected (e.g., protein–protein interaction), depending on the nature of the data source and the specific task. Compared to molecular graphs that describe chemical structures, KGs offer greater semantic richness and topological diversity. The following sections outline the key biological entity types and their associated data sources used in biomedical KG construction, while the related datasets are summarized in **Table 23**.

Table 23. Datasets related in the knowledge graph construction.

| Category | Dataset | Keywords |
| --- | --- | --- |
| Gene | NCBI Gene[508] | Primary source for gene and transcripts. |
|  | Ensembl[509] | Primary source for gene and transcripts. |
|  | UniProtKB[510] | Primary protein resources. |
| Protein | BioGRID[511] | Interactions between gene, protein, and chemicals |
|  | IntAct[512] | Interactions between gene, protein, and chemicals |
|  | STRING[513] | Physical and functional PPIs |
|  | DrugBank[514] | Drug attributes and drug-drug, drug-gene relations |
| Drug | DrugCentral[515] | Drug attributes and drug-gene, drug-disease relations |
|  | CHEMBL[206] | Drug attributes and drug-gene relations |
| Anatomy | BTO[516] | Ontology of anatomical structures in the BRENDA enzyme |
| Process, | GO[517] | Cover biological process and molecular functions |
| Function | KEGG BRITE[518] | Functional subset of KEGG, protein functions… |
|  | KEGG Pathway[519] | Primary source of pathway, gene-pathway, drug-pathway… |
| Pathway | Reactome[520] | Expert-curated, providing fine-grained pathway info |
|  | Pathway Commons[521] | Cross-database pathway info, including BioCyc, PID, etc. |
|  | MeSH[522] | Aid in the indexing of PubMed articles |
|  | DO[523] | Help link different datasets |
|  | HPO[524] | Describe the phenotypes of disease |
| Disease | Mondo[525] | Harmonize disease definitions between ontologies |
| Symptom | DisGeNET[526] | Primary disease resources, experimental and text-mined data |
|  | OMIM[527] | Focus on mendelian disorders, gene disease relations |
|  | KEGG DISEASE[528] | Subset of KEGG, linking disease to drugs and pathways |
|  | ICD[529] | Issued by WHO, providing fine-grained disease descriptions |

#### 6.1.1.1 Entities

#### 6.1.1.1.1 Gene

Genes serve as the fundamental units of genetic information and play pivotal roles in both normal physiology and disease development. Many diseases, particularly cancer, are closely associated with specific gene mutations or aberrant gene expression patterns. In cancer treatment, genes are often categorized into *driver genes*, which directly contribute to cancer progression (e.g., oncogenes and tumor suppressors), and *passenger genes*, which may harbor mutations but are not causally implicated. In a KG, gene nodes typically connect to disease nodes through relationships such as *activates*, *inhibits*, or *associated with*. Genes may also interact with other genes or protein entities via regulatory relationships. For example, in cancer research, tumor suppressor genes such as TP53 and BRCA1 are commonly associated with different cancers through mutation-induced activation[530].

Comprehensive genomic databases provide the foundational data for constructing gene-related nodes and relationships in KG. **NCBI Gene**[508], maintained by the U.S. National Center for Biotechnology Information, and **Ensembl**[509], a collaborative effort led by the European Bioinformatics Institute, offer detailed annotations such as gene sequences, genomic loci, transcript variants, and functional descriptions. These resources also support entity alignment within KGs through standardized gene identifiers, synonyms, and homology mappings. **The Cancer Genome Atlas (TCGA)**[6], a large-scale initiative jointly led by the National Cancer Institute (NCI) and the National Human Genome Research Institute (NHGRI), provides multi-omics data such as genomics, transcriptomics, proteomics, and clinical metadata across diverse cancer typesm, allowing the discovery of driver mutations and genotype-phenotype correlations for graph-based modeling.

### 6.1.1.1.2 Protein

Proteins are the functional products of gene transcription and translation, serving as the direct executors of cellular functions. Because proteins can be viewed as gene products, many biomedical KGs combine genes and proteins into a unified entity type. However, in more fine-grained representations, genes and proteins are modeled as distinct nodes, linked by relations such as *encodes* or *translated into*.

Protein-Protein Interaction (PPI) is critically important in elucidating disease pathways and mechanisms of drug action. Key signaling proteins such as MAPK and AKT frequently appear in the pathways of various diseases and are often modeled explicitly in KGs to capture their essential biological roles[531]. **UniProtKB**[510] serves as the principal protein reference database, offering sequences, functional annotations, and gene–protein mappings. Complementary resources such as **BioGRID**[511], **IntAct**[512], and **STRING**[513] provide experimentally validated or predicted PPI data, including interaction types and confidence scores, facilitating high-quality PPI subgraph construction.

### 6.1.1.1.3 Drug

Drug nodes typically represent compounds that are either approved for clinical use, under research, or identified as natural products with potential pharmacological activity. In a KG, drugs can be linked to disease nodes via relationships such as *treats* or *alleviates*, to protein nodes through relations like *inhibits* or *activates*, and to other drug nodes via interactions that capture synergy or adverse effects. These connections support applications such as drug repurposing, drug-target interaction prediction, and drug combination prediction. **DrugBank**[514] is a comprehensive drug database that includes chemical structures (e.g., SMILES and InChI), drug classifications, molecular targets, mechanisms of action, and clinical trial phases. It also records curated associations between drugs, proteins, and diseases, making it a primary data source for KG construction. **DrugCentral**[515] integrates drug-disease, drug-target, and drug-drug interaction data, annotated by therapeutic indications, contraindications, and other clinical factors, and is particularly suited for drug repurposing or interaction prediction tasks. **ChEMBL**[206], which has also been featured in earlier sections on ADMET prediction, provides detailed annotations for drug-protein bioactivity, offering high-quality information for labeling drug-target relationships in KGs.

**6.1.1.1.4 Anatomy**

Since many diseases exhibit tissue or organ specificity, and drug distribution and metabolism vary across anatomical sites, anatomical structures are often modeled as distinct entities in biomedical KGs. Examples include organs such as liver and kidney, as well as barriers like blood-brain barrier. These anatomical nodes are typically connected to disease nodes via relationships like *site of pathogenesis* or to drug nodes via *distribution* or *metabolism*, enabling spatial reasoning relevant to pharmacokinetics and pharmacodynamics. **BTO (BRENDA Tissue Ontology)**[516] provides standardized terminology and hierarchical classifications for a wide range of tissues and organs. It helps unify anatomical terminology across diverse biomedical databases and facilitates consistent integration of entities like *liver*, *kidney*, *brain*, and *blood-brain barrier* within KG semantics.

**6.1.1.1.5 Biological Processes and Molecular Functions**

Biological processes refer to coordinated sequences of biochemical or physiological events that occur within living organisms, such as cell growth, differentiation, or immune responses. Molecular functions, on the other hand, describe specific biochemical activities of genes or proteins, such as *transcription factor activity* or *metal ion binding*. Including such abstract entities in a KG supports the mechanistic interpretation of how genes and proteins contribute to disease phenotypes and signaling pathways. The **Gene Ontology**[517] (GO) provides comprehensive annotations across both biological processes and molecular functions. It enables knowledge graphs to capture relationships like "gene $X$ participates in process $Y$" or "protein $X$ exhibits function $Z$." **KEGG BRITE**[518], a functional hierarchy system within the KEGG database, adds further structure by offering layered information about protein functions, enzymatic activities, and metabolic roles, supporting systematic functional enrichment and visualization in graph-based analyses.

**6.1.1.1.6 Pathway**

A biological pathway is a series of actions among molecules in a cell that leads to a certain product or a change in the cell, including gene regulation, protein modifications, and metabolic cascades. In KG, pathways can be represented as abstract entities such as the Wnt/β-catenin and PI3K/Akt pathways. These nodes are typically linked to genes or proteins through relations like *participates in*, *regulates*, or *phosphorylates*, capturing upstream and downstream molecular relationships. **KEGG Pathway**[519] is one of the most widely used pathway databases, offering graphical representations of metabolic and signaling networks. These resources can be directly mapped into KGs through relationships such as "gene or protein *X* is part of pathway *Y*" or "drug *Z* affects pathway *Y*". **Reactome**[520] is a curated pathway knowledgebase that provides fine-grained representations of signaling steps, including molecular modifications and complex formation events. Pathway Commons[521] aggregates data from multiple sources including BioCyc[532], PID[533], and HumanCyc[534], making it a valuable integrative resource for pathway-related annotations and evidence in KGs.

**6.1.1.1.7 Disease**

Disease nodes represent pathological conditions characterized by molecular abnormalities or clinical symptoms, such as hyperglycemia in diabetes or amyloid deposition in Alzheimer's disease. In a knowledge graph, diseases are commonly linked to genes or proteins through relationships like "mutation causes disease" or "protein dysfunction leads to pathology". They can also be connected to pathway nodes to indicate aberrant signaling associated with disease mechanisms. Resources like **MeSH**[522], **Human Phenotype Ontology (HPO)**[524], and **Monarch Disease Ontology (Mondo)**[525] provide standardized disease nomenclature and semantic hierarchies. These ontologies facilitate consistent disease representation and hierarchical structuring in KGs. Databases such as **DisGeNET**[526], **OMIM (Online Mendelian Inheritance in Man)**[527], and **KEGG DISEASE**[528] offer curated gene-disease associations along with supporting literature. For example, DisGeNET provides quantitative scores for gene-disease associations, OMIM focuses on inherited disorders, and KEGG DISEASE contributes additional links between diseases, pathways, and therapeutic compounds.

**6.1.1.1.8 Symptoms**

Symptoms represent the clinical or phenotypic manifestations of diseases, such as insomnia or fever. In KG, symptom nodes are usually connected to disease nodes through relationships like *presents as* or *leads to*. Some studies also incorporate patient-specific features into KGs to support personalized medicine applications. **MeSH**, in addition to disease classification, includes clinical terms that cover many symptoms and signs. Researchers can use its hierarchical structure to assess the taxonomic proximity of symptoms and diseases. **International Classification of Diseases (ICD)**[529], issued by the World Health Organization, primarily focuses on disease taxonomy but also includes symptom-related entries in its subcategories, offering finer granularity for symptom representation.

**6.1.1.2 Ontology**

Concrete biological entities such as genes, proteins, and drugs are typically well-characterized by standardized identifiers, for instance, Ensembl IDs for genes and UniProt IDs for proteins. However, incorporating **abstract concepts** such as biological processes, molecular functions, pathways, diseases, or symptoms into a knowledge graph is more challenging due to their **ambiguous boundaries and variable definitions** across databases. In this context, **ontologies** provide a unified semantic framework that enables consistent classification, naming, and interpretation of such abstract entities across diverse datasets and research communities[535].

Ontologies define both the **hierarchical structure** of concepts and the **semantic relationships** between them. This allows knowledge graphs to capture biologically meaningful interactions between abstract entities and concrete elements. For example, by integrating ontologies such as **Gene Ontology (GO)** or **Disease Ontology (DO)**, one can represent semantic relations such as *participates in*, *localized to*, *causes*, or *associated with*, thereby linking genes and proteins to biological processes, functions, or disease states in an interpretable manner[536]. Ontological integration not only ensures semantic consistency across KG but also supports downstream tasks such as automated reasoning, semantic querying, and graph-based visualization.

**6.1.1.3 Relations in Knowledge Graph**

By integrating diverse entity types such as genes, proteins, drugs, anatomical structures, biological processes, molecular functions, signaling pathways, diseases, and symptoms, along with their corresponding databases, KG provides a unified and systematic representation of heterogeneous biomedical data. For each entity class, associated literature and databases contribute annotations, synonyms, hierarchical classifications, and interaction evidence, which can be explicitly encoded in the KG as node attributes or edge relationships. This enables the construction of large-scale, multi-layered, and interdisciplinary networks. Bonner et al.[537] compiled a comprehensive figure of biological entities and their related data sources, serving as a valuable reference for understanding the full scope of biomedical KG representations.

**6.1.2 Knowledge Graph Construction**

Constructing a biomedical KG typically begins with defining the types of entities and relations to be integrated. Based on these specifications, researchers select and combine reliable data sources. The raw data must undergo cleaning and deduplication, ensuring that different aliases or identifiers (e.g., gene or protein name) across databases are mapped to a unified ID or ontology term. Semantic consistency checks are also essential; for instance, if a protein is annotated as participating in "Pathway X" in UniProt but in "Pathway Y" in KEGG, the graph constructor must determine whether these are semantically equivalent or represent distinct, overlapping processes. Once entity and relation alignment is complete, ontologies such as GO, DO, or EFO can be used to provide hierarchical labeling and semantic abstraction. For example, to build a protein-drug interaction network enriched with functional and pathway annotations, researchers may extract protein IDs from UniProt and annotate their functions using GO terms. These proteins can then be mapped to signaling pathways via KEGG, linked to drug interactions via DrugBank, and connected to disease associations using OMIM. The resulting KG integrates multidimensional biological contexts, enabling graph-based modeling approaches to uncover drug repurposing opportunities or elucidate disease-relevant protein mechanisms.

To reduce the complexity and redundancy of integrating data from multiple sources, both academia and industry have developed pre-integrated, unified biomedical KGs. Notable examples include Hetionet[538], DRKG[539], BioKG[540], PharmKG[541], and OpenBioLink[542], which consolidate information from major databases such as DrugBank, UniProt, KEGG, OMIM, and DisGeNET. These KGs encompass a broad spectrum of biological entities, allowing researchers to bypass the burdensome data integration pipeline and focus on algorithm development for tasks like link prediction or node classification. However, a direct adoption of such general-purpose KGs remains subject to debate. These integrated KGs often reflect the design assumptions and biases of their creators, making them well-suited for some applications but suboptimal for others, particularly in certain niche domains. Many downstream tasks, such as disease-specific modeling or pathway-focused prediction, require customized graph construction, where entities, relations, and metadata are tailored to the research objective. As a result, researchers frequently use these general-purpose KGs as a starting point, augmenting or pruning them to produce task-aligned subgraphs. This hybrid approach reduces preprocessing overhead while maintaining flexibility to incorporate domain-specific knowledge and new biological evidence, thereby strengthening the foundation for downstream learning and inference.

KGs offer a unified framework for integrating multi-entity, multi-relation biomedical data across omics layers, biological processes, and clinical observations. Within this high-dimensional structure, researchers can zoom in on local subgraphs for drug-protein interaction analysis or scale out to examine global networks linking drugs, targets, and diseases. They can investigate mechanisms of disease through pathway-level molecular activity or consider macro-level influences such as anatomical location and clinical symptoms. However, the complexity of such graphs also introduces methodological challenges. Traditional GCN are better suited for homogeneous graphs, whereas heterogeneous and relational-rich KGs require more advanced models such as knowledge graph embedding methods or relational graph convolutional networks[543] (R-GCNs) to fully capture multi-relational information. In this chapter, we will explore how KG can be adapted and specialized for key tasks in drug discovery, including target identification, drug interaction prediction, adverse effect estimation, and drug repurposing. We will illustrate how heterogeneous information can be integrated and mined through tailored GNN architectures, and how these models can be validated and interpreted in real-world biomedical scenarios.

## 6.2 Link Prediction: Completing Knowledge

Link prediction[544], which aims to predict the existence of links between entities in a network, is among the most crucial and widely applied task in KG. From a graph perspective, its objective is to predict potential connections given a subset of known nodes and their existing relationships. A typical modeling approach involves first learning vector representations $h_i$, $h_j$, for nodes $i$ and $j$, respectively, then computing their dot product $o_{ij} = h_i \cdot h_j$. The resulting value $o_{ij}$ serves as a predictive indicator of whether a link exists between nodes $i$ and $j$.

As previously discussed, biomedical knowledge graphs are inherently heterogeneous, containing a wide variety of entity types (e.g., genes, proteins, drugs, diseases) and relation types. Performing link prediction in such complex graphs necessitates the use of specialized graph learning methods, such as knowledge graph embeddings (e.g., TransE[545]) or heterogeneous GNNs (e.g., R-GCN[543], R-GAT[546], or TransE[545]) o capture both the diversity of entity types and the semantics of multi-relational edges. Considering the specific interaction requirements between different node types in biomedical applications, this section categorizes link prediction tasks based on node homogeneity or heterogeneity. For homogeneous nodes (e.g., drug-drug interactions), we focus on drug combination prediction and drug-drug adverse effects, as summarized in **Table 24**; for heterogeneous nodes (e.g., drug-disease or drug-protein interactions), we highlight drug repositioning and target discovery, as summarized in **Table 25**. An illustration of link prediction in KGs is provided in **Figure 14A**.

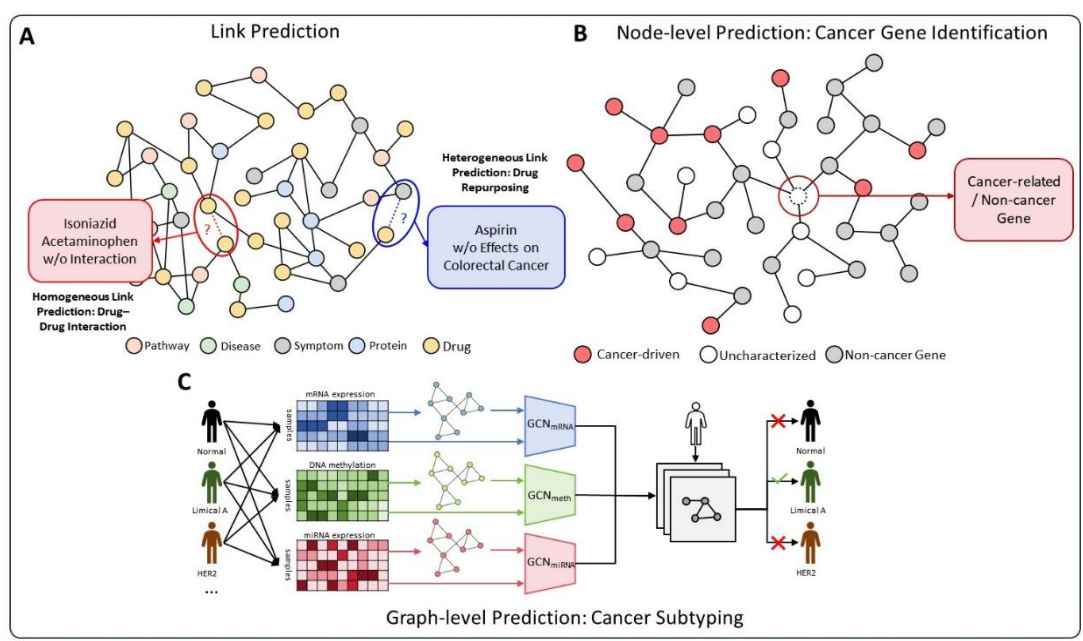

**Figure 14 A)** Link prediction. Homogeneous link prediction aims to identify missing links between entities of the same type, such as drug-drug interactions; while heterogeneous link prediction addresses interactions between different types of entities, such as drug repurposing, based on biomedical knowledge graphs composed of pathways, diseases, symptoms, proteins, and drugs. **B)** Node-level prediction: Cancer gene identification aims to classify whether a gene is cancer-related or not, based on its position and connectivity within a gene-gene interaction network. **C).** Graph-level prediction: Cancer subtyping involves integrating multi-omics features into patient-patient knowledge graph, where the final goal is to classify patients into subtypes such as Luminal A, HER2, or Normal.

**Table 24**. Overview of GNN-based KG methods in link prediction (homogeneous nodes).

| Category | Methods | Keywords |
| --- | --- | --- |
| Drug Combination | DeepDDS[547] | GNN is used as molecular fingerprint extractor |
| | MOOMIN[548] | GCN on DCI for informative molecule node |
| | HyperGraphSynergy[549] | Reconstructing drug- and CL- similarity matrices |
| | DTSyn[550] | Dual-Transformer for atom-gene and mol-gene |
| | SDCNet[551] | AutoEncoding CL-specific DDI using R-GCN |
| | Jiang et al.[552] | GCN on PPI+DTI for informative molecule node |
| | GraphSynergy[553] | Extract drug and cell info from GNN-embedded PPI |
| | KGANSynergy[554] | Construct drug-protein-cell-line-tissue knowledge graph |
| | DCMGCN[555] | Appending six drug-x relations for drug fingerprints |
| | Hu et al.[556] | Use pre-trained embedding and unsupervised learning |
| | HGVAE[557] | Use RL for generating possible drug combinations |
| Dide Effects | MK-GNN[558] | Based on medical knowledge and patient status |
| | Decagon[559] | First GNN work in side effect prediction |
| | SkipGNN[560] | Build second-order GNN (skip similarity) |
| | KGNN[561] | Similar to SkipGNN |
| | SumGNN[562] | Radius-based local graph sampling and GAT |
| | EmerGNN[563] | Path-based local graph model; inverse relation edges |

DCI: Drug Cell-line Interaction; DTI: Drug-Target Interaction; RL: Reinforcement Learning.

Table 25. Overview of GNN-based KG methods in link prediction (heterogeneous nodes).

| Category | Methods | Keywords |
| --- | --- | --- |
| Drug Repurposing | biFusion[564] | Drug-protein and disease-protein bipartite graph |
| | deepDR[565] | Multi-model info for drug-drug similarity; VAE for drug-dis |
| | DREAMwalk[566] | Introduce inter-network traversing based on random walk |
| | KG-Predict[567] | CompGCN on multi-relations KG |
| | LAGCN[568] | Layer attention GCN on drug-dis sim-association matrix |
| | DRWBNCF[569] | 2-nd order message passing |
| | DRHGCN[570] | GCN on sim matrix then on drug-disease associations |
| | DRGCC[571] | SVD for interaction learning |
| | AdaDR[572] | Replace SVD in DRGCC with GNN |
| | DRAGNN[573] | Attention on hetero info AVG pool on homo info |
| | MSSL2Drug[574] | Local and global SSL boost drug repurposing |
| | TxGNN[575] | Pre-train on other relations and explainable GNN |
| Target Identification | deepDTNet[576] | Matrix completion and had wet-lab experiments |
| | Progeni[577] | Build probabilistic-KG based on literature evidence |

PPI, DTI, DCI, DDI refers to protein-protein, drug-target, drug-cell line, drug-drug interaction network; EHR refers to electronic health records.

### 6.2.1 Homologous Node Relations: Drug Synergy, Side Effects

#### 6.2.1.1 Drug Synergy Prediction

One major application of link prediction is drug combination prediction, which seeks to determine whether two drugs exhibit synergy, antagonism, or no interaction when used together. This is particularly important for complex diseases involving multiple genes or pathways, where monotherapy may be insufficient. Combining drugs with complementary mechanisms can enhance therapeutic efficacy while reducing toxicity through dose reduction[578].

Mathematically, drug combination prediction can be formulated as $s = f(h_{d_a}, h_{d_b})$, where $s = \{0,1\}$ indicates binary synergy (or a continuous synergy score), and $h_{d_a}$ and $h_{d_b}$ are the learned representations of two drugs, respectively. While in theory one might predict combination effects from chemical structures alone, this approach may suffer from two key limitations: (1) molecular structure alone often omits biological context (e.g., signaling pathways, mutation background); and (2) the number of drugs with tested combinations is small (typically <10000), and the synergy labels are highly sparse. Thus, integrating **biomedical knowledge from KGs** becomes critical for improving prediction accuracy and interpretability[579].

**6.2.1.1.1 Cell Line-Drug Integrated KGs**

Given that most drug combination data originates from in vitro cell line experiments, explicitly incorporating cell line information into models is a logical approach. **DeepDDS**[547] is an early example: it extracts gene expression profiles from **CCLE**[580] to represent cell lines using an MLP, learns GNN-based molecular embeddings for drugs, and concatenates these features ($h_c||h_{d_a}||h_{d_b}$) before feeding into a second MLP to predict synergy scores. In this model, the GNN acts as a molecular fingerprint generator.

However, researchers have observed in DrugCombDB[581] that approximately one-third of drug combinations exhibit entirely opposite effects in different cell lines, being synergistic in some but antagonistic in others[548]. This observation motivated newer approaches that treat cell lines as nodes in a graph, allowing graph-based message passing to capture how drug responses vary with cellular context. In **MMOMIN**[548], researchers construct a **drug-cell line interaction (DCI) graph** and apply GCNs to allow information flow between similar cell lines. **HyperGraphSynergy**[549] further enriches this by adding reconstruction tasks for the similarity matrices of drugs and cell lines, enhancing embedding quality.

Building upon these foundations, subsequent studies have further explored detailed modeling of cell line-drug interactions. For example, **DTSyn**[550] (Dual-Transformer Synergy) supplements CCLE gene expression with ∼1000 **landmark genes** from LINCS. It uses a GCN to encode drug structures and applies multi-head attention in a transformer to capture substructure-gene interactions. Additionally, global cell line features are integrated to better represent compound-cell interactions. **SDCNet**[551] (Synergistic Drug Combination Network) constructs a distinct Drug-Drug Interaction (DDI) network for each cell line, treating these interactions as different relations, and applies R-GCN for representation learning. Each relational graph aggregates neighbor information, which is then combined into a general representation $h_i'$:

$$h_i' = \sum_{r \in R} \sum_{j \in N(i)^R} h_i$$

where $R$ is the relation type, $N(i)^R$ is the neighborhood of $i$ under the relation $R$. This vector is then decoded to predict synergy scores per cell line. **MGAE-DC**[582] improves upon this by treating reconstructed DDI graphs as auxiliary inputs rather than direct outputs. It constructs three DDI graphs per cell line to representing synergy, additivity, and antagonism and trains a multi-channel self-supervised model to learn both cell-line-specific and shared embeddings. These are combined with ECFP fingerprints and gene expression data into an MLP for the final prediction:

$$S = \text{MLP}(h'_{d_a} \| |h'_{d_b} \| |h_c)$$

$$h'_{d_a} = [\text{ECFP} \| h_c^{spec} \| h_c^{general}]$$

where ECFP denotes molecular fingerprints, and $h_c^{spec}$ and $h_c^{general}$ are embeddings capturing cell-line-specific and generalizable knowledge, respectively. This approach effectively integrates traditional molecular fingerprints with high-level embeddings $h_c$, accurately modeling cell line-specific knowledge and interactions.

### 6.2.1.1.2 Protein-Drug Integrated KGs

While incorporating cell line information can partially address data sparsity, its relatively coarse granularity often limits the ability to capture deeper biological mechanisms. Consequently, researchers have shifted focus to PPI networks, driven by the intuition that synergistic drug combinations frequently share common interacting proteins within these networks. **Jiang et al.**[552] pioneered the use of drug-protein mixed graphs enriched with PPI context. For each cell line, they constructed a cell-specific PPI subnetwork and embedded drug nodes into it, forming a heterogeneous graph with both protein and drug entities. A GCN was then applied to this cell line-specific drug-target interaction (DTI) graph, propagating protein-derived information into the drug embeddings:

$$h'_{d_a}, h'_{d_b} = \text{GNN}(\mathcal{G}_D TI)$$

$$s = MLP(h'_{d_a}||h'_{d_b})$$

where $h'_{d_a}$ and $h'_{d_b}$ represent the embeddings of drugs $a$ and $b$, respectively, implicitly encoding information from their proximal protein neighbors. The model thus becomes a function of both drugs and their biological context grounded in the given PPI subnetwork:

$$s = f(h_{d_a}, h_{d_b}, \{p_i\}_{i=1}^{n_p}, \{h_{d_j}\}_{j=1}^{n_d},)$$

where $\{p_i\}_{i=1}^{n_p}$ and $\{h_{d_j}\}_{j=1}^{n_d}$ denote the sets of proteins and drugs perceived by drugs $a$ and $b$, respectively, during message passing. This design allows the model to integrate drug-protein interactions at a fine-grained, cell line-specific level.

Similarly, **GraphSynergy**[553] utilizes a universal PPI network and maps drug and cell line entities to their respective proteins using CCLE data. Drug and cell line features are then aggregated from their associated protein embedding:

$$h_{p_i} = \text{GNN}(G_{\text{PPI}})$$

$$h'_{d_a}, h'_{d_b}, h'_c = \text{AGG}(\{h_{p_i}\}_{i=1}^{n_{d_a}}, \{h_{p_j}\}_{j=1}^{n_{d_b}}, \{h_{p_k}\}_{l=1}^{n_c})$$

where $h'_{d_a}$, $h'_{d_b}$ and $h'_c$ represent the features of drug $a$, drug $b$, and cell line, respectively, derived from the PPI network; $n_{d_a}, n_{d_b}$ and $n_c$ denote the number of associated proteins for drug $a$, drug $b$, and cell line, respectively; and Agg is an aggregation function, typically a simple summation. Notably, GraphSynergy's synergy score prediction differentiates between therapeutic efficacy and toxicity, formally represented as:

$$s = \sigma(s_{therapy} - s_{toxic}), \qquad s_{therapy} = MLP(h'_{d_a}, h'_{d_b}, h'_c), \qquad s_{toxic} = MLP(h'_{d_a}, h'_{d_b})$$

where $s_{therapy}$ and $s_{toxic}$ quantify therapeutic efficacy and toxicity levels, respectively, and the final synergy score $s$ combines both measures to evaluate drug synergy predictions. This dual-objective formulation enhances biological interpretability by explicitly modeling both treatment benefits and adverse effects.

### 6.2.1.1.3 Multi-entity KGs

In addition to PPI and cell-line information, researchers are increasingly incorporating other biomedical and clinical modalities, such as tissue types, genes, diseases, and even patient-level data, to improve the predictive power and interpretability of drug combination models.

**KGANSynergy**[554] constructed a multi-relational KG including proteins, drugs, cell lines, and tissues for synergistic interaction prediction. Unlike typical drug-drug combination tasks, KGANSynergy specifically evaluates whether "drug a and drug b exhibit synergistic effects within a particular cell line of a specific tissue", thereby capturing both tissue specificity and similarity. **DCMGCN**[555] adopted an alternative approach by incorporating additional relationships (e.g., drug-target, drug-side effect, drug-disease, drug-structure, drug-clinical response, and drug-event) as drug features, encoding these relationships into drug vectors using Singular Value Decomposition (SVD), and decoding the synergy matrix via GCN on the DDI network. **Hu et al.**[556] integrate diseases directly into the drug-cell line network and adopt modality-specific pretraining strategies: drug features are generated using KPGT[583], cell lines are represented using weighted ESM embeddings[350] of their associated proteins, and diseases are encoded using RotatE[584] trained on PrimeKG[505]. Furthermore, they apply **self-supervised learning** through masking and reconstructing features to enhance the utilization of multimodal biological information. **HVGAE**[557] (Hierarchical Variational Graph Auto-Encoders) adopts a generative modeling approach to predict synergistic drug combinations. Unlike traditional link prediction methods, HVGAE employed RL to iteratively grow drug fragments, thus exploring molecular structures suitable for synergistic combinations. Its primary goal is to design a "drug combination evaluation function" to determine if a drug set is more effective for a target disease. Practically, HVGAE used Variational Auto-Encoders (VAEs) to train GG-VGAE and DD-VGAE on gene and disease information, respectively, injecting their variational encodings into an RL module for sequential molecular growth. Due to its complexity and less intuitive nature, subsequent expansions of this generative method remain limited.

Beyond biological modalities such as tissues, genes, and diseases, some studies have explored integrating clinical and patient-specific biological data directly in networks to enhance drug synergy prediction. For instance, KG-CombPred[585] introduces an end-to-end embedding-based framework that scores drug pairs by integrating large-scale omics information, such as ATC codes and Gene Ontology terms, into a KG. A more clinically grounded approach is exemplified by **MK-GNN**[558] (Medical-Knowledge GNN) , which extracts diagnostic histories, treatment records, and prior health data from Electronic Health Records (EHR). These clinical features are encoded into a latent representation $h_{med}$ via attention mechanisms, and incorporated into the synergy prediction model as follows:

$$s = f(h_{d_a}, h_{d_b}, h_{med})$$

Here, $h_{med}$ represents the extracted clinical diagnostic data. Compared to mechanistic models based on biological data (e.g., PPI networks) or in vitro experiments (e.g., cell lines), MK-GNN shifts the modeling paradigm toward phenotype-level abstraction, emphasizing the influence of physician decision-making and individual patient characteristics on drug response. This represents a key step in bridging laboratory research with clinical applicability in combination therapy.

### 6.2.1.2 Drug Side Effects Prediction

Drug-drug side effect prediction, a subset of drug-drug interactions, differs from typical numeric synergy predictions as it requires handling a large and diverse array of adverse effects. For example, the SIDER[201] database lists 5,868 side effect types, including common ones such as nausea, vomiting, headaches, diarrhea, and rash. Consequently, drug-drug side effect prediction is commonly modeled as a multi-label link prediction task. Although most drug synergy prediction models can theoretically adapt to this multi-class setting by modifying their output layers, we focus here on research explicitly designed for side effect prediction.

**Decagon**[559] pioneered the use of GNNs for drug-drug side effect prediction in 2018, notably inspiring subsequent research that leveraged PPI networks for drug synergy tasks. In Decagon, protein nodes serve as intermediaries within the drug-drug network, incorporating richer biological mechanisms. It employs an R-GCN-like graph convolution structure for feature extraction and utilizes matrix factorization in the decoding phase to predict side effects through the scoring functions:

$$f(h_{d_a}, h_{d_b}, r) = z_a^T D_r R D_r z_b$$

$$f(h_i, h_j, r) = z_i^T M_r z_j$$

where $z_a$ and $z_b$ represent the node embeddings for drugs $a$ and $b$ learned by R-GCN, respectively, and $r$ denotes learned relationship embeddings. $z_i$ and $z_j$ are the embeddings for nodes involved in other relationships, respectively. The first scoring function specifically addresses drug-drug side effect relations, where $D_r$ is a relationship-specific diagonal matrix highlighting crucial dimensions for different side effect types. The second scoring function captures broader biological knowledge concerning other types of relationships.

Following Decagon, **SkipGNN**[560] and **KGNN**[561] introduced "second-order neighbor" strategies into graph convolution updates to capture higher-order interactions among biological entities. Biologically, this means explicitly accounting for the influence of protein $p_b$ on protein $p_a$ when updating the relationship between drug $a$ and protein $p_a$, enhancing the representation of protein interactions underlying side effects. Moreover, **EmerGNN**[563] and **SumGNN**[562] extended this concept further by introducing large-scale biological entities and employing local subgraphs to predict side effects. EmerGNN uses a path-based approach, selecting all nodes within paths of length less than $L$ from drug $a$ to drug $b$, forming a path subgraph. It then uses a **flow-based GNN** to propagate information from $a$ to $b$, resulting in a pairwise representation $h_{ab}$ for side effect prediction. In contrast, SumGNN constructs a radius-based subgraph centered on drugs $a$ and $b$, encompassing nodes within a shortest-path radius $L$. Then, it integrates molecular fingerprints into drug node features in its radius-based subgraph using GAT before predicting side effects. Both models localize learning to relevant biological regions of the graph and encode distinct prior assumptions: EmerGNN emphasizes biological paths, while SumGNN prioritizes proximity-based biological context. A notable design in EmerGNN is its use of **inverse relations**: if gene $a$ regulates gene $b$, a reverse edge is added to preserve directionality. This contrasts with most undirected KG models, which treat regulation symmetrically, potentially leading to semantic dilution. Ablation studies confirm that preserving edge directionality improves prediction performance.

### 6.2.2 Heterogeneous Node Relations: Drug Repurposing and Target Identification

### 6.2.2.1 Drug Repurposing

Drug repositioning is a heterogeneous link prediction task, aiming to infer connections between drugs and disease entities. Compared to developing new drugs, drug repositioning significantly reduces cost by leveraging known safety and pharmacokinetic profiles, requiring only efficacy validation[586].

Drug repositioning strategies typically follow the drug-target-disease paradigm and can broadly be categorized into three main approaches[587]. The **Drug-Centric** approach directly links approved drugs to known targets or pathological pathways through protein-drug interaction prediction, analogous to virtual screening within a limited drug space. Techniques such as molecular docking and QSAR modeling are employed to evaluate drug-target interactions. The **Target-Centric** strategy identifies novel drug repositioning opportunities by discovering new target-disease pathways. For instance, researchers identified an association between Parkinson's disease and the ABL tyrosine kinase, suggesting that nilotinib, an ABL inhibitor, might have therapeutic potential for this neurodegenerative condition[588]. However, this approach is often hindered by the complexity of molecular mechanisms and experimental limitations in target identification, which constrain the scalability of computational methods. The

**Disease-Centric** approach assumes direct links between drugs and diseases, simplifying molecular complexity to phenotype-based associations. By leveraging disease similarities or shared characteristics, this strategy repurposes known drugs for related conditions. For example, nilotinib, initially approved for chronic myeloid leukemia, was later proposed for gastrointestinal stromal tumors due to overlapping "hallmarks of cancer"[589]. Nonetheless, Disease-Centric approaches tend to prioritize repositioning within closely related diseases and may face patent-related barriers.

In KG-based repurposing, disease-centric strategies are prevalent due to their ability to uncover latent high-level associations between diseases and drugs. The prediction task is often framed as:

$$s = f(h_{drug}, h_{dis}, h_c)$$

where $h_{drug}$ encodes fine-grained drug information, $h_{dis}$ captures broad disease attributes, and $h_c$ encodes transitional information from fine-grained to coarse-grained levels within the KG, encompassing protein interactions, disease phenotypes, and side effects.

Traditional KG-based drug repositioning often relies on classical graph embedding techniques to represent nodes, followed by training scoring functions to evaluate drug-disease associations. For example, TriModel[590] integrates multi-scale information into embeddings for diseases and drugs. Another conventional approach involves designing heuristic scoring functions directly applied to KGs. For example, Al-Saleem et al.[591] developed a scoring function based on assumptions such as "drugs targeting more proteins tend to exhibit greater side effects"", thereby evaluating drug-disease relationships within KGs. While these methods can partially reveal potential associations, they struggle to model complex, multilayered interactions between drugs and diseases. Consequently, contemporary research increasingly leverages the expressive power of GNNs to overcome these limitations.

**6.2.2.1.1 Integration of Biological Mechanisms**

To enrich KGs for drug repurposing, biological entities such as **proteins, genes, and side effects** are often added. **biFusion**[564] constructed bipartite drug-protein and disease-protein graphs to allow PPI networks to mediate drug-disease inference. Additionally, **deepDR**[565] built a comprehensive KG integrating drugs, diseases, proteins, and side effects, defining seven drug similarities based on GO. It generates a probabilistic co-occurrence matrix (PCO) and pointwise mutual information (PPMI), which are used to train a VAE for decoding latent drug-disease relations. Extending this strategy, **DREAMwalk**[566] implemented cross-network random walks to capture more flexible heterogeneous information. **KG-Predict**[567] extensively mined additional biological knowledge, defining seven node types (e.g., drug, disease, gene, tissue, gene expression, and disease phenotypes) and nine edge types. To effectively manage multi-relation message passing, KG-Predict developed CompGCN, which uses relation-specific aggregation functions:

$$h_u^r = \text{AGG}_r(\{h_v\}), h_u = \text{CONCAT}(\{h_u^r\})$$

This approach enables KG-Predict to extract **mechanism-level insights** across heterogeneous biomedical graphs.

**6.2.2.1.2 Integration of Clinical Data**

Given that drug repositioning tasks are inherently disease-related, incorporating clinical data into KGs has become an increasingly popular strategy. Generally, such methods construct three distinct networks: a drug-drug similarity network, a disease-disease similarity network, and a drug-disease interaction network. Disease similarity typically refers to semantic similarity, derived from one-hot encoding of symptoms, while drug similarity is often computed based on molecular fingerprints. Due to limited experimental data on drug-drug interactions, other biological attributes (e.g., drug-gene associations) can also be leveraged to infer drug similarities[592].

**LAGCN**[568] (Layer Attention GCN) represents a prototypical example following this paradigm. It constructs an integrated adjacency matrix $A_H$, embedding drug and disease similarity information along the diagonal blocks and drug-disease interactions in the off-diagonal blocks:

$$A_H = \begin{bmatrix} \sim S^r & A \\ A^T & \sim S^d \end{bmatrix}$$

LAGCN then applies graph convolution directly to this large matrix and introduces layer-wise attention mechanisms to weigh the contributions of embeddings from different layers:

$$L_i = \sum_j \alpha_{ij} h_j$$

where $\alpha_{ij}$ is the layer-wise attention coefficient. This design has inspired numerous subsequent KG-based GNN studies. Another representative work, **DRWBNCF**[569], introduces a "weighted bilinear aggregation" strategy, similar in spirit to SkipGNN and KGNN, to capture second-order neighbor interactions:

$$h_r = \sum_i \sum_j (W_i h_i \odot W_j h_j) \cdot A^r_{ij}$$

where $W_i, h_i$ and $W_j, h_j$ are the weighted representations of two neighboring nodes, $\odot$ denotes element-wise multiplication, and $A^r_{ij}$ is the similarity (often derived from random walk co-occurrence probabilities) between nodes $i$ and $j$.

Subsequent research has further refined the integration of drug-disease network relations. For example, **DRHGCN**[593] first aggregates information via GCN within drug and disease similarity graphs, and then repeats the aggregation on drug-disease interaction graph to integrate multi-graph information. **DRGCC**[571] uses SVD to fuse representations from drug and disease similarity graphs, subsequently followed by GraphSAGE to extract similarity features separately from drug-drug and disease-disease graphs. **AdaDR**[572] separately applies GCN to three subgraphs (i.e., drug-drug, disease-disease, and drug-disease) and aggregates them using attention. **DRAGNN**[573] further distinguishes between intra-modal and cross-modal neighbors:

$$r_i = r_i^d + r_r^r$$

where $r_i^d$ aggregates cross-modality neighbors (e.g., drug-disease relations) using an attention mechanism, while $r_r^r$ aggregates homogeneous neighbors (e.g., drug-drug relations) using simple mean-pooling. Such hierarchical aggregation not only maintains cross-modality diversity but also efficiently consolidates homogeneous information.

**6.2.2.1.3 Large Models and Explainability**

With advancements in large-scale models and self-supervised pretraining, recent drug repositioning research has explored integrating large-scale heterogeneous KGs with self-supervised strategies to handle increasingly complex real-world scenarios. **MSSL2Drug**[574] and **TxGNN**[575] exemplify this direction, leveraging multimodal or multi-relational KG information to derive expressive embeddings for drug and disease nodes, thereby enhancing downstream predictive performance.

**MSSL2Drug** introduces three categories of self-supervised tasks (i.e., structural, semantic, and attribute-based), with each containing local and global structure-oriented pretraining tasks. These tasks operate jointly across multiple relation types (e.g., drug-disease, drug-target, drug-drug), allowing multi-task pretraining and improving generalization, particularly in **cold-start scenarios** where nodes (drugs or diseases) lack abundant connectivity. In contrast, **TxGNN** proposes a zero-shot drug repurposing framework built on a GNN backbone. It emphasizes the challenge of predicting treatments for diseases **with few or no known drug associations.** During pretraining, TxGNN traverses a large-scale heterogeneous medical KG, encompassing various node and edge types, to learn generalized representations. To address the limitations of self-supervised learning in zero-shot settings, TxGNN incorporates metric learning by transferring knowledge from similar diseases:

$$\mathbf{h}_i^{\text{sim}} = \sum_{j \in D_{\text{sim}}} \text{sim}(i,j) \times \mathbf{h}_j$$

where $D_{sim}$ is the set of diseases similar to disease $i$, $sim(i,j)$ is a similarity score, and $h_j$ is the embedding of disease $j$. The final embedding of disease $i$ combines the GNN-extracted representation and metric-learned representation:

$$h_i = w_i \cdot h_i^{GNN} + (1 - w_i) \cdot h_i^{sim}$$

where $w_i$ is a learnable weighting vector. Moreover, TxGNN integrates a GraphMASK module (similar to GNNExplainer) for explainability, visualizing the neighboring nodes influencing predictions. This interpretability component clarifies the biological rationale behind predictions, offering deeper insight into drug repositioning outcomes from a biological perspective.

**6.2.2.2 Target Identification**

Target discovery, as known as target fishing[594], can be categorized into three primary approaches: drug-centric, target-centric, and disease-centric. In the drug-centric strategy, known drugs are used as probes to screen protein databases via computational methods such as molecular docking or QSAR models, identifying proteins with potential high binding affinity to the drugs. The target-centric strategy, in contrast, begins with a disease-relevant protein/gene and searches for compounds that bind to it from existing drug libraries. The disease-centric strategy involves identifying diseases sharing common phenotypic or mechanistic features, such as identical mutations or biomarkers, and subsequently linking the diseases to new proteins or pathways[595].

From the perspective of KGs, target discovery and drug repositioning share conceptual similarities, as both tasks can be viewed as predicting potential links between heterogeneous entities, such as "drug-target" or "compound-protein" relationships. However, they differ in their central objectives: drug repositioning typically adopts a disease-centric approach, iterating over candidate drugs for a given disease:

$$h'_{drug} = \arg\max_{h_{drug}} f(h_{dis}, h_{drug}, h_c)$$

While target identification often base on the drug-centric approach, i.e., starting from a drug and seeks its potential binding targets:

$$h'_{dis} = \arg\max_{h_{tgt}} f(h_{tgt}, h_{drug}, h_c)$$

While, in principle, repurposing frameworks can be directly adapted for target discovery, some models are specifically tailored for this task. For example, **deepDTNet**[576] learns drug and protein embeddings $X_i$ and $Y_j$ from respective networks using graph embeddings, while decomposing the drug-protein interaction matrix via factorization

$$z = WH^T$$

The scoring function predicting drug-target associations is defined as:

$$S_{ij} = X_i Z Y_j^T$$

where $X_i$ and $Y_j$ capture interactions within the drug and protein networks separately, while $Z$ captures their cross-network interactions. Interestingly, despite deepDTNet being explicitly designed for target discovery, its experimental validation involved drug repositioning for the RORγ target, highlighting the inherent conceptual overlap between these two tasks.

Beyond classical graph embedding methods, several studies have incorporated GNN into target discovery. For instance, **Progeni**[577] introduced the concept of a probabilistic KG (prob-KG) by calculating the literature co-occurrence frequency from established databases as probabilities assigned to each edge, thus reflecting biological reliability or uncertainty at the KG. Based on this prob-KG, A relation-specific R-GCN is trained to reconstruct these edge weights via autoencoders. The model then outputs candidate disease-target associations. Notably, Progeni validated its predictions experimentally by identifying new target candidates for human melanoma and colorectal cancer (CRC). Subsequent knockdown experiments on the predicted target genes demonstrated significant inhibition of tumor cell proliferation, confirming the biological relevance of the inferred targets.

### 6.3 Node and Graph Prediction: Leveraging Knowledge

Node-level prediction and graph-level prediction aim to learn informative representations at either the node or entire graph level within a KG graph and utilize these representations to make downstream predictions[596]. Unlike link prediction, which focuses on inferring unknown relationships between nodes, these tasks assume the graph topology is fully known and seek to update either node-level or graph-level representations to support downstream inference. Formally, they can be expressed as:

$$o_i = f(h_i) \quad \text{or} \quad o_i = \sum_i f(h_i)$$

where $h_i$ represents node-level features, $f$ is the prediction model, and $o_i$ represents the output prediction. In drug discovery and biomedical research, KG-based node and graph predictions are widely applied across multiple scenarios, ranging from cancer gene discovery and biomarker identification to drug response prediction and cancer classification. These methods are summarized in **Table 26**. Below, we briefly introduce the modeling strategies for these tasks and highlight the roles of GNNs.

Table 26. Overview of GNN-based KG methods in graph and node prediction.

| Category | Methods | Keywords |
|---|---|---|
| Cancer-Driven Gene Identification | EMOGI[597] | Gene network with multi-omics; rule-based negative samples |
| | MTGCN[598] | Introducing unsupervised link masking task into EMOGI |
| | MODIG[599] | Incorporating 4 more gene edges in the gene network |
| | SPEOS[600] | Introducing GWAS data and PU learning |
| Cancer Subtypes Classfication | ERGCN[601] | Patients as nodes, genes as features and edges (patient network) |
| | OmicsGAT[602] | GAT on genetic patient network |
| | CGGA[603] | GAE on genetic patient network |
| | MoGCN[604] | GCN on multi-omics patient graph for breast cancer |
| | RRGCN[605] | Residual GCN on multi-omics patient graph for gastric cancer |
| | pDenseGCN[606] | DenseGCN on multi-omics patient graph for liver cancer |
| | MOGONET[607] | VCDN to fuse multi-omics GNN-output features |
| | GCNN[608] | GCN on patient-wise KG |

EHR refers to electronic health records.

#### 6.3.1 Cancer Gene Discovery

Cancer is often driven by aberrant gene behavior. Identifying cancer-associated genes is pivotal for unraveling molecular disease mechanisms and advancing precision medicine[609]. Large-scale sequencing projects, such as The Cancer Genome Atlas (TCGA)[6], have collected comprehensive genomic and multi-omics datasets from thousands of cancer patients across numerous cancer types. Researchers aim to uncover critical genes driving tumorigenesis, as illustrated in **Figure 14B**, thereby identifying potential therapeutic targets and designing targeted interventions.

Early approaches, such as MutSigCV[610], relied on statistical deviations in mutation frequency to detect driver genes. However, these straightforward statistical methods have inherent limitations. First, experimental constraints may lead to missing genomic data, especially for genes with low expression that are not detected by microarrays or sequencing. Second, cancer-driving factors extend beyond genetic mutations alone, involving multi-dimensional molecular features such as DNA methylation, copy number variation, protein post-translational modifications (e.g., phosphorylation, acetylation), and cellular microenvironment interactions. Lastly, gene expression profiles and epigenetic modifications vary significantly across tissues, influencing cancer susceptibility and progression. To overcome these limitations, multi-omics integration has emerged as a promising strategy to enhance cancer gene representation[611]. **EMOGI**[597] (Explainable Multi-Omics Graph Integration) is among the earliest models to formalize cancer gene discovery as a node classification task on a KG that incorporates PPI networks, formalized as:

$$c_i = f(h_i, h_j, h_c)$$

where $c_i$ indicates whether the gene is cancer-associated, $h_i$ and $h_j$ represent node features, and $h_c$ includes additional multi-omics information. Specifically, EMOGI constructs a PPI-enhanced gene KG in which each node represents a gene (protein) and each edge corresponds to a PPI. Multi-omics data from TCGA (single nucleotide variants, copy number alterations, differential methylation, and differential expression across cancer types) are appended to gene nodes, forming initial node features of dimension 16×4 (16 cancer types × 4 omics data types). During training, positive samples are derived directly from known cancer-associated genes, whereas negative samples are carefully curated using domain knowledge. Specifically, negative samples exclude genes closely linked to cancer, such as (1) known cancer-associated genes (positive samples), (2) genes in cancer-related KEGG pathways, (3) genes documented in OMIM, (4) genes predicted by MutSigDB as cancer-related, and (5) genes strongly co-expressed with known cancer genes. This ensures negative samples represent genuinely non-cancer-related genes. After training on these carefully curated positive and negative gene samples, EMOGI infers predictions across unlabeled genes within the entire KG. Furthermore, EMOGI leverages the Layer-wise Relevance Propagation (LRP) explainability method (discussed in Chapter 3) to attribute prediction outcomes to specific multi-omics evidence, elucidating biological mechanisms underlying cancer gene predictions.

Building on EMOGI, **MTGCN**[598] enhances the multi-omics graph by introducing a self-supervised learning objective that masks edge features and reconstructs them during training, thereby improving generalizability. **MODIG**[599] further expands this framework by integrating additional biological features, including gene sequence similarity, tissue-level coexpression, pathway co-occurrence, and semantic similarity, to augment edge features within the gene graph. The model adopts a GAT to assign adaptive weights to neighboring nodes and modulate the influence of different modalities via attention scores, yielding improved performance in cancer driver gene identification. **SPEOS**[600] incorporates genome-wide association studies (GWAS)[612] to identify candidate genes from a population-level perspective. Unlike single-patient mutation or CNA data, GWAS offers statistical insights into loci associated with disease risk. SPEOS integrates GWAS data with other multi-omics modalities (e.g., SNV, CNA, differential methylation, and gene expression), and applies graph representation learning and feature fusion to pinpoint critical nodes within molecular networks. Crucially, it adopts a Positive-Unlabeled (PU) learning[613] to handle open-world assumptions in datasets. By training an ensemble of 11 base classifiers, including GCNs and feature-based models like Node2Vec[614], it derives a consensus score to quantify prediction confidence across models[615]. This consensus-based uncertainty measure is conceptually aligned with the uncertainty quantification strategies used in molecular property prediction.

### 6.3.2 Cancer Subtype Classification

Cancer is a highly complex biological system characterized by extensive heterogeneity, resulting in substantial genetic, molecular, morphological, and functional variability among patients even within the same cancer type. Clinically, this heterogeneity manifests through diverse presentations, disease progression rates, and treatment responses[616]. Conversely, cancers also display notable homogeneity: tumors from distinct origins may share similar mutation patterns or gene expression profiles, leading to similar therapeutic responses. The coexistence of heterogeneity and homogeneity significantly complicates cancer diagnosis and treatment, underscoring the critical importance of precise cancer subtyping. For instance, breast cancer commonly comprises four molecular subtypes (Luminal A, Luminal B, Basal, and HER2) that differ markedly in histopathological features and treatment sensitivity[617]. Integrating genomic alterations with phenotypic data for subtyping facilitates personalized therapeutic strategies, ultimately improving clinical outcomes.

### 6.3.2.1 Patient-patient KGs

Omics data is widely used in cancer subtype classification, typically represented as a set of matrices with shape $n_s \times n_f$, where $n_s$ is the number of samples and $n_f$ is the feature dimension for a given omics type. In the case of using a single genomic modality (e.g., gene expression data), the input reduces to one $n_s \times n_f$ matrix. Traditional models, such as MLP or CNN, operate directly on these matrices to extract sample-level representations for classification. However, these models ignore the relational structure between patients. GNNs emphasize the inter-sample relationships by first computing sample-to-sample similarities, typically using the Pearson correlation coefficient, to construct a sample similarity graph. Each sample becomes a node in this graph, characterized by multi-omics feature vectors. GNNs are then applied for node- or graph-level classification. Intuitively, the label prediction for a given patient can benefit from "supportive evidence" offered by similar patients in the sample similarity graph. This modeling paradigm is often referred to as patient-patient association networks.

Early methods, such as ERGCN[601], adopted this strategy by first constructing patient similarity networks based on genomic features, then applying GCN for node classification tasks, i.e., patient cancer subtype classification. **OmicsGAT**[602] extended this approach by substituting GCN with the more expressive GAT, achieving improved prediction accuracy. **CGGA**[603] (Consensus Guided Graph Autoencoder) employed graph autoencoders within this framework, simultaneously predicting node types and reconstructing network topology to enhance model robustness against structural perturbations. Subsequent research extended the patient-patient network by integrating multiple omics data types to better capture the full complexity of cancer. Studies such as **MoGCN**[604], **RRGCN**[605], and **pDenseGCN**[606] represent different instantiations of this idea. These models ingest multiple omics modalities, such as transcriptomic profiles, copy number variation (CNV), and somatic mutations, to construct an individual similarity graph for each modality. These graphs are then fused into a **multi-omics patient graph**, where node features from different modalities are concatenated. A GNN is applied to perform patient classification based on subtype. The main distinction across these models lies in their architecture choices and target cancer types: MoGCN applies standard GCNs to breast cancer, while RRGCN and pDenseGCN employ residual GCNs or densely connected GCNs for gastrointestinal and liver cancers, respectively. Similarly, **MOGONet**[607] (Multi-Omics Graph Convolutional Networks) follows a similar design. Separate similarity graphs are constructed for mRNA expression, DNA methylation, and miRNA expression data. GCNs are applied independently to each graph to produce subtype-specific probability distributions. These are then integrated using **VCDN**[618] (View Correlation Discovery Network), which captures inter-modality correlations and fuses predictions into a final classification output from a multi-view learning perspective.

#### 6.3.2.2 Gene-gene KGs

Distinct from the macro-level **sample similarity graph** approach, another class of models formulates cancer subtyping as a **graph classification task**, where a gene-protein interaction graph is constructed individually for each patient. In this setup, the subtyping decision is made at the graph level. GCNN[608] is a representative of this direction: it uses a PPI network to define gene-gene edges, applies ChebNet to extract structural features, and classifies each patient's individual network to infer the molecular subtype.

### 6.3.3 Biomarker Identification and Drug Response

Beyond cancer subtype classification, KGs have also shown promise in biomarker identification and drug response prediction. The identification of biomarkers is often closely coupled with subtyping. By applying interpretable models to analyze the contribution of input features across different omics layers, one can pinpoint genes or molecular signatures that are pivotal within a specific cancer subtype, thus guiding precision therapy[619]. For example, **MOGONet** employs a masking strategy on input omics features and evaluates the perturbation in subtype predictions to estimate the relative importance of each feature, thereby highlighting candidate biomarkers.

In contrast, **drug response prediction** focuses more directly on personalized therapy: models aim to estimate the likely efficacy, toxicity, or resistance a patient may exhibit when exposed to a given drug. While traditional link prediction or node classification tasks in KGs focus on relational inference, drug response prediction typically uses the graph structure to encode patient-specific genomic contexts and integrate this with molecular representations of drugs. For instance, **DrugGCN**[620] constructs patient-specific gene networks using genes as nodes and PPI as edges, employs ChebNet to extract graph-level features, and ultimately predicts patient-specific drug responses. **KGDRP**[621] (Knowledge-Guided Drug Relational Predictor) extends this approach by integrating biological networks, gene expression profiles, and molecular structure-derived sequence data into a heterogeneous graph framework, leveraging GNNs to learn biologically informed representations for drug response prediction.

## 6.4 Challenges of KGs: Imperfect Data, XAI, and Multi-omics

### 6.4.1 The Data-Centric Paradigm

In the context of KG, emphasis must be placed not only on model architectures or learning algorithms but also on data curation and integration, which form the foundation of the data-centric paradigm. Unlike other tasks in computational biology that often rely on a single data type or a relatively homogeneous dataset, biomedical KGs handle heterogeneous entities, complex relationships, and contextual metadata spanning genomics, transcriptomics, proteomics, clinical pathology, and literature-derived annotations. These data sources often differ in format, semantic standards, quantity, and quality, posing challenges for integration. To handle such complexity, reusable pipelines must be established for KG construction, encompassing data cleaning, standardization, entity normalization and disambiguation, relation extraction and annotation, and ontology alignment. Given the intricacies involved, this process cannot be fully automated; domain expertise is crucial for manual or semi-automated verification. In this light, the design of KG systems is less about building a single model and more about engineering a complete KG infrastructure that enables reliable downstream inference.

In future work, large language models (LLMs) may assist with data curation by extracting biomedical entities and relations from unstructured texts, facilitating ontology mapping, and enhancing coverage by linking back to structured databases[622]. LLMs can also serve as tools for data augmentation and automated validation, resolving entity duplication, correcting typographical errors, and inferring missing relationships based on contextual knowledge.

### 6.4.2 Incompleteness and Noise

Biomedical KGs are inherently incomplete and often noisy. The open-world assumption[623] acknowledges that our current observational data cannot capture all valid biomedical facts; many associations, such as novel driver genes or unknown protein interactions, remain unobserved or unverified. Limitations in experimental detection further contribute to missing data, necessitating techniques like rule-based filtering (e.g., in EMOGI) or positive-unlabeled (PU) learning (e.g., in SPEOS) to model confidence over unlabeled entities.

Noise, on the other hand, arises from inconsistent data collection, text mining errors, and database versioning, manifesting as duplicate or conflicting entity names, mismatched relations, and poorly contextualized labels. To address this, many methods have been developed, such as removing evidently abnormal data during cleansing stages or employing noise-tolerant modeling strategies (e.g., autoencoders or probabilistic KG). For instance, MoGCN and RRGCN incorporate autoencoders to mitigate noise in cancer subtype classification, while Progeni constructs a probabilistic KG weighted by literature co-occurrence frequencies to explicitly account for uncertainty during learning. With ongoing advances in sequencing technologies and algorithmic modeling, future frameworks may better quantify and trace the provenance of noise, improving the reliability and robustness of KG–based prediction

### 6.4.3 Explainability

KGs offer a unique advantage in supporting interpretability due to their inherent structural and semantic richness. When GNNs are used to infer properties of biological entities, it becomes possible to trace the contributions of specific nodes and edges to model predictions. For example, in adverse drug reaction prediction, attention-based mechanisms (e.g., GAT or edge masking techniques like GraphMASK in TxGNN) can identify potential pathways or mutations underlying adverse reactions or distinctive phenotypes. In addition to structural interpretability, feature attribution plays a critical role, especially when multi-omics data are incorporated. In EMOGI, the model applies layer-wise relevance propagation (LRP) to analyze how predictions depend on features such as gene expression and methylation. Similarly, MoGCN uses feature masking to compare classification outputs before and after masking, identifying molecular biomarkers that most effectively distinguish cancer subtypes.

Such dual interpretability, which addresses both structural and feature-level aspects within biomedical KGs, enhances biological credibility and offers transparent theoretical foundations for mechanistic investigations, clinical decision-making, and drug development. Interpretability thus transitions from merely an "auxiliary" feature to a critical bridge linking predictive models with clinical insights, laying a solid foundation for precision medicine and personalized therapy.

### 6.4.4 Multi-Omics Integration

According to the central dogma of molecular biology, biological hierarchies ranging from DNA to protein expression and modifications, as well as signaling pathways and metabolic networks, are highly interconnected and dynamically coupled. For instance, mutations in the BRCA1 gene (e.g., point mutations, deletions, rearrangements) disrupt the function of the BRCA1 protein, impairing DNA repair mechanisms, particularly homologous recombination repair (HRR). Loss of BRCA1 function further dysregulates cell-cycle control, allowing cell division to proceed despite DNA damage, thus facilitating tumorigenesis. Such molecular-level changes may subsequently activate critical signaling networks, such as the PI3K/Akt/mTOR pathways, promoting tumor proliferation, survival, and metastasis. Consequently, integrating PPI networks and multi-omics datasets (e.g., transcriptomics, proteomics, metabolomics) requires attention not only to interactions at the same biological level but also to cross-layer coupling between omics modalities. For instance, MODIG exemplifies this by constructing separate networks across omics layers, then employing GAT to fuse features from multiple omics dimensions for each gene node, uncovering intra-layer relationships. MOGONet utilizes VCDN to discover deeper complementary associations across omics layers. However, generalizable and interpretable methods for systematically extracting multi-layer and multi-node interactions within KG remain limited. Exploring these hidden cross-omics interactions holds substantial significance for elucidating molecular alterations and their complex interplay in diseases or physiological states.

# 7. Chemical Synthesis

Organic synthesis endows human with an extraordinary capacity to reshape the natural world, forming a foundational pillar across numerous modern scientific disciplines. In drug discovery, organic synthesis plays a central role in the "Make" stage of the Design-Make-Test-Analyze (DMTA) cycle[624], where evaluating the synthetic accessibility of molecules and selecting which designs to pursue critically influences the efficiency and cost of translating computational designs into experimental validation. To address the complexity of retrosynthetic analysis, Computer-Aided Synthesis Planning (CASP) has emerged as a powerful tool for chemists[625]. Concurrently, rapid advancement of machine learning and cheminformatics has led to the curation and exploitation of large-scale, high-quality reaction databases, paving the way for data-driven approaches to synthesis planning. GNNs, which offer a natural framework for molecular representation, have attracted significant attention in recent years. They have been widely applied across a spectrum of synthesis-related tasks, including synthesis pathway prediction, reaction condition optimization, and yield prediction. This chapter would focus on the applications and challenges of GNNs in synthesis planning.

## 7.1 Synthesis Planning Tasks and Datasets

### 7.1.1 Overview of Data for Synthesis Planning

Chemical reaction data are typically structured to capture the essential details of a chemical transformation, including the structures of reactants and products, reaction conditions (e.g., solvents, catalysts, reagents, temperature), reaction class, yield, and procedural annotations. Broadly, such data can be categorized into two types: (1) reaction transformation data, which describe the conversion itself such as reaction pathways and single-step transformations; and (2) reaction property data, which include auxiliary information such as reaction class, conditions, and yield.

As showing in **Figure 15**, reaction transformation data are commonly encoded using reaction SMILES, where the dot (.) separates different molecular species, and the greater-than symbol (>) partitions reactants, reagents, and products. To capture atom-level correspondence between reactants and products, atom mapping is often applied by assigning numerical tags to atoms within the SMILES representation. These mappings enable automated identification of reaction centers and facilitate the extraction of reaction templates, which are typically encoded using reaction SMARTS or SMIRKS, the generalized extensions of the SMILES syntax. These formats allow for more flexible specification of atomic environments at the reaction center, such as atomic valence, bond order, formal charge, and local structural context (see https://www.daylight.com/dayhtml/doc/theory/theory.smarts.html for details). It is worth noting that most public datasets omit byproduct information, resulting in incomplete atom mapping, particularly for leaving groups or auxiliary fragments.

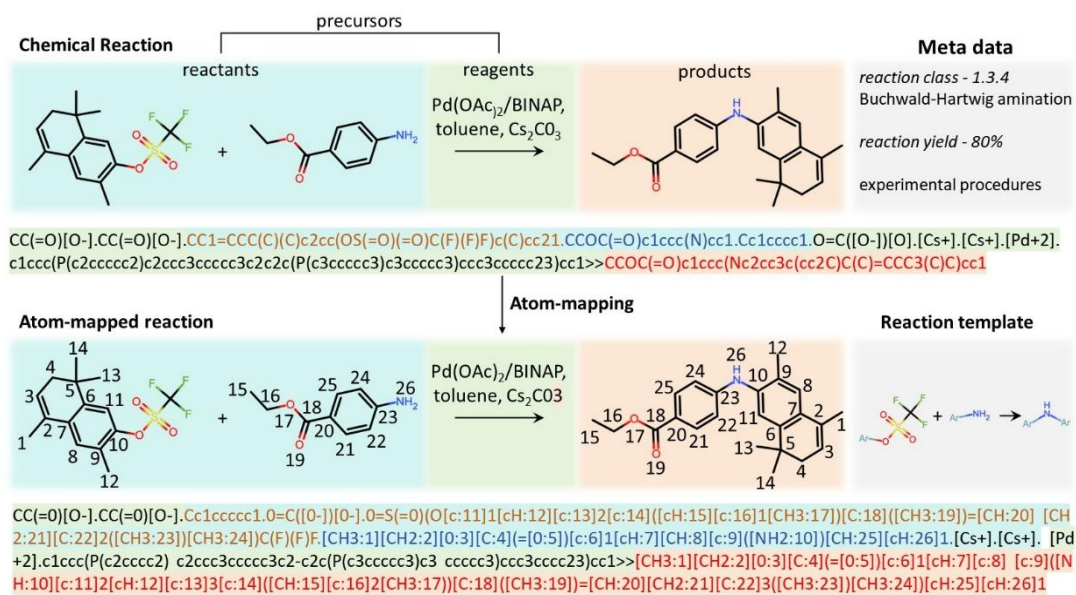

Figure 15. Machine-readable chemical reaction data format[3].

Reaction property data serve as auxiliary labels annotating chemical transformation. These include experimental conditions required for a reaction to proceed (e.g., reagents, catalysts, solvents, and temperature), as well as yields and categorical classifications. Such information is often scattered across unstructured text within experimental procedures or patent documents, necessitating extraction via specialized data cleaning and structuring workflows. While reaction classes are relatively discrete and well-defined, reaction conditions and yields are frequently subject to experimental variability and reporting bias, posing additional challenges for accurate predictive modeling.

### 7.1.2 Chemical Reaction Databases

Contemporary reaction databases can be broadly classified into two categories: general-purpose reaction datasets and high-throughput experimentation (HTE) datasets, as summarized in **Table 27**. General-purpose datasets are further divided into open-access and commercial resources, differing in reaction coverage, data provenance, and record structure. Widely used open-access datasets include the USPTO dataset[626] and its various curated subsets, as well as the more recent Open Reaction Database (ORD)[627], which is primarily derived from the USPTO data. Commercial databases, such as Reaxys[628] and Pistachio[629], provide access to larger and more richly annotated datasets but require paid licenses. In contrast, HTE datasets are typically focused on a single reaction class or catalytic system, where reactant and product scaffolds vary only slightly. These datasets are generated through automated or semi-automated platforms designed to systematically explore a matrix of reaction conditions. Compared to general-purpose datasets, HTE data are more standardized and exhibit lower noise, but they cover a much narrower region of chemical space, limiting their utility for large-scale synthesis planning and generalization.

Table 27. Overview of reaction datasets.

| Datebase | No. Of the reations | Availability |
|---|---|---|
| USPTO-MIT[630] | 479,000 | Open Access |
| USPTO-50k[631] | 50,000 | Open Access |
| USPTO-Full[632] | 1,000,000 | Open Access |
| Reaxys[628] | 40,000,000 | Proprietary |
| Pistachio[629] | 19,170,000 | Proprietary |
| Buchwald-Hartwig[633] | 3955 | Open Access |
| Suzuki-Miyaura[634] | 5760 | Open Access |

**The USPTO Dataset**: The USPTO dataset comprises reaction data extracted via text mining from U.S. patents published between 1976 and September 2016. It includes approximately 1.8 million reactions, many of which feature detailed procedural information and yield annotations. Owing to its accessibility and scale, it has become a cornerstone for benchmarking CASP models, including forward reaction prediction and retrosynthesis tasks. Commonly used subsets include:

**USPTO-MIT**: Curated by Jin et al.[630], this subset contains ~479,000 reactions and is frequently used for forward reaction prediction benchmarking.

**USPTO-50K**: Curated by Schneider et al.[631], this subset includes 50,000 stereochemically annotated reactions across ten reaction classes. It is the most widely adopted benchmark for single-step retrosynthesis prediction.

**USPTO-Full**: Curated by Dai et al.[632], it comprises ~1 million reactions. Entries with multiple products were split into separate reactions to better support model training and evaluation for retrosynthesis.

**The Reaxys Dataset:** Reaxys[635], maintained by Elsevier, is a commercial database that includes over 40 million reactions, annotated with reaction SMILES, solvents, reagents, catalysts, temperatures, yields, and detailed procedural steps. The dataset offers higher annotation quality and richer metadata than most public datasets. However, academic access is typically limited to subscription-based querying, and large-scale data export for machine learning requires a dedicated license. Despite these limitations, Reaxys has been leveraged in both industrial and academic settings to train high-precision reaction prediction models, underscoring its value as a professionally curated data source.

**The Pistachio Dataset:** Pistachio[629], curated by NextMove Software, contains over 19.1 million reactions, of which ~9.1 million are unique. The dataset primarily draws from U.S. and European patents and, like Reaxys, is commercially licensed. As such, it is less commonly used for benchmarking open-source algorithms but serves as a powerful training resource for those with licensed access.

**High-Throughput Experimentation (HTE) Datasets:** Beyond these general-purpose datasets, HTE datasets provide high-quality, narrowly scoped reaction data, typically focused on specific transformations or catalytic systems. Two representative examples are:

**Buchwald–Hartwig dataset** (Ahneman et al.[633]): This dataset focuses on Pd-catalyzed C–N coupling reactions, combining 15 aryl halides, 4 ligands, 3 bases, and 23 additives, yielding **3,955 unique reactions** with experimentally measured yields.

**Suzuki–Miyaura dataset** (Perera et al.[634]): This dataset includes 5,760 reactions, exploring 15 electrophile/nucleophile pairs in combination with 12 ligands, 8 bases, and 4 solvents.

Compared to general datasets, HTE data exhibit higher consistency and reproducibility, making them ideal for studying reaction mechanism and condition optimization. However, due to their limited coverage of chemical space, they are less suited for training or benchmarking models designed for global synthesis planning tasks.

### 7.1.3 Tasks in Synthesis Planning

Tasks related to synthesis planning can be broadly grouped into two major categories. The first involves synthesis pathway planning, which encompasses both forward and retrosynthetic prediction. The second focuses on reaction property prediction, which includes tasks such as reaction classification, reaction condition prediction, and yield prediction.

**Synthesis Pathway Planning Tasks**

**Forward Reaction Prediction**: Given a set of reactants, along with optional reagents and reaction conditions, the goal is to predict the most likely product(s). This task is fundamental for exploring novel synthetic pathways and discovering new molecular scaffolds.

**Retrosynthesis Prediction:** This task can be further divided into single-step and multi-step retrosynthesis. In single-step retrosynthesis, the model is tasked with predicting plausible reactant(s) for a given product molecule. Multi-step retrosynthesis involves chaining single-step predictions using a specific search strategy to generate a complete synthesis route from the target molecule to commercially available precursors. This class of problems lies at the heart of computer-aided synthesis planning and reflects how chemists design synthesis routes in practice.

**Reaction Property Prediction Tasks**

**Reaction Classification:** The task e is to assign a given reaction, defined by its reactants, products, and possibly conditions, to a predefined class, such as addition, redox, or coupling reactions. Accurate classification is valuable for applications such as patent mining, reaction indexing, and literature retrieval.

**Reaction Condition Prediction:** Given the structures of the reactants and products, the goal is to predict optimal reaction conditions, typically including catalysts, solvents, reagents (discrete variables), and temperature (a continuous variable). This task requires a nuanced understanding of reaction contexts and constraints.

**Reaction Yield Prediction:** The objective is to estimate the actual yield (typically ranging from 0% to 100%) of a given reaction under specified conditions. This task demands a deeper grasp of the underlying reaction mechanism and subtle experimental factors, making it a rigorous benchmark for assessing a model's ability to extract fine-grained chemical knowledge. Compared to reaction classification, yield prediction poses a more complex and quantitative challenge.

## 7.2 Synthesis Pathway Planning Tasks
### 7.2.1 Retrosynthesis Prediction

The core objective of retrosynthesis prediction is to identify a feasible sequence of chemical reactions that can synthesize a target molecule from simpler or more readily available starting materials[636]. An efficient and accurate retrosynthetic route can significantly reduce experimental workload and resource consumption. Traditional retrosynthetic analysis relies heavily on the chemist's expertise and intuition, grounded in mechanistic reasoning and accumulated experience. However, with the exponential growth of chemical space and the increasing scale of reaction databases, manual synthesis planning has become increasingly insufficient for meeting the demands of modern chemical research[637]. To address this challenge, researchers have begun integrating deep learning techniques with classical principles of organic synthesis, giving rise to a variety of data-driven retrosynthesis strategies, as shown in **Table 28**. For single-step retrosynthesis, common GNN-based approaches include template-based selection and reactant graph generation; while multi-step retrosynthesis typically builds upon these single-step models as basic components, combining them with RL strategies to explore full synthetic pathways.

Table 28. Overview of retrosynthesis models (the Top-1 and Top-5 performances are the results on USTPO-50k reported by the original reference).

| Category | Methods | Top-1 | Top-5 | Keywords |
|---|---|---|---|---|
| Template-based | GLN[632] | 52.5 | 75.6 | Edge-GCN for joint prob learning of p(R,T\|P) |
| | LocalRetro[638] | 53.4 | 85.9 | Attention-based GNN for learning of p(T\|a/ab) |
| | G2Gs[639] | 48.9 | 72.5 | Completing S with atom-wise AR generation |
| Semi-Template | GraphRetro[640] | 53.7 | 72.2 | Completing S with SMILES transformation |
| | RetroXpert[641] | 50.4 | 62.3 | Completing S with multi-class LG classification |
| | MARS[642] | 54.6 | 76.4 | Completing S with motif-wise AR generation |
| | G2GT[643] | 54.1 | 69.9 | AR decode graphs from the scratch |
| | NAG2G[644] | 55.1 | 83.4 | G2GT style but with 3D Graphomer |
| | MEAN[645] | 48.1 | 78.4 | AR graph edits transforming products to reactants |
| Template-free | Graph2Edits[646] | 55.1 | 83.4 | MEAN's strategy with leaving group edit action |
| | GNN-Retro[647] | - | - | Searching on synthesis knowledge graph, GNN is used for cost estimation |
| | RetroGraph[648] | - | - | And-or graph search, GNN is used for cost estimation |

S is synthon, LG is Leaving groups, AR is autoregressive.

### 7.2.1.1 Single-step Retrosynthesis Prediction Methods

#### 7.2.1.1.1 Template-based Methods

Based on the extent to which reaction rules are utilized, single-step retrosynthesis approaches can be broadly categorized into template-based and template-free methods. In the template-based paradigm, a reaction template refers to a generalized substructure pattern that captures the essential features of the reaction center and mechanism while abstracting away non-reactive parts of the molecule. For example, from the USPTO-MIT dataset comprising 479,000 reactions, approximately 21,080 distinct reaction templates can be extracted[638].

In template-based models (**Figure 16A**), retrosynthesis is performed by subgraph matching between the target product and a library of known templates, which are then applied to generate potential reactant candidates. However, brute-force template iteration requires thousands of subgraph matchings per prediction, making it computationally infeasible. To address this, deep learning techniques have been introduced to optimize template selection and ranking. For instance, RetroSim[649] ranks templates based on fingerprint similarity between the product and templates, while Segler and Waller[650] trained a multi-class classifier to predict the most likely template given a product fingerprint, enabling efficient node expansion within a MCTS framework.

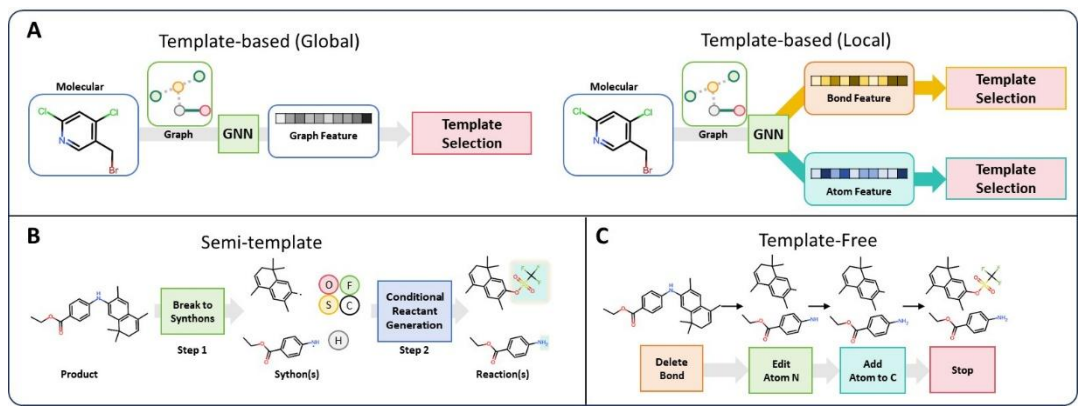

**Figure 16**. Overview of GNN-based retrosynthesis strategies. A) Template-based: Global models extract whole-graph features for template selection; local models focus on atom- or bond-level features around reaction centers. B) Semi-template: Retrosynthesis is divided into synthon decomposition (Step 1) and conditional reactant generation (Step 2). C) Template-free: Molecules are modified step-by-step via bond deletion, atom editing, and addition, without relying on predefined templates.

The use of GNNs further enhances molecular representation by capturing fine-grained atomic or bond-level features, helping to suppress interference from non-reactive regions. **GLN**[651] (Graph Logic Netwrok) is a representative GNN-based template method, which formulates the template selection as a probabilistic inference problem using an Energy-Based Model. Specifically, it defines a joint probability distribution $p(R,T|O)$, where $O$ is the product, $T$ is a retrosynthetic template, and $R$ is the set of predicted reactants:

$$p(R,T|O) \propto \exp(w_1(T,O) + w_2(R,T,O)) \cdot \phi_O(T)\phi_{O,T}(R)$$

where $w_1(T,O)$ is a template scoring function, modeling the compatibility between the template and target product; $w_2(R,T,O)$ scores the reactants conditioned on the product–template pair. Both terms are parameterized by GNNs, enabling an end-to-end learning of reaction structure and logic.

Compared to GLN, **LocalRetro**[638] introduces finer-grained modeling by scoring templates at the atom and bond levels. The model first encodes the product graph via a GNN, and then proceeds in two stepss: 1). Reaction center identification, where the model predicts which atoms or bonds are likely to participate in the reaction (denoted as $a$ or $ab$); 2); Template classification, where models classifies the appropriate reaction template based on local features. Formally, the template scoring functions in LocalRetro are defined as:

$$s(T|a) = \text{softmax}(x_a), \quad s(T|ab) = \text{softmax}(e_{ab})$$

where $x_a$ and $e_{ab}$ denote the GNN-encoded features of the atom $a$ and bond $ab$, respectively. These localized features enable the model to more accurately capture the subtle structural signals relevant to the reaction mechanism.

#### 7.2.1.1.2 Semi-Template-based Methods

The development of GNN-based single-step retrosynthesis models has given rise to a class of semi-template methods, which decompose the modeling process into two stages:

$$p(R \mid P) = p(R \mid S) \cdot p(S \mid P)$$

Here, the model first predicts an intermediate synthon $S$, and then modifies it at the predicted disconnection site to generate the final reactants $R$. The semi-template-based method is illustrated in **Figure 16B**. This approach mirrors how organic chemists identify the reactive center of a molecule, break it into synthons, and reconstruct precursor molecules through appropriate modifications. While these methods appear template-free, they rely on atom-mapped reactions during dataset construction to identify the reaction center, hence the term semi-template. From a modeling perspective, the first step does not introduce new atoms and can be framed as a bond-level classification task, while the second step involves adding atoms, making it suitable for a generation-based model starting from the synthons.

**G2Gs**[639] (Graph2Graphs) is a representative model of this strategy. It consists of two primary modules: (1) Reaction center identification, which isolates the synthons by segmenting the original molecular graph; (2) Synthons completion, which translates the synthons into the final reactants. To be more specific, for the reaction center identification module, G2Gs first extracts node-level representations using a GNN:

$$h_i = \text{GNN}(G_p), h_G = \text{ReadOut}(\{h_i\})$$

where $G_p$ is the molecular graph of the product, $h_i$ are the node features, and $h_G$ is the graph-level embedding. The edge features are then computed as:

$$e_{ij} = (h_i \| h_j \| A_{ij} \| h_G)$$

where $e_{ij}$ denotes the feature of edge $e_{ij}$, $A_{ij}$ is the edge type, and $\|$ represents concatenation. Based on these features, the reactivity score is calculated as:

$$s_{ij} = \sigma(\text{FFN}(e_{ij}))$$

where $FFN$ is a feedforward neural network and $\sigma$ is the sigmoid activation function. During inference, G2Gs computes the reactivity scores for all edges and selects the top-$k$ scoring edges as the reaction center, which are then used to fragment the product into synthons $S$.

The synthons completion module in G2Gs formulates the task as a graph editing problem, modeling the editing actions as: (1) termination prediction, (2) atom growth point selection, (3) atom prediction, and (4) bond prediction. Inspired by techniques in graph optimization, G2Gs adopts a VAE to model the probabilistic distribution of graph edits.

Observant readers may notice that the synthons completion module in G2Gs is similar to autoregressive generative models such as GraphAF, but it leverages a VAE instead of an autoregressive decoder. Subsequent models have refined this module to enhance performance. **RetroXpert**[641], for example, uses direct SMILES-based translation for synthons completion. Due to its simplified seq2seq generation process, RetroXpert can incorporate incorrect synthons into the training of its second-stage model to improve robustness. Another approach, **GraphRetro**[640], observes that the conversion from synthons to reactants is typically straightforward and often involves common leaving groups such as halide ions, methanesulfonate (MsO⁻), or acetate (AcO⁻). Remarkably, just 170 substructures cover 99.7% of the USPTO-50K dataset. Based on this insight, GraphRetro formulates the synthons completion step as a multi-class classification task over leaving groups. More recently, **MARS**[642] (Motif-based Autoregressive model for RetroSynthesis) has introduced a fragment-based generation strategy that operates between the atomic and leaving-group levels. Compared to G2Gs and GraphRetro, MARS reduces the generation path length while offering greater flexibility in synthons completion.

**7.2.1.1.3 Template-free Methods**

Despite their inherent interpretability, template-based retrosynthesis methods are fundamentally constrained by their reliance on a fixed reaction template library, limiting their ability to generalize to novel or rare reactions outside the training set. Furthermore, their multi-class classification framework often struggles to capture transferable knowledge across diverse reaction types, reducing their effectiveness in open-ended synthetic planning. To overcome these limitations, researchers have developed template-free methods, which aim to directly generate precursor molecules from a given target product by learning more fundamental chemical transformation patterns. A straightforward realization of the template-free paradigm involves representing molecular graphs using chemical languages such as SMILES or SMARTS, thereby framing retrosynthesis as a sequence-to-sequence machine translation task. While intuitive and computationally efficient, such approaches often suffer from model hallucination: generating chemically implausible transformations, and exhibit limited diversity in predicted reactions[652]. These shortcomings have led to a growing interest in graph-based template-free methods, which operate directly on molecular graphs to better preserve chemical validity.

Compared to semi-template approaches, template-free models are more end-to-end in nature, treating retrosynthesis as a graph transformation task: given a product graph, the model predicts a set of reactant graphs. This transformation can be implemented either via one-shot generation or an autoregressive decoding process, with the latter aligning more closely with the classical "arrow-pushing" formalism of organic chemistry[653]. Consequently, most contemporary models adopt autoregressive strategies, as illustrated in **Figure 16C**. Representative methods include MEGAN[654] and Graph2Edits[646].

**MEGAN**[654] predicts a sequence of graph edit operations to iteratively transform a target product into reactants. It defines five types of edit actions: atom editing, bond editing, atom addition, benzene ring addition, and termination. To train the edit sequences, the authors experimented with various orderings and ultimately settled on the sequence: bond editing → atom editing → ring addition → atom addition. **Graph2Edits**[646] extends this approach by not only modifying the model architecture but also replacing benzene ring addition with leaving group addition, making the generated transformations more chemically plausible. Beyond graph editing strategies, other models such as **G2GT**[643] and **NAG2G**[644] borrow concepts from language translation, framing retrosynthesis as a graph-to-graph translation task. G2GT[643] leverages the Graphormer architecture to autoregressively decode reactants from scratch, while NAG2G[644] incorporates 3D molecular information via a 3D Graphormer to better capture geometric features. Both models can be trained using masked token prediction strategies, analogous to those used in language modeling.

**7.2.1.2 Multi-step Retrosynthesis Prediction Methods**

In practical synthesis planning, a single-step retrosynthesis model alone is insufficient. Multi-step retrosynthesis prediction seeks to construct complete synthetic routes from the target molecule to commercially available precursors, an example is **Figure 17**. To this end, RL-based generative methods have emerged as powerful tools, framing retrosynthesis as a sequential decision-making problem over a molecule-reaction graph. These methods do not directly generate new molecular structures; instead, they operate within a known molecule-reaction database to search for the lowest-cost synthetic pathway.

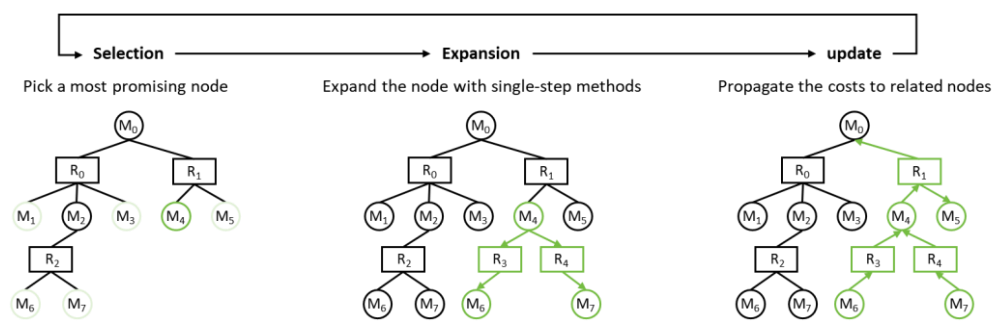

**Figure 17.** MCTS-based multi-step retrosynthesis prediction, which employs a single-step retrosynthesis model as the simulation policy to iteratively explore synthesis routes through a MCTS framework.

**GNN-Retro**[647] formulates multi-step retrosynthesis as a search over candidate intermediates generated by a single-step model. A GNN is then employed to estimate the synthetic cost of each intermediate node, guiding the policy network to select steps that minimize overall cost. A molecular similarity graph, constructed using structural fingerprints or learned embeddings, serves as the search space. During training, this graph remains static; however, during inference, it becomes dynamic as newly generated intermediates are connected to existing nodes, thereby enhancing generalization to unseen chemical structures. Similarly, **RetroGraph**[648] combines graph search with GNNs to perform multi-step planning. Unlike traditional tree-based methods, RetroGraph represents the retrosynthetic space as a directed AND-OR graph consisting of molecule nodes and reaction nodes, explicitly modeling the relationships between target molecules, intermediates, and reactions. This formulation eliminates redundancy by ensuring that each intermediate appears only once in the graph, improving search efficiency. Furthermore, RetroGraph uses a GNN to estimate the success probability of each node as a proxy for synthetic feasibility. The framework also supports batch retrosynthesis, enabling the identification of shared intermediates across different targets and further reducing computational overhead.

### 7.2.2 Forward Reaction Prediction

Forward reaction prediction, also known as reaction outcomes prediction, is the inverse of retrosynthesis: given a set of reactants, the goal is to predict the most likely products. Similar to retrosynthesis, forward reaction prediction can be approached using both template-based and template-free methods, as summarized in **Table 29**.

Table 29. Overview of forward synthesis.

| Methods | Keywords |
|---|---|
| WLDN[630] | Identify reaction center and rank all possible products |
| LocalTransform[655] | WLDN's strategy with the general local template |
| MEGAN[645] | Change the order of graph edits for forward synthesis |
| ELECTRO[656] | Not involved atom changes, only for forward synthesis |
| GTPN[657] | RL trained graph edits from reactants to products |
| NERF[658] | Graph cVAE for generating products given reactants |

In template-based approaches, **Coley**[659] extracted 140,284 reaction templates from a dataset of 1.1 million reactions. These templates were applied to reactants to generate candidate products, which were then ranked to identify the major product. However, this global template matching approach results in an excessively large template library, increasing computational cost and inference complexity. To address this issue, **WLDN**[630] (Weisfeiler–Lehman Difference Network) introduced the concept of localized templates. The model first uses a GNN to identify the reaction center, then generates and ranks product candidates based on this localized region, significantly reducing both the template set size and the search space. Building on this idea, **LocalTransform**[655] further optimized local template construction, requiring only ~100 generic templates to cover 94.6% of the reactions in the training set. Additionally, it leveraged a more modern MPNN architecture to improve model performance.

In contrast, template-free forward reaction prediction directly models reactions as graph transformation or editing tasks, closely resembling template-free retrosynthesis prediction. Notably, forward synthesis typically involves modifications to existing bonds within a molecule, without introducing many new atoms. As such, forward prediction is generally considered slightly less complex than retrosynthesis. Most graph-editing models designed for retrosynthesis can be adapted to forward prediction with minor adjustments, although the reverse is not always true. A representative example is **MEGAN**, a graph editing-based retrysynthesis method that defines five graph editing operations: atom editing, bond editing, atom addition, ring addition, and termination. In forward reaction prediction, these operations can be reused with minor modifications, such as reordering the sequence to begin with bond editing followed by atom editing, ring addition, and finally atom addition. In contrast, **ELECTRO**[656] (Electron Path Prediction Model), an earlier graph-editing method, only defines operations such as reaction center identification, bond transformation, and continuation decision, without any atom addition capabilities. As a result, ELECTRO is only applicable to forward synthesis tasks. Another notable model is **GTPN**[657] (Graph Transformation Policy Network), which employs a RL-guided autoregressive strategy for graph editing. At each step, GTPN receives immediate feedback (action-level correctness) and terminal feedback (whether the final predicted product is correct), and optimizes the editing sequence through RL to construct a path from reactants to the desired product.

In addition to autoregressive frameworks, recent studies have explored more efficient one-shot graph transformation methods. One such example is **NERF**[658] (Non-autoregressive Electron Redistribution Framework). Since forward synthesis primarily involves bond rearrangement without atom addition or deletion, NERF focuses solely on predicting changes in bond connectivity. It adopts a conditional VAE to directly learn the probability distribution $p(G_p|G_r)$, where $G_r$ is the reactant graph and $G_p$ is the product graph. Thanks to its non-autoregressive nature, NERF can predict changes for all bonds in parallel, significantly boosting computational efficiency and inference speed.

## 7.3 Reaction Property Prediction Tasks

In contrast to synthesis planning tasks, reaction property prediction is relatively simple and can be regarded as an extension of molecular property prediction. The main tasks related to reaction properties include reaction classification, reaction condition prediction, and reaction yield prediction, each of which will be discussed in detail below. **Table 30** summarizes these reaction property prediction methods.

Table 30. Overview of reaction property prediction methods.

| Task | Methods | Keywords |
|---|---|---|
| Reaction Classification | Wei et al.[660] | Early GNN fingerprint extraction |
|  | RxnRep[661] | Contrastive learning with graph augmentation (MolCLR) |
| Condition Prediction | Ryou et al.[662] | Early work modeling condition as a multi-class task |
|  | AR-GCN[663] | Attentively transform component to reaction embedding |
|  | ReactionVAE[664] | VAE for condition combination vector prediction |
|  | Reacon[665] | ConScore + condition clustering for diverse predictions |
| Yield Prediction | GraphRXN[666] | Aggregate all mol graphs as the reaction fingerprint |
|  | YieldMPNN[667] | Introduce UQ in GNN to improve performance |
|  | MPNN+Trans[668] | Graph Transformer + MolCLR-style contrastive learning |
|  | MolDescPred[669] | Psedue label pre-training + yield prediction finetune |

### 7.3.1 Reaction Classification and Reaction Fingerprints

Reaction classification is the simplest task within the domain of chemical synthesis. Given a reaction system, the model is required to predict which reaction family it belongs to. Categorizing reactions into families not only facilitates efficient indexing of reactions in literature and databases[670] but also supports subsequent reactivity prediction tasks[671].

Due to the relative simplicity of this task, GNNs are primarily employed to construct learnable reaction fingerprints. These fingerprints serve as abstract vector representations of entire reaction systems, typically expressed as fixed-length numerical vectors. Unlike conventional molecular fingerprints, reaction fingerprints must account for the complexity of multi-component systems (e.g., reactants, products, catalysts, and solvents), which introduces additional design challenges. A well-known example is the rule-based Differential Reaction Fingerprint (DRFP)[672], which generates binary vectors by computing the symmetric difference between the circular substructure sets of reactants and products, effectively serving as a reaction version of the Extended Connectivity Fingerprintt (ECFP).

As early as 2016, **Wei et al.**[660] explored the use of GNNs to represent reactants and reagents for reaction classification. Later, **RxnRep**[661] (Reaction Representation) was introduced to address the challenge of data scarcity, where traditional supervised learning models often suffer from limited generalization capabilities. RxnRep proposes an unsupervised pretraining strategy inspired by MolCLR, using contrastive learning techniques on augmented graph pairs. These augmentations include atom masking, bond masking, atom dropping, bond dropping, and extraction of subgraphs centered on the reaction core. The empirical results show that the resulting reaction fingerprints outperform both rule-based and language model-derived representations on reaction classification tasks, even without task-specific fine-tuning. This highlights the strong transferability and generalizability of self-supervised reaction embeddings.

### 7.3.2 Reaction Condition Prediction

In synthetic planning, even with an appropriate combination of reactants, selecting suitable reaction conditions, such as solvent, catalyst, reagent, and temperature, is essential to ensure the reaction proceeds successfully. As such, reaction condition prediction becomes an indispensable component of computational synthesis.

#### 7.3.2.1 Multi-label Classification

A straightforward strategy is to formulate condition prediction as a multi-label classification task. Training data can be derived from large reaction databases such as Reaxys, where each reaction is annotated with the corresponding condition labels. The model takes a reaction input and outputs a probability distribution across multiple condition categories. These predictions are typically represented as a $k \times d$ matrix, where $k$ denotes condition categories (e.g., catalyst, solvent, temperature) and $d$ represents the set of possible values within each category (e.g., different catalyst or solvent types). **Ryou et al.**[662] initially explored this direction by evaluating various GNNs on four types of coupling reactions. Their follow-up work reported by **Maser et al.**[663] explicitly addressed the presence of disconnected molecular graphs in real-world reaction systems. AR-GCN employs an attention mechanism to integrate information across all molecular components into a unified reaction fingerprint, which is subsequently passed through a multi-layer perceptron to perform classification over each condition category.

### 7.3.2.2 Generative Modeling

While classification-based approaches like AR-GCN can predict the relevance of individual components, they lack the ability to generate coherent and complete condition sets. In practice, a single reaction may proceed under multiple feasible condition combinations. Independent classification over condition categories fails to capture such flexibility and does not resolve how many additives or solvents are actually required.

To address this, **ReactionVAE**[664] formulates condition prediction as a generative task. Specifically, given a reaction pair $(P, R)$, the dataset may provide one or more valid condition sets $\{c_1, c_2, ..., c_n\}$. Each condition set $c_i$ is modeled as a k-dimensional vector corresponding to values across $k$ condition categories. ReactionVAE utilizes a graph-based variational autoencoder to model the conditional distribution $p(c|P,R)$, allowing the model to sample diverse condition combinations that are consistent with the reaction context.

**Reacon**[665] offers another probabilistic approach to reaction condition prediction. It trains individual binary classifiers for each condition component (e.g., catalyst, solvent1, solvent2, reagent1, reagent2, reagent3), learning the probability that each component appears in a given reaction context. During inference, Reacon combines the probabilities of all components to form a composite condition score:

$$\text{ConScore}(x) = \prod_{i \in \{\text{cat, solv1, solv2, reag1, reag2, reag3}\}} P_i(x)$$

To improve prediction diversity and practical applicability, Reacon incorporates a condition clustering strategy during inference. Reactions are grouped according to features such as functional group presence, metal elements, or redox status. In addition, reaction template guidance is applied, leveraging the insight that reactions sharing the same template often exhibit similar mechanistic characteristics and condition requirements. This approach enables the model to select high-quality candidates across top-k entries in each condition category, enhancing both robustness and coverage.

### 7.3.3 Reaction Yield Prediction

In organic chemistry, predicting the yield of a chemical reaction is a critical component of synthetic planning. Accurate yield prediction not only helps assess the overall efficiency of a synthetic route but also aids in identifying bottleneck steps that may lead to poor outcomes[673]. Similar to reaction classification, GNN-based yield prediction models typically encode the entire reaction system into a vector representation followed by a regression layer. However, unlike classification, yield prediction involves modeling fine-grained, continuous variables, which requires higher model sensitivity and expressiveness.

Early approaches often relied on combining handcrafted physicochemical descriptors with classical machine learning algorithms. For example, **Doyle et al.**[633] used features such as NMR parameters and HOMO-LUMO gaps alongside a random forest model to successfully predict yields for Buchwald–Hartwig amination reactions, achieving a coefficient of determination ($R^2$) of 0.92. Notably, they observed that more complex models like DNNs did not perform better, likely due to their reliance on manually defined descriptors. To fully exploit the capabilities of deep learning, end-to-end frameworks are needed to automatically learn rich reaction representations directly from data.

**GraphRXN**[666] proposed a direct approach by extracting molecular features from all components in the reaction system using GNNs. These features are then aggregated either by summation or concatenation, and then fed into a regression model for yield prediction. **YieldMPNN**[667], an earlier but conceptually richer model, introduced uncertainty modeling into the prediction framework. It applies separate GNNs to the reactants and products. To handle variable numbers of reactants, YieldMPNN sums the embeddings of all reactants as $h_R = \sum(h_{R_1}, h_{R_2}, \ldots, h_{R_m})$ and then concatenates this with the product representation to form $h_{PR}$, which is then used for the final yield score. To enhance robustness, YieldMPNN employs heteroscedastic uncertainty regression to capture aleatoric noise, and uses MC-Dropout to estimate epistemic uncertainty. To further improve model expressiveness, **MPNN+Transformer**[668] combines inductive biases from MPNNs with the long-range modeling capabilities of Transformers, forming a GraphTransformer-like architecture. This model also incorporates RxnRep-style graph augmentations for unsupervised pretraining, yielding significant performance gains. Another model, **MolDescPred**[669], uses molecular fingerprints as pseudo-labels for pre-training first and finetuning on the yield prediction latter. MolDescPred also adopts a similar framework to YieldMPNN.

It is important to note that most reaction yield prediction models are trained and evaluated on HTE datasets, which represent only a narrow slice of the chemical space. Some efforts, such as **YieldGNN**[117], have attempted to scale up using large pharmaceutical electronic laboratory notebook (ELN) data. When applied to large, heterogeneous datasets, model performance typically deteriorates, with $R^2$ values rarely exceeding 0.2. Furthermore, when predicting reaction yields from general-purpose databases such as Reaxys, even for a coarse binary classification task (e.g., predicting whether yield is above 65%), the best models only achieve around 65 ± 5% accuracy[674]. These observations underscore a critical limitation of current models: poor generalizability across diverse reaction classes and chemical contexts. Addressing data sparsity and improving domain coverage remains a major challenge for future research.

## 7.4 Perspectives of Chemical Synthesis

### 7.4.1 Conformations are Crucial

Overall, CASP remains a field with numerous open questions worthy of deeper exploration. One fundamental limitation is that current models almost only take 2D molecular graph as input, neglecting the chemical reality that reactions are governed by 3D conformations and dynamic processes. Features such as nucleophilic attack sites, spatial orientations, and transition-state geometries strongly influence reactivity and selectivity. While models like NAG2G have begun incorporating geometric information, a comprehensive framework for learning in 3D chemical space has yet to be established. The integration of geometric deep learning with temporal dynamics could offer more accurate modeling of stereoselectivity and transition-state features.

### 7.4.2 Molecular Generation and Expert Synthesis System

In the realm of reaction pathway planning, the availability of plausible synthetic routes is a crucial criterion for assessing molecular synthesizability. Many molecules generated by de novo molecular design models are often difficult to synthesize, making experimental validation highly challenging. Integrating forward and retrosynthetic prediction directly into the molecular generation process could enable models to assess synthetic accessibility in tandem with structural innovation, thus dramatically enhancing their relevance for real-world drug discovery. Moreover, since reaction pathway planning heavily depends on domain expertise, improving the interpretability of graph-based models is essential. Interpretability not only allows chemists to better understand the rationale behind model predictions but also increases trust in automated suggestions, moving towards expert-system-level platforms, exemplified by tools like Chematica[675].

### 7.4.3 Cross-Reaction Prediction is the Future

From the perspective of reaction property prediction, the primary bottleneck remains the limited scope of training data. Most models are trained on HTE datasets that represent only a narrow fraction of reaction types and chemical diversity. As a result, models often perform well within known reaction regimes but fail to generalize to novel mechanisms. This issue is especially evident when models trained on curated datasets are applied to real-world sources such as Reaxys. In such settings, performance degradation is significant, mirroring the data sparsity problem observed in ADMET prediction. To overcome this, future models may benefit from the integration of external knowledge sources such as chemical knowledge graphs. These resources can inject biochemical and mechanistic priors into the learning process, potentially overcoming the limitations of molecular graph-based-only representations and yielding more robust, generalizable, and interpretable prediction frameworks.

# 8. Conclusions and Perspectives

The chapters above have examined how graph neural networks address the diverse tasks of modern drug discovery, including molecular property prediction, virtual screening, reaction modeling, molecular generation, and biomedical knowledge graph construction. In each context we identified recurring obstacles, such as data sparsity, physical validity, overfitting, metric bias, and the tension among multiple design objectives. A synthesis of these insights points to a broader trajectory. Although graph neural networks have become highly effective at learning expressive representations from individual molecular graphs, their greater potential lies in acting as relational bridges that connect chemical structures to the many biological modalities that govern therapeutic response. Progress in this direction will require message-passing schemes that extend beyond intramolecular neighborhoods to protein interfaces, signaling pathways, and clinically observed phenotypes, the incorporation of symmetry-aware architectures with differentiable physical constraints that ensure chemical plausibility during training, and foundation-scale pretraining on billions of compounds and large collections of macromolecular complexes to create transferable inductive priors for data-limited settings. Equally important is the integration of causal explanation, calibrated uncertainty, and expert feedback within data-centric pipelines that can evolve alongside experimental workflows. Under these conditions, graph neural networks can move from isolated predictors to comprehensive integrators, providing the connective framework that links molecular design decisions with biological outcomes and thereby accelerating the rational discovery of safe, efficacious therapeutics.

# Author Contributions

Odin Zhang led the overall conceptualization, formal analysis, investigation, project administration, validation, and completed both the original draft and subsequent revisions. Haitao Lin contributed to the investigation and review of the entire manuscript, with a focus on the methodology section. Xujun Zhang was involved in the investigation and revision of the full text, particularly Chapter 4. Xiaorui Wang contributed to the investigation and editing of the manuscript, with emphasis on Chapter 7. Zhengxing Wu focused on the investigation and revision of Chapter 2. Qingye contributed to the investigation and review of Chapter 6. Weibo Zhao was responsible for figure design, preparation, and refinement. Jike Wang contributed to Chapter 6 through investigation and review. Kejun Ying was involved in the review of Chapter 6 and participated in the overall discussion. Yu Kang provided scientific input and writing suggestions across all sections. Chang-yu Hsieh reviewed the manuscript comprehensively, with a focus on verifying mathematical formulations. Tingjun Hou supervised the entire work, guided scientific direction, and contributed to critical manuscript revisions.

# Biographies

**Odin Zhang** received his dual B.S. degrees in Physics and Pharmaceutics from Zhejiang University in 2022, and M.S. degree in Pharmaceutics in 2024, where he worked with Prof. Tingjun Hou on computer-aided drug design. He will pursue his Ph.D. in Computer Science at the University of Washington, advised by Prof. David Baker, focusing on general-purpose molecular design. His research spans a wide range of topics in AI-driven drug discovery, including molecular docking, ADMET prediction, and molecular design.

**Haitao Lin** received his B.S. degree in Solid State Physics from Sichuan University. Since 2015, he has been pursuing a Ph.D. degree in Computer Science at Zhejiang University and Westlake University. His primary research interests lie in generative artificial intelligence and its applications in drug discovery, including small molecule and protein generation.

**Xujun Zhang** received his Ph.D. degree (2024) from Zhejiang University, where he focused on methodologies in machine learning-based molecular docking under Prof. Tingjun Hou. After that, he joined the postdoctoral program at the College of Pharmaceutical Sciences, Zhejiang University, under the continued supervision of Prof. Tingjun Hou. His current research interest focuses on integrating co-folding methodologies (such as AlphaFold3) with virtual screening applications, with the goal of advancing computational drug discovery through improved protein-ligand interaction predictions.

**Xiaorui Wang** received his Ph.D. degree in 2024 from Macau University of Science and Technology, under the supervision of Prof. Xiaojun Yao, with a focus on algorithmic approaches for drug synthesis planning. Upon graduation, he joined the research group of Prof. Tingjun Hou at the College of Pharmaceutical Sciences, Zhejiang University, as a postdoctoral fellow. His current research interests include hybrid organic-enzymatic synthesis planning algorithms and the application of natural language technologies to advance autonomous and intelligent synthesis planning.

**Zhenxing Wu** received his Ph.D. degree in 2023 from Zhejiang University, where he focused on artificial intelligence–based drug discovery under the supervision of Prof. Tingjun Hou. He is currently a postdoctoral researcher at the College of Pharmaceutical Sciences, Zhejiang University. His research interests center around AI-driven molecular property prediction, molecular optimization, and interpretable machine learning for drug design.

**Qing Ye** received her Ph.D. in Electronic and Information Engineering from Zhejiang University in 2024, where she focused on graph representation learning for the discovery of therapeutic drug targets. Her research was conducted under the co-supervision of Prof. Tingjun Hou, Prof. Shibo He, and Prof. Chang-Yu Hsieh.

**Weibo Zhao** received his B.S.(2024) from the College of Pharmacy at Zhejiang University, where he performed research on the molecular representation based on Electron Cloud Models. In 2025, he received an offer and was preparing to continue his postgraduate studies at Johns Hopkins University.

**Jike Wang** received his Ph.D. degree (2022) from School of Computer Science, Wuhan University, where he focused on artificial intelligence drug design. After that, he joined the group of Prof. Tingjun Hou at College of Pharmaceutical Sciences, Zhejiang University. His current research interests focus on the design of small molecules, peptides, and antibodies using artificial intelligence.

**Kejun Ying** received his Ph.D. in Biology (2025) and M.S. in Computational Science & Engineering (2024) from Harvard University, where he focused on aging systems biology under Prof. Vadim Gladyshev. He is currently a postdoctoral researcher at Stanford University (Wyss-Coray lab) and the Institute of Protein Design (Baker lab), supported by an NIH/NIA K00 fellowship. His current research focuses on developing computational approaches to understand aging mechanisms, with notable contributions including the first causal inference-based aging clock and ClockBase, a comprehensive biological age data platform. His goal is to leverage systems biology and machine learning to advance our understanding of aging and develop interventions for healthy longevity.

**Yu Kang** received her Ph.D. in 2011 from Zhejiang University, where she focused on molecular simulation under the guidance of Prof. Qi Wang. After that, she joined the group of Dr. Mark Santer and Prof. Reinhard Lipowsky as a postdoctoral researcher at Max Planck Institute of Colloids and Interfaces, focusing on multiscale simulation and computer-aided design of biomolecules. She is currently an assosciate professor at College of Pharmaceutical Sciences, Zhejiang University. Her research interests primarily encompass computer-aided drug design and molecular modeling.

**Changyu Hsieh** received his Ph.D. in Physics from the University of Ottawa and completed his postdoctoral training in theoretical chemistry at the University of Toronto and MIT. He previously earned his B.Eng. in Engineering Physics from the University of Toronto. From 2018 to 2022, he worked at Tencent Quantum Lab, focusing on quantum computing and AI for Science. Since 2022, he has been a faculty member at the College of Pharmaceutical Sciences, Zhejiang University, where his research explores the intersection of computational theory, algorithm development, and drug discovery.

**Tingjun Hou** received his Ph.D. in Computational Chemistry from Peking University in 2002. Currently, he is a full professor in College of Pharmaceutical Sciences at Zhejiang University. His research focuses on molecular modeling and AI-driven drug discovery, including development of structure-based virtual screening methodologies, theoretical predictions of ADMET and drug-likeness, discovery of small molecule inhibitors towards important drug targets, and multiscale simulations of target-ligand recognition. Prof. Hou has co-authored more than 600 publications in peer-reviewed journals with an h-index of 93.

# Acknowledgements

This work was financially supported by National Key R&D Program of China (2024YFA1300051), National Science Foundation of China (92370130, 22220102001).

# Reference

(1) Dickson, M.; Gagnon, J. P. The cost of new drug discovery and development. *Discov. Med.* **2009**, *4* (22), 172-179.

(2) Sacan, A.; Ekins, S.; Kortagere, S. Applications and limitations of in silico models in drug discovery. *Methods Mol. Biol.* **2012**, *910*, 87-124.

(3) Mandal, S.; Mandal, S. K. Rational drug design. *Eur. J. Pharmacol.* **2009**, *625* (1-3), 90-100.

(4) Lu, P.; Liu, L. Computational protein design and structure prediction—The 2024 Nobel Prize in Chemistry. *Sci. China Chem.* **2025**, *68* (3), 812-814.

(5) Sussman, J. L.; Lin, D.; Jiang, J.; Manning, N. O.; Prilusky, J.; Ritter, O.; Abola, E. E. Protein Data Bank (PDB): database of three-dimensional structural information of biological macromolecules. *Acta Crystallogr., Sec. D: Biol. Crystallogr.* **1998**, *54* (6), 1078-1084.

(6) Weinstein, J. N.; Collisson, E. A.; Mills, G. B.; Shaw, K. R.; Ozenberger, B. A.; Ellrott, K.; Shmulevich, I.; Sander, C.; Stuart, J. M. The cancer genome atlas pan-cancer analysis project. *Nat. Genet.* **2013**, *45* (10), 1113-1120.

(7) Blanco-Gonzalez, A.; Cabezon, A.; Seco-Gonzalez, A.; Conde-Torres, D.; Antelo-Riveiro, P.; Pineiro, A.; Garcia-Fandino, R. The role of AI in drug discovery: challenges, opportunities, and strategies. *Pharmaceuticals* **2023**, *16* (6), 891.

(8) Vijayan, R.; Kihlberg, J.; Cross, J. B.; Poongavanam, V. Enhancing preclinical drug discovery with artificial intelligence. *Drug Discov. Today* **2022**, *27* (4), 967-984. Mak, K.-K.; Wong, Y.-H.; Pichika, M. R. Artificial intelligence in drug discovery and development. In *Drug Discovery and Evaluation: Safety and Pharmacokinetic Assays*, Hock, F. J., Pugsley, M. K. Eds.; Springer Cham, 2024; pp 1461-1498.

(9) Weininger, D. SMILES, a chemical language and information system. 1. Introduction to methodology and encoding rules. *J. Chem. Inf. Comput. Sci.* **1988**, *28* (1), 31-36.

(10) Krenn, M.; Häse, F.; Nigam, A.; Friederich, P.; Aspuru-Guzik, A. Self-referencing embedded strings (SELFIES): A 100% robust molecular string representation. *Mach. Learn.: Sci. Technol.* **2020**, *1* (4), 045024.

(11) Capecchi, A.; Probst, D.; Reymond, J.-L. One molecular fingerprint to rule them all: drugs, biomolecules, and the metabolome. *J. Cheminformatics* **2020**, *12*, 1-15. Batra, R.; Tran, H. D.; Kim, C.; Chapman, J.; Chen, L.; Chandrasekaran, A.; Ramprasad, R. General atomic neighborhood fingerprint for machine learning-based methods. *J. Phys. Chem. C* **2019**, *123* (25), 15859-15866.

(12) Seo, S.; Kim, W. Y. PharmacoNet: Accelerating Large-Scale Virtual Screening by Deep Pharmacophore Modeling. *arXiv preprint arXiv:2310.00681* **2023**.

(13) O'shea, K.; Nash, R. An introduction to convolutional neural networks. *arXiv preprint arXiv:1511.08458* **2015**.

(14) Vaswani, A.; Shazeer, N.; Parmar, N.; Uszkoreit, J.; Jones, L.; Gomez, A. N.; Kaiser, Ł.; Polosukhin, I. Attention is all you need. *Adv. Neural Inf. Process. Syst* **2017**, *31*, 6000-6010.

(15) Li, Z.; Jiang, M.; Wang, S.; Zhang, S. Deep learning methods for molecular representation and property prediction. *Drug Discov. Today* **2022**, *27* (12), 103373.

(16) Eckert, H.; Bajorath, J. Molecular similarity analysis in virtual screening: foundations, limitations and novel approaches. *Drug Discov. Today* **2007**, *12* (5-6), 225-233.

(17) Mokaya, M.; Imrie, F.; van Hoorn, W. P.; Kalisz, A.; Bradley, A. R.; Deane, C. M. Testing the limits of SMILES-based de novo molecular generation with curriculum and deep reinforcement learning. *Nat. Mach. Intell.* **2023**, *5* (4), 386-394.

(18) Hu, W.; Fey, M.; Zitnik, M.; Dong, Y.; Ren, H.; Liu, B.; Catasta, M.; Leskovec, J. Open graph benchmark: Datasets for machine learning on graphs. *Adv. Neural Inf. Process. Syst* **2020**, *33*, 22118-22133.

(19) Axelrod, S.; Gomez-Bombarelli, R. GEOM, energy-annotated molecular conformations for property prediction and molecular generation. *Sci. Data* **2022**, *9* (1), 185.

(20) Shuman, D. I.; Narang, S. K.; Frossard, P.; Ortega, A.; Vandergheynst, P. The emerging field of signal processing on graphs: Extending high-dimensional data analysis to networks and other irregular domains. *IEEE Signal Process. Mag.* **2013**, *30* (3), 83-98.

(21) Bruna, J.; Zaremba, W.; Szlam, A.; LeCun, Y. Spectral networks and locally connected networks on graphs. *arXiv preprint arXiv:1312.6203* **2013**.

(22) Defferrard, M.; Bresson, X.; Vandergheynst, P. Convolutional neural networks on graphs with fast localized spectral filtering. *Adv. Neural Inf. Process. Syst* **2016**, *30*, 3844-3852.

(23) Kipf, T. N.; Welling, M. Semi-supervised classification with graph convolutional networks. *arXiv preprint arXiv:1609.02907* **2016**.

(24) Wu, Z.; Pan, S.; Chen, F.; Long, G.; Zhang, C.; Yu, P. S. A comprehensive survey on graph neural networks. *IEEE Trans. Neural Networks Learn. Syst.* **2020**, *32* (1), 4-24.

(25) Gilmer, J.; Schoenholz, S. S.; Riley, P. F.; Vinyals, O.; Dahl, G. E. Neural message passing for quantum chemistry. In *International Conference on Machine Learning*, 2017; PMLR: pp 1263-1272.

(26) Hamilton, W.; Ying, Z.; Leskovec, J. Inductive representation learning on large graphs. *Adv. Neural Inf. Process. Syst* **2017**, *31*.

(27) Xu, K.; Hu, W.; Leskovec, J.; Jegelka, S. How powerful are graph neural networks? *arXiv preprint arXiv:1810.00826* **2018**.

(28) Veličković, P.; Cucurull, G.; Casanova, A.; Romero, A.; Lio, P.; Bengio, Y. Graph attention networks. *arXiv preprint arXiv:1710.10903* **2017**.

(29) Fey, M.; Lenssen, J. E. Fast graph representation learning with PyTorch Geometric. *arXiv preprint arXiv:1903.02428* **2019**.

(30) Wang, M.; Zheng, D.; Ye, Z.; Gan, Q.; Li, M.; Song, X.; Zhou, J.; Ma, C.; Yu, L.; Gai, Y. Deep graph library: A graph-centric, highly-performant package for graph neural networks. *arXiv preprint arXiv:1909.01315* **2019**.

(31) Dong, X.; Thanou, D.; Toni, L.; Bronstein, M.; Frossard, P. Graph signal processing for machine learning: A review and new perspectives. *IEEE Signal Process. Mag.* **2020**, *37* (6), 117-127.

(32) Min, Y.; Wenkel, F.; Wolf, G. Scattering gcn: Overcoming oversmoothness in graph convolutional networks. *Adv. Neural Inf. Process. Syst* **2020**, *33*, 14498-14508.

(33) Xiong, Z.; Wang, D.; Liu, X.; Zhong, F.; Wan, X.; Li, X.; Li, Z.; Luo, X.; Chen, K.; Jiang, H. Pushing the boundaries of molecular representation for drug discovery with the graph attention mechanism. *J. Med. Chem.* **2019**, *63* (16), 8749-8760.

(34) Chen, J.; Chen, H. Edge-featured graph attention network. *arXiv preprint arXiv:2101.07671* **2021**.

(35) Villar, S.; Hogg, D. W.; Storey-Fisher, K.; Yao, W.; Blum-Smith, B. Scalars are universal: Equivariant machine learning, structured like classical physics. *Adv. Neural Inf. Process. Syst* **2021**, *34*, 28848-28863.

(36) Schütt, K.; Kindermans, P.-J.; Sauceda Felix, H. E.; Chmiela, S.; Tkatchenko, A.; Müller, K.-R. Schnet: A continuous-filter convolutional neural network for modeling quantum interactions. *Adv. Neural Inf. Process. Syst* **2017**, *31*, 992-1002.

(37) Klicpera, J.; Groß, J.; Günnemann, S. Directional message passing for molecular graphs. *arXiv preprint arXiv:2003.03123* **2020**.

(38) Gasteiger, J.; Becker, F.; Günnemann, S. Gemnet: Universal directional graph neural networks for molecules. *Adv. Neural Inf. Process. Syst* **2021**, *34*, 6790-6802.

(39) Liu, Y.; Wang, L.; Liu, M.; Zhang, X.; Oztekin, B.; Ji, S. Spherical message passing for 3d graph networks. *arXiv preprint arXiv:2102.05013* **2021**.

(40) Wang, L.; Liu, Y.; Lin, Y.; Liu, H.; Ji, S. ComENet: Towards complete and efficient message passing for 3D molecular graphs. *Adv. Neural Inf. Process. Syst* **2022**, *35*, 650-664.

(41) Thomas, N.; Smidt, T.; Kearnes, S.; Yang, L.; Li, L.; Kohlhoff, K.; Riley, P. Tensor field networks: Rotation-and translation-equivariant neural networks for 3d point clouds. *arXiv preprint arXiv:1802.08219* **2018**.

(42) Fuchs, F.; Worrall, D.; Fischer, V.; Welling, M. Se (3)-transformers: 3d roto-translation equivariant attention networks. *Adv. Neural Inf. Process. Syst* **2020**, *33*, 1970-1981.

(43) Anderson, B.; Hy, T. S.; Kondor, R. Cormorant: Covariant molecular neural networks. *Adv. Neural Inf. Process. Syst* **2019**, *32*.

(44) Brandstetter, J.; Hesselink, R.; van der Pol, E.; Bekkers, E. J.; Welling, M. Geometric and physical quantities improve e (3) equivariant message passing. *arXiv preprint arXiv:2110.02905* **2021**.

(45) Liao, Y.-L.; Smidt, T. Equiformer: Equivariant graph attention transformer for 3d atomistic graphs. *arXiv preprint arXiv:2206.11990* **2022**.

(46) Batatia, I.; Kovacs, D. P.; Simm, G.; Ortner, C.; Csányi, G. MACE: Higher order equivariant message passing neural networks for fast and accurate force fields. *Adv. Neural Inf. Process. Syst* **2022**, *35*, 11423-11436.

(47) Jing, B.; Eismann, S.; Suriana, P.; Townshend, R. J.; Dror, R. Learning from protein structure with geometric vector perceptrons. *arXiv preprint arXiv:2009.01411* **2020**.

(48) Satorras, V. G.; Hoogeboom, E.; Welling, M. E (n) equivariant graph neural networks. In *International conference on machine learning*, 2021; PMLR: pp 9323-9332.

(49) Haghighatlari, M.; Li, J.; Guan, X.; Zhang, O.; Das, A.; Stein, C. J.; Heidar-Zadeh, F.; Liu, M.; Head-Gordon, M.; Bertels, L. NewtonNet: a Newtonian message passing network for deep learning of interatomic potentials and forces. *Digit. Discov.* **2022**, *1* (3), 333-343.

(50) Le, T.; Noé, F.; Clevert, D.-A. Equivariant graph attention networks for molecular property prediction. *arXiv preprint arXiv:2202.09891* **2022**.

(51) Schütt, K.; Unke, O.; Gastegger, M. Equivariant message passing for the prediction of tensorial properties and molecular spectra. In *International Conference on Machine Learning*, 2021; PMLR: pp 9377-9388.

(52) Bonomi, M.; Branduardi, D.; Bussi, G.; Camilloni, C.; Provasi, D.; Raiteri, P.; Donadio, D.; Marinelli, F.; Pietrucci, F.; Broglia, R. A. PLUMED: A portable plugin for free-energy calculations with molecular dynamics. *Comput. Phys. Commun.* **2009**, *180* (10), 1961-1972.

(53) Baek, M.; DiMaio, F.; Anishchenko, I.; Dauparas, J.; Ovchinnikov, S.; Lee, G. R.; Wang, J.; Cong, Q.; Kinch, L. N.; Schaeffer, R. D. Accurate prediction of protein structures and interactions using a three-track neural network. *Science* **2021**, *373* (6557), 871-876.

(54) Jumper, J.; Evans, R.; Pritzel, A.; Green, T.; Figurnov, M.; Ronneberger, O.; Tunyasuvunakool, K.; Bates, R.; Žídek, A.; Potapenko, A. Highly accurate protein structure prediction with AlphaFold. *Nature* **2021**, *596* (7873), 583-589.

(55) Worrall, D. E.; Garbin, S. J.; Turmukhambetov, D.; Brostow, G. J. Harmonic networks: Deep translation and rotation equivariance. In *Proceedings of the IEEE conference on computer vision and pattern recognition*, 2017; pp 5028-5037.

(56) Cohen, T. S.; Geiger, M.; Köhler, J.; Welling, M. Spherical cnns. *arXiv preprint arXiv:1801.10130* **2018**.

(57) Deng, C.; Litany, O.; Duan, Y.; Poulenard, A.; Tagliasacchi, A.; Guibas, L. J. Vector neurons: A general framework for so (3)-equivariant networks. In *Proceedings of the IEEE/CVF International Conference on Computer Vision*, 2021; pp 12200-12209.

(58) Zhang, X.; Zhang, O.; Shen, C.; Qu, W.; Chen, S.; Cao, H.; Kang, Y.; Wang, Z.; Wang, E.; Zhang, J. Efficient and accurate large library ligand docking with KarmaDock. *Nat. Comput. Sci.* **2023**, *3* (9), 789-804.

(59) Rosenfeld, J. S.; Rosenfeld, A.; Belinkov, Y.; Shavit, N. A constructive prediction of the generalization error across scales. *arXiv preprint arXiv:1909.12673* **2019**. Abnar, S.; Dehghani, M.; Neyshabur, B.; Sedghi, H. Exploring the limits of large scale pre-training. *arXiv preprint arXiv:2110.02095* **2021**. Kaplan, J.; McCandlish, S.; Henighan, T.; Brown, T. B.; Chess, B.; Child, R.; Gray, S.; Radford, A.; Wu, J.; Amodei, D. Scaling laws for neural language models. *arXiv preprint arXiv:2001.08361* **2020**.

(60) Brown, T.; Mann, B.; Ryder, N.; Subbiah, M.; Kaplan, J. D.; Dhariwal, P.; Neelakantan, A.; Shyam, P.; Sastry, G.; Askell, A. Language models are few-shot learners. *Adv. Neural Inf. Process. Syst* **2020**, *33*, 1877-1901.

(61) *Dataset: Instagram Images*. https://service.tib.eu/ldmservice/dataset/instagram-images (accessed 2024 Nov, 25).

(62) Sun, C.; Shrivastava, A.; Singh, S.; Gupta, A. Revisiting unreasonable effectiveness of data in deep learning era. In *Proceedings of the IEEE international conference on computer vision*, 2017; pp 843-852.

(63) Radford, A.; Kim, J. W.; Hallacy, C.; Ramesh, A.; Goh, G.; Agarwal, S.; Sastry, G.; Askell, A.; Mishkin, P.; Clark, J. Learning transferable visual models from natural language supervision. In *International conference on machine learning*, 2021; PMLR: pp 8748-8763.

(64) Zhang, W.; Sheng, Z.; Yin, Z.; Jiang, Y.; Xia, Y.; Gao, J.; Yang, Z.; Cui, B. Model degradation hinders deep graph neural networks. In *Proceedings of the 28th ACM SIGKDD conference on knowledge discovery and data mining*, 2022; pp 2493-2503.

(65) Wang, Z.; Zhang, Z.; Zhang, C.; Ye, Y. Tackling Negative Transfer on Graphs. *arXiv preprint arXiv:2402.08907* **2024**.

(66) Papp, P. A.; Martinkus, K.; Faber, L.; Wattenhofer, R. DropGNN: Random dropouts increase the expressiveness of graph neural networks. *Adv. Neural Inf. Process. Syst* **2021**, *34*, 21997-22009.

(67) Zhang, S.; Zhu, F.; Yan, J.; Zhao, R.; Yang, X. DOTIN: Dropping task-irrelevant nodes for GNNs. *arXiv preprint arXiv:2204.13429* **2022**.

(68) Rong, Y.; Huang, W.; Xu, T.; Huang, J. Dropedge: Towards deep graph convolutional networks on node classification. *arXiv preprint arXiv:1907.10903* **2019**.

(69) Guo, H.; Sun, S. Softedge: regularizing graph classification with random soft edges. *arXiv preprint arXiv:2204.10390* **2022**.

(70) Li, G.; Muller, M.; Thabet, A.; Ghanem, B. Deepgcns: Can gcns go as deep as cnns? In *Proceedings of the IEEE/CVF international conference on computer vision*, 2019; pp 9267-9276.

(71) Chen, M.; Wei, Z.; Huang, Z.; Ding, B.; Li, Y. Simple and deep graph convolutional networks. In *International conference on machine learning*, 2020; PMLR: pp 1725-1735.

(72) Xu, K.; Li, C.; Tian, Y.; Sonobe, T.; Kawarabayashi, K.-i.; Jegelka, S. Representation learning on graphs with jumping knowledge networks. In *International conference on machine learning*, 2018; PMLR: pp 5453-5462.

(73) Liu, M.; Gao, H.; Ji, S. Towards deeper graph neural networks. In *Proceedings of the 26th ACM SIGKDD international conference on knowledge discovery & data mining*, 2020; pp 338-348.

(74) Topping, J.; Di Giovanni, F.; Chamberlain, B. P.; Dong, X.; Bronstein, M. M. Understanding over-squashing and bottlenecks on graphs via curvature. *arXiv preprint arXiv:2111.14522* **2021**.

(75) Liu, Y.; Zhou, C.; Pan, S.; Wu, J.; Li, Z.; Chen, H.; Zhang, P. Curvdrop: A ricci curvature based approach to prevent graph neural networks from over-smoothing and over-squashing. In *Proceedings of the ACM Web Conference 2023*, 2023; pp 221-230.

(76) Shi, D.; Guo, Y.; Shao, Z.; Gao, J. How Curvature Enhance the Adaptation Power of Framelet GCNs. *arXiv preprint arXiv:2307.09768* **2023**.

(77) Gutteridge, B.; Dong, X.; Bronstein, M. M.; Di Giovanni, F. Drew: Dynamically rewired message passing with delay. In *International Conference on Machine Learning*, 2023; PMLR: pp 12252-12267.

(78) Luzhnica, E.; Day, B.; Lio, P. Clique pooling for graph classification. *arXiv preprint arXiv:1904.00374* **2019**.

(79) He, K.; Zhang, X.; Ren, S.; Sun, J. Deep residual learning for image recognition. In *Proceedings of the IEEE conference on computer vision and pattern recognition*, 2016; pp 770-778.

(80) Nair, V.; Hinton, G. E. Rectified linear units improve restricted boltzmann machines. In *Proceedings of the 27th international conference on machine learning (ICML-10)*, 2010; pp 807-814.

(81) Ioffe, S.; Szegedy, C. Batch normalization: Accelerating deep network training by reducing internal covariate shift. In *International conference on machine learning*, 2015; pmlr: pp 448-456.

(82) He, K.; Zhang, X.; Ren, S.; Sun, J. Delving deep into rectifiers: Surpassing human-level performance on imagenet classification. In *Proceedings of the IEEE international conference on computer vision*, 2015; pp 1026-1034.

(83) Chen, D.; Lin, Y.; Li, W.; Li, P.; Zhou, J.; Sun, X. Measuring and relieving the over-smoothing problem for graph neural networks from the topological view. In *Proceedings of the AAAI conference on artificial intelligence*, 2020; Vol. 34, pp 3438-3445.

(84) Li, Q.; Han, Z.; Wu, X.-M. Deeper insights into graph convolutional networks for semi-supervised learning. In *Proceedings of the AAAI conference on artificial intelligence*, 2018; Vol. 32.

(85) Taubin, G. A signal processing approach to fair surface design. In *Proceedings of the 22nd annual conference on Computer graphics and interactive techniques*, 1995; pp 351-358.

(86) Hochreiter, S. The vanishing gradient problem during learning recurrent neural nets and problem solutions. *Int. J. Uncertain. Fuzz.* **1998**, *6* (02), 107-116.

(87) Hu, K.; Fredrikson, M. Revisiting Exploding Gradient: A Ghost That Never Leaves. **2016**.

(88) Zhou, X.; Wu, O. Drop "noise" edge: an approximation of the Bayesian GNNs. In *Asian Conference on Pattern Recognition*, 2021; Springer: pp 59-72.

(89) Thürlemann, M.; Riniker, S. Anisotropic message passing: Graph neural networks with directional and long-range interactions. In *The Eleventh International Conference on Learning Representations*, 2023.

(90) Huang, G.; Liu, Z.; Van Der Maaten, L.; Weinberger, K. Q. Densely connected convolutional networks. In *Proceedings of the IEEE conference on computer vision and pattern recognition*, 2017; pp 4700-4708.

(91) Li, G.; Xiong, C.; Thabet, A.; Ghanem, B. Deepergcn: All you need to train deeper gcns. *arXiv preprint arXiv:2006.07739* **2020**.

(92) Gong, S.; Bahri, M.; Bronstein, M. M.; Zafeiriou, S. Geometrically principled connections in graph neural networks. In *Proceedings of the IEEE/CVF conference on computer vision and pattern recognition*, 2020; pp 11415-11424. Xu, K.; Zhang, M.; Jegelka, S.; Kawaguchi, K. Optimization of graph neural networks: Implicit acceleration by skip connections and more depth. In *International Conference on Machine Learning*, 2021; PMLR: pp 11592-11602.

(93) Alon, U.; Yahav, E. On the bottleneck of graph neural networks and its practical implications. *arXiv preprint arXiv:2006.05205* **2020**.

(94) Xu, B.; Wang, N.; Chen, T.; Li, M. Empirical evaluation of rectified activations in convolutional network. *arXiv preprint arXiv:1505.00853* **2015**.

(95) Clevert, D.-A.; Unterthiner, T.; Hochreiter, S. Fast and accurate deep network learning by exponential linear units (elus). *arXiv preprint arXiv:1511.07289* **2015**.

(96) Zhang, W.; Sheng, Z.; Jiang, Y.; Xia, Y.; Gao, J.; Yang, Z.; Cui, B. Evaluating deep graph neural networks. *arXiv preprint arXiv:2108.00955* **2021**.

(97) Rosenfeld, J. S. Scaling laws for deep learning. *arXiv preprint arXiv:2108.07686* **2021**.

(98) Irwin, J. J.; Shoichet, B. K. ZINC− a free database of commercially available compounds for virtual screening. *J. Chem. Inf. Model.* **2005**, *45* (1), 177-182.

(99) Radford, A.; Narasimhan, K.; Salimans, T.; Sutskever, I. Improving language understanding by generative pre-training. OpenAI: 2018.

(100) Devlin, J.; Chang, M.-W.; Lee, K.; Toutanova, K. Bert: Pre-training of deep bidirectional transformers for language understanding. *arXiv preprint arXiv:1810.04805* **2018**.

(101) Rasley, J.; Rajbhandari, S.; Ruwase, O.; He, Y. Deepspeed: System optimizations enable training deep learning models with over 100 billion parameters. In *Proceedings of the 26th ACM SIGKDD International Conference on Knowledge Discovery & Data Mining*, 2020; pp 3505-3506.

(102) Ke, G.; He, D.; Liu, T.-Y. Rethinking positional encoding in language pre-training. *arXiv preprint arXiv:2006.15595* **2020**.

(103) Shaw, P.; Uszkoreit, J.; Vaswani, A. Self-attention with relative position representations. *arXiv preprint arXiv:1803.02155* **2018**.

(104) Zhang, J.; Zhang, H.; Xia, C.; Sun, L. Graph-bert: Only attention is needed for learning graph representations. *arXiv preprint arXiv:2001.05140* **2020**.

(105) Dwivedi, V. P.; Bresson, X. A generalization of transformer networks to graphs. *arXiv preprint arXiv:2012.09699* **2020**.

(106) Wu, Z.; Jain, P.; Wright, M.; Mirhoseini, A.; Gonzalez, J. E.; Stoica, I. Representing long-range context for graph neural networks with global attention. *Adv. Neural Inf. Process. Syst* **2021**, *34*, 13266-13279.

(107) Ying, C.; Cai, T.; Luo, S.; Zheng, S.; Ke, G.; He, D.; Shen, Y.; Liu, T.-Y. Do transformers really perform badly for graph representation? *Adv. Neural Inf. Process. Syst* **2021**, *34*, 28877-28888.

(108) Mialon, G.; Chen, D.; Selosse, M.; Mairal, J. Graphit: Encoding graph structure in transformers. *arXiv preprint arXiv:2106.05667* **2021**.

(109) Kreuzer, D.; Beaini, D.; Hamilton, W.; Létourneau, V.; Tossou, P. Rethinking graph transformers with spectral attention. *Adv. Neural Inf. Process. Syst* **2021**, *34*, 21618-21629.

(110) Chen, D.; O'Bray, L.; Borgwardt, K. Structure-aware transformer for graph representation learning. In *International Conference on Machine Learning*, 2022; PMLR: pp 3469-3489.

(111) Hussain, M. S.; Zaki, M. J.; Subramanian, D. Global self-attention as a replacement for graph convolution. In *Proceedings of the 28th ACM SIGKDD Conference on Knowledge Discovery and Data Mining*, 2022; pp 655-665.

(112) Park, W.; Chang, W.; Lee, D.; Kim, J.; Hwang, S.-w. Grpe: Relative positional encoding for graph transformer. *arXiv preprint arXiv:2201.12787* **2022**.

(113) Rampášek, L.; Galkin, M.; Dwivedi, V. P.; Luu, A. T.; Wolf, G.; Beaini, D. Recipe for a general, powerful, scalable graph transformer. *Adv. Neural Inf. Process. Syst* **2022**, *35*, 14501-14515.

(114) Bo, D.; Shi, C.; Wang, L.; Liao, R. Specformer: Spectral graph neural networks meet transformers. *arXiv preprint arXiv:2303.01028* **2023**.

(115) Beaini, D.; Passaro, S.; Létourneau, V.; Hamilton, W.; Corso, G.; Liò, P. Directional graph networks. In *International Conference on Machine Learning*, 2021; PMLR: pp 748-758.

(116) Shi, H.; Gao, J.; Xu, H.; Liang, X.; Li, Z.; Kong, L.; Lee, S.; Kwok, J. T. Revisiting over-smoothing in bert from the perspective of graph. *arXiv preprint arXiv:2202.08625* **2022**. Wang, P.; Zheng, W.; Chen, T.; Wang, Z. Anti-oversmoothing in deep vision transformers via the fourier domain analysis: From theory to practice. *arXiv preprint arXiv:2203.05962* **2022**.

(117) Li, P.; Wang, Y.; Wang, H.; Leskovec, J. Distance encoding: Design provably more powerful neural networks for graph representation learning. *Adv. Neural Inf. Process. Syst* **2020**, *33*, 4465-4478.

(118) Fung, V.; Zhang, J.; Juarez, E.; Sumpter, B. G. Benchmarking graph neural networks for materials chemistry. *Npj Comput. Mater.* **2021**, *7* (1), 84.

(119) Bouritsas, G.; Frasca, F.; Zafeiriou, S.; Bronstein, M. M. Improving graph neural network expressivity via subgraph isomorphism counting. *IEEE Trans. Pattern Anal. Mach. Intell.* **2022**, *45* (1), 657-668.

(120) Dwivedi, V. P.; Luu, A. T.; Laurent, T.; Bengio, Y.; Bresson, X. Graph neural networks with learnable structural and positional representations. *arXiv preprint arXiv:2110.07875* **2021**.

(121) Wang, R.; Fang, X.; Lu, Y.; Yang, C.-Y.; Wang, S. The PDBbind database: methodologies and updates. *J. Med. Chem.* **2005**, *48* (12), 4111-4119.

(122) Landrum, G. A.; Riniker, S. Combining IC50 or K i Values from Different Sources Is a Source of Significant Noise. *J. Chem. Inf. Model.* **2024**.

(123) Glorot, X.; Bengio, Y. Understanding the difficulty of training deep feedforward neural networks. In *Proceedings of the thirteenth international conference on artificial intelligence and statistics*, 2010; JMLR Workshop and Conference Proceedings: pp 249-256.

(124) Hu, Z.; Dong, Y.; Wang, K.; Chang, K.-W.; Sun, Y. Gpt-gnn: Generative pre-training of graph neural networks. In *Proceedings of the 26th ACM SIGKDD international conference on knowledge discovery & data mining*, 2020; pp 1857-1867.

(125) Liu, S.; Wang, H.; Liu, W.; Lasenby, J.; Guo, H.; Tang, J. Pre-training Molecular Graph Representation with 3D Geometry. In *ICLR 2022 Workshop on Geometrical and Topological Representation Learning*, 2022.

(126) Zhu, J.; Xia, Y.; Wu, L.; Xie, S.; Qin, T.; Zhou, W.; Li, H.; Liu, T.-Y. Unified 2d and 3d pre-training of molecular representations. In *Proceedings of the 28th ACM SIGKDD conference on knowledge discovery and data mining*, 2022; pp 2626-2636.

(127) Armitage, J.; Spalek, L. J.; Nguyen, M.; Nikolka, M.; Jacobs, I. E.; Marañón, L.; Nasrallah, I.; Schweicher, G.; Dimov, I.; Simatos, D. Fragment graphical variational autoencoding for screening molecules with small data. *arXiv preprint arXiv:1910.13325* **2019**.

(128) Benjamin, R.; Singer, U.; Radinsky, K. Graph neural networks pretraining through inherent supervision for molecular property prediction. In *Proceedings of the 31st ACM International Conference on Information & Knowledge Management*, 2022; pp 2903-2912.

(129) Hu, W.; Liu, B.; Gomes, J.; Zitnik, M.; Liang, P.; Pande, V.; Leskovec, J. Strategies for pre-training graph neural networks. *arXiv preprint arXiv:1905.12265* **2019**.

(130) Rong, Y.; Bian, Y.; Xu, T.; Xie, W.; Wei, Y.; Huang, W.; Huang, J. Self-supervised graph transformer on large-scale molecular data. *Adv. Neural Inf. Process. Syst* **2020**, *33*, 12559-12571.

(131) Zhang, Z.; Liu, Q.; Wang, H.; Lu, C.; Lee, C.-K. Motif-based graph self-supervised learning for molecular property prediction. *Adv. Neural Inf. Process. Syst* **2021**, *34*, 15870-15882.

(132) Xia, J.; Zhao, C.; Hu, B.; Gao, Z.; Tan, C.; Liu, Y.; Li, S.; Li, S. Z. Mole-bert: Rethinking pre-training graph neural networks for molecules. In *The Eleventh International Conference on Learning Representations*, 2022.

(133) Wang, Y.; Zhang, J.; Jin, J.; Wei, L. MolCAP: Molecular Chemical reActivity Pretraining and prompted-finetuning enhanced molecular representation learning. *Comput. Biol. Med.* **2023**, *167*, 107666.

(134) Godwin, J.; Schaarschmidt, M.; Gaunt, A.; Sanchez-Gonzalez, A.; Rubanova, Y.; Veličković, P.; Kirkpatrick, J.; Battaglia, P. Simple gnn regularisation for 3d molecular property prediction & beyond. *arXiv preprint arXiv:2106.07971* **2021**.

(135) Liu, S.; Guo, H.; Tang, J. Molecular geometry pretraining with se (3)-invariant denoising distance matching. *arXiv preprint arXiv:2206.13602* **2022**.

(136) Zaidi, S.; Schaarschmidt, M.; Martens, J.; Kim, H.; Teh, Y. W.; Sanchez-Gonzalez, A.; Battaglia, P.; Pascanu, R.; Godwin, J. Pre-training via denoising for molecular property prediction. *arXiv preprint arXiv:2206.00133* **2022**.

(137) Feng, S.; Ni, Y.; Lan, Y.; Ma, Z.-M.; Ma, W.-Y. Fractional denoising for 3d molecular pre-training. In *International Conference on Machine Learning*, 2023; PMLR: pp 9938-9961.

(138) You, Y.; Chen, T.; Sui, Y.; Chen, T.; Wang, Z.; Shen, Y. Graph contrastive learning with augmentations. *Adv. Neural Inf. Process. Syst* **2020**, *33*, 5812-5823.

(139) Sun, F.-Y.; Hoffmann, J.; Verma, V.; Tang, J. Infograph: Unsupervised and semi-supervised graph-level representation learning via mutual information maximization. *arXiv preprint arXiv:1908.01000* **2019**.

(140) Li, S.; Zhou, J.; Xu, T.; Dou, D.; Xiong, H. Geomgcl: Geometric graph contrastive learning for molecular property prediction. In *Proceedings of the AAAI conference on artificial intelligence*, 2022; Vol. 36, pp 4541-4549.

(141) Stärk, H.; Beaini, D.; Corso, G.; Tossou, P.; Dallago, C.; Günnemann, S.; Liò, P. 3d infomax improves gnns for molecular property prediction. In *International Conference on Machine Learning*, 2022; PMLR: pp 20479-20502.

(142) Zhang, S.; Hu, Z.; Subramonian, A.; Sun, Y. Motif-driven contrastive learning of graph representations. *arXiv preprint arXiv:2012.12533* **2020**.

(143) You, Y.; Chen, T.; Shen, Y.; Wang, Z. Graph contrastive learning automated. In *International Conference on Machine Learning*, 2021; PMLR: pp 12121-12132.

(144) Li, P.; Wang, J.; Qiao, Y.; Chen, H.; Yu, Y.; Yao, X.; Gao, P.; Xie, G.; Song, S. An effective self-supervised framework for learning expressive molecular global representations to drug discovery. *Brief. Bioinform.* **2021**, *22* (6), bbab109.

(145) Kaufman, B.; Williams, E. C.; Underkoffler, C.; Pederson, R.; Mardirossian, N.; Watson, I.; Parkhill, J. Coati: Multimodal contrastive pretraining for representing and traversing chemical space. *J. Chem. Inf. Model.* **2024**, *64* (4), 1145-1157.

(146) Pinheiro, G. A.; Da Silva, J. L.; Quiles, M. G. Smiclr: Contrastive learning on multiple molecular representations for semisupervised and unsupervised representation learning. *J. Chem. Inf. Model.* **2022**, *62* (17), 3948-3960.

(147) Xiang, H.; Jin, S.; Liu, X.; Zeng, X.; Zeng, L. Chemical structure-aware molecular image representation learning. *Brief. Bioinform.* **2023**, *24* (6), bbad404.

(148) Chen, J.; Zhang, X.; Ma, Z.-M.; Liu, S. Molecule Joint Auto-Encoding: Trajectory Pretraining with 2D and 3D Diffusion. *Adv. Neural Inf. Process. Syst* **2024**, *36*.

(149) Jin, W.; Barzilay, R.; Jaakkola, T. Junction tree variational autoencoder for molecular graph generation. In *International conference on machine learning*, 2018; PMLR: pp 2323-2332.

(150) Rosenstein, M. T.; Marx, Z.; Kaelbling, L. P.; Dietterich, T. G. To transfer or not to transfer. In *NIPS 2005 workshop on transfer learning*, 2005; Vol. 898.

(151) Klopman, G.; Wang, S.; Balthasar, D. M. Estimation of aqueous solubility of organic molecules by the group contribution approach. Application to the study of biodegradation. *J. Chem. Inf. Comput. Sci.* **1992**, *32* (5), 474-482. Hou, T.; Xia, K.; Zhang, W.; Xu, X. ADME evaluation in drug discovery. 4. Prediction of aqueous solubility based on atom contribution approach. *J. Chem. Inf. Comput. Sci.* **2004**, *44* (1), 266-275.

(152) Van Den Oord, A.; Vinyals, O. Neural discrete representation learning. *Adv. Neural Inf. Process. Syst* **2017**, *30*.

(153) Tian, Y.; Sun, C.; Poole, B.; Krishnan, D.; Schmid, C.; Isola, P. What makes for good views for contrastive learning? *Adv. Neural Inf. Process. Syst* **2020**, *33*, 6827-6839.

(154) Chen, T.; Kornblith, S.; Norouzi, M.; Hinton, G. A simple framework for contrastive learning of visual representations. In *International conference on machine learning*, 2020; PMLR: pp 1597-1607.

(155) Hjelm, R. D.; Fedorov, A.; Lavoie-Marchildon, S.; Grewal, K.; Bachman, P.; Trischler, A.; Bengio, Y. Learning deep representations by mutual information estimation and maximization. *arXiv preprint arXiv:1808.06670* **2018**.

(156) Nigam, A.; Pollice, R.; Krenn, M.; dos Passos Gomes, G.; Aspuru-Guzik, A. Beyond generative models: superfast traversal, optimization, novelty, exploration and discovery (STONED) algorithm for molecules using SELFIES. *Chem. Sci.* **2021**, *12* (20), 7079-7090.

(157) Foster, D. *Generative deep learning*; " O'Reilly Media, Inc.", 2022.

(158) Van Den Oord, A.; Kalchbrenner, N.; Kavukcuoglu, K. Pixel recurrent neural networks. In *International conference on machine learning*, 2016; PMLR: pp 1747-1756.

(159) Dinh, L.; Krueger, D.; Bengio, Y. Nice: Non-linear independent components estimation. In *arXiv preprint arXiv:1410.8516*, 2014.

(160) LeCun, Y.; Chopra, S.; Hadsell, R.; Ranzato, M.; Huang, F. A tutorial on energy-based learning. *Predicting structured data* **2006**, *1* (0).

(161) Kingma, D. P.; Welling, M. Auto-encoding variational bayes. In *arXiv preprint arXiv:1312.6114*, 2013.

(162) Goodfellow, I.; Pouget-Abadie, J.; Mirza, M.; Xu, B.; Warde-Farley, D.; Ozair, S.; Courville, A.; Bengio, Y. Generative adversarial nets. *Adv. Neural Inf. Process. Syst* **2014**, *27*.

(163) Song, Y.; Sohl-Dickstein, J.; Kingma, D. P.; Kumar, A.; Ermon, S.; Poole, B. Score-based generative modeling through stochastic differential equations. *arXiv preprint arXiv:2011.13456* **2020**.

(164) Zhou, L.; Lou, A.; Khanna, S.; Ermon, S. Denoising diffusion bridge models. *arXiv preprint arXiv:2309.16948* **2023**.

(165) Lipman, Y.; Chen, R. T.; Ben-Hamu, H.; Nickel, M.; Le, M. Flow matching for generative modeling. *arXiv preprint arXiv:2210.02747* **2022**.

(166) Dinh, L.; Sohl-Dickstein, J.; Bengio, S. Density estimation using real nvp. *arXiv preprint arXiv:1605.08803* **2016**.

(167) Kingma, D. P.; Dhariwal, P. Glow: Generative flow with invertible 1x1 convolutions. *Adv. Neural Inf. Process. Syst* **2018**, *31*.

(168) Du, Y.; Li, S.; Tenenbaum, J.; Mordatch, I. Improved contrastive divergence training of energy based models. *arXiv preprint arXiv:2012.01316* **2020**.

(169) Nijkamp, E.; Hill, M.; Zhu, S.-C.; Wu, Y. N. Learning non-convergent non-persistent short-run mcmc toward energy-based model. *Adv. Neural Inf. Process. Syst* **2019**, *32*.

(170) Song, Y.; Ermon, S. Generative modeling by estimating gradients of the data distribution. *Adv. Neural Inf. Process. Syst* **2019**, *32*.

(171) Ho, J.; Jain, A.; Abbeel, P. Denoising diffusion probabilistic models. *Adv. Neural Inf. Process. Syst* **2020**, *33*, 6840-6851.

(172) Williams, R. J. Simple statistical gradient-following algorithms for connectionist reinforcement learning. *Mach. Learn.* **1992**, *8*, 229-256.

(173) Schulman, J.; Levine, S.; Abbeel, P.; Jordan, M.; Moritz, P. Trust region policy optimization. In *International conference on machine learning*, 2015; PMLR: pp 1889-1897.

(174) Schulman, J.; Wolski, F.; Dhariwal, P.; Radford, A.; Klimov, O. Proximal policy optimization algorithms. *arXiv preprint arXiv:1707.06347* **2017**.

(175) Watkins, C. J.; Dayan, P. Q-learning. *Mach. Learn.* **1992**, *8*, 279-292.

(176) Mnih, V.; Kavukcuoglu, K.; Silver, D.; Graves, A.; Antonoglou, I.; Wierstra, D.; Riedmiller, M. Playing atari with deep reinforcement learning. *arXiv preprint arXiv:1312.5602* **2013**.

(177) Konda, V.; Tsitsiklis, J. Actor-critic algorithms. *Adv. Neural Inf. Process. Syst* **1999**, *12*.

(178) Torabi, F.; Warnell, G.; Stone, P. Behavioral cloning from observation. *arXiv preprint arXiv:1805.01954* **2018**.

(179) Abbeel, P.; Ng, A. Y. Apprenticeship learning via inverse reinforcement learning. In *Proceedings of the twenty-first international conference on Machine learning*, 2004; p 1.

(180) Silver, D.; Schrittwieser, J.; Simonyan, K.; Antonoglou, I.; Huang, A.; Guez, A.; Hubert, T.; Baker, L.; Lai, M.; Bolton, A. Mastering the game of go without human knowledge. *nature* **2017**, *550* (7676), 354-359.

(181) Clark, D. E.; Pickett, S. D. Computational methods for the prediction of 'drug-likeness'. *Drug Discov. Today* **2000**, *5* (2), 49-58.

(182) Hessel, M.; Modayil, J.; Van Hasselt, H.; Schaul, T.; Ostrovski, G.; Dabney, W.; Horgan, D.; Piot, B.; Azar, M.; Silver, D. Rainbow: Combining improvements in deep reinforcement learning. In *Proceedings of the AAAI conference on artificial intelligence*, 2018; Vol. 32.

(183) Mnih, V.; Badia, A. P.; Mirza, M.; Graves, A.; Lillicrap, T.; Harley, T.; Silver, D.; Kavukcuoglu, K. Asynchronous methods for deep reinforcement learning. In *International conference on machine learning*, 2016; PMLR: pp 1928-1937.

(184) Meldgaard, S. A.; Köhler, J.; Mortensen, H. L.; Christiansen, M.-P. V.; Noé, F.; Hammer, B. Generating stable molecules using imitation and reinforcement learning. *Mach. Learn.: Sci. Technol.* **2021**, *3* (1), 015008.

(185) Agyemang, B.; Wu, W.-P.; Addo, D.; Kpiebaareh, M. Y.; Nanor, E.; Roland Haruna, C. Deep inverse reinforcement learning for structural evolution of small molecules. *Brief. Bioinform.* **2021**, *22* (4), bbaa364.

(186) Dowden, H.; Munro, J. Trends in clinical success rates and therapeutic focus. *Nat. Rev. Drug Discov.* **2019**, *18* (7), 495-496.

(187) Walters, W. P.; Barzilay, R. Applications of deep learning in molecule generation and molecular property prediction. *Acc. Chem. Res.* **2020**, *54* (2), 263-270.

(188) Xia, J.; Zhang, L.; Zhu, X.; Liu, Y.; Gao, Z.; Hu, B.; Tan, C.; Zheng, J.; Li, S.; Li, S. Z. Understanding the limitations of deep models for molecular property prediction: Insights and solutions. *Adv. Neural Inf. Process. Syst* **2023**, *36*, 64774-64792.

(189) Currie, G. M. Pharmacology, part 2: introduction to pharmacokinetics. *J. Nucl. Med. Technol.* **2018**, *46* (3), 221-230.

(190) Hou, T.; Zhang, W.; Xia, K.; Qiao, X.; Xu, X. ADME evaluation in drug discovery. 5. Correlation of Caco-2 permeation with simple molecular properties. *J. Chem. Inf. Comput. Sci.* **2004**, *44* (5), 1585-1600.

(191) Chen, L.; Li, Y.; Zhao, Q.; Peng, H.; Hou, T. ADME evaluation in drug discovery. 10. Predictions of P-glycoprotein inhibitors using recursive partitioning and naive Bayesian classification techniques. *Mol. Pharm.* **2011**, *8* (3), 889-900.

(192) Gibaldi, M.; Koup, J. Pharmacokinetic concepts—drug binding, apparent volume of distribution and clearance. *Eur. J. Clin. Pharmacol.* **1981**, *20*, 299-305.

(193) Martins, I. F.; Teixeira, A. L.; Pinheiro, L.; Falcao, A. O. A Bayesian approach to in silico blood-brain barrier penetration modeling. *J. Chem. Inf. Model.* **2012**, *52* (6), 1686-1697.

(194) Meyer, U. A. Overview of enzymes of drug metabolism. *J. Pharmacokinet. Biopharm.* **1996**, *24*, 449-459.

(195) Martinez, M. N. Article II: Volume, clearance, and half-life. *J. Am. Vet. Med. Assoc.* **1998**, *213* (8), 1122-1127.

(196) Gaulton, A.; Bellis, L. J.; Bento, A. P.; Chambers, J.; Davies, M.; Hersey, A.; Light, Y.; McGlinchey, S.; Michalovich, D.; Al-Lazikani, B. ChEMBL: a large-scale bioactivity database for drug discovery. *Nucleic Acids Res.* **2012**, *40* (D1), D1100-D1107.

(197) Dong, J.; Wang, N.-N.; Yao, Z.-J.; Zhang, L.; Cheng, Y.; Ouyang, D.; Lu, A.-P.; Cao, D.-S. ADMETlab: a platform for systematic ADMET evaluation based on a comprehensively collected ADMET database. *J. Cheminformatics* **2018**, *10*, 1-11.

(198) Hou, T.; Wang, J.; Zhang, W.; Xu, X. ADME evaluation in drug discovery. 6. Can oral bioavailability in humans be effectively predicted by simple molecular property-based rules? *J. Chem. Inf. Model.* **2007**, *47* (2), 460-463.

(199) NCATS. *Tox21 Data Challenge*. 2014. https://tripod.nih.gov/tox21/challenge/ (accessed.

(200) Richard, A. M.; Judson, R. S.; Houck, K. A.; Grulke, C. M.; Volarath, P.; Thillainadarajah, I.; Yang, C.; Rathman, J.; Martin, M. T.; Wambaugh, J. F. ToxCast chemical landscape: paving the road to 21st century toxicology. *Chem. Res. Toxicol.* **2016**, *29* (8), 1225-1251.

(201) Kuhn, M.; Letunic, I.; Jensen, L. J.; Bork, P. The SIDER database of drugs and side effects. *Nucleic Acids Res.* **2016**, *44* (D1), D1075-D1079.

(202) Gayvert, K. M.; Madhukar, N. S.; Elemento, O. A data-driven approach to predicting successes and failures of clinical trials. *Cell Chem. Biol.* **2016**, *23* (10), 1294-1301.

(203) Zheng, J. W. a. L.-J., Olivier. IUPAC Digitized pKa Dataset, v2.0. 2024.

(204) Delaney, J. S. ESOL: estimating aqueous solubility directly from molecular structure. *J. Chem. Inf. Comput. Sci.* **2004**, *44* (3), 1000-1005.

(205) Mobley, D. L.; Guthrie, J. P. FreeSolv: a database of experimental and calculated hydration free energies, with input files. *J. Comput. Aided Mol. Des.* **2014**, *28*, 711-720.

(206) Mendez, D.; Gaulton, A.; Bento, A. P.; Chambers, J.; De Veij, M.; Félix, E.; Magariños, M. P.; Mosquera, J. F.; Mutowo, P.; Nowotka, M. ChEMBL: towards direct deposition of bioassay data. *Nucleic Acids Res.* **2019**, *47* (D1), D930-D940.

(207) Dwivedi, V. P.; Joshi, C. K.; Luu, A. T.; Laurent, T.; Bengio, Y.; Bresson, X. Benchmarking graph neural networks. *J. Mach. Learn. Res.* **2023**, *24* (43), 1-48.

(208) Hoja, J.; Medrano Sandonas, L.; Ernst, B. G.; Vazquez-Mayagoitia, A.; DiStasio Jr, R. A.; Tkatchenko, A. QM7-X, a comprehensive dataset of quantum-mechanical properties spanning the chemical space of small organic molecules. *Sci. Data* **2021**, *8* (1), 43.

(209) Ramakrishnan, R.; Hartmann, M.; Tapavicza, E.; Von Lilienfeld, O. A. Electronic spectra from TDDFT and machine learning in chemical space. *J. Chem. Phys.* **2015**, *143* (8).

(210) Ramakrishnan, R.; Dral, P. O.; Rupp, M.; Von Lilienfeld, O. A. Quantum chemistry structures and properties of 134 kilo molecules. *Sci. Data* **2014**, *1* (1), 1-7.

(211) Chmiela, S.; Tkatchenko, A.; Sauceda, H. E.; Poltavsky, I.; Schütt, K. T.; Müller, K.-R. Machine learning of accurate energy-conserving molecular force fields. *Sci. Adv.* **2017**, *3* (5), e1603015.

(212) Nakata, M.; Shimazaki, T. PubChemQC project: a large-scale first-principles electronic structure database for data-driven chemistry. *J. Chem. Inf. Model.* **2017**, *57* (6), 1300-1308.

(213) Wang, Y.; Xiao, J.; Suzek, T. O.; Zhang, J.; Wang, J.; Zhou, Z.; Han, L.; Karapetyan, K.; Dracheva, S.; Shoemaker, B. A. PubChem's BioAssay database. *Nucleic Acids Res.* **2012**, *40* (D1), D400-D412.

(214) Rohrer, S. G.; Baumann, K. Maximum unbiased validation (MUV) data sets for virtual screening based on PubChem bioactivity data. *J. Chem. Inf. Model.* **2009**, *49* (2), 169-184.

(215) NCI. AIDS antiviral screen data.

(216) Subramanian, G.; Ramsundar, B.; Pande, V.; Denny, R. A. Computational modeling of β-secretase 1 (BACE-1) inhibitors using ligand based approaches. *J. Chem. Inf. Model.* **2016**, *56* (10), 1936-1949.

(217) Wale, N.; Watson, I. A.; Karypis, G. Comparison of descriptor spaces for chemical compound retrieval and classification. *Knowl. Inf. Syst.* **2008**, *14*, 347-375.

(218) Xiong, J.; Li, Z.; Wang, G.; Fu, Z.; Zhong, F.; Xu, T.; Liu, X.; Huang, Z.; Liu, X.; Chen, K. Multi-instance learning of graph neural networks for aqueous p K a prediction. *Bioinformatics* **2022**, *38* (3), 792-798.

(219) Pan, X.; Wang, H.; Li, C.; Zhang, J. Z.; Ji, C. MolGpka: A web server for small molecule p K a prediction using a graph-convolutional neural network. *J. Chem. Inf. Model.* **2021**, *61* (7), 3159-3165.

(220) Johnston, R. C.; Yao, K.; Kaplan, Z.; Chelliah, M.; Leswing, K.; Seekins, S.; Watts, S.; Calkins, D.; Chief Elk, J.; Jerome, S. V. Epik: p K a and Protonation State Prediction through Machine Learning. *J. Chem. Theory Comput.* **2023**, *19* (8), 2380-2388.

(221) Wu, J.; Wan, Y.; Wu, Z.; Zhang, S.; Cao, D.; Hsieh, C.-Y.; Hou, T. MF-SuP-pKa: Multi-fidelity modeling with subgraph pooling mechanism for pKa prediction. *Acta Pharm. Sin. B* **2023**, *13* (6), 2572-2584.

(222) Sander, T.; Freyss, J.; Von Korff, M.; Rufener, C. DataWarrior: an open-source program for chemistry aware data visualization and analysis. *J. Chem. Inf. Model.* **2015**, *55* (2), 460-473.

(223) Yang, K.; Swanson, K.; Jin, W.; Coley, C.; Eiden, P.; Gao, H.; Guzman-Perez, A.; Hopper, T.; Kelley, B.; Mathea, M. Analyzing learned molecular representations for property prediction. *J. Chem. Inf. Model.* **2019**, *59* (8), 3370-3388.

(224) Zhang, Z.; Guan, J.; Zhou, S. FraGAT: a fragment-oriented multi-scale graph attention model for molecular property prediction. *Bioinformatics* **2021**, *37* (18), 2981-2987.

(225) Zhu, W.; Zhang, Y.; Zhao, D.; Xu, J.; Wang, L. HiGNN: A hierarchical informative graph neural network for molecular property prediction equipped with feature-wise attention. *J. Chem. Inf. Model.* **2022**, *63* (1), 43-55.

(226) Kong, Y.; Zhao, X.; Liu, R.; Yang, Z.; Yin, H.; Zhao, B.; Wang, J.; Qin, B.; Yan, A. Integrating concept of pharmacophore with graph neural networks for chemical property prediction and interpretation. *J. Cheminformatics* **2022**, *14* (1), 52.

(227) Adams, K.; Pattanaik, L.; Coley, C. W. Learning 3d representations of molecular chirality with invariance to bond rotations. *arXiv preprint arXiv:2110.04383* **2021**.

(228) Liu, Y. L.; Wang, Y.; Vu, O.; Moretti, R.; Bodenheimer, B.; Meiler, J.; Derr, T. Interpretable chirality-aware graph neural network for quantitative structure activity relationship modeling in drug discovery. In *Proceedings of the AAAI Conference on Artificial Intelligence*, 2023; Vol. 37, pp 14356-14364.

(229) Withnall, M.; Lindelöf, E.; Engkvist, O.; Chen, H. Building attention and edge message passing neural networks for bioactivity and physical–chemical property prediction. *J. Cheminformatics* **2020**, *12* (1), 1.

(230) Han, X.; Jia, M.; Chang, Y.; Li, Y.; Wu, S. Directed message passing neural network (D-MPNN) with graph edge attention (GEA) for property prediction of biofuel-relevant species. *Energy and AI* **2022**, *10*, 100201.

(231) Heid, E.; Greenman, K. P.; Chung, Y.; Li, S.-C.; Graff, D. E.; Vermeire, F. H.; Wu, H.; Green, W. H.; McGill, C. J. Chemprop: a machine learning package for chemical property prediction. *J. Chem. Inf. Model.* **2023**, *64* (1), 9-17.

(232) Degen, J.; Wegscheid-Gerlach, C.; Zaliani, A.; Rarey, M. On the Art of Compiling and Using'Drug-Like'Chemical Fragment Spaces. *ChemMedChem: Chemistry Enabling Drug Discovery* **2008**, *3* (10), 1503-1507.

(233) Schütt, K. T.; Sauceda, H. E.; Kindermans, P.-J.; Tkatchenko, A.; Müller, K.-R. Schnet–a deep learning architecture for molecules and materials. *J. Chem. Phys.* **2018**, *148* (24), 241722. Moon, K.; Im, H.-J.; Kwon, S. 3D graph contrastive learning for molecular property prediction. *Bioinformatics* **2023**, *39* (6), btad371.

(234) Cremer, J.; Medrano Sandonas, L.; Tkatchenko, A.; Clevert, D.-A.; De Fabritiis, G. Equivariant graph neural networks for toxicity prediction. *Chem. Res. Toxicol.* **2023**, *36* (10), 1561-1573.

(235) Fang, X.; Liu, L.; Lei, J.; He, D.; Zhang, S.; Zhou, J.; Wang, F.; Wu, H.; Wang, H. Geometry-enhanced molecular representation learning for property prediction. *Nat. Mach. Intell.* **2022**, *4* (2), 127-134.

(236) Axelrod, S.; Gomez-Bombarelli, R. Molecular machine learning with conformer ensembles. *Mach. Learn.: Sci. Technol.* **2023**, *4* (3), 035025.

(237) H. Brooks, W.; C. Guida, W.; G. Daniel, K. The significance of chirality in drug design and development. *Curr. Top. Med. Chem.* **2011**, *11* (7), 760-770.

(238) Ridings, J. E. The thalidomide disaster, lessons from the past. In *Teratogenicity Testing: Methods and Protocols*, Springer, 2012; pp 575-586.

(239) Wu, Z.; Jiang, D.; Wang, J.; Hsieh, C.-Y.; Cao, D.; Hou, T. Mining toxicity information from large amounts of toxicity data. *J. Med. Chem.* **2021**, *64* (10), 6924-6936.

(240) Liu, Z.; Lin, L.; Jia, Q.; Cheng, Z.; Jiang, Y.; Guo, Y.; Ma, J. Transferable multilevel attention neural network for accurate prediction of quantum chemistry properties via multitask learning. *J. Chem. Inf. Model.* **2021**, *61* (3), 1066-1082.

(241) Spyromitros-Xioufis, E.; Tsoumakas, G.; Groves, W.; Vlahavas, I. Multi-target regression via input space expansion: treating targets as inputs. *Mach. Learn.* **2016**, *104*, 55-98.

(242) Du, B.-X.; Xu, Y.; Yiu, S.-M.; Yu, H.; Shi, J.-Y. MTGL-ADMET: a novel multi-task graph learning framework for ADMET prediction enhanced by status-theory and maximum flow. In *International Conference on Research in Computational Molecular Biology*, 2023; Springer: pp 85-103.

(243) Nguyen, C. Q.; Kreatsoulas, C.; Branson, K. M. Meta-learning GNN initializations for low-resource molecular property prediction. *arXiv preprint arXiv:2003.05996* **2020**.

(244) Guo, Z.; Zhang, C.; Yu, W.; Herr, J.; Wiest, O.; Jiang, M.; Chawla, N. V. Few-shot graph learning for molecular property prediction. In *Proceedings of the web conference 2021*, 2021; pp 2559-2567.

(245) Zhuang, X.; Zhang, Q.; Wu, B.; Ding, K.; Fang, Y.; Chen, H. Graph sampling-based meta-learning for molecular property prediction. *arXiv preprint arXiv:2306.16780* **2023**.

(246) Yu, T.; Kumar, S.; Gupta, A.; Levine, S.; Hausman, K.; Finn, C. Gradient surgery for multi-task learning. *Adv. Neural Inf. Process. Syst* **2020**, *33*, 5824-5836.

(247) Zhang, L.; Yang, Q.; Liu, X.; Guan, H. Rethinking hard-parameter sharing in multi-domain learning. In *2022 IEEE International Conference on Multimedia and Expo (ICME)*, 2022; IEEE: pp 01-06.

(248) Han, C.; Wang, H.; Zhu, J.; Liu, Q.; Zhu, W. Comparison of multi-task approaches on molecular property prediction. *Chin. J. Chem. Phys.* **2023**, *36* (4), 443-452.

(249) Vanschoren, J. Meta-learning. *Automated machine learning: methods, systems, challenges* **2019**, 35-61.

(250) Li, H.; Dong, W.; Mei, X.; Ma, C.; Huang, F.; Hu, B.-G. LGM-Net: Learning to generate matching networks for few-shot learning. In *International conference on machine learning*, 2019; PMLR: pp 3825-3834.

(251) Finn, C.; Abbeel, P.; Levine, S. Model-agnostic meta-learning for fast adaptation of deep networks. In *International conference on machine learning*, 2017; PMLR: pp 1126-1135.

(252) Nichol, A.; Achiam, J.; Schulman, J. On first-order meta-learning algorithms. *arXiv preprint arXiv:1803.02999* **2018**.

(253) Raghu, A.; Raghu, M.; Bengio, S.; Vinyals, O. Rapid learning or feature reuse? towards understanding the effectiveness of maml. *arXiv preprint arXiv:1909.09157* **2019**.

(254) Li, Y.; Tarlow, D.; Brockschmidt, M.; Zemel, R. Gated graph sequence neural networks. *arXiv preprint arXiv:1511.05493* **2015**.

(255) de Ocáriz Borde, H. S.; Barbero, F. Graph neural network expressivity and meta-learning for molecular property regression. In *The First Learning on Graphs Conference*, 2022.

(256) OpenAI. *ChatGPT*. 2023. https://chat.openai.com/chat (accessed.

(257) Buterez, D.; Janet, J. P.; Kiddle, S. J.; Oglic, D.; Lió, P. Transfer learning with graph neural networks for improved molecular property prediction in the multi-fidelity setting. *Nat. Commun.* **2024**, *15* (1), 1517.

(258) He, H.; Garcia, E. A. Learning from imbalanced data. *IEEE Trans. Knowledge Data Eng.* **2009**, *21* (9), 1263-1284.

(259) Landrum, G. A.; Riniker, S. Combining IC50 or K i Values from Different Sources Is a Source of Significant Noise. *Journal of Chemical Information and Modeling* **2024**, *64* (5), 1560-1567.

(260) Hirschfeld, L.; Swanson, K.; Yang, K.; Barzilay, R.; Coley, C. W. Uncertainty quantification using neural networks for molecular property prediction. *J. Chem. Inf. Model.* **2020**, *60* (8), 3770-3780.

(261) Rutemiller, H. C.; Bowers, D. A. Estimation in a heteroscedastic regression model. *J. Am. Stat. Assoc.* **1968**, *63* (322), 552-557.

(262) Gal, Y.; Ghahramani, Z. Dropout as a bayesian approximation: Representing model uncertainty in deep learning. In *international conference on machine learning*, 2016; PMLR: pp 1050-1059.

(263) Lakshminarayanan, B.; Pritzel, A.; Blundell, C. Simple and scalable predictive uncertainty estimation using deep ensembles. *Adv. Neural Inf. Process. Syst* **2017**, *30*.

(264) Mooney, C. Z.; Duval, R. D.; Duvall, R. *Bootstrapping: A nonparametric approach to statistical inference*; sage, 1993.

(265) Welling, M.; Teh, Y. W. Bayesian learning via stochastic gradient Langevin dynamics. In *Proceedings of the 28th international conference on machine learning (ICML-11)*, 2011; Citeseer: pp 681-688.

(266) Blundell, C.; Cornebise, J.; Kavukcuoglu, K.; Wierstra, D. Weight uncertainty in neural network. In *International conference on machine learning*, 2015; PMLR: pp 1613-1622.

(267) Sensoy, M.; Kaplan, L.; Kandemir, M. Evidential deep learning to quantify classification uncertainty. *Adv. Neural Inf. Process. Syst* **2018**, *31*.

(268) Guo, C.; Pleiss, G.; Sun, Y.; Weinberger, K. Q. On calibration of modern neural networks. In *International conference on machine learning*, 2017; PMLR: pp 1321-1330.

(269) Fontana, M.; Zeni, G.; Vantini, S. Conformal prediction: a unified review of theory and new challenges. *Bernoulli* **2023**, *29* (1), 1-23.

(270) Gal, Y.; Hron, J.; Kendall, A. Concrete dropout. *Adv. Neural Inf. Process. Syst* **2017**, *30*.

(271) Duvenaud, D.; Maclaurin, D.; Adams, R. Early stopping as nonparametric variational inference. In *Artificial intelligence and statistics*, 2016; PMLR: pp 1070-1077. Pearce, T.; Zaki, M.; Brintrup, A.; Anastassacos, N.; Neely, A. Uncertainty in neural networks: Bayesian ensembling. *stat* **2018**, *1050*, 12.

(272) Liu, Q.; Wang, D. Stein variational gradient descent: A general purpose bayesian inference algorithm. *Adv. Neural Inf. Process. Syst* **2016**, *29*.

(273) Zhang, Y. Bayesian semi-supervised learning for uncertainty-calibrated prediction of molecular properties and active learning. *Chem. Sci.* **2019**, *10* (35), 8154-8163.

(274) Ryu, S.; Kwon, Y.; Kim, W. Y. A Bayesian graph convolutional network for reliable prediction of molecular properties with uncertainty quantification. *Chem. Sci.* **2019**, *10* (36), 8438-8446.

(275) Scalia, G.; Grambow, C. A.; Pernici, B.; Li, Y.-P.; Green, W. H. Evaluating scalable uncertainty estimation methods for deep learning-based molecular property prediction. *J. Chem. Inf. Model.* **2020**, *60* (6), 2697-2717.

(276) Li, Y.; Kong, L.; Du, Y.; Yu, Y.; Zhuang, Y.; Mu, W.; Zhang, C. Muben: Benchmarking the uncertainty of molecular representation models. **2023**.

(277) Soleimany, A. P.; Amini, A.; Goldman, S.; Rus, D.; Bhatia, S. N.; Coley, C. W. Evidential deep learning for guided molecular property prediction and discovery. *ACS Cent. Sci.* **2021**, *7* (8), 1356-1367.

(278) Busk, J.; Jørgensen, P. B.; Bhowmik, A.; Schmidt, M. N.; Winther, O.; Vegge, T. Calibrated uncertainty for molecular property prediction using ensembles of message passing neural networks. *Mach. Learn.: Sci. Technol.* **2021**, *3* (1), 015012.

(279) Svensson, F.; Norinder, U.; Bender, A. Modelling compound cytotoxicity using conformal prediction and PubChem HTS data. *Toxicol. Res.* **2017**, *6* (1), 73-80. Norinder, U.; Carlsson, L.; Boyer, S.; Eklund, M. Introducing conformal prediction in predictive modeling. A transparent and flexible alternative to applicability domain determination. *J. Chem. Inf. Model.* **2014**, *54* (6), 1596-1603. Alvarsson, J.; McShane, S. A.; Norinder, U.; Spjuth, O. Predicting with confidence: using conformal prediction in drug discovery. *J. Pharm. Sci.* **2021**, *110* (1), 42-49.

(280) Laghuvarapu, S.; Lin, Z.; Sun, J. CoDrug: Conformal Drug Property Prediction with Density Estimation under Covariate Shift. *Adv. Neural Inf. Process. Syst* **2024**, *36*.

(281) Huang, K.; Jin, Y.; Candes, E.; Leskovec, J. Uncertainty quantification over graph with conformalized graph neural networks. *Adv. Neural Inf. Process. Syst* **2024**, *36*.

(282) Ribeiro, M. T.; Singh, S.; Guestrin, C. " Why should i trust you?" Explaining the predictions of any classifier. In *Proceedings of the 22nd ACM SIGKDD international conference on knowledge discovery and data mining*, 2016; pp 1135-1144.

(283) Shrikumar, A.; Greenside, P.; Kundaje, A. Learning important features through propagating activation differences. In *International conference on machine learning*, 2017; PMlR: pp 3145-3153.

(284) Lundberg, S. M.; Lee, S.-I. A unified approach to interpreting model predictions. *Adv. Neural Inf. Process. Syst* **2017**, *30*.

(285) Simonyan, K.; Vedaldi, A.; Zisserman, A. Deep inside convolutional networks: Visualising image classification models and saliency maps. *arXiv preprint arXiv:1312.6034* **2013**.

(286) Zhang, J.; Bargal, S. A.; Lin, Z.; Brandt, J.; Shen, X.; Sclaroff, S. Top-down neural attention by excitation backprop. *Int. J. Comput. Vis.* **2018**, *126* (10), 1084-1102.

(287) Selvaraju, R. R.; Cogswell, M.; Das, A.; Vedantam, R.; Parikh, D.; Batra, D. Grad-cam: Visual explanations from deep networks via gradient-based localization. In *Proceedings of the IEEE international conference on computer vision*, 2017; pp 618-626.

(288) Sundararajan, M.; Taly, A.; Yan, Q. Axiomatic attribution for deep networks. In *International conference on machine learning*, 2017; PMLR: pp 3319-3328.

(289) Ying, Z.; Bourgeois, D.; You, J.; Zitnik, M.; Leskovec, J. Gnnexplainer: Generating explanations for graph neural networks. *Adv. Neural Inf. Process. Syst* **2019**, *32*.

(290) Luo, D.; Cheng, W.; Xu, D.; Yu, W.; Zong, B.; Chen, H.; Zhang, X. Parameterized explainer for graph neural network. *Adv. Neural Inf. Process. Syst* **2020**, *33*, 19620-19631.

(291) Schlichtkrull, M. S.; De Cao, N.; Titov, I. Interpreting graph neural networks for NLP with differentiable edge masking. *arXiv preprint arXiv:2010.00577* **2020**.

(292) Yuan, H.; Yu, H.; Wang, J.; Li, K.; Ji, S. On explainability of graph neural networks via subgraph explorations. In *International conference on machine learning*, 2021; PMLR: pp 12241-12252.

(293) Wu, Z.; Wang, J.; Du, H.; Jiang, D.; Kang, Y.; Li, D.; Pan, P.; Deng, Y.; Cao, D.; Hsieh, C.-Y. Chemistry-intuitive explanation of graph neural networks for molecular property prediction with substructure masking. *Nat. Commun.* 2023, *14* (1), 2585.

(294) Lin, W.; Lan, H.; Li, B. Generative causal explanations for graph neural networks. In *International Conference on Machine Learning*, 2021; PMLR: pp 6666-6679.

(295) Huang, Q.; Yamada, M.; Tian, Y.; Singh, D.; Chang, Y. Graphlime: Local interpretable model explanations for graph neural networks. *IEEE Trans. Knowledge Data Eng.* 2022.

(296) Zhang, Y.; Defazio, D.; Ramesh, A. Relex: A model-agnostic relational model explainer. In *Proceedings of the 2021 AAAI/ACM Conference on AI, Ethics, and Society*, 2021; pp 1042-1049.

(297) Duval, A.; Malliaros, F. D. Graphsvx: Shapley value explanations for graph neural networks. In *Machine Learning and Knowledge Discovery in Databases. Research Track: European Conference, ECML PKDD 2021, Bilbao, Spain, September 13–17, 2021, Proceedings, Part II 21*, 2021; Springer: pp 302-318.

(298) Miao, S.; Liu, M.; Li, P. Interpretable and generalizable graph learning via stochastic attention mechanism. In *International Conference on Machine Learning*, 2022; PMLR: pp 15524-15543.

(299) Zhang, Z.; Liu, Q.; Wang, H.; Lu, C.; Lee, C. Protgnn: Towards self-explaining graph neural networks. In *Proceedings of the AAAI Conference on Artificial Intelligence*, 2022; Vol. 36, pp 9127-9135.

(300) Lucic, A.; Ter Hoeve, M. A.; Tolomei, G.; De Rijke, M.; Silvestri, F. Cf-gnnexplainer: Counterfactual explanations for graph neural networks. In *International Conference on Artificial Intelligence and Statistics*, 2022; PMLR: pp 4499-4511.

(301) Tan, J.; Geng, S.; Fu, Z.; Ge, Y.; Xu, S.; Li, Y.; Zhang, Y. Learning and evaluating graph neural network explanations based on counterfactual and factual reasoning. In *Proceedings of the ACM Web Conference 2022*, 2022; pp 1018-1027.

(302) Bajaj, M.; Chu, L.; Xue, Z. Y.; Pei, J.; Wang, L.; Lam, P. C.-H.; Zhang, Y. Robust counterfactual explanations on graph neural networks. *Adv. Neural Inf. Process. Syst* 2021, *34*, 5644-5655.

(303) Numeroso, D.; Bacciu, D. Meg: Generating molecular counterfactual explanations for deep graph networks. In *2021 International Joint Conference on Neural Networks (IJCNN)*, 2021; IEEE: pp 1-8.

(304) Wellawatte, G. P.; Seshadri, A.; White, A. D. Model agnostic generation of counterfactual explanations for molecules. *Chem. Sci.* 2022, *13* (13), 3697-3705.

(305) Huang, Z.; Kosan, M.; Medya, S.; Ranu, S.; Singh, A. Global counterfactual explainer for graph neural networks. In *Proceedings of the Sixteenth ACM International Conference on Web Search and Data Mining*, 2023; pp 141-149.

(306) Ma, J.; Guo, R.; Mishra, S.; Zhang, A.; Li, J. Clear: Generative counterfactual explanations on graphs. *Adv. Neural Inf. Process. Syst* 2022, *35*, 25895-25907.

(307) Kindermans, P.-J.; Hooker, S.; Adebayo, J.; Alber, M.; Schütt, K. T.; Dähne, S.; Erhan, D.; Kim, B. The (un) reliability of saliency methods. *Explainable AI: Interpreting, explaining and visualizing deep learning* 2019, 267-280.

(308) Sanchez-Lengeling, B.; Wei, J.; Lee, B.; Reif, E.; Wang, P.; Qian, W.; McCloskey, K.; Colwell, L.; Wiltschko, A. Evaluating attribution for graph neural networks. *Adv. Neural Inf. Process. Syst* 2020, *33*, 5898-5910.

(309) Jiménez-Luna, J.; Skalic, M.; Weskamp, N. Benchmarking molecular feature attribution methods with activity cliffs. *J. Chem. Inf. Model.* 2022, *62* (2), 274-283.

(310) Rao, J.; Zheng, S.; Lu, Y.; Yang, Y. Quantitative evaluation of explainable graph neural networks for molecular property prediction. *Patterns* 2022, *3* (12).

(311) Jang, E.; Gu, S.; Poole, B. Categorical reparameterization with gumbel-softmax. *arXiv preprint arXiv:1611.01144* **2016**.

(312) Kuhn, H. W.; Tucker, A. W. *Contributions to the Theory of Games*; Princeton University Press, 1953.

(313) Hu, Y.; Stumpfe, D.; Bajorath, J. r. Computational exploration of molecular scaffolds in medicinal chemistry: Miniperspective. *J. Med. Chem.* **2016**, *59* (9), 4062-4076.

(314) Yuan, H.; Tang, J.; Hu, X.; Ji, S. Xgnn: Towards model-level explanations of graph neural networks. In *Proceedings of the 26th ACM SIGKDD international conference on knowledge discovery & data mining*, 2020; pp 430-438.

(315) Bressler, S. L.; Seth, A. K. Wiener–Granger causality: a well established methodology. *Neuroimage* **2011**, *58* (2), 323-329.

(316) Gao, H.; Ji, S. Graph u-nets. In *international conference on machine learning*, 2019; PMLR: pp 2083-2092.

(317) Lee, J.; Lee, I.; Kang, J. Self-attention graph pooling. In *International conference on machine learning*, 2019; PMLR: pp 3734-3743.

(318) Serrano, S.; Smith, N. A. Is attention interpretable? *arXiv preprint arXiv:1906.03731* **2019**.

(319) Meylan, W. M.; Howard, P. H. Atom/fragment contribution method for estimating octanol–water partition coefficients. *J. Pharm. Sci.* **1995**, *84* (1), 83-92.

(320) Crawford, V.; Kuhnle, A.; Thai, M. Submodular cost submodular cover with an approximate oracle. In *International Conference on Machine Learning*, 2019; PMLR: pp 1426-1435.

(321) Pemantle, R. Vertex-reinforced random walk. *Probab. Theory Rel. Fields* **1992**, *92* (1), 117-136.

(322) Ryu, S.; Lim, J.; Hong, S. H.; Kim, W. Y. Deeply learning molecular structure-property relationships using attention-and gate-augmented graph convolutional network. *arXiv preprint arXiv:1805.10988* **2018**.

(323) Preuer, K.; Klambauer, G.; Rippmann, F.; Hochreiter, S.; Unterthiner, T. Interpretable deep learning in drug discovery. *Explainable AI: interpreting, explaining and visualizing deep learning* **2019**, 331-345.

(324) Hochuli, J.; Helbling, A.; Skaist, T.; Ragoza, M.; Koes, D. R. Visualizing convolutional neural network protein-ligand scoring. *J. Mol. Graph. Model.* **2018**, *84*, 96-108.

(325) Coley, C. W.; Jin, W.; Rogers, L.; Jamison, T. F.; Jaakkola, T. S.; Green, W. H.; Barzilay, R.; Jensen, K. F. A graph-convolutional neural network model for the prediction of chemical reactivity. *Chem. Sci.* **2019**, *10* (2), 370-377. Ishida, S.; Terayama, K.; Kojima, R.; Takasu, K.; Okuno, Y. Prediction and interpretable visualization of retrosynthetic reactions using graph convolutional networks. *J. Chem. Inf. Model.* **2019**, *59* (12), 5026-5033.

(326) Debnath, A. K.; Lopez de Compadre, R. L.; Debnath, G.; Shusterman, A. J.; Hansch, C. Structure-activity relationship of mutagenic aromatic and heteroaromatic nitro compounds. correlation with molecular orbital energies and hydrophobicity. *J. Med. Chem.* **1991**, *34* (2), 786-797.

(327) Zhou, G.; Gao, Z.; Ding, Q.; Zheng, H.; Xu, H.; Wei, Z.; Zhang, L.; Ke, G. Uni-Mol: A Universal 3D Molecular Representation Learning Framework. In *The Eleventh International Conference on Learning Representations*, 2022.

(328) Mayor, M. J. B.; Ivars, A. J. E-Learning for interpreting. *Babel: International Journal of Translation/Revue Internationale de la Traduction* **2007**, *53* (4). Du, M.; Liu, N.; Hu, X. Techniques for interpretable machine learning. *Commun. ACM* **2019**, *63* (1), 68-77.

(329) Zhang, Q. C.; Petrey, D.; Norel, R.; Honig, B. H. Protein interface conservation across structure space. *Proc. Natl. Acad. Sci.* **2010**, *107* (24), 10896-10901. Yang, J.; Roy, A.; Zhang, Y. Protein–ligand binding site recognition using complementary binding-specific substructure comparison and sequence profile alignment. *Bioinformatics* **2013**, *29* (20), 2588-2595.

(330) Yan, J.; Kurgan, L. DRNApred, fast sequence-based method that accurately predicts and discriminates DNA-and RNA-binding residues. *Nucleic Acids Res.* **2017**, *45* (10), e84-e84. Taherzadeh, G.; Yang, Y.; Zhang, T.; Liew, A. W. C.; Zhou, Y. Sequence-based prediction of protein–peptide binding sites using support vector machine. *J. Comput. Chem.* **2016**, *37* (13), 1223-1229.

(331) Yang, J.; Roy, A.; Zhang, Y. BioLiP: a semi-manually curated database for biologically relevant ligand–protein interactions. *Nucleic Acids Res.* **2012**, *41* (D1), D1096-D1103.

(332) Buttenschoen, M.; Morris, G. M.; Deane, C. M. PoseBusters: AI-based docking methods fail to generate physically valid poses or generalise to novel sequences. *Chem. Sci.* **2024**, *15* (9), 3130-3139.

(333) Hu, L.; Benson, M. L.; Smith, R. D.; Lerner, M. G.; Carlson, H. A. Binding MOAD (mother of all databases). *Proteins* **2005**, *60* (3), 333-340.

(334) Bauer, M. R.; Ibrahim, T. M.; Vogel, S. M.; Boeckler, F. M. Evaluation and optimization of virtual screening workflows with DEKOIS 2.0–a public library of challenging docking benchmark sets. *J. Chem. Inf. Model.* **2013**, *53* (6), 1447-1462.

(335) Mysinger, M. M.; Carchia, M.; Irwin, J. J.; Shoichet, B. K. Directory of useful decoys, enhanced (DUD-E): better ligands and decoys for better benchmarking. *J. Med. Chem.* **2012**, *55* (14), 6582-6594.

(336) Tran-Nguyen, V.-K.; Jacquemard, C.; Rognan, D. LIT-PCBA: an unbiased data set for machine learning and virtual screening. *J. Chem. Inf. Model.* **2020**, *60* (9), 4263-4273.

(337) Schindler, C. E.; Baumann, H.; Blum, A.; Böse, D.; Buchstaller, H.-P.; Burgdorf, L.; Cappel, D.; Chekler, E.; Czodrowski, P.; Dorsch, D. Large-scale assessment of binding free energy calculations in active drug discovery projects. *J. Chem. Inf. Model.* **2020**, *60* (11), 5457-5474.

(338) Liu, T.; Lin, Y.; Wen, X.; Jorissen, R. N.; Gilson, M. K. BindingDB: a web-accessible database of experimentally determined protein–ligand binding affinities. *Nucleic Acids Res.* **2007**, *35* (suppl_1), D198-D201.

(339) Shi, W.; Singha, M.; Pu, L.; Srivastava, G.; Ramanujam, J.; Brylinski, M. Graphsite: ligand binding site classification with deep graph learning. *Biomolecules* **2022**, *12* (8), 1053.

(340) Smith, Z.; Strobel, M.; Vani, B. P.; Tiwary, P. Graph attention site prediction (grasp): Identifying druggable binding sites using graph neural networks with attention. *J. Chem. Inf. Model.* **2024**, *64* (7), 2637-2644.

(341) Gainza, P.; Sverrisson, F.; Monti, F.; Rodola, E.; Boscaini, D.; Bronstein, M.; Correia, B. Deciphering interaction fingerprints from protein molecular surfaces using geometric deep learning. *Nat. Methods* **2020**, *17* (2), 184-192.

(342) Zhang, O.; Wang, T.; Weng, G.; Jiang, D.; Wang, N.; Wang, X.; Zhao, H.; Wu, J.; Wang, E.; Chen, G. Learning on topological surface and geometric structure for 3D molecular generation. *Nat. Comput. Sci.* **2023**, 1-11.

(343) Li, P.; Liu, Z.-P. GeoBind: segmentation of nucleic acid binding interface on protein surface with geometric deep learning. *Nucleic Acids Res.* **2023**, *51* (10), e60-e60.

(344) Krapp, L. F.; Abriata, L. A.; Cortés Rodriguez, F.; Dal Peraro, M. PeSTo: parameter-free geometric deep learning for accurate prediction of protein binding interfaces. *Nat. Commun.* **2023**, *14* (1), 2175.

(345) Tubiana, J.; Schneidman-Duhovny, D.; Wolfson, H. J. ScanNet: an interpretable geometric deep learning model for structure-based protein binding site prediction. *Nat. Methods* **2022**, *19* (6), 730-739.

(346) Yuan, Q.; Tian, C.; Yang, Y. Genome-scale annotation of protein binding sites via language model and geometric deep learning. *Elife* **2024**, *13*, RP93695.

(347) Wang, Y.; Xia, Y.; Yan, J.; Yuan, Y.; Shen, H.-B.; Pan, X. ZeroBind: a protein-specific zero-shot predictor with subgraph matching for drug-target interactions. *Nat. Commun.* **2023**, *14* (1), 7861.

(348) Puny, O.; Atzmon, M.; Ben-Hamu, H.; Misra, I.; Grover, A.; Smith, E. J.; Lipman, Y. Frame averaging for invariant and equivariant network design. *arXiv preprint arXiv:2110.03336* **2021**.

(349) Elnaggar, A.; Heinzinger, M.; Dallago, C.; Rehawi, G.; Wang, Y.; Jones, L.; Gibbs, T.; Feher, T.; Angerer, C.; Steinegger, M. Prottrans: Toward understanding the language of life through self-supervised learning. *IEEE Trans. Pattern Anal. Mach. Intell.* **2021**, *44* (10), 7112-7127.

(350) Lin, Z.; Akin, H.; Rao, R.; Hie, B.; Zhu, Z.; Lu, W.; dos Santos Costa, A.; Fazel-Zarandi, M.; Sercu, T.; Candido, S. Language models of protein sequences at the scale of evolution enable accurate structure prediction. *BioRxiv* **2022**, *2022*, 500902.

(351) Morris, G. M.; Goodsell, D. S.; Huey, R.; Hart, W. E.; Halliday, S.; Belew, R.; Olson, A. J. AutoDock. *Automated docking of flexible ligands to receptor-User Guide* **2001**.

(352) Liu, N.; Xu, Z. Using LeDock as a docking tool for computational drug design. In *IOP Conference Series: Earth and Environmental Science*, 2019; IOP Publishing: Vol. 218, p 012143.

(353) Méndez-Lucio, O.; Ahmad, M.; del Rio-Chanona, E. A.; Wegner, J. K. A geometric deep learning approach to predict binding conformations of bioactive molecules. *Nat. Mach. Intell.* **2021**, *3* (12), 1033-1039.

(354) Masters, M.; Mahmoud, A. H.; Wei, Y.; Lill, M. A. Deep learning model for flexible and efficient protein-ligand docking. In *ICLR2022 Machine Learning for Drug Discovery*, 2022.

(355) Lu, W.; Wu, Q.; Zhang, J.; Rao, J.; Li, C.; Zheng, S. Tankbind: Trigonometry-aware neural networks for drug-protein binding structure prediction. *Adv. Neural Inf. Process. Syst* **2022**, *35*, 7236-7249.

(356) Cai, H.; Shen, C.; Jian, T.; Zhang, X.; Chen, T.; Han, X.; Yang, Z.; Dang, W.; Hsieh, C.-Y.; Kang, Y. CarsiDock: a deep learning paradigm for accurate protein–ligand docking and screening based on large-scale pre-training. *Chem. Sci.* **2024**, *15* (4), 1449-1471.

(357) Stärk, H.; Ganea, O.; Pattanaik, L.; Barzilay, R.; Jaakkola, T. Equibind: Geometric deep learning for drug binding structure prediction. In *International Conference on Machine Learning*, 2022; PMLR: pp 20503-20521.

(358) Zhang, J.; He, K.; Dong, T. Accurate Protein-Ligand Complex Structure Prediction using Geometric Deep Learning. **2022**.

(359) Corso, G.; Jing, B.; Barzilay, R.; Jaakkola, T. DiffDock: Diffusion Steps, Twists, and Turns for Molecular Docking. In *International Conference on Learning Representations (ICLR 2023)*, 2023.

(360) Liu, L.; He, D.; Ye, X.; Zhang, S.; Zhang, X.; Zhou, J.; Li, J.; Chai, H.; Wang, F.; He, J. Pre-Training on Large-Scale Generated Docking Conformations with HelixDock to Unlock the Potential of Protein-ligand Structure Prediction Models. *arXiv preprint arXiv:2310.13913* **2023**.

(361) Pei, Q.; Gao, K.; Wu, L.; Zhu, J.; Xia, Y.; Xie, S.; Qin, T.; He, K.; Liu, T.-Y.; Yan, R. Fabind: Fast and accurate protein-ligand binding. *Adv. Neural Inf. Process. Syst* **2024**, *36*.

(362) Gao, K.; Pei, Q.; Zhu, J.; Qin, T.; He, K.; Liu, T.-Y.; Wu, L. FABind+: Enhancing Molecular Docking through Improved Pocket Prediction and Pose Generation. *arXiv preprint arXiv:2403.20261* **2024**.

(363) Krishna, R.; Wang, J.; Ahern, W.; Sturmfels, P.; Venkatesh, P.; Kalvet, I.; Lee, G. R.; Morey-Burrows, F. S.; Anishchenko, I.; Humphreys, I. R. Generalized biomolecular modeling and design with RoseTTAFold All-Atom. *Science* **2024**, eadl2528.

(364) AlQuraishi, M. AlphaFold at CASP13. *Bioinformatics* **2019**, *35* (22), 4862-4865.

(365) Bryant, P.; Kelkar, A.; Guljas, A.; Clementi, C.; Noé, F. Structure prediction of protein-ligand complexes from sequence information with Umol. *Nat. Commun.* **2024**, *15* (1), 4536.

(366) Qiao, Z.; Nie, W.; Vahdat, A.; Miller III, T. F.; Anandkumar, A. State-specific protein–ligand complex structure prediction with a multiscale deep generative model. *Nat. Mach. Intell.* **2024**, *6* (2), 195-208.

(367) Dong, T.; Yang, Z.; Zhou, J.; Chen, C. Y.-C. Equivariant flexible modeling of the protein–ligand binding pose with geometric deep learning. *J. Chem. Theory Comput.* **2023**, *19* (22), 8446-8459.

(368) Zhu, J.; Gu, Z.; Pei, J.; Lai, L. DiffBindFR: an SE (3) equivariant network for flexible protein–ligand docking. *Chem. Sci.* **2024**, *15* (21), 7926-7942.

(369) Lu, W.; Zhang, J.; Huang, W.; Zhang, Z.; Jia, X.; Wang, Z.; Shi, L.; Li, C.; Wolynes, P. G.; Zheng, S. DynamicBind: predicting ligand-specific protein-ligand complex structure with a deep equivariant generative model. *Nat. Commun.* **2024**, *15* (1), 1071.

(370) Huang, Y.; Zhang, O.; Wu, L.; Tan, C.; Lin, H.; Gao, Z.; Li, S.; Li, S. Re-Dock: Towards Flexible and Realistic Molecular Docking with Diffusion Bridge. *arXiv preprint arXiv:2402.11459* **2024**.

(371) Burley, S. K.; Berman, H. M.; Kleywegt, G. J.; Markley, J. L.; Nakamura, H.; Velankar, S. Protein Data Bank (PDB): the single global macromolecular structure archive. *Protein Crystallogr.* **2017**, 627-641.

(372) Krivák, R.; Hoksza, D. P2Rank: machine learning based tool for rapid and accurate prediction of ligand binding sites from protein structure. *J. Cheminformatics* **2018**, *10*, 1-12.

(373) Alcaide, E.; Gao, Z.; Ke, G.; Li, Y.; Zhang, L.; Zheng, H.; Zhou, G. Uni-Mol Docking V2: Towards Realistic and Accurate Binding Pose Prediction. *arXiv preprint arXiv:2405.11769* **2024**.

(374) Abramson, J.; Adler, J.; Dunger, J.; Evans, R.; Green, T.; Pritzel, A.; Ronneberger, O.; Willmore, L.; Ballard, A. J.; Bambrick, J. Accurate structure prediction of biomolecular interactions with AlphaFold 3. *Nature* **2024**, 1-3.

(375) Evans, R.; O'Neill, M.; Pritzel, A.; Antropova, N.; Senior, A.; Green, T.; Žídek, A.; Bates, R.; Blackwell, S.; Yim, J. Protein complex prediction with AlphaFold-Multimer. *biorxiv* **2021**, 2021.2010.2004.463034.

(376) Aggarwal, R.; Gupta, A.; Priyakumar, U. Apobind: a dataset of ligand unbound protein conformations for machine learning applications in de novo drug design. *arXiv preprint arXiv:2108.09926* **2021**.

(377) Nguyen, T.; Le, H.; Quinn, T. P.; Nguyen, T.; Le, T. D.; Venkatesh, S. GraphDTA: predicting drug–target binding affinity with graph neural networks. *Bioinformatics* **2021**, *37* (8), 1140-1147.

(378) Yang, Z.; Zhong, W.; Zhao, L.; Chen, C. Y.-C. MGraphDTA: deep multiscale graph neural network for explainable drug–target binding affinity prediction. *Chem. Sci.* **2022**, *13* (3), 816-833.

(379) Feinberg, E. N.; Sur, D.; Wu, Z.; Husic, B. E.; Mai, H.; Li, Y.; Sun, S.; Yang, J.; Ramsundar, B.; Pande, V. S. PotentialNet for molecular property prediction. *ACS Cent. Sci.* **2018**, *4* (11), 1520-1530.

(380) Lim, J.; Ryu, S.; Park, K.; Choe, Y. J.; Ham, J.; Kim, W. Y. Predicting drug–target interaction using a novel graph neural network with 3D structure-embedded graph representation. *J. Chem. Inf. Model.* **2019**, *59* (9), 3981-3988.

(381) Jiang, D.; Hsieh, C.-Y.; Wu, Z.; Kang, Y.; Wang, J.; Wang, E.; Liao, B.; Shen, C.; Xu, L.; Wu, J. Interactiongraphnet: A novel and efficient deep graph representation learning framework for accurate protein–ligand interaction predictions. *J. Med. Chem.* **2021**, *64* (24), 18209-18232.

(382) Li, S.; Zhou, J.; Xu, T.; Huang, L.; Wang, F.; Xiong, H.; Huang, W.; Dou, D.; Xiong, H. Structure-aware interactive graph neural networks for the prediction of protein-ligand binding affinity. In *Proceedings of the 27th ACM SIGKDD conference on knowledge discovery & data mining*, 2021; pp 975-985.

(383) Wu, J.; Chen, H.; Cheng, M.; Xiong, H. CurvAGN: Curvature-based Adaptive Graph Neural Networks for Predicting Protein-Ligand Binding Affinity. *BMC Bioinf.* 2023, 24 (1), 378.

(384) Yang, Z.; Zhong, W.; Lv, Q.; Dong, T.; Yu-Chian Chen, C. Geometric interaction graph neural network for predicting protein–ligand binding affinities from 3d structures (gign). *J. Phys. Chem. Lett.* 2023, 14 (8), 2020-2033.

(385) Yang, Z.; Zhong, W.; Lv, Q.; Dong, T.; Chen, G.; Chen, C. Y.-C. Interaction-Based Inductive Bias in Graph Neural Networks: Enhancing Protein-Ligand Binding Affinity Predictions From 3D Structures. *IEEE Trans. Pattern Anal. Mach. Intell.* 2024.

(386) Moon, S.; Zhung, W.; Yang, S.; Lim, J.; Kim, W. Y. PIGNet: a physics-informed deep learning model toward generalized drug–target interaction predictions. *Chem. Sci.* 2022, 13 (13), 3661-3673.

(387) Guo, J. Improving structure-based protein-ligand affinity prediction by graph representation learning and ensemble learning. *PLoS One* 2024, 19 (1), e0296676.

(388) Jones, D.; Kim, H.; Zhang, X.; Zemla, A.; Stevenson, G.; Bennett, W. D.; Kirshner, D.; Wong, S. E.; Lightstone, F. C.; Allen, J. E. Improved protein–ligand binding affinity prediction with structure-based deep fusion inference. *J. Chem. Inf. Model.* 2021, 61 (4), 1583-1592.

(389) Kyro, G. W.; Brent, R. I.; Batista, V. S. Hac-net: A hybrid attention-based convolutional neural network for highly accurate protein–ligand binding affinity prediction. *J. Chem. Inf. Model.* 2023, 63 (7), 1947-1960.

(390) Mqawass, G.; Popov, P. GraphLambda: fusion graph neural networks for binding affinity prediction. *J. Chem. Inf. Model.* 2024, 64 (7), 2323-2330.

(391) Dong, L.; Shi, S.; Qu, X.; Luo, D.; Wang, B. Ligand binding affinity prediction with fusion of graph neural networks and 3D structure-based complex graph. *Phys. Chem. Chem. Phys.* 2023, 25 (35), 24110-24120.

(392) Yu, J.; Li, Z.; Chen, G.; Kong, X.; Hu, J.; Wang, D.; Cao, D.; Li, Y.; Huo, R.; Wang, G. Computing the relative binding affinity of ligands based on a pairwise binding comparison network. *Nat. Comput. Sci.* 2023, 3 (10), 860-872.

(393) Shen, C.; Zhang, X.; Deng, Y.; Gao, J.; Wang, D.; Xu, L.; Pan, P.; Hou, T.; Kang, Y. Boosting Protein–Ligand Binding Pose Prediction and Virtual Screening Based on Residue–Atom Distance Likelihood Potential and Graph Transformer. *J. Med. Chem.* 2022, 65 (15), 10691-10706.

(394) Moon, S.; Hwang, S.-Y.; Lim, J.; Kim, W. Y. PIGNet2: a versatile deep learning-based protein–ligand interaction prediction model for binding affinity scoring and virtual screening. *Digit. Discov.* 2024, 3 (2), 287-299.

(395) Shen, C.; Zhang, X.; Hsieh, C.-Y.; Deng, Y.; Wang, D.; Xu, L.; Wu, J.; Li, D.; Kang, Y.; Hou, T. A generalized protein–ligand scoring framework with balanced scoring, docking, ranking and screening powers. *Chem. Sci.* 2023, 14 (30), 8129-8146.

(396) Cao, D.; Chen, G.; Jiang, J.; Yu, J.; Zhang, R.; Chen, M.; Zhang, W.; Chen, L.; Zhong, F.; Zhang, Y. Generic protein–ligand interaction scoring by integrating physical prior knowledge and data augmentation modelling. *Nat. Mach. Intell.* 2024, 1-13.

(397) Luo, D.; Liu, D.; Qu, X.; Dong, L.; Wang, B. Enhancing Generalizability in Protein–Ligand Binding Affinity Prediction with Multimodal Contrastive Learning. *J. Chem. Inf. Model.* 2024, 64 (6), 1892-1906.

(398) Wang, Z.; Wang, S.; Li, Y.; Guo, J.; Wei, Y.; Mu, Y.; Zheng, L.; Li, W. A new paradigm for applying deep learning to protein–ligand interaction prediction. *Brief. Bioinform.* **2024**, *25* (3), bbae145.

(399) Scantlebury, J.; Vost, L.; Carbery, A.; Hadfield, T. E.; Turnbull, O. M.; Brown, N.; Chenthamarakshan, V.; Das, P.; Grosjean, H.; Von Delft, F. A small step toward generalizability: training a machine learning scoring function for structure-based virtual screening. *J. Chem. Inf. Model.* **2023**, *63* (10), 2960-2974.

(400) Zhang, X.; Gao, H.; Wang, H.; Chen, Z.; Zhang, Z.; Chen, X.; Li, Y.; Qi, Y.; Wang, R. Planet: a multi-objective graph neural network model for protein–ligand binding affinity prediction. *J. Chem. Inf. Model.* **2023**, *64* (7), 2205-2220.

(401) Liu, J.; Wang, R. Classification of current scoring functions. *J. Chem. Inf. Model.* **2015**, *55* (3), 475-482.

(402) Li, Y.; Han, L.; Liu, Z.; Wang, R. Comparative assessment of scoring functions on an updated benchmark: 2. Evaluation methods and general results. *J. Chem. Inf. Model.* **2014**, *54* (6), 1717-1736.

(403) Wang, L.; Chambers, J.; Abel, R. Protein-ligand binding free energy calculations with FEP+. *Methods Mol. Biol.* **2019**, *2022*, 201-232.

(404) Ballester, P. J.; Mitchell, J. B. A machine learning approach to predicting protein–ligand binding affinity with applications to molecular docking. *Bioinformatics* **2010**, *26* (9), 1169-1175.

(405) Jusoh, S.; Almajali, S. A systematic review on fusion techniques and approaches used in applications. *IEEE Access* **2020**, *8*, 14424-14439.

(406) Jiménez-Luna, J.; Pérez-Benito, L.; Martinez-Rosell, G.; Sciabola, S.; Torella, R.; Tresadern, G.; De Fabritiis, G. DeltaDelta neural networks for lead optimization of small molecule potency. *Chem. Sci.* **2019**, *10* (47), 10911-10918.

(407) Wang, Z.; Zheng, L.; Wang, S.; Lin, M.; Wang, Z.; Kong, A. W.-K.; Mu, Y.; Wei, Y.; Li, W. A fully differentiable ligand pose optimization framework guided by deep learning and a traditional scoring function. *Brief. Bioinform.* **2023**, *24* (1), bbac520.

(408) Wu, J.; Leng, D.; Pan, L. ParaVS: A Simple, Fast, Efficient and Flexible Graph Neural Network Framework for Structure-Based Virtual Screening. *arXiv preprint arXiv:2102.06086* **2021**.

(409) Koes, D. R.; Baumgartner, M. P.; Camacho, C. J. Lessons learned in empirical scoring with smina from the CSAR 2011 benchmarking exercise. *J. Chem. Inf. Model.* **2013**, *53* (8), 1893-1904.

(410) Imrie, F.; Bradley, A. R.; Deane, C. M. Generating property-matched decoy molecules using deep learning. *Bioinformatics* **2021**, *37* (15), 2134-2141.

(411) Sastry, G. M.; Dixon, S. L.; Sherman, W. Rapid shape-based ligand alignment and virtual screening method based on atom/feature-pair similarities and volume overlap scoring. *J. Chem. Inf. Model.* **2011**, *51* (10), 2455-2466.

(412) Sieg, J.; Flachsenberg, F.; Rarey, M. In need of bias control: evaluating chemical data for machine learning in structure-based virtual screening. *J. Chem. Inf. Model.* **2019**, *59* (3), 947-961.

(413) Volkov, M.; Turk, J.-A.; Drizard, N.; Martin, N.; Hoffmann, B.; Gaston-Mathé, Y.; Rognan, D. On the frustration to predict binding affinities from protein–ligand structures with deep neural networks. *J. Med. Chem.* **2022**, *65* (11), 7946-7958.

(414) Mastropietro, A.; Pasculli, G.; Bajorath, J. Learning characteristics of graph neural networks predicting protein–ligand affinities. *Nat. Mach. Intell.* **2023**, *5* (12), 1427-1436.

(415) Liu, Q.; Allamanis, M.; Brockschmidt, M.; Gaunt, A. Constrained graph variational autoencoders for molecule design. *Adv. Neural Inf. Process. Syst* **2018**, *31*.

(416) Liu, J.; Kumar, A.; Ba, J.; Kiros, J.; Swersky, K. Graph normalizing flows. *Adv. Neural Inf. Process. Syst* **2019**, *32*.

(417) Madhawa, K.; Ishiguro, K.; Nakago, K.; Abe, M. Graphnvp: An invertible flow model for generating molecular graphs. *arXiv preprint arXiv:1905.11600* **2019**.

(418) Simonovsky, M.; Komodakis, N. Graphvae: Towards generation of small graphs using variational autoencoders. In *International conference on artificial neural networks*, 2018; Springer: pp 412-422.

(419) De Cao, N.; Kipf, T. MolGAN: An implicit generative model for small molecular graphs. *arXiv preprint arXiv:1805.11973* **2018**.

(420) Jo, J.; Lee, S.; Hwang, S. J. Score-based generative modeling of graphs via the system of stochastic differential equations. In *International conference on machine learning*, 2022; PMLR: pp 10362-10383.

(421) Vignac, C.; Krawczuk, I.; Siraudin, A.; Wang, B.; Cevher, V.; Frossard, P. Digress: Discrete denoising diffusion for graph generation. *arXiv preprint arXiv:2209.14734* **2022**.

(422) Popova, M.; Shvets, M.; Oliva, J.; Isayev, O. MolecularRNN: Generating realistic molecular graphs with optimized properties. *arXiv preprint arXiv:1905.13372* **2019**.

(423) Shi, C.; Xu, M.; Zhu, Z.; Zhang, W.; Zhang, M.; Tang, J. Graphaf: a flow-based autoregressive model for molecular graph generation. *arXiv preprint arXiv:2001.09382* **2020**.

(424) Lippe, P.; Gavves, E. Categorical normalizing flows via continuous transformations. *arXiv preprint arXiv:2006.09790* **2020**.

(425) Luo, Y.; Yan, K.; Ji, S. Graphdf: A discrete flow model for molecular graph generation. In *International conference on machine learning*, 2021; PMLR: pp 7192-7203.

(426) Kwon, Y.; Yoo, J.; Choi, Y.; Son, W.; Lee, D.; Kang, S. Efficient learning of non-autoregressive graph variational autoencoders for molecular graph generation. J Cheminf 11 (1): 1–10. 2019.

(427) Kwon, Y.; Lee, D.; Choi, Y.-S.; Shin, K.; Kang, S. Compressed graph representation for scalable molecular graph generation. *J. Cheminformatics* **2020**, *12*, 1-8.

(428) Arjovsky, M.; Chintala, S.; Bottou, L. Wasserstein generative adversarial networks. In *International conference on machine learning*, 2017; PMLR: pp 214-223.

(429) Li, Z. *MolGAN without Mode Collapse*. 2020. https://github.com/ZiyaoLi/molgan-without-mode-collapse (accessed 2020.

(430) Jin, W.; Barzilay, R.; Jaakkola, T. Hierarchical generation of molecular graphs using structural motifs. In *International conference on machine learning*, 2020; PMLR: pp 4839-4848.

(431) Xie, Y.; Shi, C.; Zhou, H.; Yang, Y.; Zhang, W.; Yu, Y.; Li, L. Mars: Markov molecular sampling for multi-objective drug discovery. *arXiv preprint arXiv:2103.10432* **2021**.

(432) Bradshaw, J.; Paige, B.; Kusner, M. J.; Segler, M.; Hernández-Lobato, J. M. A model to search for synthesizable molecules. *Adv. Neural Inf. Process. Syst* **2019**, *32*.

(433) Kong, X.; Huang, W.; Tan, Z.; Liu, Y. Molecule generation by principal subgraph mining and assembling. *Adv. Neural Inf. Process. Syst* **2022**, *35*, 2550-2563.

(434) Geng, Z.; Xie, S.; Xia, Y.; Wu, L.; Qin, T.; Wang, J.; Zhang, Y.; Wu, F.; Liu, T.-Y. De novo molecular generation via connection-aware motif mining. *arXiv preprint arXiv:2302.01129* **2023**.

(435) Simm, G. N.; Hernández-Lobato, J. M. A generative model for molecular distance geometry. *arXiv preprint arXiv:1909.11459* **2019**.

(436) Xu, M.; Luo, S.; Bengio, Y.; Peng, J.; Tang, J. Learning neural generative dynamics for molecular conformation generation. *arXiv preprint arXiv:2102.10240* **2021**.

(437) Xu, M.; Wang, W.; Luo, S.; Shi, C.; Bengio, Y.; Gomez-Bombarelli, R.; Tang, J. An end-to-end framework for molecular conformation generation via bilevel programming. In *International Conference on Machine Learning*, 2021; PMLR: pp 11537-11547.

(438) Zhang, H.; Li, S.; Zhang, J.; Wang, Z.; Wang, J.; Jiang, D.; Bian, Z.; Zhang, Y.; Deng, Y.; Song, J. SDEGen: learning to evolve molecular conformations from thermodynamic noise for conformation generation. *Chem. Sci.* **2023**, *14* (6), 1557-1568.

(439) Zhu, J.; Xia, Y.; Liu, C.; Wu, L.; Xie, S.; Wang, T.; Wang, Y.; Zhou, W.; Qin, T.; Li, H. Direct molecular conformation generation. *arXiv preprint arXiv:2202.01356* **2022**.

(440) Xu, M.; Yu, L.; Song, Y.; Shi, C.; Ermon, S.; Tang, J. Geodiff: A geometric diffusion model for molecular conformation generation. *arXiv preprint arXiv:2203.02923* **2022**.

(441) Jing, B.; Corso, G.; Chang, J.; Barzilay, R.; Jaakkola, T. Torsional Diffusion for Molecular Conformer Generation. *arXiv preprint arXiv:2206.01729* **2022**.

(442) Riniker, S.; Landrum, G. A. Better informed distance geometry: using what we know to improve conformation generation. *J. Chem. Inf. Model.* **2015**, *55* (12), 2562-2574.

(443) Shi, C.; Luo, S.; Xu, M.; Tang, J. Learning Gradient Fields for Molecular Conformation Generation. In Proceedings of the 38th International Conference on Machine Learning, Proceedings of Machine Learning Research; 2021.

(444) Garcia Satorras, V.; Hoogeboom, E.; Fuchs, F.; Posner, I.; Welling, M. E (n) equivariant normalizing flows. *Adv. Neural Inf. Process. Syst* **2021**, *34*, 4181-4192.

(445) Hoogeboom, E.; Satorras, V. G.; Vignac, C.; Welling, M. Equivariant diffusion for molecule generation in 3d. In *International Conference on Machine Learning*, 2022; PMLR: pp 8867-8887.

(446) Zhang, H.; Liu, Y.; Liu, X.; Wang, C.; Guo, M. Equivariant score-based generative diffusion framework for 3D molecules. *BMC Bioinf.* **2024**, *25* (1), 203.

(447) Huang, H.; Sun, L.; Du, B.; Lv, W. Learning joint 2-d and 3-d graph diffusion models for complete molecule generation. *IEEE Trans. Neural Networks Learn. Syst.* **2024**.

(448) Huang, L.; Zhang, H.; Xu, T.; Wong, K.-C. Mdm: Molecular diffusion model for 3d molecule generation. In *Proceedings of the AAAI Conference on Artificial Intelligence*, 2023; Vol. 37, pp 5105-5112.

(449) Dunn, I.; Koes, D. R. Mixed Continuous and Categorical Flow Matching for 3D De Novo Molecule Generation. *ArXiv* **2024**.

(450) Gebauer, N.; Gastegger, M.; Schütt, K. Symmetry-adapted generation of 3d point sets for the targeted discovery of molecules. *Adv. Neural Inf. Process. Syst* **2019**, *32*.

(451) Gebauer, N. W.; Gastegger, M.; Hessmann, S. S.; Müller, K.-R.; Schütt, K. T. Inverse design of 3d molecular structures with conditional generative neural networks. *Nat. Commun.* **2022**, *13* (1), 973.

(452) O'Boyle, N. M.; Banck, M.; James, C. A.; Morley, C.; Vandermeersch, T.; Hutchison, G. R. Open Babel: An open chemical toolbox. *J. Cheminformatics* **2011**, *3*, 1-14.

(453) Huang, X.; Belongie, S. Arbitrary style transfer in real-time with adaptive instance normalization. In *Proceedings of the IEEE international conference on computer vision*, 2017; pp 1501-1510.

(454) Lu, H.; Diaz, D. J.; Czarnecki, N. J.; Zhu, C.; Kim, W.; Shroff, R.; Acosta, D. J.; Alexander, B. R.; Cole, H. O.; Zhang, Y. Machine learning-aided engineering of hydrolases for PET depolymerization. *Nature* **2022**, *604* (7907), 662-667.

(455) Liu, M.; Luo, Y.; Uchino, K.; Maruhashi, K.; Ji, S. Generating 3D Molecules for Target Protein Binding. In Proceedings of the 39th International Conference on Machine Learning, Proceedings of Machine Learning Research; 2022.

(456) Luo, S.; Guan, J.; Ma, J.; Peng, J. A 3D generative model for structure-based drug design. *Adv. Neural Inf. Process. Syst* **2021**, *34*, 6229-6239.

(457) Peng, X.; Luo, S.; Guan, J.; Xie, Q.; Peng, J.; Ma, J. Pocket2mol: Efficient molecular sampling based on 3d protein pockets. In *International Conference on Machine Learning*, 2022; PMLR: pp 17644-17655.

(458) Zhang, O.; Zhang, J.; Jin, J.; Zhang, X.; Hu, R.; Shen, C.; Cao, H.; Du, H.; Kang, Y.; Deng, Y. ResGen is a pocket-aware 3D molecular generation model based on parallel multiscale modelling. *Nat. Mach. Intell.* **2023**, *5* (9), 1020-1030.

(459) Zhung, W.; Kim, H.; Kim, W. Y. 3D molecular generative framework for interaction-guided drug design. *Nat. Commun.* **2024**, *15* (1), 2688.

(460) Jiang, Y.; Zhang, G.; You, J.; Zhang, H.; Yao, R.; Xie, H.; Zhang, L.; Xia, Z.; Dai, M.; Wu, Y. Pocketflow is a data-and-knowledge-driven structure-based molecular generative model. *Nat. Mach. Intell.* **2024**, *6* (3), 326-337.

(461) Zhang, Z.; Min, Y.; Zheng, S.; Liu, Q. Molecule generation for target protein binding with structural motifs. In *The Eleventh International Conference on Learning Representations*, 2022.

(462) Zhang, O.; Huang, Y.; Cheng, S.; Yu, M.; Zhang, X.; Lin, H.; Zeng, Y.; Wang, M.; Wu, Z.; Zhao, H. FragGen: towards 3D geometry reliable fragment-based molecular generation. *Chem. Sci.* **2024**.

(463) Schneuing, A.; Du, Y.; Harris, C.; Jamasb, A.; Igashov, I.; Du, W.; Blundell, T.; Lió, P.; Gomes, C.; Welling, M. Structure-based drug design with equivariant diffusion models. In *arXiv preprint arXiv:2210.13695*, 2022.

(464) Lin, H.; Huang, Y.; Liu, M.; Li, X.; Ji, S.; Li, S. Z. DiffBP: Generative Diffusion of 3D Molecules for Target Protein Binding. *arXiv preprint arXiv:2211.11214* **2022**.

(465) Guan, J.; Qian, W. W.; Peng, X.; Su, Y.; Peng, J.; Ma, J. 3d equivariant diffusion for target-aware molecule generation and affinity prediction. *arXiv preprint arXiv:2303.03543* **2023**.

(466) Choi, S.; Seo, S.; Kim, B. J.; Park, C.; Park, S. PIDiff: Physics informed diffusion model for protein pocket-specific 3D molecular generation. *Comput. Biol. Med.* **2024**, *180*, 108865.

(467) Qian, H.; Huang, W.; Tu, S.; Xu, L. KGDiff: towards explainable target-aware molecule generation with knowledge guidance. *Brief. Bioinform.* **2024**, *25* (1), bbad435.

(468) Guan, J.; Zhou, X.; Yang, Y.; Bao, Y.; Peng, J.; Ma, J.; Liu, Q.; Wang, L.; Gu, Q. DecompDiff: diffusion models with decomposed priors for structure-based drug design. *arXiv preprint arXiv:2403.07902* **2024**.

(469) Pracht, P.; Bohle, F.; Grimme, S. Automated exploration of the low-energy chemical space with fast quantum chemical methods. *Phys. Chem. Chem. Phys.* **2020**, *22* (14), 7169-7192.

(470) Olanders, G.; Alogheli, H.; Brandt, P.; Karlén, A. Conformational analysis of macrocycles: comparing general and specialized methods. *J. Comput. Aided Mol. Des.* **2020**, *34*, 231-252.

(471) Francoeur, P. G.; Masuda, T.; Sunseri, J.; Jia, A.; Iovanisci, R. B.; Snyder, I.; Koes, D. R. Three-dimensional convolutional neural networks and a cross-docked data set for structure-based drug design. *J. Chem. Inf. Model.* **2020**, *60* (9), 4200-4215.

(472) Gürsoy, O.; Smieško, M. Searching for bioactive conformations of drug-like ligands with current force fields: how good are we? *J. Cheminformatics* **2017**, *9*, 1-13.

(473) Rogers, D.; Hahn, M. Extended-connectivity fingerprints. *J. Chem. Inf. Model.* **2010**, *50* (5), 742-754.

(474) Preuer, K.; Renz, P.; Unterthiner, T.; Hochreiter, S.; Klambauer, G. Fréchet ChemNet distance: a metric for generative models for molecules in drug discovery. *J. Chem. Inf. Model.* **2018**, *58* (9), 1736-1741.

(475) Ertl, P.; Schuffenhauer, A. Estimation of synthetic accessibility score of drug-like molecules based on molecular complexity and fragment contributions. *J. Cheminformatics* **2009**, *1* (1), 1-11.

(476) Ganesan, A. The impact of natural products upon modern drug discovery. *Curr. Opin. Chem.Biol.* **2008**, *12* (3), 306-317.

(477) Wu, K.; Zhao, Z.; Wang, R.; Wei, G. W. TopP–S: Persistent homology-based multi-task deep neural networks for simultaneous predictions of partition coefficient and aqueous solubility. *J. Comput. Chem.* **2018**, *39* (20), 1444-1454.

(478) Trott, O.; Olson, A. J. AutoDock Vina: improving the speed and accuracy of docking with a new scoring function, efficient optimization, and multithreading. *J. Comput. Chem.* **2010**, *31* (2), 455-461.

(479) Genheden, S.; Ryde, U. The MM/PBSA and MM/GBSA methods to estimate ligand-binding affinities. *Expert Opin. Drug Discov.* **2015**, *10* (5), 449-461.

(480) Lin, H.; Zhao, G.; Zhang, O.; Huang, Y.; Wu, L.; Liu, Z.; Li, S.; Tan, C.; Gao, Z.; Li, S. Z. CBGBench: Fill in the Blank of Protein-Molecule Complex Binding Graph. *arXiv preprint arXiv:2406.10840* **2024**.

(481) Schneider, P.; Walters, W. P.; Plowright, A. T.; Sieroka, N.; Listgarten, J.; Goodnow Jr, R. A.; Fisher, J.; Jansen, J. M.; Duca, J. S.; Rush, T. S. Rethinking drug design in the artificial intelligence era. *Nat. Rev. Drug Discov.* **2020**, *19* (5), 353-364.

(482) You, J.; Liu, B.; Ying, Z.; Pande, V.; Leskovec, J. Graph convolutional policy network for goal-directed molecular graph generation. *Adv. Neural Inf. Process. Syst* **2018**, *31*.

(483) Yang, S.; Hwang, D.; Lee, S.; Ryu, S.; Hwang, S. J. Hit and lead discovery with explorative rl and fragment-based molecule generation. *Adv. Neural Inf. Process. Syst* **2021**, *34*, 7924-7936.

(484) Jin, W.; Barzilay, R.; Jaakkola, T. Multi-objective molecule generation using interpretable substructures. In *International conference on machine learning*, 2020; PMLR: pp 4849-4859.

(485) Zhou, Z.; Kearnes, S.; Li, L.; Zare, R. N.; Riley, P. Optimization of molecules via deep reinforcement learning. *Sci. Rep.* **2019**, *9* (1), 10752.

(486) Jeon, W.; Kim, D. Autonomous molecule generation using reinforcement learning and docking to develop potential novel inhibitors. *Sci. Rep.* **2020**, *10* (1), 22104.

(487) Ståhl, N.; Falkman, G.; Karlsson, A.; Mathiason, G.; Bostrom, J. Deep reinforcement learning for multiparameter optimization in de novo drug design. *J. Chem. Inf. Model.* **2019**, *59* (7), 3166-3176.

(488) Bolcato, G.; Heid, E.; Boström, J. On the value of using 3D shape and electrostatic similarities in deep generative methods. *J. Chem. Inf. Model.* **2022**, *62* (6), 1388-1398.

(489) Liu, X.; Ye, K.; van Vlijmen, H. W.; IJzerman, A. P.; van Westen, G. J. DrugEx v3: scaffold-constrained drug design with graph transformer-based reinforcement learning. *J. Cheminformatics* **2023**, *15* (1), 24.

(490) Chen, Z.; Min, M. R.; Parthasarathy, S.; Ning, X. A deep generative model for molecule optimization via one fragment modification. *Nat. Mach. Intell.* **2021**, *3* (12), 1040-1049.

(491) Mukaidaisi, M.; Vu, A.; Grantham, K.; Tchagang, A.; Li, Y. Multi-objective drug design based on graph-fragment molecular representation and deep evolutionary learning. *Front. Pharmacol.* **2022**, *13*, 920747.

(492) Sun, M.; Xing, J.; Meng, H.; Wang, H.; Chen, B.; Zhou, J. Molsearch: search-based multi-objective molecular generation and property optimization. In *Proceedings of the 28th ACM SIGKDD conference on knowledge discovery and data mining*, 2022; pp 4724-4732.

(493) Devi, R. V.; Sathya, S. S.; Coumar, M. S. Multi-objective genetic algorithm for de novo drug design (MoGADdrug). *Curr. Comput.-Aid. Drug* **2021**, *17* (3), 445-457.

(494) Nicolaou, C. A.; Apostolakis, J.; Pattichis, C. S. De novo drug design using multiobjective evolutionary graphs. *J. Chem. Inf. Model.* **2009**, *49* (2), 295-307.

(495) Jensen, J. H. A graph-based genetic algorithm and generative model/Monte Carlo tree search for the exploration of chemical space. *Chem. Sci.* **2019**, *10* (12), 3567-3572.

(496) Brown, N.; McKay, B.; Gilardoni, F.; Gasteiger, J. A graph-based genetic algorithm and its application to the multiobjective evolution of median molecules. *J. Chem. Inf. Comput. Sci.* **2004**, *44* (3), 1079-1087.

(497) Virshup, A. M.; Contreras-García, J.; Wipf, P.; Yang, W.; Beratan, D. N. Stochastic voyages into uncharted chemical space produce a representative library of all possible drug-like compounds. *J. Am. Chem. Soc.* **2013**, *135* (19), 7296-7303.

(498) Grantham, K.; Mukaidaisi, M.; Ooi, H. K.; Ghaemi, M. S.; Tchagang, A.; Li, Y. Deep evolutionary learning for molecular design. *IEEE Comput. Intell. Mag.* **2022**, *17* (2), 14-28.

(499) Murcko, M. A. What Makes a Great Medicinal Chemist? A Personal Perspective: Miniperspective. *J. Med. Chem.* **2018**, *61* (17), 7419-7424.

(500) Rosenthal, S.; Borschbach, M. Design perspectives of an evolutionary process for multi-objective molecular optimization. In *International Conference on Evolutionary Multi-Criterion Optimization*, 2017; Springer: pp 529-544. Lambrinidis, G.; Tsantili-Kakoulidou, A. Multi-objective optimization methods in novel drug design. *Expert Opin. Drug Discov.* **2021**, *16* (6), 647-658.

(501) Srinivas, N.; Deb, K. Muiltiobjective optimization using nondominated sorting in genetic algorithms. *Evol. Comput.* **1994**, *2* (3), 221-248. Deb, K.; Pratap, A.; Agarwal, S.; Meyarivan, T. A fast and elitist multiobjective genetic algorithm: NSGA-II. *IEEE Trans. Evol. Comput.* **2002**, *6* (2), 182-197.

(502) Angelo, J. S.; Guedes, I. A.; Barbosa, H. J.; Dardenne, L. E. Multi-and many-objective optimization: present and future in de novo drug design. *Front. Chem.* **2023**, *11*, 1288626.

(503) Zhang, H.; Zhao, H.; Zhang, X.; Su, Q.; Du, H.; Shen, C.; Wang, Z.; Li, D.; Pan, P.; Chen, G. Delete: Deep Lead Optimization Enveloped in Protein Pocket through Unified Deleting Strategies and a Structure-aware Network. In *arXiv preprint arXiv:2308.02172*, 2023.

(504) Nicholson, D. N.; Greene, C. S. Constructing knowledge graphs and their biomedical applications. *Comput. Struct. Biotechnol. J.* **2020**, *18*, 1414-1428.

(505) Chandak, P.; Huang, K.; Zitnik, M. Building a knowledge graph to enable precision medicine. *Sci. Data* **2023**, *10* (1), 67.

(506) Zeng, X.; Tu, X.; Liu, Y.; Fu, X.; Su, Y. Toward better drug discovery with knowledge graph. *Curr. Opin. Struct. Biol.* **2022**, *72*, 114-126.

(507) Chen, X.; Jia, S.; Xiang, Y. A review: Knowledge reasoning over knowledge graph. *Expert Syst. Appl.* **2020**, *141*, 112948.

(508) Brown, G. R.; Hem, V.; Katz, K. S.; Ovetsky, M.; Wallin, C.; Ermolaeva, O.; Tolstoy, I.; Tatusova, T.; Pruitt, K. D.; Maglott, D. R. Gene: a gene-centered information resource at NCBI. *Nucleic Acids Res.* **2015**, *43* (D1), D36-D42.

(509) Howe, K. L.; Achuthan, P.; Allen, J.; Allen, J.; Alvarez-Jarreta, J.; Amode, M. R.; Armean, I. M.; Azov, A. G.; Bennett, R.; Bhai, J. Ensembl 2021. *Nucleic Acids Res.* **2021**, *49* (D1), D884-D891.

(510) Consortium, U. UniProt: a worldwide hub of protein knowledge. *Nucleic Acids Res.* **2019**, *47* (D1), D506-D515.

(511) Oughtred, R.; Rust, J.; Chang, C.; Breitkreutz, B. J.; Stark, C.; Willems, A.; Boucher, L.; Leung, G.; Kolas, N.; Zhang, F. The BioGRID database: A comprehensive biomedical resource of curated protein, genetic, and chemical interactions. *Protein Sci.* **2021**, *30* (1), 187-200.

(512) Hermjakob, H.; Montecchi-Palazzi, L.; Lewington, C.; Mudali, S.; Kerrien, S.; Orchard, S.; Vingron, M.; Roechert, B.; Roepstorff, P.; Valencia, A. IntAct: an open source molecular interaction database. *Nucleic Acids Res.* **2004**, *32* (suppl_1), D452-D455.

(513) Mering, C. v.; Huynen, M.; Jaeggi, D.; Schmidt, S.; Bork, P.; Snel, B. STRING: a database of predicted functional associations between proteins. *Nucleic Acids Res.* **2003**, *31* (1), 258-261.

(514) Knox, C.; Wilson, M.; Klinger, C. M.; Franklin, M.; Oler, E.; Wilson, A.; Pon, A.; Cox, J.; Chin, N. E.; Strawbridge, S. A. DrugBank 6.0: the DrugBank knowledgebase for 2024. *Nucleic Acids Res.* **2024**, *52* (D1), D1265-D1275.

(515) Ursu, O.; Holmes, J.; Knockel, J.; Bologa, C. G.; Yang, J. J.; Mathias, S. L.; Nelson, S. J.; Oprea, T. I. DrugCentral: online drug compendium. *Nucleic Acids Res.* **2016**, gkw993.

(516) Gremse, M.; Chang, A.; Schomburg, I.; Grote, A.; Scheer, M.; Ebeling, C.; Schomburg, D. The BRENDA Tissue Ontology (BTO): the first all-integrating ontology of all organisms for enzyme sources. *Nucleic Acids Res.* **2010**, *39* (suppl_1), D507-D513.

(517) Consortium, G. O. The gene ontology resource: 20 years and still GOing strong. *Nucleic Acids Res.* **2019**, *47* (D1), D330-D338.

(518) Aoki-Kinoshita, K. F.; Kanehisa, M. Gene annotation and pathway mapping in KEGG. *Comparative genomics* **2007**, 71-91.

(519) Ogata, H.; Goto, S.; Fujibuchi, W.; Kanehisa, M. Computation with the KEGG pathway database. *Biosystems* **1998**, *47* (1-2), 119-128.

(520) Fabregat, A.; Jupe, S.; Matthews, L.; Sidiropoulos, K.; Gillespie, M.; Garapati, P.; Haw, R.; Jassal, B.; Korninger, F.; May, B. The reactome pathway knowledgebase. *Nucleic Acids Res.* **2018**, *46* (D1), D649-D655.

(521) Cerami, E. G.; Gross, B. E.; Demir, E.; Rodchenkov, I.; Babur, Ö.; Anwar, N.; Schultz, N.; Bader, G. D.; Sander, C. Pathway Commons, a web resource for biological pathway data. *Nucleic Acids Res.* **2010**, *39* (suppl_1), D685-D690.

(522) Lipscomb, C. E. Medical subject headings (MeSH). *BMLA* **2000**, *88* (3), 265.

(523) Schriml, L. M.; Arze, C.; Nadendla, S.; Chang, Y.-W. W.; Mazaitis, M.; Felix, V.; Feng, G.; Kibbe, W. A. Disease Ontology: a backbone for disease semantic integration. *Nucleic Acids Res.* **2012**, *40* (D1), D940-D946.

(524) Robinson, P. N.; Köhler, S.; Bauer, S.; Seelow, D.; Horn, D.; Mundlos, S. The Human Phenotype Ontology: a tool for annotating and analyzing human hereditary disease. *Am. J. Hum. Genet.* **2008**, *83* (5), 610-615.

(525) Vasilevsky, N.; Essaid, S.; Matentzoglu, N.; Harris, N. L.; Haendel, M.; Robinson, P.; Mungall, C. J. Mondo Disease Ontology: harmonizing disease concepts across the world. In *CEUR Workshop Proceedings, CEUR-WS*, 2020; Vol. 2807.

(526) Piñero, J.; Bravo, À.; Queralt-Rosinach, N.; Gutiérrez-Sacristán, A.; Deu-Pons, J.; Centeno, E.; García-García, J.; Sanz, F.; Furlong, L. I. DisGeNET: a comprehensive platform integrating information on human disease-associated genes and variants. *Nucleic Acids Res.* **2016**, gkw943.

(527) Hamosh, A.; Scott, A. F.; Amberger, J. S.; Bocchini, C. A.; McKusick, V. A. Online Mendelian Inheritance in Man (OMIM), a knowledgebase of human genes and genetic disorders. *Nucleic Acids Res.* **2005**, *33* (suppl_1), D514-D517.

(528) Kanehisa, M.; Goto, S.; Furumichi, M.; Tanabe, M.; Hirakawa, M. KEGG for representation and analysis of molecular networks involving diseases and drugs. *Nucleic Acids Res.* **2010**, *38* (suppl_1), D355-D360.

(529) Organization, W. H. *International classification of diseases:[9th] ninth revision, basic tabulation list with alphabetic index*; World Health Organization, 1978.

(530) Olivier, M.; Hollstein, M.; Hainaut, P. TP53 mutations in human cancers: origins, consequences, and clinical use. *CSH Perspect. Biol.* **2010**, *2* (1), a001008. Rosen, E. M.; Fan, S.; Pestell, R. G.; Goldberg, I. D. BRCA1 gene in breast cancer. *J. Cell. Physiol.* **2003**, *196* (1), 19-41.

(531) Seger, R.; Krebs, E. G. The MAPK signaling cascade. *FASEB J.* **1995**, *9* (9), 726-735. Franke, T. PI3K/Akt: getting it right matters. *Oncogene* **2008**, *27* (50), 6473-6488.

(532) Karp, P. D.; Billington, R.; Caspi, R.; Fulcher, C. A.; Latendresse, M.; Kothari, A.; Keseler, I. M.; Krummenacker, M.; Midford, P. E.; Ong, Q. The BioCyc collection of microbial genomes and metabolic pathways. *Brief. Bioinform.* **2019**, *20* (4), 1085-1093.

(533) Schaefer, C. F.; Anthony, K.; Krupa, S.; Buchoff, J.; Day, M.; Hannay, T.; Buetow, K. H. PID: the pathway interaction database. *Nucleic Acids Res.* **2009**, *37* (suppl_1), D674-D679.

(534) Romero, P.; Wagg, J.; Green, M. L.; Kaiser, D.; Krummenacker, M.; Karp, P. D. Computational prediction of human metabolic pathways from the complete human genome. *Genome Biol.* **2005**, *6*, 1-17.

(535) Guarino, N.; Oberle, D.; Staab, S. What is an ontology? *Handbook on ontologies* **2009**, 1-17.

(536) Ehrlinger, L.; Wöß, W. Towards a definition of knowledge graphs. *SEMANTiCS (Posters, Demos, SuCCESS)* **2016**, *48* (1-4), 2.

(537) Bonner, S.; Barrett, I. P.; Ye, C.; Swiers, R.; Engkvist, O.; Bender, A.; Hoyt, C. T.; Hamilton, W. L. A review of biomedical datasets relating to drug discovery: a knowledge graph perspective. *Brief. Bioinform.* **2022**, *23* (6), bbac404.

(538) Himmelstein, D. S.; Lizee, A.; Hessler, C.; Brueggeman, L.; Chen, S. L.; Hadley, D.; Green, A.; Khankhanian, P.; Baranzini, S. E. Systematic integration of biomedical knowledge prioritizes drugs for repurposing. *Elife* **2017**, *6*, e26726.

(539) Ioannidis, V. N.; Song, X.; Manchanda, S.; Li, M.; Pan, X.; Zheng, D.; Ning, X.; Zeng, X.; Karypis, G. Drkg-drug repurposing knowledge graph for covid-19. *arXiv preprint arXiv:2010.09600* **2020**.

(540) Walsh, B.; Mohamed, S. K.; Nováček, V. Biokg: A knowledge graph for relational learning on biological data. In *Proceedings of the 29th ACM International Conference on Information & Knowledge Management*, 2020; pp 3173-3180.

(541) Zheng, S.; Rao, J.; Song, Y.; Zhang, J.; Xiao, X.; Fang, E. F.; Yang, Y.; Niu, Z. PharmKG: a dedicated knowledge graph benchmark for bomedical data mining. *Brief. Bioinform.* **2021**, *22* (4), bbaa344.

(542) Breit, A.; Ott, S.; Agibetov, A.; Samwald, M. OpenBioLink: a benchmarking framework for large-scale biomedical link prediction. *Bioinformatics* **2020**, *36* (13), 4097-4098.

(543) Schlichtkrull, M.; Kipf, T. N.; Bloem, P.; Van Den Berg, R.; Titov, I.; Welling, M. Modeling relational data with graph convolutional networks. In *The semantic web: 15th international conference, ESWC 2018, Heraklion, Crete, Greece, June 3–7, 2018, proceedings 15*, 2018; Springer: pp 593-607.

(544) Kazemi, S. M.; Poole, D. Simple embedding for link prediction in knowledge graphs. *Adv. Neural Inf. Process. Syst* **2018**, *31*.

(545) Bordes, A.; Usunier, N.; Garcia-Duran, A.; Weston, J.; Yakhnenko, O. Translating embeddings for modeling multi-relational data. *Adv. Neural Inf. Process. Syst* **2013**, *26*.

(546) Chen, M.; Zhang, Y.; Kou, X.; Li, Y.; Zhang, Y. R-gat: relational graph attention network for multi-relational graphs. *arXiv preprint arXiv:2109.05922* **2021**.

(547) Wang, J.; Liu, X.; Shen, S.; Deng, L.; Liu, H. DeepDDS: deep graph neural network with attention mechanism to predict synergistic drug combinations. *Brief. Bioinform.* **2022**, *23* (1), bbab390.

(548) Rozemberczki, B.; Gogleva, A.; Nilsson, S.; Edwards, G.; Nikolov, A.; Papa, E. Moomin: Deep molecular omics network for anti-cancer drug combination therapy. In *Proceedings of the 31st ACM international conference on information & knowledge management*, 2022; pp 3472-3483.

(549) Liu, X.; Song, C.; Liu, S.; Li, M.; Zhou, X.; Zhang, W. Multi-way relation-enhanced hypergraph representation learning for anti-cancer drug synergy prediction. *Bioinformatics* **2022**, *38* (20), 4782-4789.

(550) Hu, J.; Gao, J.; Fang, X.; Liu, Z.; Wang, F.; Huang, W.; Wu, H.; Zhao, G. DTSyn: a dual-transformer-based neural network to predict synergistic drug combinations. *Brief. Bioinform.* **2022**, *23* (5), bbac302.

(551) Zhang, P.; Tu, S.; Zhang, W.; Xu, L. Predicting cell line-specific synergistic drug combinations through a relational graph convolutional network with attention mechanism. *Brief. Bioinform.* **2022**, *23* (6), bbac403.

(552) Jiang, P.; Huang, S.; Fu, Z.; Sun, Z.; Lakowski, T. M.; Hu, P. Deep graph embedding for prioritizing synergistic anticancer drug combinations. *Comput. Struct. Biotechnol. J.* **2020**, *18*, 427-438.

(553) Yang, J.; Xu, Z.; Wu, W. K. K.; Chu, Q.; Zhang, Q. GraphSynergy: a network-inspired deep learning model for anticancer drug combination prediction. *J. Am. Med. Inform. Assoc.* **2021**, *28* (11), 2336-2345.

(554) Zhang, G.; Gao, Z.; Yan, C.; Wang, J.; Liang, W.; Luo, J.; Luo, H. KGANSynergy: knowledge graph attention network for drug synergy prediction. *Brief. Bioinform.* **2023**, *24* (3), bbad167.

(555) Chen, H.; Lu, Y.; Yang, Y.; Rao, Y. A drug combination prediction framework based on graph convolutional network and heterogeneous information. *IEEE ACM Trans. Comput. Biol. Bioinform.* **2022**, *20* (3), 1917-1925.

(556) Hu, Z.; Yu, Q.; Gao, Y. X.; Guo, L.; Song, T.; Li, Y.; King, I. Drug synergistic combinations predictions via large-scale pre-training and graph structure learning. In *Springer Nature*, 2023; Springer: Vol. 13976, p 265.

(557) Karimi, M.; Hasanzadeh, A.; Shen, Y. Network-principled deep generative models for designing drug combinations as graph sets. *Bioinformatics* **2020**, *36* (Supplement_1), i445-i454.

(558) Gao, C.; Yin, S.; Wang, H.; Wang, Z.; Du, Z.; Li, X. Medical-knowledge-based graph neural network for medication combination prediction. *IEEE Trans. Neural Networks Learn. Syst.* **2023**.

(559) Zitnik, M.; Agrawal, M.; Leskovec, J. Modeling polypharmacy side effects with graph convolutional networks. *Bioinformatics* **2018**, *34* (13), i457-i466.

(560) Huang, K.; Xiao, C.; Glass, L. M.; Zitnik, M.; Sun, J. SkipGNN: predicting molecular interactions with skip-graph networks. *Sci. Rep.* **2020**, *10* (1), 21092.

(561) Lin, X.; Quan, Z.; Wang, Z.-J.; Ma, T.; Zeng, X. KGNN: Knowledge Graph Neural Network for Drug-Drug Interaction Prediction. In *IJCAI*, 2020; Vol. 380, pp 2739-2745.

(562) Yu, Y.; Huang, K.; Zhang, C.; Glass, L. M.; Sun, J.; Xiao, C. SumGNN: multi-typed drug interaction prediction via efficient knowledge graph summarization. *Bioinformatics* **2021**, *37* (18), 2988-2995.

(563) Zhang, Y.; Yao, Q.; Yue, L.; Wu, X.; Zhang, Z.; Lin, Z.; Zheng, Y. Emerging drug interaction prediction enabled by a flow-based graph neural network with biomedical network. *Nat. Comput. Sci.* **2023**, *3* (12), 1023-1033.

(564) Wang, Z.; Zhou, M.; Arnold, C. Toward heterogeneous information fusion: bipartite graph convolutional networks for in silico drug repurposing. *Bioinformatics* **2020**, *36* (Supplement_1), i525-i533.

(565) Zeng, X.; Zhu, S.; Liu, X.; Zhou, Y.; Nussinov, R.; Cheng, F. deepDR: a network-based deep learning approach to in silico drug repositioning. *Bioinformatics* **2019**, *35* (24), 5191-5198.

(566) Bang, D.; Lim, S.; Lee, S.; Kim, S. Biomedical knowledge graph learning for drug repurposing by extending guilt-by-association to multiple layers. *Nat. Commun.* **2023**, *14* (1), 3570.

(567) Gao, Z.; Ding, P.; Xu, R. KG-Predict: A knowledge graph computational framework for drug repurposing. *J. Biomed. Inform.* **2022**, *132*, 104133.

(568) Yu, Z.; Huang, F.; Zhao, X.; Xiao, W.; Zhang, W. Predicting drug–disease associations through layer attention graph convolutional network. *Brief. Bioinform.* **2021**, *22* (4), bbaa243.

(569) Meng, Y.; Lu, C.; Jin, M.; Xu, J.; Zeng, X.; Yang, J. A weighted bilinear neural collaborative filtering approach for drug repositioning. *Brief. Bioinform.* **2022**, *23* (2), bbab581.

(570) Cai, L.; Lu, C.; Xu, J.; Meng, Y.; Wang, P.; Fu, X.; Zeng, X.; Su, Y. Drug repositioning based on the heterogeneous information fusion graph convolutional network. *Brief. Bioinform.* **2021**, *22* (6), bbab319.

(571) Zhang, Y.; Lei, X.; Pan, Y.; Wu, F.-X. Drug repositioning with GraphSAGE and clustering constraints based on drug and disease networks. *Front. Pharmacol.* **2022**, *13*, 872785.

(572) Sun, X.; Jia, X.; Lu, Z.; Tang, J.; Li, M. Drug repositioning with adaptive graph convolutional networks. *Bioinformatics* **2024**, *40* (1), btad748.

(573) Meng, Y.; Wang, Y.; Xu, J.; Lu, C.; Tang, X.; Peng, T.; Zhang, B.; Tian, G.; Yang, J. Drug repositioning based on weighted local information augmented graph neural network. *Brief. Bioinform.* **2024**, *25* (1), bbad431.

(574) Wang, X.; Cheng, Y.; Yang, Y.; Yu, Y.; Li, F.; Peng, S. Multitask joint strategies of self-supervised representation learning on biomedical networks for drug discovery. *Nat. Mach. Intell.* **2023**, *5* (4), 445-456.

(575) Huang, K.; Chandak, P.; Wang, Q.; Havaldar, S.; Vaid, A.; Leskovec, J.; Nadkarni, G. N.; Glicksberg, B. S.; Gehlenborg, N.; Zitnik, M. A foundation model for clinician-centered drug repurposing. *Nat. Med.* **2024**, 1-13.

(576) Zeng, X.; Zhu, S.; Lu, W.; Liu, Z.; Huang, J.; Zhou, Y.; Fang, J.; Huang, Y.; Guo, H.; Li, L. Target identification among known drugs by deep learning from heterogeneous networks. *Chem. Sci.* **2020**, *11* (7), 1775-1797.

(577) Liu, C.; Xiao, K.; Yu, C.; Lei, Y.; Lyu, K.; Tian, T.; Zhao, D.; Zhou, F.; Tang, H.; Zeng, J. A probabilistic knowledge graph for target identification. *PLoS Comput. Biol.* **2024**, *20* (4), e1011945.

(578) Jia, J.; Zhu, F.; Ma, X.; Cao, Z.; Cao, Z. W.; Li, Y.; Li, Y. X.; Chen, Y. Z. Mechanisms of drug combinations: interaction and network perspectives. *Nat. Rev. Drug Discov.* **2009**, *8* (2), 111-128. DOI: 10.1038/nrd2683.

(579) Besharatifard, M.; Vafaee, F. A review on graph neural networks for predicting synergistic drug combinations. *Artif. Intell. Rev.* **2024**, *57* (3), 49.

(580) Ghandi, M.; Huang, F. W.; Jané-Valbuena, J.; Kryukov, G. V.; Lo, C. C.; McDonald III, E. R.; Barretina, J.; Gelfand, E. T.; Bielski, C. M.; Li, H. Next-generation characterization of the cancer cell line encyclopedia. *Nature* **2019**, *569* (7757), 503-508.

(581) Liu, H.; Zhang, W.; Zou, B.; Wang, J.; Deng, Y.; Deng, L. DrugCombDB: a comprehensive database of drug combinations toward the discovery of combinatorial therapy. *Nucleic Acids Res.* **2020**, *48* (D1), D871-D881.

(582) Zhang, P.; Tu, S. MGAE-DC: Predicting the synergistic effects of drug combinations through multi-channel graph autoencoders. *PLoS Comput. Biol.* **2023**, *19* (3), e1010951.

(583) Li, H.; Zhao, D.; Zeng, J. KPGT: knowledge-guided pre-training of graph transformer for molecular property prediction. In *Proceedings of the 28th ACM SIGKDD Conference on Knowledge Discovery and Data Mining*, 2022; pp 857-867.

(584) Sun, Z.; Deng, Z.-H.; Nie, J.-Y.; Tang, J. Rotate: Knowledge graph embedding by relational rotation in complex space. *arXiv preprint arXiv:1902.10197* **2019**.

(585) Ye, Q.; Xu, R.; Li, D.; Kang, Y.; Deng, Y.; Zhu, F.; Chen, J.; He, S.; Hsieh, C.-Y.; Hou, T. Integrating multi-modal deep learning on knowledge graph for the discovery of synergistic drug combinations against infectious diseases. *Cell Rep. Phys. Sci.* **2023**, *4* (8).

(586) Pushpakom, S.; Iorio, F.; Eyers, P. A.; Escott, K. J.; Hopper, S.; Wells, A.; Doig, A.; Guilliams, T.; Latimer, J.; McNamee, C. Drug repurposing: progress, challenges and recommendations. *Nat. Rev. Drug Discov.* **2019**, *18* (1), 41-58.

(587) Parisi, D.; Adasme, M. F.; Sveshnikova, A.; Bolz, S. N.; Moreau, Y.; Schroeder, M. Drug repositioning or target repositioning: A structural perspective of drug-target-indication relationship for available repurposed drugs. *Comput. Struct. Biotechnol. J.* **2020**, *18*, 1043-1055.

(588) Karuppagounder, S. S.; Brahmachari, S.; Lee, Y.; Dawson, V. L.; Dawson, T. M.; Ko, H. S. The c-Abl inhibitor, nilotinib, protects dopaminergic neurons in a preclinical animal model of Parkinson's disease. *Sci. Rep.* **2014**, *4* (1), 4874.

(589) Hanahan, D.; Weinberg, R. A. Hallmarks of cancer: the next generation. *Cell* **2011**, *144* (5), 646-674. Kantarjian, H.; Giles, F.; Wunderle, L.; Bhalla, K.; O'Brien, S.; Wassmann, B.; Tanaka, C.; Manley, P.; Rae, P.; Mietlowski, W. Nilotinib in imatinib-resistant CML and Philadelphia chromosome–positive ALL. *New Eng. J. Med.* **2006**, *354* (24), 2542-2551.

(590) Mohamed, S. K.; Nováček, V.; Nounu, A. Discovering protein drug targets using knowledge graph embeddings. *Bioinformatics* **2020**, *36* (2), 603-610.

(591) Al-Saleem, J.; Granet, R.; Ramakrishnan, S.; Ciancetta, N. A.; Saveson, C.; Gessner, C.; Zhou, Q. Knowledge graph-based approaches to drug repurposing for COVID-19. *J. Chem. Inf. Model.* **2021**, *61* (8), 4058-4067.

(592) Groza, V.; Udrescu, M.; Bozdog, A.; Udrescu, L. Drug repurposing using modularity clustering in drug-drug similarity networks based on drug–gene interactions. *Pharmaceutics* **2021**, *13* (12), 2117.

(593) Cai, L.; Lu, C.; Xu, J.; Meng, Y.; Wang, P.; Fu, X.; Zeng, X.; Su, Y. Drug repositioning based on the heterogeneous information fusion graph convolutional network. *Brief. Bioinform.* **2021**, *22* (6). DOI: 10.1093/bib/bbab319.

(594) Cereto-Massagué, A.; Ojeda, M. J.; Valls, C.; Mulero, M.; Pujadas, G.; Garcia-Vallve, S. Tools for in silico target fishing. *Methods* **2015**, *71*, 98-103.

(595) Jenkins, J. L.; Bender, A.; Davies, J. W. In silico target fishing: Predicting biological targets from chemical structure. *Drug Discov. Today: Technol.* **2006**, *3* (4), 413-421. Ji, K.-Y.; Liu, C.; Liu, Z.-Q.; Deng, Y.-F.; Hou, T.-J.; Cao, D.-S. Comprehensive assessment of nine target prediction web services: which should we choose for target fishing? *Brief. Bioinform.* **2023**, *24* (2), bbad014.

(596) Chen, Z.; Wang, Y.; Zhao, B.; Cheng, J.; Zhao, X.; Duan, Z. Knowledge graph completion: A review. *Ieee Access* **2020**, *8*, 192435-192456.

(597) Schulte-Sasse, R.; Budach, S.; Hnisz, D.; Marsico, A. Integration of multiomics data with graph convolutional networks to identify new cancer genes and their associated molecular mechanisms. *Nat. Mach. Intell.* **2021**, *3* (6), 513-526.

(598) Peng, W.; Tang, Q.; Dai, W.; Chen, T. Improving cancer driver gene identification using multi-task learning on graph convolutional network. *Brief. Bioinform.* **2022**, *23* (1), bbab432.

(599) Zhao, W.; Gu, X.; Chen, S.; Wu, J.; Zhou, Z. MODIG: integrating multi-omics and multi-dimensional gene network for cancer driver gene identification based on graph attention network model. *Bioinformatics* **2022**, *38* (21), 4901-4907.

(600) Ratajczak, F.; Joblin, M.; Hildebrandt, M.; Ringsquandl, M.; Falter-Braun, P.; Heinig, M. Speos: an ensemble graph representation learning framework to predict core gene candidates for complex diseases. *Nat. Commun.* **2023**, *14* (1), 7206.

(601) Dai, W.; Yue, W.; Peng, W.; Fu, X.; Liu, L.; Liu, L. Identifying cancer subtypes using a residual graph convolution model on a sample similarity network. *Genes* **2021**, *13* (1), 65.

(602) Baul, S.; Ahmed, K. T.; Filipek, J.; Zhang, W. omicsGAT: Graph attention network for cancer subtype analyses. *Int. J. Mol. Sci.* **2022**, *23* (18), 10220.

(603) Liang, C.; Shang, M.; Luo, J. Cancer subtype identification by consensus guided graph autoencoders. *Bioinformatics* **2021**, *37* (24), 4779-4786.

(604) Li, X.; Ma, J.; Leng, L.; Han, M.; Li, M.; He, F.; Zhu, Y. MoGCN: a multi-omics integration method based on graph convolutional network for cancer subtype analysis. *Front. Genet.* **2022**, *13*, 806842.

(605) Liu, C.; Duan, Y.; Zhou, Q.; Wang, Y.; Gao, Y.; Kan, H.; Hu, J. A classification method of gastric cancer subtype based on residual graph convolution network. *Front. Genet.* **2023**, *13*, 1090394.

(606) Zhang, G.; Peng, Z.; Yan, C.; Wang, J.; Luo, J.; Luo, H. A novel liver cancer diagnosis method based on patient similarity network and DenseGCN. *Sci. Rep.* **2022**, *12* (1), 6797.

(607) Wang, T.; Shao, W.; Huang, Z.; Tang, H.; Zhang, J.; Ding, Z.; Huang, K. MOGONET integrates multi-omics data using graph convolutional networks allowing patient classification and biomarker identification. *Nat. Commun.* **2021**, *12* (1), 3445.

(608) Ramirez, R.; Chiu, Y.-C.; Hererra, A.; Mostavi, M.; Ramirez, J.; Chen, Y.; Huang, Y.; Jin, Y.-F. Classification of cancer types using graph convolutional neural networks. *Front. Phys.* **2020**, *8*, 203.

(609) Barabási, A.-L.; Gulbahce, N.; Loscalzo, J. Network medicine: a network-based approach to human disease. *Nat. Rev. Genet.* **2011**, *12* (1), 56-68. Goh, K.-I.; Cusick, M. E.; Valle, D.; Childs, B.; Vidal, M.; Barabási, A.-L. The human disease network. *Proc. Natl. Acad. Sci.* **2007**, *104* (21), 8685-8690. Li, Y.; Agarwal, P. A pathway-based view of human diseases and disease relationships. *PLoS One* **2009**, *4* (2), e4346.

(610) Lawrence, M. S.; Stojanov, P.; Polak, P.; Kryukov, G. V.; Cibulskis, K.; Sivachenko, A.; Carter, S. L.; Stewart, C.; Mermel, C. H.; Roberts, S. A. Mutational heterogeneity in cancer and the search for new cancer-associated genes. *Nature* **2013**, *499* (7457), 214-218.

(611) Chen, C.; Wang, J.; Pan, D.; Wang, X.; Xu, Y.; Yan, J.; Wang, L.; Yang, X.; Yang, M.; Liu, G. P. Applications of multi-omics analysis in human diseases. *MedComm* **2023**, *4* (4), e315.

(612) Uffelmann, E.; Huang, Q. Q.; Munung, N. S.; De Vries, J.; Okada, Y.; Martin, A. R.; Martin, H. C.; Lappalainen, T.; Posthuma, D. Genome-wide association studies. *Nat. Rev. Methods Primers* **2021**, *1* (1), 59.

(613) Li, F.; Dong, S.; Leier, A.; Han, M.; Guo, X.; Xu, J.; Wang, X.; Pan, S.; Jia, C.; Zhang, Y. Positive-unlabeled learning in bioinformatics and computational biology: a brief review. *Brief. Bioinform.* **2022**, *23* (1), bbab461.

(614) Grover, A.; Leskovec, J. node2vec: Scalable feature learning for networks. In *Proceedings of the 22nd ACM SIGKDD international conference on Knowledge discovery and data mining*, 2016; pp 855-864.

(615) Parvandeh, S.; Yeh, H.-W.; Paulus, M. P.; McKinney, B. A. Consensus features nested cross-validation. *Bioinformatics* **2020**, *36* (10), 3093-3098.

(616) Janku, F. Tumor heterogeneity in the clinic: is it a real problem? *Ther. Adv. Med. Oncol.* **2014**, *6* (2), 43-51.

(617) Ritchie, M. E.; Phipson, B.; Wu, D.; Hu, Y.; Law, C. W.; Shi, W.; Smyth, G. K. limma powers differential expression analyses for RNA-sequencing and microarray studies. *Nucleic Acids Res.* **2015**, *43* (7), e47-e47.

(618) Wang, L.; Ding, Z.; Tao, Z.; Liu, Y.; Fu, Y. Generative multi-view human action recognition. In *Proceedings of the IEEE/CVF International Conference on Computer Vision*, 2019; pp 6212-6221.

(619) Goncalves, M.; Cohen-Setton, J.; Kagiampakis, I.; Sidders, B.; Bulusu, K. Multi-modal knowledge graphs enhance patient stratification & biomarker discovery. *Cancer Res.* **2024**, *84* (6_Supplement), 4888-4888.

(620) Nguyen, T.; Nguyen, G. T. T.; Nguyen, T.; Le, D. H. Graph Convolutional Networks for Drug Response Prediction. *IEEE/ACM Trans. Comput. Biol. Bioinform.* **2022**, *19* (1), 146-154. DOI: 10.1109/TCBB.2021.3060430.

(621) Ye, Q.; Zeng, Y.; Jiang, L.; Kang, Y.; Pan, P.; Chen, J.; Deng, Y.; Zhao, H.; He, S.; Hou, T.; et al. A Knowledge-Guided Graph Learning Approach Bridging Phenotype- and Target-Based Drug Discovery. *Adv. Sci.* **2025**, e2412402. DOI: 10.1002/advs.202412402.

(622) Tan, Z.; Li, D.; Wang, S.; Beigi, A.; Jiang, B.; Bhattacharjee, A.; Karami, M.; Li, J.; Cheng, L.; Liu, H. Large language models for data annotation and synthesis: A survey. *arXiv preprint arXiv:2402.13446* **2024**.

(623) Drummond, N.; Shearer, R. The open world assumption. In *eSI Workshop: The Closed World of Databases meets the Open World of the Semantic Web*, 2006; Vol. 15, p 1.

(624) Cook, A.; Johnson, A. P.; Law, J.; Mirzazadeh, M.; Ravitz, O.; Simon, A. Computer-aided synthesis design: 40 years on. *Wires Comput. Mol. Sci.* **2012**, *2* (1), 79-107.

(625) Coley, C. W.; Green, W. H.; Jensen, K. F. Machine learning in computer-aided synthesis planning. *Acc. Chem. Res.* **2018**, *51* (5), 1281-1289.

(626) *USPTO*. https://www.uspto.gov/ (accessed 2025 Mar, 25).

(627) Kearnes, S. M.; Maser, M. R.; Wleklinski, M.; Kast, A.; Doyle, A. G.; Dreher, S. D.; Hawkins, J. M.; Jensen, K. F.; Coley, C. W. The open reaction database. *J. Am. Chem. Soc.* **2021**, *143* (45), 18820-18826.

(628) *Reaxys*. https://www.reaxys.com/ (accessed 2025 Mar, 25).

(629) *Nextmove Software Pistachio*. https://www.nextmovesoftware.com/pistachio.html (accessed 2025 Mar, 25).

(630) Jin, W.; Coley, C.; Barzilay, R.; Jaakkola, T. Predicting organic reaction outcomes with weisfeiler-lehman network. *Adv. Neural Inf. Process. Syst* **2017**, *30*.

(631) Schneider, N.; Stiefl, N.; Landrum, G. A. What's what: The (nearly) definitive guide to reaction role assignment. *J. Chem. Inf. Model.* **2016**, *56* (12), 2336-2346.

(632) Dai, H.; Li, C.; Coley, C.; Dai, B.; Song, L. Retrosynthesis prediction with conditional graph logic network. *Adv. Neural Inf. Process. Syst* **2019**, *32*.

(633) Ahneman, D. T.; Estrada, J. G.; Lin, S.; Dreher, S. D.; Doyle, A. G. Predicting reaction performance in C–N cross-coupling using machine learning. *Science* **2018**, *360* (6385), 186-190.

(634) Perera, D.; Tucker, J. W.; Brahmbhatt, S.; Helal, C. J.; Chong, A.; Farrell, W.; Richardson, P.; Sach, N. W. A platform for automated nanomole-scale reaction screening and micromole-scale synthesis in flow. *Science* **2018**, *359* (6374), 429-434.

(635) Goodman, J. Computer Software Review: Reaxys. *J. Chem. Inf. Model.* **2009**, *49* (12), 2897-2898. DOI: 10.1021/ci900437n.

(636) Corey, E. J. The logic of chemical synthesis: multistep synthesis of complex carbogenic molecules (nobel lecture). *Angew. Chem. Int. Ed.* **1991**, *30* (5), 455-465.

(637) Hoffmann, R. W. *Elements of synthesis planning*; Springer, 2009.

(638) Chen, S.; Jung, Y. Deep retrosynthetic reaction prediction using local reactivity and global attention. *JACS Au* **2021**, *1* (10), 1612-1620.

(639) Shi, C.; Xu, M.; Guo, H.; Zhang, M.; Tang, J. A graph to graphs framework for retrosynthesis prediction. In *International conference on machine learning*, 2020; PMLR: pp 8818-8827.

(640) Somnath, V. R.; Bunne, C.; Coley, C.; Krause, A.; Barzilay, R. Learning graph models for retrosynthesis prediction. *Adv. Neural Inf. Process. Syst* **2021**, *34*, 9405-9415.

(641) Yan, C.; Ding, Q.; Zhao, P.; Zheng, S.; Yang, J.; Yu, Y.; Huang, J. Retroxpert: Decompose retrosynthesis prediction like a chemist. *Adv. Neural Inf. Process. Syst* **2020**, *33*, 11248-11258.

(642) Liu, J.; Yan, C.; Yu, Y.; Lu, C.; Huang, J.; Ou-Yang, L.; Zhao, P. Mars: a motif-based autoregressive model for retrosynthesis prediction. *Bioinformatics* **2024**, *40* (3), btae115.

(643) Lin, Z.; Yin, S.; Shi, L.; Zhou, W.; Zhang, Y. J. G2GT: retrosynthesis prediction with graph-to-graph attention neural network and self-training. *J. Chem. Inf. Model.* **2023**, *63* (7), 1894-1905.

(644) Yao, L.; Guo, W.; Wang, Z.; Xiang, S.; Liu, W.; Ke, G. Node-aligned graph-to-graph: elevating template-free deep learning approaches in single-step retrosynthesis. *JACS Au* **2024**, *4* (3), 992-1003.

(645) Sacha, M.; Błaz, M.; Byrski, P.; Dabrowski-Tumanski, P.; Chrominski, M.; Loska, R.; Włodarczyk-Pruszynski, P.; Jastrzebski, S. Molecule edit graph attention network: modeling chemical reactions as sequences of graph edits. *J. Chem. Inf. Model.* **2021**, *61* (7), 3273-3284.

(646) Zhong, W.; Yang, Z.; Chen, C. Y.-C. Retrosynthesis prediction using an end-to-end graph generative architecture for molecular graph editing. *Nat. Commun.* **2023**, *14* (1), 3009.

(647) Han, P.; Zhao, P.; Lu, C.; Huang, J.; Wu, J.; Shang, S.; Yao, B.; Zhang, X. Gnn-retro: Retrosynthetic planning with graph neural networks. In *Proceedings of the AAAI conference on artificial intelligence*, 2022; Vol. 36, pp 4014-4021.

(648) Xie, S.; Yan, R.; Han, P.; Xia, Y.; Wu, L.; Guo, C.; Yang, B.; Qin, T. Retrograph: Retrosynthetic planning with graph search. In *Proceedings of the 28th ACM SIGKDD Conference on Knowledge Discovery and Data Mining*, 2022; pp 2120-2129.

(649) Coley, C. W.; Rogers, L.; Green, W. H.; Jensen, K. F. Computer-assisted retrosynthesis based on molecular similarity. *ACS Cent. Sci.* **2017**, *3* (12), 1237-1245.

(650) Segler, M. H.; Preuss, M.; Waller, M. P. Planning chemical syntheses with deep neural networks and symbolic AI. *Nature* **2018**, *555* (7698), 604-610.

(651) Dai, H.; Li, C.; Coley, C. W.; Dai, B.; Song, L. Retrosynthesis prediction with conditional graph logic network. *arXiv* **2020**, (NeurIPS), 1–15.

(652) Schwaller, P.; Petraglia, R.; Zullo, V.; Nair, V. H.; Haeuselmann, R. A.; Pisoni, R.; Bekas, C.; Iuliano, A.; Laino, T. Predicting retrosynthetic pathways using transformer-based models and a hyper-graph exploration strategy. *Chem. Sci.* **2020**, *11* (12), 3316-3325.

(653) Clayden, J.; Greeves, N.; Warren, S. *Organic chemistry*; Oxford university press, 2012.

(654) Sacha, M.; Błaż, M.; Byrski, P.; Dąbrowski-Tumański, P.; Chromiński, M.; Loska, R.; Włodarczyk-Pruszyński, P.; Jastrzębski, S. Molecule Edit Graph Attention Network: Modeling Chemical Reactions as Sequences of Graph Edits. *J. Chem. Inf. Model.* **2021**, *61* (7), 3273-3284. DOI: 10.1021/acs.jcim.1c00537 (acccessed 2024-01-16 13:40:05).DOI.org (Crossref).

(655) Chen, S.; Jung, Y. A generalized-template-based graph neural network for accurate organic reactivity prediction. *Nat. Mach. Intell.* **2022**, *4* (9), 772-780.

(656) Bradshaw, J.; Kusner, M. J.; Paige, B.; Segler, M. H.; Hernández-Lobato, J. M. A generative model for electron paths. *arXiv preprint arXiv:1805.10970* **2018**.

(657) Do, K.; Tran, T.; Venkatesh, S. Graph transformation policy network for chemical reaction prediction. In *Proceedings of the 25th ACM SIGKDD international conference on knowledge discovery & data mining*, 2019; pp 750-760.

(658) Bi, H.; Wang, H.; Shi, C.; Coley, C.; Tang, J.; Guo, H. Non-autoregressive electron redistribution modeling for reaction prediction. In *International Conference on Machine Learning*, 2021; PMLR: pp 904-913.

(659) Coley, C. W.; Barzilay, R.; Jaakkola, T. S.; Green, W. H.; Jensen, K. F. Prediction of organic reaction outcomes using machine learning. *ACS Cent. Sci.* **2017**, *3* (5), 434-443.

(660) Wei, J. N.; Duvenaud, D.; Aspuru-Guzik, A. Neural networks for the prediction of organic chemistry reactions. *ACS Cent. Sci.* **2016**, *2* (10), 725-732.

(661) Wen, M.; Blau, S. M.; Xie, X.; Dwaraknath, S.; Persson, K. A. Improving machine learning performance on small chemical reaction data with unsupervised contrastive pretraining. *Chem. Sci.* **2022**, *13* (5), 1446-1458.

(662) Ryou, S.; Maser, M. R.; Cui, A. Y.; DeLano, T. J.; Yue, Y.; Reisman, S. E. Graph neural networks for the prediction of substrate-specific organic reaction conditions. *arXiv preprint arXiv:2007.04275* **2020**.

(663) Maser, M. R.; Cui, A. Y.; Ryou, S.; DeLano, T. J.; Yue, Y.; Reisman, S. E. Multilabel classification models for the prediction of cross-coupling reaction conditions. *J. Chem. Inf. Model.* **2021**, *61* (1), 156-166.

(664) Kwon, Y.; Kim, S.; Choi, Y.-S.; Kang, S. Generative modeling to predict multiple suitable conditions for chemical reactions. *J. Chem. Inf. Model.* **2022**, *62* (23), 5952-5960.

(665) Wang, Z.; Lin, K.; Pei, J.; Lai, L. Reacon: a template-and cluster-based framework for reaction condition prediction. *Chem. Sci.* **2025**, *16* (2), 854-866.

(666) Li, B.; Su, S.; Zhu, C.; Lin, J.; Hu, X.; Su, L.; Yu, Z.; Liao, K.; Chen, H. A deep learning framework for accurate reaction prediction and its application on high-throughput experimentation data. *J. Cheminformatics* **2023**, *15* (1), 72.

(667) Kwon, Y.; Lee, D.; Choi, Y.-S.; Kang, S. Uncertainty-aware prediction of chemical reaction yields with graph neural networks. *J. Cheminformatics* **2022**, *14*, 1-10.

(668) Sato, A.; Asahara, R.; Miyao, T. Chemical Graph-Based Transformer Models for Yield Prediction of High-Throughput Cross-Coupling Reaction Datasets. *ACS Omega* **2024**, *9* (39), 40907-40919.

(669) Han, J.; Kwon, Y.; Choi, Y.-S.; Kang, S. Improving chemical reaction yield prediction using pre-trained graph neural networks. *J. Cheminformatics* **2024**, *16* (1), 25.

(670) Kraut, H.; Eiblmaier, J.; Grethe, G.; Löw, P.; Matuszczyk, H.; Saller, H. Algorithm for reaction classification. *J. Chem. Inf. Model.* **2013**, *53* (11), 2884-2895.

(671) Bell, R. P. The theory of reactions involving proton transfers. *Proc. R. soc. Lond. Ser. A Math. Phys. Sci.* **1936**, *154* (882), 414-429.

(672) Probst, D.; Schwaller, P.; Reymond, J.-L. Reaction classification and yield prediction using the differential reaction fingerprint DRFP. *Digit. Discov.* **2022**, *1* (2), 91-97.

(673) Meuwly, M. Machine learning for chemical reactions. *Chem. Rev.* **2021**, *121* (16), 10218-10239.

(674) Skoraczyński, G.; Dittwald, P.; Miasojedow, B.; Szymkuć, S.; Gajewska, E. P.; Grzybowski, B. A.; Gambin, A. Predicting the outcomes of organic reactions via machine learning: are current descriptors sufficient? *Sci. Rep.* **2017**, *7* (1), 3582.

(675) Grzybowski, B. A.; Szymkuć, S.; Gajewska, E. P.; Molga, K.; Dittwald, P.; Wołos, A.; Klucznik, T. Chematica: a story of computer code that started to think like a chemist. *Chem* **2018**, *4* (3), 390-398.